# Nonlinear Quantum Optics in an Atomic Cavity

*Towards a quantum fluid of light*

Simon Panyella Pedersen

# Abstract

The idea of making photons effectively interact has attracted a lot of interest in recent years, for several reasons. Firstly, since photons do not naturally interact with each other, it is of fundamental physical interest to see whether we can make them do so. To find out what kind of medium can mediate interactions between these fundamental and non-interacting particles, and to what extent. Secondly, photonics is a major candidate for future quantum technology, where light, rather than electricity, is employed as the signal carrier. The easy manipulation, readout, and transport of photons makes them ideal for quantum information processing. Finally, achieving strong and tunable interactions among photons would open up an avenue for exploring the many-body physics of a fluid of light.

In this thesis, we will see how a cavity formed of subwavelength ordered lattices of two-level atoms can confine photons to a nonlinear environment for a long time, such that emitted photons have accumulated strong correlations both among their momenta and in their temporal statistics. This speaks of a strong photon-photon interaction within the cavity. The nonlinearity originates in the saturability of individual atoms, and the lattice structure results in a strong and low-loss collective interaction with light. While a single atomic lattice has a largely linear nature, as the effect of individual atoms washes out exactly due to the collective interaction, the confining geometry of the cavity means the photons are exposed to the underlying saturability of the atoms for such a long time that the nonlinearity is revived.

We will analyse this system both using a standard approach based on an input-output formalism, where the nonlinear physics of the system is handled numerically, and a powerful Green's function-based approach that allows for exact analytical results with no additional approximations compared to the numerical calculations. This analytical description has the potential to lead to an exact study of the many-body physics of interacting photons in a two-dimensional setting.

# Resumé (Dansk)

Idéen om at få fotoner til effektivt at interagere har tiltrukket megen interesse i de seneste år af adskillige årsager. For det første, siden fotoner ikke interagerer naturligt med hinanden, er det af fundamental fysisk interesse at undersøge, hvorvidt vi kan få dem til at gøre det. At undersøge hvilket slags medium kan mediere interaktioner imellem disse fundamentale og ikke-interagerende partikler, og i hvilket omfang. For det andet er fotonik en central kandidat for fremtidig kvanteteknologi, hvor lys, i stedet for elektricitet, bliver brugt som signalbærer. Den nemme manipulation, aflæsning og transport af fotoner gør dem ideele til kvanteinformationsprocessering. Endeligt ville opnåelsen af stærke og justerbare interaktioner imellem fotoner åbne op for udforskningen af mange-legeme fysik i lysfluider.

I denne afhandling skal vi se, hvordan en kavitet udformet af subbølgelængde regulære gitre af to-niveau atomer kan begrænse fotoner til et ikke-lineært miljø i lang tid, således at udsendte fotoner har akkumuleret stærke korrelationer både imellem deres impulser og i deres tidslige statistik. Dette bevidner om en stærk foton-foton interaktion i kaviteten. Denne ikke-linearitet oprinder i mætbarheden af de individuelle atomer, og gitterstrukturen resulterer i en stærk kollektiv interaktion med lys med lavt tab. Mens et enkelt atomgitter har en overvejende lineær natur, da effekten af individuelle atomer bliver udvasket præcis på grund af den kollektive interaktion, betyder kavitetens indgrænsende geometri, at fotoner bliver udsat for den underliggende mætbarhed af atomerne i så lang tid, at ikke-lineariteten bliver genoplivet.

Vi kommer til at analysere systemet både med en standard tilgang baseret på en input-output formalisme, hvor systemets ikke-lineære fysik bliver beregnet numerisk, og med en kraftfuld tilgang baseret på Greens funktioner, som tillader eksakte analytiske resulter uden yderligere approksimationer end i den numeriske tilgang. Denne analytiske beskrivelse har potentialet til at lede til et eksakt studie af mange-legeme fysikken i interagerende fotoner i to-dimensionelle omgivelser.

# Contents

## Appendix

# Preface

This thesis represents the culmination of my Ph.D. studies conducted from March 2020 to August 2023 under the supervision of Thomas Pohl. The work was carried out at the Center for Complex Quantum Systems at the Department of Physics and Astronomy, Aarhus University. In accordance with the rules of GSNS (the Graduate School of Natural Sciences, Aarhus University), some of the content of this thesis was featured in the progress report for the qualifying examination.

❉

I think it was sometime in the sixth grade that I decided I wanted to do a Ph.D. in physics. This would have been around the time my oldest brother started his, so inspiration for the idea would have been immediate. A far-sighted plan that I didn't question until sometime during my graduate studies, where it suddenly hit me that single-sentence, potentially life-defining plans made at the age of twelve perhaps deserve some scrutiny. However, I quickly concluded that proceeding with The Plan seemed a good choice, and so here we are.

I have decided to write my dissertation in the so-called monograph style, the canonical format of a dissertation, as I like to be thorough and complete in any explanation. Thus, the text might read like a lecture or a singular derivation, and be a bit long or technical at times, but I have a romantic idea that future students of the group, or elsewhere, can pick up this thesis and read it for its attention to detail and stringency. A reader experienced with the field may therefore be able to simply skim through some of the more technical sections.

I started my Ph.D. project a couple of weeks before the first national shut-down due to the Covid-19 pandemic. In other words, I had barely moved into my office or met any of my colleagues before I was sent home to work and had to become familiar with my supervisor via Zoom. Fortunately, this state of affairs did not last forever, and I am very grateful for the supervisor and colleagues I have had throughout these years. I would like to thank Thomas for offering me this Ph.D. in the first place, and for his skillful and ambitious supervision. I have always been impressed and pleased by the fact that he will actually have specific and concrete input to derivations and calculations, even doing some himself occasionally. I would like to thank my colleagues for their invigorating company during lunch breaks and our regular foosball games. I would in particular like to thank Lida Zhang and Jan Kumlin, my postdoc office mates, who have often helped me. The work presented in this thesis has its roots in Lida's own work, and some of the code used is a direct continuation of code he wrote.

A significant part of my Ph.D. has been the hours involved in teaching. An obligation to some, it has been a part I have thoroughly enjoyed, as I like to teach and I have had the privilege of being able to instruct the same class of people for all my teaching hours, and now count them as my friends. It has been comforting and joyful to have had continued contact with the student body of the department, and to have been invited to re-experience some of the social life only bestowed upon students.

Following that thought, I would also like to thank all the friends I have made here at uni for their continued and delightful friendship. From the now old friends met on the first year, to new friends found among first year students. The people and the revue of that excellent student organization TÅGEKAMMERET, who have been a constant source of company.

Finally, I would like to thank Jakob Lysgaard Rørsted for creating the beautiful template used for this document, and the friends who took the time to proofread and give feedback on the text.

# Introduction

Photons only interact directly with charged particles, thus excluding self-interactions. They do have an indirect interaction through very improbable charge quantum fluctuations which, as such, results in a very weak coupling. It is therefore fundamentally interesting to see whether we can engineer a physical system in which photons effectively interact strongly. In such a system, the presence of one photon affects the perceived environment of another, such that its dynamics are indirectly affected by the first. This may find application in future quantum technologies that employ photons as the active component and as carriers of information [1, 2]. Likewise, interacting photons may also find application in the analogue simulation of complex quantum systems [3–5], where we wish to study some complicated system by constructing an analogue that is more accessible and workable. In particular, it might open a way for the exploration of the complex and challenging many-body physics of interacting photons, a so-called quantum fluid of light [6].

A system which mediates photon-photon interactions has a nonlinear optical response. That is, the optical response of the system, the light that is emitted from the system, depends on the number of incident photons in a more complicated fashion than simply adding more photons to the output. Since the photons interact, the presence of more photons changes the outcome of the dynamics. We thus speak of linear or nonlinear systems, or media, and we can loosely speak of "how nonlinear" a system is, as the response may be more or less close to being linear. This reflects how strongly the photons are interacting. Intuitively, the degree of nonlinearity may itself also depend on the number of photons, as a weak field would probe without itself causing a change, while a strong field may actively change the response of a system. In this thesis I will be considering very few photons, indeed no more than two, which is also the

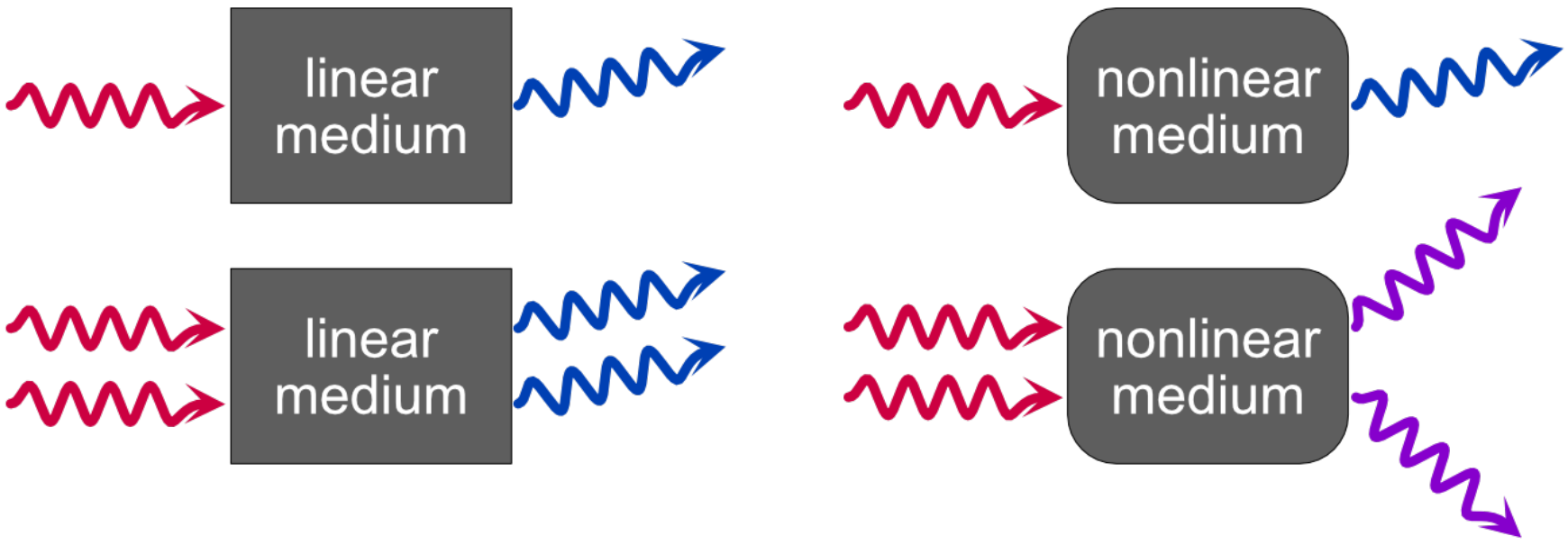

**A linear and a nonlinear medium.** In a linear medium, the optical response to a number of incident photons is the same irrespective of the number of photons. A nonlinear medium, on the other hand, changes its response as soon as there is more than one photon present. Thus, the nonlinear medium mediates an effective interaction between photons, and results in the outgoing photons being correlated.

working regime of many applications (for example constructing quantum two-qubit logic gates, or simulating few-body interactions).

A central example of an optically nonlinear system is a two-level atom with a dipole transition. Such a system only absorbs or emits a single photon. Hence, if a photon is absorbed, a second photon perceives a completely different situation, namely an empty environment as the atom becomes invisible to it. Thus, at the level of one or two photons the two-level atom is highly nonlinear. However, the coupling of light to a single atom is so weak that while the emitted light is highly correlated there would only be a very small signal of it, which in turn corresponds to a weak interaction between photons. This weak coupling is a result of, on the one hand, the small cross section for absorption, and on the other hand, the fact that there is a fundamental limit, the so-called diffraction limit [7], to how narrowly light can be focused in free space.

To exploit the nonlinearity of a two-level atom something must therefore be done to either increase the cross section or to change the conditions for the light. The latter is done by changing the environment, for example by placing atoms in a waveguide [8–10] or an optical cavity [11–13], where particular guided modes or cavity modes can be focused to less than the squared wavelength or the same light can couple to the atoms repeatedly, resulting in a strong interaction. A simple approach to increase the cross section is to have many atoms, so that there are many possibilities for photons to be absorbed, thus making it more likely to happen. A strong light-matter coupling can thus be observed in a gas of atoms, which can

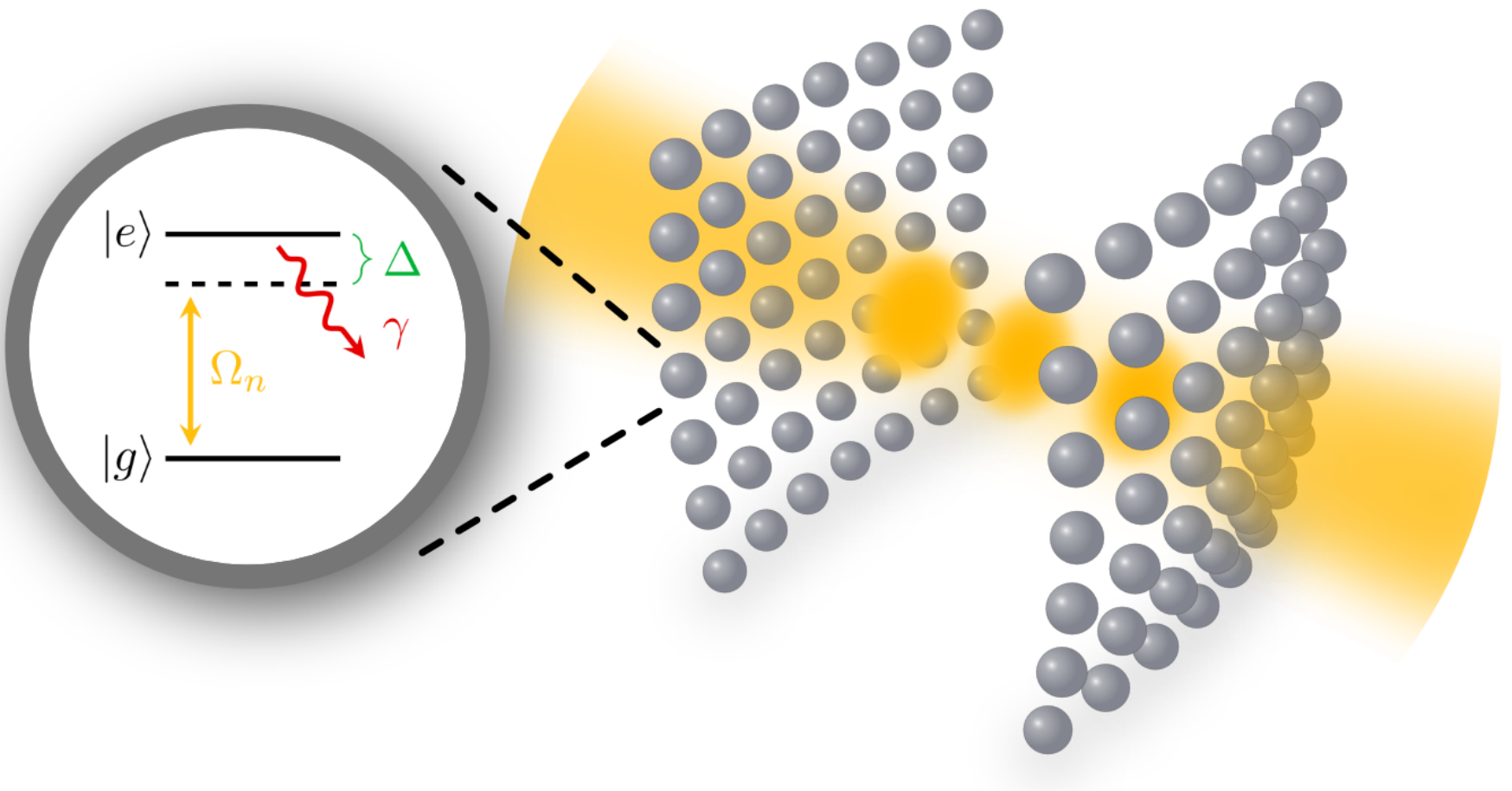

**Schematic of the atomic cavity.** Two two-dimensional lattices of two-level atoms are placed in parallel, forming a cavity of atomic mirrors. Photons are confined for a long time in this system, and while the individual lattices are only weakly nonlinear, the long exposure to this weakly nonlinear environment results in the accumulation of strong nonlinearities for the photons.

act collectively and absorb the photon as a delocalized atomic excitation [14–16]. However, there are two issues with this: Firstly, a disordered and moving gas of atoms causes scattering in all directions and induces decoherence. Secondly, increasing the number of atoms decreases the nonlinearity, as this originated in the saturability of a single two-level system, but a gas of many atoms can absorb many photons before reaching saturation.

To take advantage of the strong coupling between light and many atoms, but circumventing the lossy scattering, the atoms can be arranged in regular two-dimensional lattices [17]. The discrete translational symmetry results in the conservation of the photons' momentum up to Bragg scattering, i.e. momentum is conserved up to the addition of a reciprocal lattice vector. Reducing the real space lattice spacing to the point where the reciprocal lattice vectors become so large that their addition always takes the photon momentum outside the light cone, such that any scattered light becomes evanescent, results in a situation where light is only emitted with the same momentum as it initially had. We say that only the zeroth Bragg channel (or diffraction order) is open, and thus it becomes possible to have single-mode dynamics. This is a similar effect to having a waveguide, but in free space. These arrays then behave as atomic mirrors, being capable of perfect reflection of single photons [18].

Just as with the gas, such an extended array of atoms is also mostly linear, due to the suppression of saturability. To circumvent this, I will consider a particular configuration of two atomic mirrors in parallel that form a cavity-like system. This system shows extremely narrow single-photon transmission and reflection resonances. That is, very narrow frequency ranges of sudden perfect, or near-perfect, transmission or reflection. Following the intuitive picture that a second photon sees a system that is perturbed by the presence of the first, these might result in strong nonlinearities as a narrow feature can be shifted away or otherwise vanish due to even a small perturbation. More stringently, I will show how a recovery of strong nonlinearities arises as a consequence of long confinement of the photons in this atomic cavity around the transmission resonances. While the individual lattices are still mostly linear, the long exposure to a weakly nonlinear environment results in the photons accumulating strong correlations, implying a strong interaction among photons.

At first I will study this system using analytics for the linear physics and numerics for the nonlinear, but then I will develop a method based on Green's function propagators, as known from quantum field theory, to analytically determine both the linear and nonlinear behaviour. Using a Feynman diagrammatic approach to determine the propagators, I analytically perform time-evolution of the photonic wave functions, and from them extract all correlations. The highly involved evaluation of integrals required to do this, becomes feasible due to the relatively simple form of the propagators in the system, and due to the fact that I consider propagating photons in the steady state, i.e. evanescent fields near the atomic array are neglected, and I do not consider the transient dynamics before the system settles into its steady state. In this regime, and for a single-frequency initial state of light, I thus derive exact expressions for the nonlinear response of both the atomic mirror and the atomic cavity. This is exactly the same regime as considered in the numerical approach, and so the analytical results require no additional restrictions.

The thesis is structured in two main parts. In the first part I introduce the physical system and the formalism used to analyse such atomic-optical systems. In Chapter 1 I will derive, in as general terms as possible, the formulas that will subsequently be used to describe and study atomic arrays. Chapter 2 then moves on to explore the linear physics of this system. I will consider the single-photon transmission and reflection, as well as the delay time (i.e. confinement time) of photons, comparing the results with the corresponding ones for a single atomic mirror and for a single atom along the way. This is done analytically for infinite lattices of atoms, and the chapter ends with considerations regarding finite-size arrays, which must be used when performing numerics for the nonlinear physics. Chapter 3 then concerns the nonlinear physics of the atomic cavity. I start by introducing the correlation functions used to study the nonlinearity and the numerical approach to calculating these. In particular, I will look at momentum and temporal correlations, and with these show how the atomic cavity produces strong and long-lasting correlations, which correspond to photons strongly interacting and exchanging momentum before being emitted from the system.

The second part of the thesis is dedicated to the development of the propagator-based approach to analytically determining the nonlinear physics of the atomic arrays. In Chapter 4 the method is presented and exemplified by applying it to the atomic mirror. We perform the derivation in full detail, and arrive at analytical results for the two-photon correlations of the atomic mirror, from which we make a few new observations compared to the numerical study. We use the atomic mirror to show the method, as the derivation is identical in form to that of the atomic cavity, but less cluttered. In Chapter 5 we therefore simply quote the analytical results pertaining to the atomic cavity, giving details only on the parts of the derivation, which are different from the atomic mirror due to the presence of a second atomic lattice. These results are briefly discussed. A conclusion and outlook for the entire thesis is given in Chapter 6.

The thesis is written using the pronoun *we*, rather than *I*. This is to avoid the confusion of switching between “I, the author”, “we, the reader and the author”, and “we, the author’s colleagues and the author”. It also highlights the fact that most of the work done in thesis took place with the help and collaboration of my colleagues, and, of course, on the basis of the existing literature built up over the years by many other physicists.

Thus prepared, let us begin!

# Part I

## *The Atomic Cavity*

# 1

# Basic formalism

In this, the first part of the thesis, the basic formalism to describe interacting two-level atoms and photons will be introduced. Initially, the presented formulas will be valid and relevant for any configuration of atoms in vacuum, and then we specialize to a particular class of atom geometry that will be used for the remainder of the thesis. Namely, we consider a (possibly infinite) number of two-level atoms arranged in an array of either a single or two parallel square lattices, interacting with the electromagnetic field. We will be focusing on the physics of the latter. Our goal with analysing this system is to show how it induces strong interactions among photons. We will see that the atomic arrays couple strongly with individual photons, and how the saturability of the underlying two-level atoms enables pairs of photons to effectively interact. Indeed, we will see how narrow features of transmission or reflection are associated with a long confinement of the photons, and subsequently how the long exposure to the nonlinear saturability of the atoms generates strong correlations among photons, indicating a strong interaction.

Taking advantage of the fact that the dynamics of the electromagnetic field can be expressed in terms of the atoms' behaviour, we focus in this part on the atomic sector of the system, either analytically or numerically solving their dynamics and from it characterize the electromagnetic field. In the next part we take a different perspective, using an analytical approach that treats the atoms and photons on a more equal basis.

Most of the results presented in this part are published in Ref. [19].

In this chapter, the atomic arrays are first introduced in Section 1.1. In particular we will look at the mentioned single and dual two-dimensional lattices of atoms, where the former has already received much attention in the field of quantum optics. Section 1.2 will go through the derivation of the input-output relation for the electric field, and the effective atomic Hamiltonian and Lindbladian, which are the basis for the analysis performed in this part of the thesis. These are found by integration of the electromagnetic degrees of freedom, i.e. solving the photonic sector of the system in terms of the atomic dynamics. We then take time in Section 1.3 to analyse and understand the dyadic Green's function of the electromagnetic wave equation, as this function and the manipulation of it are central to the analysis at hand. Finally, in Section 1.4, we reformulate the input-output relation in terms of given modes, and in Section 1.5 the effective atomic Hamiltonian and Lindbladian in terms of quasimomentum modes. This chapter will thus be rather technical, with many preparatory and general derivations performed to set up the stage for the following chapters, where we dive into the physics of the system at hand.

## 1.1 Interacting dipolar atoms and photons

We consider an ensemble of identical two-level atoms interacting with the electromagnetic (EM) field in the electric dipole approximation. This is described by the following Hamiltonian (with Planck's reduced constant $\hbar = 1$) [20]

$$H = \sum_{\boldsymbol{k},\nu} \omega_{\boldsymbol{k}} b^\dagger_{\boldsymbol{k}\nu} b_{\boldsymbol{k}\nu} + \sum_n \omega_a \sigma^\dagger_n \sigma_n - \sum_n \boldsymbol{E}^\dagger(\boldsymbol{r}_n)\boldsymbol{p}_n \quad , \tag{1.1}$$

where $\omega_{\boldsymbol{k}} = |\boldsymbol{k}|$ is the photonic dispersion (with the speed of light $c = 1$), $b_{\boldsymbol{k}\nu}$ is the bosonic annihilation operator for photons with wave vector (or momentum) $\boldsymbol{k}$ and polarization indexed by $\nu = 1, 2$. Furthermore, $\omega_a = k_a = 2\pi/\lambda_a$ is the transition frequency of the two-level atoms[1], $\sigma_n$ is the lowering operator for the $n$'th atom positioned at $\boldsymbol{r}_n$, and $\boldsymbol{p}_n \equiv \boldsymbol{d}\sigma_n + \boldsymbol{d}^*\sigma^\dagger_n$ is the corresponding dipole operator, where $\boldsymbol{d}$ is the dipole moment of the atoms. Finally, $\boldsymbol{E}(\boldsymbol{r})$ is the electric field operator given by

$$\boldsymbol{E}(\boldsymbol{r}) = \frac{i}{\sqrt{2\epsilon_0 V}} \sum_{\boldsymbol{k},\nu} \sqrt{\omega_{\boldsymbol{k}}}\, \hat{\boldsymbol{e}}_{\boldsymbol{k}\nu} (e^{i\boldsymbol{k}\cdot\boldsymbol{r}} b_{\boldsymbol{k}\nu} - e^{-i\boldsymbol{k}\cdot\boldsymbol{r}} b^\dagger_{\boldsymbol{k}\nu}) \quad . \tag{1.2}$$

1: We will be using frequencies and wave numbers interchangeably, going from one to the other depending on what notation is the most intuitive.

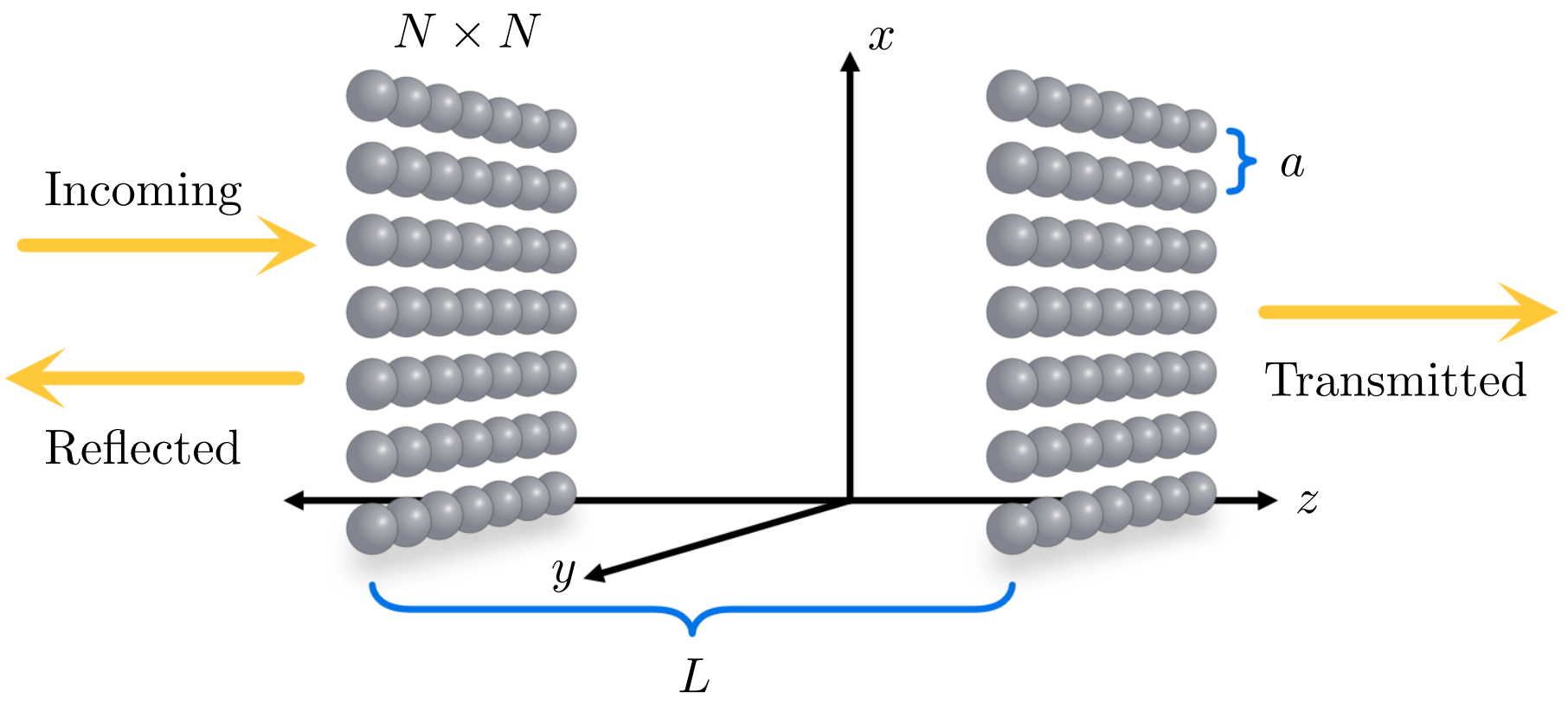

**Figure 1.1 | Dual array atomic configuration.** Two parallel planar lattices of atoms with lattice spacing $a$ are placed a distance $L$ apart. Light is incident on the system from the left, and gets transmitted and reflected.

Here $V$ is the quantization volume, which we eventually take to be infinite, and $\hat{\boldsymbol{e}}_{\boldsymbol{k}\nu}$ are two polarization vectors orthogonal to $\boldsymbol{k}$, indexed by $\nu$. The front factor is chosen such that the classical expression for the electromagnetic energy yields the free photonic Hamiltonian (i.e. the first term of $H$) upon using Eq. (1.2) [21]. Notice that we use the †-symbol to indicate both Hermitian conjugation of operators and to indicate complex conjugation plus transposition of vectors (or matrices).

These equations are completely general for identical atoms with a single dipolar transition in the presence of the EM field, but we will specifically be considering atoms arranged in a regular square lattice in the $xy$-plane, or two such lattices at $z = \pm L/2$. We refer to these configurations as the *single array* or *atomic mirror*, and the *dual array* or *atomic cavity*, respectively. In Fig. 1.1 a diagram of the atomic cavity is shown for the case of finite lattices with $N \times N$ atoms. We will also be considering the infinite lattice case, $N \to \infty$. We furthermore consider an EM field that is incident from the left, propagating in the positive $z$-direction, driving the atoms. This field may either be a single plane wave, with some frequency and transverse momentum (i.e. $\boldsymbol{k}_\perp = (k_x, k_y)^T$), or a more complicated mode of light. In particular, we will consider a continuous-wave Gaussian mode driving the system.

The atomic mirror [17, 22–24], and other ordered configurations of quantum emitters [25, 26], have been the subject of a wealth of papers in recent years, due to their strong coupling to light while suppressing unwanted scattering [18, 27–29], as well as the related fact that they can host highly super- and subradiant states (i.e. states with increased or suppressed emission) [30, 31]. These phenomena are based in the cooperative behaviour of the arrays [32, 33], where the configuration of emitters acts or responds as a single collective entity, rather than as individuals.

As mentioned in the Introduction, and as we shall derive for the atomic cavity, the discrete translational symmetry of regular emitters results in the well-ordered Bragg scattering of incoming light (a collective response as opposed to the dipole scattering of individual atoms), which for subwavelength structures means only a single forward and backward channel are open for emission. This is a decided advantage over atoms coupled to gasses of atoms or waveguides [8], where a significant part of the incoming light is lost due to scattering into undesired modes. The relative fraction of emission into the desired guided mode, designated $\beta$, may be on the order of $\beta \sim 10^{-1}$-$10^{-2}$ for waveguides [8, 34, 35]. Superconducting circuits or quantum dots feature very low loss similar to the atomic arrays ($\beta \simeq 1$) [8, 36], but compared to using real atoms, or arrays of these, it is very challenging to achieve identical emitters in these platforms. That is, since each emitter is a solid state object constructed from a great number of constituents, their properties will generally not be aligned without a great effort of tuning them. State-of-the-art experiments have reported on the order of 10 or fewer identical emitters for these systems [8], while real atoms intrinsically are identical such that experiments with waveguides can feature in the thousands of identical emitters (but at the cost of a very weak coupling to the guided mode, as mentioned).

However, while the light-matter coupling of subwavelength atomic arrays is high, and the emitters are intrinsically identical, each array requires many atoms precisely located and kept in place. Due to the collective, single emitter-like behaviour of the atomic mirror at the linear level, one could argue that a single two-dimensional array of atoms corresponds to one emitter in the waveguide [37–40], such that it would require a great deal of real atoms to achieve just a few effective ones (again, this may be outweighed by the strong coupling to light). Arrays will generally suffer from imperfections in the position or filling of emitter sites, which deteriorates the low-loss collective response [17, 41–44], and it may be difficult to implement the desired subwavelength spacing. However, a great effort is being put into the construction of arbitrary and defect-free atomic lattices using tweezer arrays [45–49] or optical lattices [5, 28, 49] to implement experimental realizations of the physics we will discuss here.

We now derive the input-output formalism and effective spin-description commonly used to study the atomic arrays [17, 18, 42, 50–55]. The idea is that rather than attempt to solve the dynamics of the full Hamiltonian, Eq. (1.1), we will formally solve the photonic dynamics in terms of the atomic one, resulting in an input-output formula for the E-field and an open-system description of the atomic dynamics.

## 1.2 Integration of the photonic degrees of freedom

As Eq. (1.1) is quadratic in the photonic operators, and as these are bosonic, the dynamics of the photons can be formally solved in terms of the remaining degrees of freedom (the dynamics of bosonic degrees of freedom with a quadratic Hamiltonian can generally be solved). This procedure was first used for interacting photons and atoms in the seminal work by Lehmberg [56]. We can write Heisenberg's equation of motion for the photonic annihilation operator

$$
\begin{aligned}
i\dot{b}_{\boldsymbol{k}\nu} &= [b_{\boldsymbol{k}\nu}, H] \\
&= \omega_{\boldsymbol{k}} b_{\boldsymbol{k}\nu} - \sum_n \left( -\frac{i}{\sqrt{2\epsilon_0 V}} \sqrt{\omega_{\boldsymbol{k}}} \hat{\boldsymbol{e}}^\dagger_{\boldsymbol{k}\nu} e^{-i\boldsymbol{k}\cdot\boldsymbol{r}} \right) \boldsymbol{p}_n \\
&= \omega_{\boldsymbol{k}} b_{\boldsymbol{k}\nu} + i\sqrt{\frac{\omega_{\boldsymbol{k}}}{2\epsilon_0 V}} \hat{\boldsymbol{e}}^\dagger_{\boldsymbol{k}\nu} \sum_n \boldsymbol{p}_n e^{-i\boldsymbol{k}\cdot\boldsymbol{r}_n} \quad .
\end{aligned}
\tag{1.3}
$$

By inspection we find that this is formally solved by

$$
b_{\boldsymbol{k}\nu}(t) = b^{\text{free}}_{\boldsymbol{k}\nu}(t) + \sqrt{\frac{\omega_{\boldsymbol{k}}}{2\epsilon_0 V}} \hat{\boldsymbol{e}}^\dagger_{\boldsymbol{k}\nu} \sum_n \int_{t'}^{t} \mathrm{d}\tau \boldsymbol{p}_n(\tau) e^{-i\boldsymbol{k}\cdot\boldsymbol{r}_n} e^{-i\omega_{\boldsymbol{k}}(t-\tau)} \quad , \tag{1.4}
$$

where $b^{\text{free}}_{\boldsymbol{k}\nu}(t)$ evolves according to the free photonic Hamiltonian, and $t' < t$ is some time in the past. Thus, the photonic field is given by a freely evolving component and a component stemming from the atomic dynamics. We can use this to write the E-field dynamics in terms of the atomic dynamics, and to write an effective Hamiltonian for just the atomic degrees of freedom driven by the free component of the E-field.

### 1.2.1 Derivation of the input-output relation

Inserting the above formal solution for $b_{\boldsymbol{k}\nu}(t)$ in the expression for the E-field, Eq. (1.2), we find

$$\begin{aligned}\boldsymbol{E}(\boldsymbol{r},t) = \boldsymbol{E}_{\text{free}}(\boldsymbol{r},t) + \frac{i}{\sqrt{2\epsilon_0 V}}\sum_{\boldsymbol{k},\nu}\sqrt{\omega_{\boldsymbol{k}}}\hat{\boldsymbol{e}}_{\boldsymbol{k}\nu}\sqrt{\frac{\omega_{\boldsymbol{k}}}{2\epsilon_0 V}}\hat{\boldsymbol{e}}^{\dagger}_{\boldsymbol{k}\nu} \\ \times\sum_n\int_{t'}^{t}\mathrm{d}\tau\boldsymbol{p}_n(\tau)\left(e^{i\boldsymbol{k}\cdot\boldsymbol{r}}e^{-i\boldsymbol{k}\cdot\boldsymbol{r}_n}e^{-i\omega_{\boldsymbol{k}}(t-\tau)}-\text{c.c.}\right) \\ = \boldsymbol{E}_{\text{free}}(\boldsymbol{r},t) \\ + \frac{i}{2\epsilon_0 V}\sum_{n,\boldsymbol{k}}\omega_{\boldsymbol{k}}\boldsymbol{Q}e^{i\boldsymbol{k}\cdot(\boldsymbol{r}-\boldsymbol{r}_n)}\int_{t'}^{t}\mathrm{d}\tau\boldsymbol{p}_n(\tau)e^{-i\omega_{\boldsymbol{k}}(t-\tau)} + \text{H.c.}\ ,\end{aligned} \tag{1.5}$$

where $\boldsymbol{E}_{\text{free}}(\boldsymbol{r},t)$ is the freely evolving E-field, consisting of the externally controlled input field (the driving field) and of quantum noise [55], and c.c. and H.c. are the complex and Hermitian conjugates respectively. Furthermore, we have introduced $\boldsymbol{Q}\equiv\sum_\nu\hat{\boldsymbol{e}}_{\boldsymbol{k}\nu}\hat{\boldsymbol{e}}^{\dagger}_{\boldsymbol{k}\nu}=\boldsymbol{I}-\hat{\boldsymbol{k}}\hat{\boldsymbol{k}}^{\dagger}$, which is the projection matrix onto the plane spanned by the polarization vectors. The second identity follows from the fact that $\hat{\boldsymbol{e}}_{\boldsymbol{k}\nu}$ are unit vectors orthogonal to each other and to $\boldsymbol{k}$. Hence, $\{\hat{\boldsymbol{e}}_{\boldsymbol{k}1},\hat{\boldsymbol{e}}_{\boldsymbol{k}2},\hat{\boldsymbol{k}}\}$ is a spanning basis, such that $\hat{\boldsymbol{e}}_{\boldsymbol{k}1}\hat{\boldsymbol{e}}^{\dagger}_{\boldsymbol{k}1}+\hat{\boldsymbol{e}}_{\boldsymbol{k}2}\hat{\boldsymbol{e}}^{\dagger}_{\boldsymbol{k}2}+\hat{\boldsymbol{k}}\hat{\boldsymbol{k}}^{\dagger}=\boldsymbol{I}$.

We now let $V\to\infty$, such that $\frac{(2\pi)^3}{V}\sum_{\boldsymbol{k}}\to\int\mathrm{d}^3k=\int_0^\infty\mathrm{d}k\int_{S_k}\mathrm{d}s_k$, where we perform the integration of $\boldsymbol{k}$-space using spherical coordinates with radial component $k$, and $S_k$ a sphere in $\boldsymbol{k}$-space with radius $k$, and differential surface element $\mathrm{d}s_k=k^2\sin\theta\mathrm{d}\theta\mathrm{d}\phi$ (with the usual polar and azimuthal angles). With this, and using $\omega_{\boldsymbol{k}}=k$, we find

$$\begin{aligned}\boldsymbol{E}(\boldsymbol{r},t) &= \boldsymbol{E}_{\text{free}}(\boldsymbol{r},t) \\ &+ \frac{i}{16\pi^3\epsilon_0}\sum_n\int_0^\infty\mathrm{d}kk\int_{S_k}\mathrm{d}s_k\boldsymbol{Q}e^{i\boldsymbol{k}\cdot(\boldsymbol{r}-\boldsymbol{r}_n)}\int_{t'}^{t}\mathrm{d}\tau\boldsymbol{p}_n(\tau)e^{-ik(t-\tau)} \\ &+ \text{H.c.}\end{aligned} \tag{1.6}$$

We now introduce the function

$$\boldsymbol{G}(\boldsymbol{r},\boldsymbol{r}',k) = \frac{i}{8\pi^2 k}\int_{H_k^{\text{sgn}(z-z')}}\mathrm{d}s_k\boldsymbol{Q}e^{i\boldsymbol{k}\cdot(\boldsymbol{r}-\boldsymbol{r}')}\ , \tag{1.7}$$

where $H_k^{\pm}$ is the upper or lower hemisphere of $S_k$ with respect to the $k_z$-direction. As we will show in Section 1.3 this is the (vacuum) dyadic

Green's function of the EM wave equation. This function in principle contains all information about how an E-field propagates given some source. The Green's function written here is in fact only for the propagating part of the E-field (as opposed to evanescent components), which stems from the fact that Eq. (1.2) only contains propagating field components. The expression for the E-field operator, we are deriving, thus also only pertains to the propagating part of the full field. In Section 1.3 we will study the Green's function further and make the generalization to include evanescent fields as well. With this definition we see

$$\begin{aligned}\int_{S_k} & \mathrm{d}s_k \boldsymbol{Q} e^{i\boldsymbol{k}\cdot(\boldsymbol{r}-\boldsymbol{r}_n)} \\ &= \int_{H_k^{\mathrm{sgn}(z-z_n)}} \mathrm{d}s_k \boldsymbol{Q} e^{i\boldsymbol{k}\cdot(\boldsymbol{r}-\boldsymbol{r}_n)} + \int_{H_k^{-\mathrm{sgn}(z-z')}} \mathrm{d}s_k \boldsymbol{Q} e^{i\boldsymbol{k}\cdot(\boldsymbol{r}-\boldsymbol{r}_n)} \\ &= \int_{H_k^{\mathrm{sgn}(z-z_n)}} \mathrm{d}s_k \boldsymbol{Q} e^{i\boldsymbol{k}\cdot(\boldsymbol{r}-\boldsymbol{r}_n)} + \int_{H_k^{\mathrm{sgn}(z-z')}} \mathrm{d}s_k \boldsymbol{Q} e^{-i\boldsymbol{k}\cdot(\boldsymbol{r}-\boldsymbol{r}_n)} \\ &= -8i\pi^2 k(\boldsymbol{G}(\boldsymbol{r},\boldsymbol{r}_n,k) - \boldsymbol{G}^*(\boldsymbol{r},\boldsymbol{r}_n,k)) \\ &= 16\pi^2 k \Im\left[\boldsymbol{G}(\boldsymbol{r},\boldsymbol{r}_n,k)\right] \quad ,\end{aligned} \tag{1.8}$$

where $\Im[\cdot]$ indicates the imaginary part. We have here flipped the sign of $\boldsymbol{k}$, causing $H_k^{\pm} \to H_k^{\mp}$, while we can freely flip the sign of $k_x$ and $k_y$ in the integral $\int_{H_k^{\pm}} \mathrm{d}^2k$, and $\boldsymbol{Q}$ is even in $\boldsymbol{k}$. With this we have

$$\begin{aligned}\boldsymbol{E}(\boldsymbol{r},t) = {}& \boldsymbol{E}_{\mathrm{free}}(\boldsymbol{r},t) \\ &+ \frac{i}{\pi\epsilon_0}\sum_n \int_0^\infty \mathrm{d}k k^2 \Im\left[\boldsymbol{G}(\boldsymbol{r},\boldsymbol{r}_n,k)\right] \int_{t'}^{t} \mathrm{d}\tau \boldsymbol{p}_n(\tau) e^{-ik(t-\tau)} \\ &+ \mathrm{H.c.}\end{aligned} \tag{1.9}$$

Let us deal with the temporal integral first.

We write $\sigma_n(t) = \tilde{\sigma}_n(t) e^{-ik_a t}$, such that the integral becomes

$$\int_{t'}^{t} \mathrm{d}\tau \boldsymbol{p}_n(\tau) e^{-ik(t-\tau)} = \int_{t'}^{t} \mathrm{d}\tau \left(\boldsymbol{d}\tilde{\sigma}_n(\tau) e^{-ik_a\tau} + \boldsymbol{d}^*\tilde{\sigma}_n^\dagger(\tau) e^{ik_a\tau}\right) e^{-ik(t-\tau)} \quad . \tag{1.10}$$

Here we have extracted the fast part of the time-dependence of atomic operators corresponding to free evolution, $e^{-ik_a t}$. If we now assume that the scale $t_c$ of temporal correlations in the photonic dynamics is $t_c \ll t - t'$, we may let $t' \to \infty$. That is, we assume that the dynamics are only affected by the immediate past so that the part of the integral which looks at the far past, or indeed all the past, does not contribute anyway. Furthermore, we assume that the atomic dynamics are much slower than $t_c$, such that $\tilde{\sigma}_n(\tau)$ can be taken at time $t$, i.e. $\tilde{\sigma}_n(\tau) \simeq \tilde{\sigma}_n(t) = \sigma_n(t)e^{ik_a t}$, and moved outside the integral. This is the Markov approximation (the system has no memory of the past), and the steps we have taken here are similar to those taken in Wigner-Weisskopf theory [7, 55, 57]. In total we get

$$\begin{aligned}
&\int_{t'}^{t} \mathrm{d}\tau \boldsymbol{p}_n(\tau) e^{-ik(t-\tau)} \\
&\qquad \simeq \boldsymbol{d}\sigma_n(t) \int_{-\infty}^{t} \mathrm{d}\tau e^{i(k-k_a)(\tau-t)} + \boldsymbol{d}^* \sigma_n^\dagger(t) \int_{-\infty}^{t} \mathrm{d}\tau e^{i(k+k_a)(\tau-t)} \\
&\qquad = \boldsymbol{d}\sigma_n(t) \int_{0}^{\infty} \mathrm{d}\tau e^{-i(k-k_a)\tau} + \boldsymbol{d}^* \sigma_n^\dagger(t) \int_{0}^{\infty} \mathrm{d}\tau e^{-i(k+k_a)\tau} \\
&\qquad = \boldsymbol{d}\sigma_n(t) i\zeta(k_a - k) + \boldsymbol{d}^* \sigma_n^\dagger(t) i\zeta(-k-k_a) \ ,
\end{aligned} \tag{1.11}$$

where we have identified the Heitler zeta-function $\zeta(x) \equiv -i\int_0^\infty \mathrm{d}\tau e^{ix\tau} = \mathrm{P}\frac{1}{x} - i\pi\delta(x) = \frac{1}{x+i\eta}$ [58], where $\eta$ is a positive infinitesimal, and P denotes the principal value (we will in general use the last expression of the function when doing contour integrations). In Lehmberg's original derivation [56], rather than directly performing the Markov approximation, he imposes a UV cut-off on the $k$-integral in Eq. (1.9) at $k = 1/a_B$ (i.e. he replaces the integration limit $+\infty$ with $1/a_B$), where $a_B$ is the Bohr radius, arguing that the dipole approximation breaks down at frequencies above $1/a_B$. He uses this cut-off to justify essentially the same steps as we have taken, arriving at the same result.

In total we have

$$
\begin{aligned}
\boldsymbol{E}(\boldsymbol{r},t) &= \boldsymbol{E}_{\text{free}}(\boldsymbol{r},t) \\
&\quad - \frac{1}{\pi\epsilon_0}\sum_n \int_0^\infty \mathrm{d}k k^2 \Im\left[\boldsymbol{G}(\boldsymbol{r},\boldsymbol{r}_n,k)\right] \\
&\qquad\qquad \times \Big(\boldsymbol{d}\sigma_n(t)\zeta(k_a-k) + \boldsymbol{d}^*\sigma_n^\dagger(t)\zeta(-k-k_a)\Big) \\
&\quad + \text{H.c.} \\
&= \boldsymbol{E}_{\text{free}}(\boldsymbol{r},t) \\
&\quad - \frac{1}{\pi\epsilon_0}\sum_n \int_0^\infty \mathrm{d}k k^2 \Im\left[\boldsymbol{G}(\boldsymbol{r},\boldsymbol{r}_n,k)\right] \boldsymbol{d}\sigma_n(t) \\
&\qquad\qquad \times \left(\zeta(k_a-k) + \zeta^*(-k-k_a)\right) \\
&\quad - \frac{1}{\pi\epsilon_0}\sum_n \int_0^\infty \mathrm{d}k k^2 \Im\left[\boldsymbol{G}(\boldsymbol{r},\boldsymbol{r}_n,k)\right] \boldsymbol{d}^*\sigma_n^\dagger(t) \\
&\qquad\qquad \times \left(\zeta(-k-k_a) + \zeta^*(k_a-k)\right) \quad ,
\end{aligned}
\tag{1.12}
$$

where we have written out the Hermitian conjugate, and rearranged the terms, according to whether they contain $\sigma_n$ or $\sigma_n^\dagger$. From this we identify the so-called positive frequency part of the E-field (the second term of the final expression in the above equation)

$$
\begin{aligned}
\boldsymbol{E}^+(\boldsymbol{r},t) &= \boldsymbol{E}^+_{\text{free}}(\boldsymbol{r},t) \\
&\quad - \frac{1}{\pi\epsilon_0}\sum_n \int_0^\infty \mathrm{d}k k^2 \Im\left[\boldsymbol{G}(\boldsymbol{r},\boldsymbol{r}_n,k)\right] \boldsymbol{d}\sigma_n(t) \\
&\qquad\qquad \times \left(\zeta(k_a-k) + \zeta^*(-k-k_a)\right) \quad .
\end{aligned}
\tag{1.13}
$$

The full field is then $\boldsymbol{E}(\boldsymbol{r},t) = \boldsymbol{E}^+(\boldsymbol{r},t) + \boldsymbol{E}^-(\boldsymbol{r},t)$ where the negative frequency part of the field, $\boldsymbol{E}^-(\boldsymbol{r},t)$, is the Hermitian conjugate of $\boldsymbol{E}^+(\boldsymbol{r},t)$. Using the Schwarz reflection principle [59] (which can be seen directly from Eq. (1.7))

$$
\boldsymbol{G}^*(\boldsymbol{r},\boldsymbol{r}_n,z) = \boldsymbol{G}(\boldsymbol{r},\boldsymbol{r}_n,-z^*) \quad , \tag{1.14}
$$

we find

$$\begin{aligned}
\boldsymbol{E}^{+}(\boldsymbol{r},t) &= \boldsymbol{E}^{+}_{\text{free}}(\boldsymbol{r},t) \\
&\quad - \frac{1}{\pi\epsilon_0}\sum_n \int_0^{\infty} \mathrm{d}k k^2 \frac{1}{2i}\left(\boldsymbol{G}(\boldsymbol{r},\boldsymbol{r}_n,k) - \boldsymbol{G}^*(\boldsymbol{r},\boldsymbol{r}_n,k)\right)\boldsymbol{d}\sigma_n(t) \\
&\qquad\qquad \times \left(\zeta(k_a - k) + \zeta^*(-k - k_a)\right) \\
&= \boldsymbol{E}^{+}_{\text{free}}(\boldsymbol{r},t) \\
&\quad + \frac{i}{2\pi\epsilon_0}\sum_n \int_0^{\infty} \mathrm{d}k k^2 \boldsymbol{G}(\boldsymbol{r},\boldsymbol{r}_n,k)\boldsymbol{d}\sigma_n(t) \\
&\qquad\qquad \times \left(\frac{1}{k_a - k + i\eta} + \frac{1}{-k - k_a - i\eta}\right) \\
&\quad - \frac{i}{2\pi\epsilon_0}\sum_n \int_{-\infty}^{0} \mathrm{d}k k^2 \boldsymbol{G}(\boldsymbol{r},\boldsymbol{r}_n,k)\boldsymbol{d}\sigma_n(t) \\
&\qquad\qquad \times \left(\frac{1}{k_a + k + i\eta} + \frac{1}{k - k_a - i\eta}\right) \\
&= \boldsymbol{E}^{+}_{\text{free}}(\boldsymbol{r},t) \\
&\quad - \frac{i}{2\pi\epsilon_0}\sum_n \int_{-\infty}^{\infty} \mathrm{d}k k^2 \boldsymbol{G}(\boldsymbol{r},\boldsymbol{r}_n,k)\boldsymbol{d}\sigma_n(t) \\
&\qquad\qquad \times \left(\frac{1}{k - k_a - i\eta} + \frac{1}{k + k_a + i\eta}\right) \quad .
\end{aligned} \tag{1.15}$$

We will now calculate the above integral, called $I$, including only the $k$-dependent factors, using complex contour integration. We will use a contour consisting of the real axis (yielding the integral $I$) and a large semicircle in upper half plane called $\Gamma$ (this curve is centred at 0, and has radius $R \to \infty$), see Fig. 1.2. The integrand of $I$ has a pole within this contour at $k_a + i\eta$ (we will show in Section 1.3 that $k^2\boldsymbol{G}(\boldsymbol{r},\boldsymbol{r}_n,k)$ has no poles in terms of $k$), such that

$$I + \int_{\Gamma} \mathrm{d}z z^2 \frac{\boldsymbol{G}(\boldsymbol{r},\boldsymbol{r}_n,z)}{z - k_a} = 2\pi i k_a^2 \boldsymbol{G}(\boldsymbol{r},\boldsymbol{r}_n,k_a) \quad , \tag{1.16}$$

where we have taken the orientation of the curves into account and the right hand side of the equation is $2\pi i$ times the residue at the pole. It is known that $\lim_{|z|\to\infty} z^2 \boldsymbol{G}(\boldsymbol{r}, \boldsymbol{r}', z) = -\delta(\boldsymbol{r} - \boldsymbol{r}')$ [59], such that

$$\begin{aligned}
\int_\Gamma \mathrm{d}z z^2 \frac{\boldsymbol{G}(\boldsymbol{r}, \boldsymbol{r}_n, z)}{z - k_a} &= \lim_{R\to\infty} \int_0^\pi iRe^{i\phi} \mathrm{d}\phi (Re^{i\phi})^2 \frac{\boldsymbol{G}(\boldsymbol{r}, \boldsymbol{r}_n, Re^{i\phi})}{Re^{i\phi} - k_a} \\
&= i \int_0^\pi \mathrm{d}\phi \lim_{R\to\infty} \frac{(Re^{i\phi})^2 \boldsymbol{G}(\boldsymbol{r}, \boldsymbol{r}_n, Re^{i\phi})}{1 - e^{-i\phi} k_a / R} \\
&= -i \int_0^\pi \mathrm{d}\phi \delta(\boldsymbol{r} - \boldsymbol{r}_n) \\
&= -i\pi\delta(\boldsymbol{r} - \boldsymbol{r}_n) \ .
\end{aligned} \tag{1.17}$$

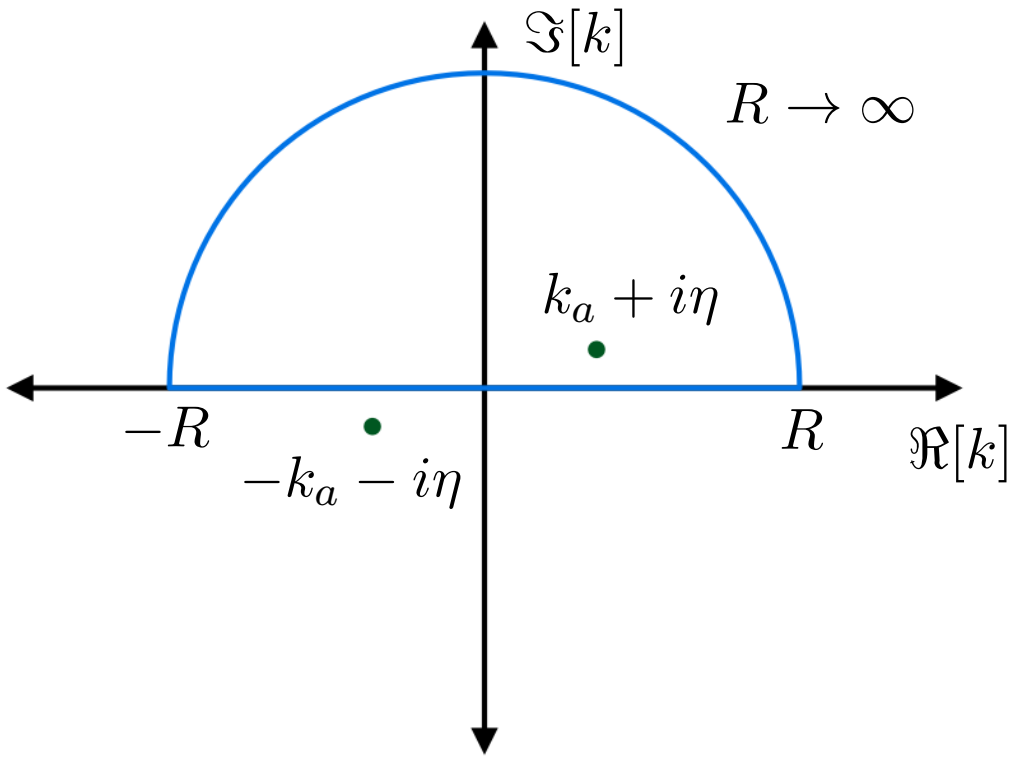

**Figure 1.2 | Contour for $k$-integral in Eq. (1.15).** The contour is shown in blue in the complex $k$-plane, with the two poles of the integrand marked with green dots.

In total

$$I = 2i\pi k_a^2 \boldsymbol{G}(\boldsymbol{r}, \boldsymbol{r}_n, k_a) + i\pi\delta(\boldsymbol{r} - \boldsymbol{r}_n) \ . \tag{1.18}$$

Ignoring the delta-function in the above result (we will not be evaluating the E-field at the position of the atoms, where we expect it to be ill-defined in any case), we can conclude the calculation of the E-field, finding the input-output relation

$$\boldsymbol{E}^+(\boldsymbol{r}, t) = \boldsymbol{E}^+_{\text{in}}(\boldsymbol{r}, t) + \mu_0 \omega_a^2 \sum_n \boldsymbol{G}(\boldsymbol{r}, \boldsymbol{r}_n, k_a) \boldsymbol{d} \sigma_n(t) \ , \tag{1.19}$$

where we have used $1/\epsilon_0 = \mu_0$. Furthermore, we have replaced the free-field operator $\boldsymbol{E}_{\text{free}}$ with the classical field $\boldsymbol{E}_{\text{in}}$. Here, $\boldsymbol{E}_{\text{in}}$ is the externally controlled driving field, which we take to be in a coherent state, i.e. it behaves classically and we can describe it with a c-number vector (as opposed to an operator). This procedure also means we are neglecting quantum noise in the free field. We can do so, as we will ultimately be interested in calculating expectation values of normal-ordered products of the E-field, and it can be shown that for a zero-temperature vacuum (which is what we consider[2]) the noise operators do not contribute to such expectation values [55]. We have thus arrived at the familiar input-output relation [17, 18, 50], giving the full E-field (the output) in terms of the externally controlled field (the input) and the atomic dynamics, which results in a field radiated from the atoms according the EM Green's function. This equation will be central in the analysis performed in this part of the thesis, as it will allow us to focus on the atomic dynamics, and from that calculate properties of the E-field.

### 1.2.2 Derivation of the effective atomic Hamiltonian and Lindbladian

Using the input-output relation of the E-field operator, Eq. (1.19), we can derive a Heisenberg-Langevin equation for the atomic dynamics that only involves the incoming (driving) field and atomic degrees of freedom. From this we can extract the effective atomic Hamiltonian and Lindbladian, and write down a master equation for the density matrix. To do so, we consider the general Hamiltonian, Eq. (1.1), within a rotating wave approximation (RWA). This means we remove terms containing $\boldsymbol{E}^+\sigma_n$ and $\boldsymbol{E}^-\sigma_n^\dagger$, as these will be shown to quickly oscillate in time, such that their contribution to the dynamics averages out and they can be neglected. Furthermore, we will order the E-field operators and the atomic operators, such that when the E-field operators are replaced according to Eq. (1.19), we get a normal-ordered expression (at this point the operators commute and we can order them freely). In total, we will use the Hamiltonian

$$\begin{aligned} H &= H_{\text{ph}} + H_{\text{at}} + H_{\text{int}} \\ &= \sum_{\boldsymbol{k},\nu} \omega_{\boldsymbol{k}} b_{\boldsymbol{k}\nu}^\dagger b_{\boldsymbol{k}\nu} + \sum_n \omega_a \sigma_n^\dagger \sigma_n - \sum_n \left[ (\boldsymbol{E}^+(\boldsymbol{r}_n))^\dagger \boldsymbol{d} \sigma_n + \sigma_n^\dagger \boldsymbol{d}^\dagger \boldsymbol{E}^+(\boldsymbol{r}_n) \right] \ , \end{aligned} \tag{1.20}$$

where we have used the properties of $\boldsymbol{E}^\pm$. Consider now an operator $A$, which depends only on the atomic dynamics. We may then write its

2: Thus far we have not included temperature or thermodynamics in our calculations, and such an approach corresponds to taking the temperature to be zero.

equation of motion as

$$i\dot{A} = [A, H] = [A, H_{\text{at}}] + [A, H_{\text{int}}] \ . \tag{1.21}$$

Let us focus on the second term. We have

$$[A, H_{\text{int}}] = -\sum_n \Big( (\boldsymbol{E}^+(\boldsymbol{r}_n))^\dagger \boldsymbol{d}[A, \sigma_n] + [A, \sigma_n^\dagger]\boldsymbol{d}^\dagger \boldsymbol{E}^+(\boldsymbol{r}_n) \Big) \ . \tag{1.22}$$

We then express the E-field in terms of the input-output relation, Eq. (1.19), suppressing the time-dependence of all operators, except the free E-field, for a less cluttered notation, which yields

$$\begin{aligned}
[A, H_{\text{int}}] = &- \sum_n \Big[ \boldsymbol{d}^\dagger \boldsymbol{E}_{\text{in}}^+(\boldsymbol{r}_n, t)[A, \sigma_n^\dagger] + (\boldsymbol{E}_{\text{in}}^+(\boldsymbol{r}_n, t))^\dagger \boldsymbol{d}[A, \sigma_n] \Big] \\
&- \mu_0 \omega_a^2 \sum_{n,m} \Big[ \boldsymbol{d}^\dagger \boldsymbol{G}^*(\boldsymbol{r}_n, \boldsymbol{r}_m, k_a)\boldsymbol{d}\sigma_m^\dagger [A, \sigma_n] \\
&\qquad + \boldsymbol{d}^\dagger \boldsymbol{G}(\boldsymbol{r}_n, \boldsymbol{r}_m, k_a)\boldsymbol{d}[A, \sigma_n^\dagger]\sigma_m \Big] \ .
\end{aligned} \tag{1.23}$$

Defining $J_{nm} + i\Gamma_{nm} = \mu_0 \omega_a^2 \boldsymbol{d}^\dagger \boldsymbol{G}(\boldsymbol{r}_n, \boldsymbol{r}_m, k_a)\boldsymbol{d}$, and suppressing the "+"-superscript on the incoming positive frequency E-field, we have

$$\begin{aligned}
[A, H_{\text{int}}] = &- \sum_n \Big( \boldsymbol{E}_{\text{in}}^\dagger(\boldsymbol{r}_n)\boldsymbol{d}[A, \sigma_n] + \boldsymbol{d}^\dagger \boldsymbol{E}_{\text{in}}(\boldsymbol{r}_n, t)[A, \sigma_n^\dagger] \Big) \\
&- \sum_{n,m} \Big( (J_{nm} - i\Gamma_{nm})\sigma_m^\dagger [A, \sigma_n] + (J_{nm} + i\Gamma_{nm})[A, \sigma_n^\dagger]\sigma_m \Big) \\
= &- \sum_n \Big( \boldsymbol{E}_{\text{in}}^\dagger(\boldsymbol{r}_n)\boldsymbol{d}[A, \sigma_n] + \boldsymbol{d}^\dagger \boldsymbol{E}_{\text{in}}(\boldsymbol{r}_n, t)[A, \sigma_n^\dagger] \Big) \\
&- \sum_{n,m} J_{nm}[A, \sigma_n^\dagger \sigma_m] \\
&+ \sum_{n,m} i\Gamma_{nm} \Big( 2\sigma_n^\dagger A \sigma_m - \Big\{ \sigma_n^\dagger \sigma_m, A \Big\} \Big) \ ,
\end{aligned} \tag{1.24}$$

where we have used the fact that $\boldsymbol{G}(\boldsymbol{r}, \boldsymbol{r}', k_a)$ is a symmetric matrix, which depends only on $\boldsymbol{r} - \boldsymbol{r}'$. With this calculation we may now say the operator $A$ evolves according to the Heisenberg-Langevin equation

$i\dot{A} = [A, H] + i\mathcal{L}[A]$, with

$$H = \sum_n \omega_a \sigma_n^\dagger \sigma_n - \sum_n \left( \boldsymbol{E}_{\text{in}}^\dagger(\boldsymbol{r}_n) \boldsymbol{d} \sigma_n + \boldsymbol{d}^\dagger \boldsymbol{E}_{\text{in}}(\boldsymbol{r}_n, t) \sigma_n^\dagger \right) - \sum_{\substack{n,m \\ n \neq m}} J_{nm} \sigma_n^\dagger \sigma_m \quad , \tag{1.25a}$$

$$\mathcal{L}[A] = \sum_{n,m} \Gamma_{nm} \left( 2\sigma_n^\dagger A \sigma_m - \left\{ \sigma_n^\dagger \sigma_m, A \right\} \right) \quad . \tag{1.25b}$$

Here the Hamiltonian describes the coherent part of the dynamics, while the Lindbladian $\mathcal{L}$ describes dissipative processes. We have omitted the $J_{nn}$ terms from the dipole-dipole interaction, which give a shift to the atom frequency (the Lamb shift), as this shift is divergent in this nonrelativistic setting[3] [55, 58] (see also Section 1.3). We assume the (correctly calculated and finite) shift is absorbed in the definition of $\omega_a$. Thus, we have arrived at an open-system description of operators pertaining to the atomic degrees of freedom. The description is simplified by the absence of the photonic degrees of freedom, but at the price of introducing interactions among atoms and loss into the environment. This physically corresponds respectively to photons being emitted and reabsorbed by the atoms, and to being emitted and lost into the vacuum. We now do some work to simplify the Hamiltonian and to derive a Lindblad master equation for the density matrix of the atoms.

To do so, we assume that the incoming free field has only a single frequency component, as, for example, is the case for plane waves and continuous-wave beams. That is, we assume $\boldsymbol{E}_{\text{in}}(\boldsymbol{r}_n, t) = \boldsymbol{E}_{\text{in}}(\boldsymbol{r}_n) e^{-i\omega t}$, where $\omega = k = 2\pi/\lambda$ is the frequency of the incoming light. We then go into a rotating frame by transforming all states via $|\psi(t)\rangle \to U(t)\,|\psi(t)\rangle$, where $U(t) = e^{i\omega t \sum_n \sigma_n^\dagger \sigma_n}$. By taking the time derivative of the transformed state, one finds that the Hamiltonian describing its time-evolution is $H_R = UHU^\dagger + i(\partial_t U)U^\dagger$. Using $i(\partial_t U)U^\dagger = -\omega \sum_n \sigma_n^\dagger \sigma_n$, $U\sigma_n U^\dagger = \sigma_n e^{-i\omega t}$ and $U\sigma_n^\dagger \sigma_n U^\dagger = \sigma_n^\dagger \sigma_n$, we then find the following Hamiltonian

3: In principle we should, of course, have omitted this infinite term from the beginning of the derivation, but it would not have made a difference to the derivation and simply made the notation denser.

in the rotating frame[4]

$$H = -\sum_n \Delta\sigma_n^\dagger\sigma_n - \sum_n \left(\boldsymbol{E}_{\text{in}}^\dagger(\boldsymbol{r}_n)\boldsymbol{d}\sigma_n + \boldsymbol{d}^\dagger\boldsymbol{E}_{\text{in}}(\boldsymbol{r}_n)\sigma_n^\dagger\right) - \sum_{\substack{n,m\\ n\neq m}} J_{nm}\sigma_n^\dagger\sigma_m \quad , \tag{1.26}$$

where we have defined the detuning of the incoming field $\Delta \equiv \omega - \omega_a$. The Lindbladian $\mathcal{L}$ is unchanged.

The expectation value of $A(t)$ is

$$\langle A(t)\rangle = \text{Tr}_{\text{ph+at}}(W(0)A(t)) \quad , \tag{1.27}$$

where trace is over both the photonic and atomic sector of the system, and $W(0) = \rho_{\text{ph}}(0) \otimes \rho_{\text{at}}(0)$ is some initial density matrix of the system, which is constituted by the free photonic sector and the atomic sector. Using the cyclic property of the trace, we can find

$$\langle A(t)\rangle = \text{Tr}_{\text{ph+at}}(W(t)A(0)) \quad . \tag{1.28}$$

Going to the rotating frame, as above, we thus have

$$\langle A(t)\rangle = \text{Tr}_{\text{ph+at}}(\rho_{\text{ph}}(0) \otimes \rho_{\text{at}}(t)A(0)) = \text{Tr}_{\text{at}}(\rho_{\text{at}}(t)A(0)) \quad , \tag{1.29}$$

where have used that $A$ depends only on the atomic dynamics to take the trace of the free photonic sector. Taking the time-derivative of Eq. (1.27), and expressing the left-hand side in terms of the time-derivative of the density matrix via Eq. (1.29), we can again use the cyclic properties of the trace, and the fact that the master equation we have derived for $A$ holds for all $A$, to conclude that $\rho_{\text{at}}$ evolves according to the Lindblad master equation $i\dot{\rho_{\text{at}}} = [H, \rho_{\text{at}}] + i\mathcal{L}[\rho_{\text{at}}]$. Here the Hamiltonian is given by Eq. (1.26), and the Lindbladian for the density matrix is essentially given by the Hermitian conjugate of Eq. (1.25b)

$$\mathcal{L}[\rho] = \sum_{n,m} \Gamma_{nm}\left(2\sigma_n\rho\sigma_m^\dagger - \left\{\sigma_n^\dagger\sigma_m, \rho\right\}\right) \quad . \tag{1.30}$$

We have thus derived an effective open-system description of the atomic dynamics, while the photonic dynamics is then given by the input-output relation, Eq. (1.19). These are general in terms of atomic configuration,

4: The terms removed in the RWA performed at the beginning of this section would at this point have been $\boldsymbol{d}^\dagger\boldsymbol{E}_{\text{in}}(\boldsymbol{r}_n)\sigma_n e^{-2i\omega t} + \boldsymbol{E}_{\text{in}}^\dagger(\boldsymbol{r}_n)\boldsymbol{d}\sigma_n^\dagger e^{2i\omega t}$, which oscillate quickly with the frequency $2\omega_a$.

and for the remainder of this chapter, we will spend some time studying both general properties and properties specific to the described atomic arrays of the EM dyadic Green's function, the input-output relation, and the above Hamiltonian and Lindbladian.

Before moving on to, let us here define the decay rate of a single free-space atom $\gamma = \Gamma_{nn} = \mu_0 \omega_a^3 d^2/6\pi$ (this can shown via the explicit expression for $\boldsymbol{G}(\boldsymbol{r}, \boldsymbol{r}', k)$ that we find in Section 1.3.1). Conventionally, in the literature, the collective energies would are defined via $J_{nm} + i\Gamma_{nm}/2 = \mu_0 \omega_a^2 \boldsymbol{d}^\dagger \boldsymbol{G}(\boldsymbol{r}_n, \boldsymbol{r}_m, k_a)\boldsymbol{d}$, such that the decay rate would have been $\gamma = \mu_0 \omega_a^3 d^2/3\pi$ [18, 56, 60]. This would result in the excited state population of a single free-space atom decaying in time as $e^{-\gamma t}$. We choose to not include this factor of 2 for the sake of a less cluttered notation, such that for us it is the complex amplitude of the excited state which decays as $e^{-\gamma t}$ (while the population would decay as $e^{-2\gamma t}$). We will assume the free atomic decay rate $\gamma$ to be extremely small compared to $\omega_a$, corresponding to a narrow excited state, as this is the case in experiments. For example, in Ref. [28] $\gamma/\omega_a \sim 10^{-7}$. This will be relevant for our analytical approach in the second part of the thesis, but has in fact already been used implicitly. Upon making the Markov approximation we used that the time scale of atomic dynamics was much larger than the time scale of the photonic correlations. The former is given by $\gamma^{-1}$, such that the smallness of $\gamma$ is what allowed us to do this approximation.

## 1.3 The electromagnetic dyadic Green's function

Let us again write the dyadic Green's function of the EM field (Eq. (1.7))

$$\boldsymbol{G}(\boldsymbol{r}, \boldsymbol{r}', \omega) = \frac{i}{8\pi^2\omega} \int_{H^{\mathrm{sgn}(z-z')}_{|\boldsymbol{k}|=\omega}} \mathrm{d}s_k \boldsymbol{Q} e^{i\boldsymbol{k}\cdot(\boldsymbol{r}-\boldsymbol{r}')} \ , \tag{1.31}$$

where, again, the integral is over the surface area of $H_k^{\pm}$, which is the upper or lower hemisphere, with respect to the $k_z$-direction, of a sphere of radius $k$ in $\boldsymbol{k}$-space. Dyadic means $\boldsymbol{G}$ is a tensor of order two (which we simply consider as a matrix), and the fact that it is the (vacuum) EM Green's function means it fulfils the (vacuum) EM wave equation in frequency space

$$\nabla \times \nabla \times \boldsymbol{E}(\boldsymbol{r}, \omega) - \omega^2 \boldsymbol{E}(\boldsymbol{r}, \omega) = \boldsymbol{f}(\boldsymbol{r}) \ , \tag{1.32}$$

with the source term $\boldsymbol{f}(\boldsymbol{r})$ replaced by $\boldsymbol{\delta}(\boldsymbol{r} - \boldsymbol{r}') = \boldsymbol{I}\delta(\boldsymbol{r} - \boldsymbol{r}')$. That is, the Green's function is defined by the equation

$$\nabla \times \nabla \times \boldsymbol{G}(\boldsymbol{r}, \boldsymbol{r}', \omega) - \omega^2 \boldsymbol{G}(\boldsymbol{r}, \boldsymbol{r}', \omega) = \boldsymbol{\delta}(\boldsymbol{r} - \boldsymbol{r}') \ . \tag{1.33}$$

Being the solution to the wave equation with a delta-function source, $\boldsymbol{G}$ has the property that for any given source $\boldsymbol{f}(\boldsymbol{r})$, the following E-field would satisfy Eq. (1.32)

$$\boldsymbol{E}(\boldsymbol{r},\omega) = \int \mathrm{d}^3 r' \boldsymbol{G}(\boldsymbol{r},\boldsymbol{r}',\omega)\boldsymbol{f}(\boldsymbol{r}') \ . \tag{1.34}$$

This is the central property of Green's functions: Having solved a differential equation for the case of a delta-function source, we know the solution for the general case, or rather, we have formally solved the general case in terms of an integral, which, of course, then needs to be evaluated. We can consider the $ij$ component of $\boldsymbol{G}$ as the $i$'th entry of an E-field at $\boldsymbol{r}$ due to a point-like source at $\boldsymbol{r}'$ with polarization along the $j$'th direction. In this sense the Green's function contains all information of how E-fields behave given a collection of sources. This explains its appearance in the input-output relation Eq. (1.19), which in fact takes on the form of Eq. (1.34)[5].

As can be seen in Eq. (1.31), $\boldsymbol{G}(\boldsymbol{r},\boldsymbol{r}',\omega)$ depends only on $|\boldsymbol{r}-\boldsymbol{r}'|$, and so we will write $\boldsymbol{G}(\boldsymbol{r},\omega) = \boldsymbol{G}(\boldsymbol{r},0,\omega)$ whenever it clarifies our notation. This property is a consequence of the translational invariance of vacuum.

Due to its appearance in the input-output relation, $\boldsymbol{G}$ will be ubiquitous in our subsequent analyses and so we go through some transformations and properties of this function. We will derive its explicit real space form (evaluating the integral in Eq. (1.31)), and look at both its continuous and discrete transverse Fourier transform. Many of the properties detailed here can be found in textbooks like Refs. [7, 59].

### 1.3.1 Explicit real space expression

Remembering the definition of $\boldsymbol{Q} = \sum_\nu \hat{\boldsymbol{e}}_{\boldsymbol{k}\nu}\hat{\boldsymbol{e}}_{\boldsymbol{k}\nu}^\dagger = \boldsymbol{I} - \hat{\boldsymbol{k}}\hat{\boldsymbol{k}}^\dagger$, the projection matrix onto the space of polarization vectors pertaining to $\boldsymbol{k}$, we can

5: This is seen by dividing the source into external sources, which yields the free field, and the atoms, where the integral is replaced by a sum due to their discrete nature.

write

$$
\begin{aligned}
\boldsymbol{G}(\boldsymbol{r},k) &= \frac{i}{8\pi^2 k}\int_{H_k^{\mathrm{sgn}(z)}} \mathrm{d}s_k (\boldsymbol{I}-\hat{\boldsymbol{k}}\hat{\boldsymbol{k}}^\dagger) e^{i\boldsymbol{k}\cdot\boldsymbol{r}} \\
&= \frac{i}{8\pi^2 k}\int_{H_k^{\mathrm{sgn}(z)}} \mathrm{d}s_k \left(\boldsymbol{I}-\frac{1}{k^2}\nabla\nabla^\dagger\right) e^{i\boldsymbol{k}\cdot\boldsymbol{r}} \\
&= \frac{i}{8\pi^2 k}\left(\boldsymbol{I}-\frac{1}{k^2}\nabla\nabla^\dagger\right)\int_{H_k^{\mathrm{sgn}(z)}} \mathrm{d}s_k e^{i\boldsymbol{k}\cdot\boldsymbol{r}} \\
&= \frac{i}{8\pi^2}\left(\boldsymbol{I}-\frac{1}{k^2}\nabla\nabla^\dagger\right)\int_{k_\perp\leq k} \frac{\mathrm{d}^2 k_\perp}{k_{z,k}} e^{i\boldsymbol{k}_\perp\cdot\boldsymbol{r}_\perp + ik_{z,k}|z|} \quad ,
\end{aligned}
\tag{1.35}
$$

where in the final expression $k_{z,k} = \sqrt{k^2 - k_\perp^2}$, as we have performed a change of integration variables from a hemisphere to a disk in the transverse momentum plane, $\boldsymbol{k}_\perp = (k_x, k_y)^T$. Lastly, $\boldsymbol{r}_\perp = (x, y)^T$. The integral here is in fact the Weyl expansion of a spherical wave [7, 61]

$$
\frac{e^{ikr}}{r} = \frac{i}{2\pi}\int \frac{\mathrm{d}^2 k_\perp}{k_{z,k}} e^{i\boldsymbol{k}_\perp\cdot\boldsymbol{r}_\perp + ik_{z,k}|z|} \quad , \tag{1.36}
$$

except in our case only propagating fields ($k_\perp \leq k$) are included. This is ultimately because the E-field operator from which we started our derivations, Eq. (1.2), contains only propagating fields as mentioned in Section 1.2.1. We make the step to include evanescent fields ($k_\perp > k$) in our model by relaxing the condition $k_\perp \leq k$ in the integral expression for the Green's function, Eq. (1.35). In doing so, we can write

$$
\boldsymbol{G}(\boldsymbol{r},k) = \frac{1}{4\pi}\left(\boldsymbol{I}-\frac{1}{k^2}\nabla\nabla^\dagger\right)\frac{e^{ikr}}{r} \quad . \tag{1.37}
$$

Calculating the derivatives, one finds the explicit real space Green's function [59, 62–64]

$$
\begin{aligned}
\boldsymbol{G}(\boldsymbol{r},k) = \frac{e^{ikr}}{4\pi r}&\left[\left(1+\frac{ikr-1}{k^2r^2}\right)\boldsymbol{I} - \left(1+3\frac{ikr-1}{k^2r^2}\right)\hat{\boldsymbol{r}}\hat{\boldsymbol{r}}^\dagger\right] \\
&- \frac{1}{3k^2}\boldsymbol{\delta}(r) \quad .
\end{aligned}
\tag{1.38}
$$

As used in the derivation of the input-output relation (Eq. (1.19)), we see here that $k^2\boldsymbol{G}(\boldsymbol{r},k)$ has no poles in terms of $k$. Also, as in that derivation, we will ignore the delta-function in the above as we will not be evaluating the Green's function at $\boldsymbol{r} = 0$. As mentioned the dyadic Green's function may be thought of as the E-field given a point-like source with a particular polarization. The part of $\boldsymbol{G}(\boldsymbol{r},k)$ which goes as $1/r$ is then the far-field, and is simply a spherical wave with polarization transverse to $\hat{\boldsymbol{r}}$ (the far-field has no component in the longitudinal direction). The remaining part, which goes as a combination of $1/r^2$ and $1/r^3$, describes the near-field behaviour.

### 1.3.2 Continuous transverse Fourier transform

Performing the same change of integration variables in Eq. (1.31) as we did in Eq. (1.35), and relaxing the condition $k_\perp \leq k$, allows us to write

$$\boldsymbol{G}(\boldsymbol{r},k) = \frac{i}{8\pi^2}\int \frac{\mathrm{d}^2k_\perp}{k_{z,k}}\boldsymbol{Q}_{\mathrm{sgn}(z),k}e^{i\boldsymbol{k}_\perp\cdot\boldsymbol{r}_\perp + ik_{z,k}|z|} \quad , \tag{1.39}$$

where we have introduced $\boldsymbol{Q}_{\pm,k}$, which is the same as $\boldsymbol{Q}$, but with $k_z \to \pm k_{z,k}$. We can then easily Fourier transform with respect to $\boldsymbol{r}_\perp$

$$\begin{aligned}\boldsymbol{G}(\boldsymbol{k}_\perp,z,k) &= \int \mathrm{d}^2r_\perp \boldsymbol{G}(\boldsymbol{r},k)e^{-i\boldsymbol{k}_\perp\cdot\boldsymbol{r}_\perp} \\ &= \frac{i}{2k_{z,k}}\boldsymbol{Q}_{\mathrm{sgn}(z),k}e^{ik_{z,k}|z|} \quad .\end{aligned} \tag{1.40}$$

This is the continuos Fourier transformation of the EM Green's function.

### 1.3.3 Discrete transverse Fourier transform

As we will be working with square planar lattices of atoms, the discrete Fourier transformation of $\boldsymbol{G}(\boldsymbol{r},k)$ pertaining to such a lattice will be needed [18]. We therefore define the following for $z \neq 0$

$$\tilde{\boldsymbol{G}}(\boldsymbol{k}_\perp,z,k) = \sum_n \boldsymbol{G}(\boldsymbol{r}_{\perp,n},z,k)e^{-i\boldsymbol{k}_\perp\cdot\boldsymbol{r}_{\perp,n}} \quad , \tag{1.41}$$

where $\boldsymbol{G}(\boldsymbol{r}_\perp,z,k) = \boldsymbol{G}(\boldsymbol{r},k)$, and the sum is over the sites of a single infinite square lattice, i.e. $\boldsymbol{r}_{\perp,n} = a(i_{x,n},i_{y,n})^T$ where $i_{x,n}, i_{y,n}$ are integers, and $a$ is the lattice spacing. The case of $z = 0$ will be handled below. Using Eq. (1.39) we can start to evaluate the above sum

$$\begin{aligned}\tilde{\boldsymbol{G}}(\boldsymbol{k}_\perp,z,k) &= \frac{i}{8\pi^2}\sum_n\int \frac{\mathrm{d}^2k'_\perp}{k'_{z,k}}\boldsymbol{Q}'_{\mathrm{sgn}(z),k}e^{i\boldsymbol{k}'_\perp\cdot\boldsymbol{r}_{\perp,n}+ik'_{z,k}|z|}e^{-i\boldsymbol{k}_\perp\cdot\boldsymbol{r}_{\perp,n}} \\ &= \frac{i}{8\pi^2}\int \frac{\mathrm{d}^2k'_\perp}{k'_{z,k}}\boldsymbol{Q}'_{\mathrm{sgn}(z),k}e^{ik'_{z,k}|z|}\sum_n e^{i(\boldsymbol{k}'_\perp-\boldsymbol{k}_\perp)\cdot\boldsymbol{r}_{\perp,n}} \quad .\end{aligned} \tag{1.42}$$

Here we use the identity

$$\sum_n e^{i(\boldsymbol{k}'_\perp - \boldsymbol{k}_\perp)\cdot \boldsymbol{r}_{\perp,n}} = \frac{(2\pi)^2}{a^2} \sum_m \delta(\boldsymbol{k}'_\perp - \boldsymbol{k}_\perp - \boldsymbol{q}_m) \ , \tag{1.43}$$

where the sum on the right hand side is over the reciprocal lattices vectors $\boldsymbol{q}_m$, defined by $e^{i\boldsymbol{q}_m \cdot \boldsymbol{r}_{\perp,n}} = 1$ for all $m, n$ (a sketch of the proof of this relation is mentioned after Eq. (1.69a)). For the square lattice we specifically have $\boldsymbol{q}_m = (2\pi/a)(i_{x,m}, i_{y,m})^T$ with integer $i_{x,m}, i_{y,m}$. With this we have

$$\begin{aligned} \tilde{\boldsymbol{G}}(\boldsymbol{k}_\perp, z, k) &= \frac{i}{8\pi^2} \int \frac{\mathrm{d}^2 k'_\perp}{k'_{z,k}} \boldsymbol{Q}'_{\mathrm{sgn}(z),k} e^{ik'_{z,k}|z|} \frac{(2\pi)^2}{a^2} \sum_m \delta(\boldsymbol{k}'_\perp - \boldsymbol{k}_\perp - \boldsymbol{q}_m) \\ &= \frac{i}{2a^2} \sum_m \frac{1}{\kappa_z} \left( \boldsymbol{I} - \frac{\hat{\boldsymbol{\kappa}}\hat{\boldsymbol{\kappa}}^\dagger}{k^2} \right) e^{i\kappa_z |z|} \ , \end{aligned} \tag{1.44}$$

where in this final expression we have introduced $\boldsymbol{\kappa} = (\boldsymbol{k}_\perp + \boldsymbol{q}_m, \mathrm{sgn}(z)\kappa_z)^T$ with $\kappa_z = \sqrt{k^2 - (\boldsymbol{k}_\perp + \boldsymbol{q}_m)^2}$ for the sake of a compact notation. This infinite series must be calculated numerically. In principle, one could also have calculated the original Fourier series numerically, but the above converges more quickly and also makes some of the behaviour of $\tilde{\boldsymbol{G}}(\boldsymbol{k}_\perp, z, k)$ clearer, for example the divergence occurring for $\kappa_z = 0$ and the dependence on $z$, the effects of which we will study further in the following chapters. We note the simple relation $\tilde{\boldsymbol{G}}(\boldsymbol{k}_\perp, z, k) = \frac{1}{a^2} \sum_m \boldsymbol{G}(\boldsymbol{k}_\perp + \boldsymbol{q}_m, z, k)$.

For $z = 0$, the above sum would diverge. This originates from the fact that the real part of the Green's function evaluated at the origin, $\lim_{|\boldsymbol{r}| \to 0} \Re[\boldsymbol{G}(\boldsymbol{r}, k)]$, is divergent (while the imaginary part is convergent), as can be seen from Eq. (1.38). As mentioned below Eq. (1.25), this term corresponds to the Lamb shift of the atomic transition frequency $\omega_a$, which appears infinite if relativistic effects are not taken appropriately into account. We therefore define a discrete Fourier transform, which omits the real part of this term, assuming the shift to be included in the value of $\omega_a$, such that for $z = 0$

$$\begin{aligned} \tilde{\boldsymbol{G}}(\boldsymbol{k}_\perp, 0, k) &= \sum_{n \neq 0} \Re[\boldsymbol{G}(\boldsymbol{r}_{\perp,n}, 0, k)] e^{-i\boldsymbol{k}_\perp \cdot \boldsymbol{r}_{\perp,n}} \\ &\quad + \sum_n \Im[\boldsymbol{G}(\boldsymbol{r}_{\perp,n}, 0, k)] e^{-i\boldsymbol{k}_\perp \cdot \boldsymbol{r}_{\perp,n}} \\ &= \sum_n \boldsymbol{G}(\boldsymbol{r}_{\perp,n}, 0, k) e^{-i\boldsymbol{k}_\perp \cdot \boldsymbol{r}_{\perp,n}} - \Re[\boldsymbol{G}(0, k)] \ . \end{aligned} \tag{1.45}$$

The final expression is to be understood as a way to express $\tilde{\boldsymbol{G}}(\boldsymbol{k}_\perp, 0, k)$ formally via the sum in Eq. (1.44), evaluated at $z = 0$, and $\boldsymbol{G}(0, k)$, both of which are divergent. To calculate this quantity we must regularize the expressions, allowing us to find a finite difference between the two divergent quantities.

### 1.3.4 Regularization of the discrete Fourier transform

To regularize $\tilde{\boldsymbol{G}}(\boldsymbol{k}_\perp, 0, k)$, which is formally the difference of two infinite quantities, we introduce a factor into each quantity that depends on some parameter $\alpha$, such that for $\alpha \to 0$ the factor goes to unity (and each quantity thus returns to its original value), and for all finite $\alpha$ each quantity is convergent. Then taking the difference of the convergent expressions and letting $\alpha \to 0$ should give us the correct finite result (which can also be found by numerically calculating the first sum in Eq. (1.45)). To do this we consider again Eq. (1.39), but now introduce a Gaussian factor to make the integral convergent for all $z$ [18, 62, 65]. Specifically we write

$$\boldsymbol{G}_\alpha(\boldsymbol{r}, k) = \frac{i}{8\pi^2} \int \mathrm{d}^2 k_\perp e^{-\alpha k_\perp^2} \frac{1}{k_{z,k}} \boldsymbol{Q}_{\mathrm{sgn}(z),k} e^{i\boldsymbol{k}_\perp \cdot \boldsymbol{r}_{\perp,n} + i k_{z,k}|z|} \quad , \tag{1.46}$$

which for $\alpha \to 0$ yields $\boldsymbol{G}(\boldsymbol{r}, k)$. We then calculate each term in the final line of Eq. (1.45) separately. The sum is evaluated following exactly the same steps as led to Eq. (1.44), and results in

$$\sum_n \boldsymbol{G}_\alpha(\boldsymbol{r}_{\perp,n}, 0, k) e^{-i\boldsymbol{k}_\perp \cdot \boldsymbol{r}_{\perp,n}} = \frac{i}{2a^2} \sum_m e^{-\alpha \kappa_\perp^2} \frac{1}{\kappa_z} \left( \boldsymbol{I} - \frac{\hat{\boldsymbol{\kappa}} \hat{\boldsymbol{\kappa}}^\dagger}{k^2} \right) \quad , \tag{1.47}$$

where $\boldsymbol{\kappa}$ is defined below Eq. (1.44). For the other term we will only consider a certain component of $\Re[\boldsymbol{G}(0, k)]$, as the calculations for the full matrix are cumbersome, and we will not need them. We will later on be specifically interested in fields and sources that both are right-circularly polarized with respect to propagation in the positive $z$-direction, as given by the polarization vector $\hat{\boldsymbol{e}}_+ = (1, i, 0)^T/\sqrt{2}$. We therefore consider the following

$$\begin{aligned} \hat{\boldsymbol{e}}_+^\dagger \Re[\boldsymbol{G}(0, k)] \hat{\boldsymbol{e}}_+ &= \Re \left[ \frac{i}{8\pi^2} \int \mathrm{d}^2 k_\perp e^{-\alpha k_\perp^2} \frac{1}{\sqrt{k^2 - k_\perp^2}} \left( 1 - \frac{k_\perp^2}{2k^2} \right) \right] \\ &= \frac{1}{16\sqrt{\pi}} e^{-\alpha k^2} \frac{1}{\sqrt{\alpha}} \left( 1 - \frac{1}{2k^2 \alpha} \right) \quad , \end{aligned} \tag{1.48}$$

where the integral is performed with a change of integration variable, resulting in simple Gaussian integrals. Thus, the regularized expression for the right-circular component of the discrete Fourier transformed Green's function at $z = 0$ is

$$\begin{aligned}\hat{\boldsymbol{e}}_+^\dagger \tilde{\boldsymbol{G}}(\boldsymbol{k}_\perp, 0, k)\hat{\boldsymbol{e}}_+ = \frac{i}{2a^2}\sum_m e^{-\alpha\kappa_\perp^2}\frac{1-\kappa_\perp^2/2k^2}{\kappa_z} \\ -\frac{1}{16\sqrt{\pi}}e^{-\alpha k^2}\frac{1}{\sqrt{\alpha}}\left(1-\frac{1}{2k^2\alpha}\right) \quad .\end{aligned} \tag{1.49}$$

As prescribed, both of these terms are finite for finite $\alpha$, but individually diverge for $\alpha \to 0$. However, their difference will remain finite, and we can numerically calculate the sum and take the limit to evaluate $\hat{\boldsymbol{e}}_+^\dagger \tilde{\boldsymbol{G}}(\boldsymbol{k}_\perp, 0, k)\hat{\boldsymbol{e}}_+$.

## 1.4 Transverse-momentum and paraxial input-output relation

In the following chapters we will be studying the transverse-momentum components of light, as well as consider its overlap with a paraxial Gaussian mode of light (corresponding to a Gaussian laser beam [7]). Therefore this section will be used to derive the input-output relation in transverse momentum space and for a general paraxial mode. Physically, the input-output relation as written in Eq. (1.19) gives the full field in real space, but in an experiment one might be interested in correlating the momentum components of light, or detecting light in a certain spatial mode. By taking the input-output relation into transverse momentum space or finding its overlap with a spatial mode, we can study such cases. Here, we keep the derivations as general as possible, and then specialize in the following chapter to the specific mode of light we will actually study.

### 1.4.1 Transverse Fourier transform

We calculate the transverse Fourier transform of the input-output relation Eq. (1.19), giving us the E-field at a certain transverse momentum $\boldsymbol{k}_\perp$ and at a certain $z$. This mixture of coordinates is meaningful in the systems we consider, i.e. planar arrays parallel to the $xy$-plane, as $k_z$ is not conserved, while $\boldsymbol{k}_\perp$ is conserved up to the addition of a reciprocal lattice vector (Bragg scattering), as we will elaborate on later. Furthermore, $z$-dependencies in the quantities we consider will cancel, or become trivial, as propagation of light along this direction is free away from the arrays.

As when deriving the effective atomic Hamiltonian and Lindbladian, we assume the input E-field has only one frequency component, and we move into a frame rotating with this frequency. We can thus write

$$\boldsymbol{E}_{\text{in}}(\boldsymbol{r}) = \int \frac{\mathrm{d}^2 k_\perp}{(2\pi)^2} \boldsymbol{E}_{\text{in}}(\boldsymbol{k}_\perp) e^{i\boldsymbol{k}\cdot\boldsymbol{r}} \quad , \tag{1.50}$$

where the $z$-component of $\boldsymbol{k}$ is $k_{z,k}$, and $k$ is the single wave number of the incoming field. The mode satisfies $\boldsymbol{E}^\dagger_{\text{in}}(\boldsymbol{k}_\perp)\boldsymbol{k} = 0$, and could be written in the form $\boldsymbol{E}_{\text{in}}(\boldsymbol{k}_\perp) = \sum_\nu E_{\text{in},\nu}(\boldsymbol{k}_\perp)\hat{\boldsymbol{e}}_{\boldsymbol{k}\nu}$. As $k_{z,k} > 0$, we are assuming the incoming field is propagating in the positive $z$-direction. Furthermore, to only include propagating components (i.e. components with real $k_{z,k}$) in the incoming field, we assume $\boldsymbol{E}_{\text{in}}(\boldsymbol{k}_\perp) = 0$ for $|\boldsymbol{k}_\perp| > k$.

We calculate the transverse Fourier transform of Eq. (1.19) (suppressing time dependencies for clarity)

$$\begin{aligned} \boldsymbol{E}(\boldsymbol{k}_\perp, z) &= \int \mathrm{d}^2 r_\perp \boldsymbol{E}(\boldsymbol{r}) e^{-i\boldsymbol{k}_\perp\cdot\boldsymbol{r}_\perp} \\ &= \int \mathrm{d}^2 r_\perp \boldsymbol{E}_{\text{in}}(\boldsymbol{r}) e^{-i\boldsymbol{k}_\perp\cdot\boldsymbol{r}_\perp} \\ &\quad + \mu_0\omega_a^2 \sum_n \left( \int \mathrm{d}^2 r_\perp \boldsymbol{G}(\boldsymbol{r}, \boldsymbol{r}_n, k_a) e^{-i\boldsymbol{k}_\perp\cdot\boldsymbol{r}_\perp} \right) \boldsymbol{d}\sigma_n \quad . \end{aligned} \tag{1.51}$$

Let us deal with each integral individually. First the incoming field, using Eq. (1.50), which simply yields

$$\boldsymbol{E}_{\text{in}}(\boldsymbol{k}_\perp, z) = \int \mathrm{d}^2 r_\perp \boldsymbol{E}_{\text{in}}(\boldsymbol{r}) e^{-i\boldsymbol{k}_\perp\cdot\boldsymbol{r}_\perp} = \boldsymbol{E}_{\text{in}}(\boldsymbol{k}_\perp) e^{ik_{z,k}z} \quad . \tag{1.52}$$

Then the Fourier transform of the Green's function, using Eq. (1.40),

$$\begin{aligned} \int \mathrm{d}^2 r_\perp \boldsymbol{G}(\boldsymbol{r}, \boldsymbol{r}_n, k_a) e^{-i\boldsymbol{k}_\perp\cdot\boldsymbol{r}_\perp} &= \int \mathrm{d}^2 r'_\perp \boldsymbol{G}(\boldsymbol{r}'_\perp, z - z_n, k_a) e^{-i\boldsymbol{k}_\perp\cdot(\boldsymbol{r}'_\perp + \boldsymbol{r}_{\perp,n})} \\ &= \boldsymbol{G}(\boldsymbol{k}_\perp, z - z_n, k_a) e^{-i\boldsymbol{k}_\perp\cdot\boldsymbol{r}_{\perp,n}} \quad , \end{aligned} \tag{1.53}$$

where we have used the fact that $\boldsymbol{G}(\boldsymbol{r}, \boldsymbol{r}_n, k_a)$ depends only on $\boldsymbol{r} - \boldsymbol{r}_n$. With this we have

$$\boldsymbol{E}(\boldsymbol{k}_\perp, z) = \boldsymbol{E}_{\text{in}}(\boldsymbol{k}_\perp, z) + \mu_0\omega_a^2 \sum_n \boldsymbol{G}(\boldsymbol{k}_\perp, z - z_n, k_a) \boldsymbol{d} e^{-i\boldsymbol{k}_\perp\cdot\boldsymbol{r}_{\perp,n}} \sigma_n \quad . \tag{1.54}$$

We now assume that the atoms are arranged in identical lattices at different values of $z$, such that the sum over $n$ can be split into a sum over $\boldsymbol{r}_{\perp,i}$, the transverse coordinate vector of the $i$'th lattice site in any lattice, and $z_j$, the $z$-coordinate of the $j$'th lattice (whereas $\boldsymbol{r}_{\perp,n}$ or $z_n$ pertain to the $n$'th atom of the whole system). We then introduce the discrete Fourier transformed atomic operators

$$\tilde{\sigma}_{\boldsymbol{k}_\perp,j} = \sum_i \sigma_{ij} e^{-i\boldsymbol{k}_\perp \cdot \boldsymbol{r}_{\perp,i}} \quad , \tag{1.55}$$

where $\sigma_{ij}$ is the atomic operator pertaining to the $i$'th site in the $j$'th lattice. With this we get the final expression

$$\boldsymbol{E}(\boldsymbol{k}_\perp, z) = \boldsymbol{E}_{\text{in}}(\boldsymbol{k}_\perp, z) + \mu_0 \omega_a^2 \sum_j \boldsymbol{G}(\boldsymbol{k}_\perp, z - z_j, k_a) \boldsymbol{d} \tilde{\sigma}_{\boldsymbol{k}_\perp,j} \quad . \tag{1.56}$$

Eventually taking the detuning of the incoming field relative to $\omega_a$ to be very small, we will see how the $z$-dependencies of quantities of interest cancel. This reflects the fact the $z$-dependence in the above expressions is that of a freely propagating field, and essentially has no effect on most quantities (as most properties of a freely propagating field do not change). With Eq. (1.56) we can study the transverse momentum components of light, corresponding to detecting the outgoing light in a plane-wave basis.

### 1.4.2 The paraxial approximation

Before deriving the input-output relation for a paraxial mode, we will briefly understand what that and the paraxial approximation are. Intuitively put, a paraxial mode is a mode of light with an approximately well-defined direction of propagation, say in the positive $z$-direction. More explicitly, if the mode is written as

$$\boldsymbol{f}(\boldsymbol{r}, k) = \int \frac{\mathrm{d}^2 k_\perp}{(2\pi)^2} \boldsymbol{f}(\boldsymbol{k}_\perp) e^{i\boldsymbol{k}\cdot\boldsymbol{r}} \quad , \tag{1.57}$$

where again the $z$-component of $\boldsymbol{k}$ is $k_{z,k}$, then $\boldsymbol{f}(\boldsymbol{k}_\perp)$ is vanishingly small, except for momenta where $k_\perp$ is small compared to $k$. In other words, the direction of propagation of each plane-wave component of the mode, given by $\hat{\boldsymbol{k}}$, only diverges slightly from $\hat{\boldsymbol{z}}$. We can thus perform lowest order expansions in $k_\perp/k$, which is the paraxial approximation. Another common way of defining this, is the following. We consider the EM wave equation Eq. (1.32) for the case of a divergenceless field (no sources), such that it reduces to

$$(\nabla^2 + k^2)\boldsymbol{E} = 0 \quad , \tag{1.58}$$

where we have again replaced the frequency $\omega$ with the wave number $k$, as it will be more intuitive. Each entry of the vector is independent in this equation, so for simplicity we consider the equation for a scalar function $A$. We write the function on the form $A(\boldsymbol{r}) = u(\boldsymbol{r})e^{ikz}$, such that the equation reduces to

$$\nabla^2 u + 2ik\frac{\partial u}{\partial z} = 0 \ . \tag{1.59}$$

The paraxial approximation then consists of neglecting the term $\partial^2 u/\partial z^2$. Modes satisfying the resulting paraxial wave equation

$$\nabla_\perp^2 u + 2ik\frac{\partial u}{\partial z} = 0 \tag{1.60}$$

are called paraxial modes. Notice that $e^{ikz}$ is the spatial dependence for a wave with a well-defined direction of propagation along $+\hat{\boldsymbol{z}}$ (i.e. a plane wave with $\boldsymbol{k}_\perp = 0$ and $k_z = k$). Inserting the full mode Eq. (1.57) in the paraxial wave equation Eq. (1.60), after extracting a factor of $e^{ikz}$, results in the equation $-k_\perp^2 - 2k(k_z - k) = 0 \Rightarrow k_z = k - k_\perp^2/2k$, which is the lowest order expansion of $k_z = \sqrt{k^2 - k_\perp^2}$, thus showing that the paraxial equation indeed implies that $k_\perp/k$ is small.

### 1.4.3 Paraxial input-output

We now derive the input-output relation for a paraxial mode, by taking the overlap between such a mode, written as in Eq. (1.57), and the full input-output relation, Eq. (1.19). As described in the previous section such a mode has an approximately well-defined direction of propagation and, taking this direction to be along $\hat{\boldsymbol{z}}$, a $z$-dependence that is approximately that of free propagation. Having a fixed frequency, the transverse shape of the mode is sufficient to fully determine it, as described by Eq. (1.57). For these reasons, when calculating the overlap between a mode and the full field, it is sufficient to calculate the transverse overlap (as opposed to overlapping in all three spatial coordinates). This corresponds to having a detector, appropriately suited for the mode, placed at some fixed value of $z$.

For atomic configurations consisting of planar lattices parallel to the $xy$-plane, we in fact know beforehand that the $z$-dependence of any emitted field must be that of a freely propagating field, symmetrically propagating away from the array (when the distance to the atoms is large enough to neglect evanescent fields). Calculating only the transverse overlap allows us to easily distinguish between transmitted and reflected fields, and under the assumption that the detuning of the incoming field with respect to the atomic transition frequency is small, $z$-dependencies will turn out to cancel, leaving the result independent of real space coordinates.

In total, we find the component of the full field in a paraxial mode $\boldsymbol{f}$ (given by Eq. (1.57)) via the following calculation

$$\begin{aligned} E_{\boldsymbol{f}} &= \int \mathrm{d}^2 r_\perp \boldsymbol{f}^\dagger(\boldsymbol{r}, k)\boldsymbol{E}(\boldsymbol{r}) \\ &= \int \mathrm{d}^2 r_\perp \int \frac{\mathrm{d}^2 k_\perp}{(2\pi)^2} \boldsymbol{f}^\dagger(\boldsymbol{k}_\perp) e^{-i\boldsymbol{k}\cdot\boldsymbol{r}} \boldsymbol{E}(\boldsymbol{r}) \\ &= \int \frac{\mathrm{d}^2 k_\perp}{(2\pi)^2} \boldsymbol{f}^\dagger(\boldsymbol{k}_\perp)\boldsymbol{E}(\boldsymbol{k}_\perp, z) e^{-ik_{z,k}z} \quad . \end{aligned} \tag{1.61}$$

Taking the detuning of the incoming mode with respect to the atomic frequency to be small (such that $k_{z,k} \simeq k_{z,k_a}$), we can use Eq. (1.56) to find

$$\begin{aligned} E_{\boldsymbol{f}} = &\int \frac{\mathrm{d}^2 k_\perp}{(2\pi)^2} \boldsymbol{f}^\dagger(\boldsymbol{k}_\perp)\boldsymbol{E}_{\mathrm{in}}(\boldsymbol{k}_\perp) \\ &+ \int \frac{\mathrm{d}^2 k_\perp}{(2\pi)^2} \boldsymbol{f}^\dagger(\boldsymbol{k}_\perp)\mu_0\omega_a^2 \sum_j \theta(z-z_j)\boldsymbol{G}(\boldsymbol{k}_\perp, z-z_j, k_a)\boldsymbol{d}\tilde{\sigma}_{\boldsymbol{k}_\perp,j} e^{-ik_{z,k}z} \\ = &\int \frac{\mathrm{d}^2 k_\perp}{(2\pi)^2} \boldsymbol{f}^\dagger(\boldsymbol{k}_\perp)\boldsymbol{E}_{\mathrm{in}}(\boldsymbol{k}_\perp) \\ &+ i\mu_0\omega_a^2 \int \frac{\mathrm{d}^2 k_\perp}{(2\pi)^2} \frac{1}{k_{z,k_a}} \boldsymbol{f}^\dagger(\boldsymbol{k}_\perp)\boldsymbol{Q}_{+,k_a}\boldsymbol{d} \sum_j \theta(z-z_j) e^{-ik_{z,k_a}z_j} \tilde{\sigma}_{\boldsymbol{k}_\perp,j} \quad , \end{aligned} \tag{1.62}$$

where we have used Eqs. (1.40) and (1.52). Here, both the incoming field mode and the chosen paraxial mode, propagate in the positive $z$-direction (as $k_{z,k} > 0$). The Heaviside step-function $\theta(z - z_j)$ comes about as a consequence of the fact that the corresponding terms propagate in the negative $z$-direction. We will generally use this expression for $z > z_j$ (when calculating properties of the transmitted field) for all $j$, such that the step-functions are all unity. We can now use the approximation $k_{z,k} \simeq k_{z,k_a}$ and the fact that $\boldsymbol{Q}_{+,k_a}$ is projector on to the polarization vectors pertaining to $\boldsymbol{k} = (\boldsymbol{k}_\perp, k_{z,k_a})^T$ to say that $\boldsymbol{f}^\dagger(\boldsymbol{k}_\perp)\boldsymbol{Q}_{+,k_a} = \boldsymbol{f}^\dagger(\boldsymbol{k}_\perp)$, i.e. $\boldsymbol{f}$ is already transversely polarized ($\boldsymbol{f}^\dagger\boldsymbol{k} = 0$), such that the projection results in the full vector again. Furthermore, we transform the atomic operator back to real space. Letting $E_{\boldsymbol{f},\mathrm{at}}$ be the atomic contribution in the above (the second term), we find

$$E_{\boldsymbol{f},\mathrm{at}} = i\mu_0\omega_a^2 \int \frac{\mathrm{d}^2k_\perp}{(2\pi)^2}\frac{1}{k_{z,k_a}}\boldsymbol{f}^\dagger(\boldsymbol{k}_\perp)\boldsymbol{d} \times \sum_j \theta(z-z_j)e^{-ik_{z,k_a}z_j}\sum_i \sigma_{ij}e^{-i\boldsymbol{k}_\perp\cdot\boldsymbol{r}_{\perp,i}}$$
$$= i\mu_0\omega_a^2 \sum_{i,j}\theta(z-z_j)\int \frac{\mathrm{d}^2k_\perp}{(2\pi)^2}\frac{1}{k_{z,k_a}}\boldsymbol{f}^\dagger(\boldsymbol{k}_\perp)e^{-i(\boldsymbol{k}_\perp\cdot\boldsymbol{r}_{\perp,i}+k_{z,k_a}z_j)}\boldsymbol{d}\sigma_{ij} \quad . \tag{1.63}$$

We now approximate $1/k_{z,k_a} \simeq 1/k_a$, according to the paraxial approximation[6], and comparing with Eq. (1.57), we recover the full real-space mode

$$E_{\boldsymbol{f},\mathrm{at}} = i\mu_0\omega_a \sum_n \theta(z-z_n)\boldsymbol{f}^\dagger(\boldsymbol{r}_n,k_a)\boldsymbol{d}\sigma_{ij} \quad . \tag{1.64}$$

Thus, the full E-field in the paraxial $\boldsymbol{f}$ mode can be written as

$$E_{\boldsymbol{f}} = \int \mathrm{d}^2r_\perp \boldsymbol{f}^\dagger(\boldsymbol{r},k_a)\boldsymbol{E}_{\mathrm{in}}(\boldsymbol{r}) + i\mu_0\omega_a \sum_n \theta(z-z_n)\boldsymbol{f}^\dagger(\boldsymbol{r}_n,k_a)\boldsymbol{d}\sigma_n \quad , \tag{1.65}$$

where we have also transformed the overlap between the detection mode and the incoming E-field to real space for consistency (and performed the approximation $\boldsymbol{f}(\boldsymbol{r},k_a) = \boldsymbol{f}(\boldsymbol{r},k)$ again). This is the input-output relation for a paraxial mode moving in the positive $z$-direction.

For a paraxial mode $\boldsymbol{f}'$ propagating in the negative $z$-direction (replacing $k_{z,k}$ with $-k_{z,k}$), we would find

$$E_{\boldsymbol{f}'} = i\mu_0\omega_a \sum_n \theta(z_n-z)\boldsymbol{f}^\dagger(\boldsymbol{r}_n,k_a)\boldsymbol{d}\sigma_n \quad . \tag{1.66}$$

There is no contribution from the incoming field here, because it propagates in the opposite direction than the mode. In the following chapters we will choose a specific detection mode, and use these expressions, Eqs. (1.65) and (1.66), to calculate the E-field as detected in that mode.

6: A higher order expansion of $k_{z,k_a}$ could be made, and results in equations that are workable (each power of $k_\perp^2$ result in a corresponding power of $\nabla_\perp^2$ as usual for Fourier transforms), but it turns out this lowest order approximation is sufficient for us.

Another, perhaps more stringent, approach to finding the the component of the full light field pertaining to a certain mode would be to calculate the three-dimensional overlap. However, if we were to calculate the overlap with paraxial modes that have a free-propagation $z$-dependence, the overlap in the $z$-coordinate would in general result in diverging factors. Specifically, the overlap with the free part of the full field would result in delta-functions, and the atomic emission part would result in zeta-functions[7]. These factors stem from the fact that plane waves are normalized to delta-functions. By taking only the overlap in the transverse coordinates we avoid carrying these complicating factors along. However, by not performing the overlap in $z$, we have had to artificially remove terms that should otherwise vanish due to the field and the mode propagating in opposite directions (as we did by introducing step-functions in Eq. (1.62) and by having no contribution from the incoming field in Eq. (1.66)).

## 1.5 Momentum-space Hamiltonian

As a final technical preparation we now derive the Hamiltonian and Lindbladian in terms of atomic quasimomentum modes for the square lattices we will consider, i.e. we perform a discrete Fourier transformation of the atomic operators. We do this for the single array in detail, and then generalize to multiple arrays in the end.

### 1.5.1 Fourier transforming the Hamiltonian

Let $\boldsymbol{k}_{\perp,l}$ be a quasimomentum of the first Brillouin zone (BZ). For a finite square lattice we have $\boldsymbol{k}_{\perp,l} = \frac{2\pi}{aN}(i_{x,l}, i_{y,l})^T$ for integer $i_{x,l}, i_{y,l}$, and $N^2$ is the total number of atoms, with $N$ on each side of the lattice. Similar to Eq. (1.55), we define

$$\tilde{\sigma}_{\boldsymbol{k}_{\perp,l}} = \sum_n \sigma_n e^{-i\boldsymbol{k}_{\perp,l}\cdot\boldsymbol{r}_{\perp,n}} \quad , \tag{1.67}$$

with inverse

$$\sigma_n = N^{-2} \sum_l \tilde{\sigma}_{\boldsymbol{k}_{\perp,l},j} e^{i\boldsymbol{k}_{\perp,l}\cdot\boldsymbol{r}_{\perp,n}} \quad , \tag{1.68}$$

7: The Heitler zeta-function, $\zeta(x) = \frac{1}{x+i\eta}$, see text below Eq. (1.11)

as can be seen from

$$\sum_n e^{-i(\boldsymbol{k}_{\perp,l}-\boldsymbol{k}_{\perp,l'})\cdot\boldsymbol{r}_{\perp,n}} = N^2\delta_{l,l'} \tag{1.69a}$$

$$\sum_l e^{i\boldsymbol{k}_{\perp,l}\cdot(\boldsymbol{r}_{\perp,n}-\boldsymbol{r}_{\perp,m})} = N^2\delta_{n,m} \tag{1.69b}$$

for $\boldsymbol{k}_{\perp,l}, \boldsymbol{k}_{\perp,l'} \in \mathrm{BZ}$, where $\delta_{x,y}$ is the Kronecker delta-function. Equation (1.43) is found by taking the infinite lattice limit of Eq. (1.69a) after adding a reciprocal lattice vector to the exponent on the left hand side (thus allowing for arbitrary momenta, while the above is specifically for momenta from the first Brillouin zone). We now insert Eq. (1.68) in each term of the Hamiltonian from Eq. (1.26). First the atomic energy term

$$\begin{aligned}\sum_n \Delta\sigma_n^\dagger\sigma_n &= -\sum_n \Delta N^{-4}\sum_l \tilde{\sigma}^\dagger_{\boldsymbol{k}_{\perp,l}} e^{-i\boldsymbol{k}_{\perp,l}\cdot\boldsymbol{r}_{\perp,n}} \sum_{l'} \tilde{\sigma}_{\boldsymbol{k}_{\perp,l'}} e^{i\boldsymbol{k}_{\perp,l'}\cdot\boldsymbol{r}_{\perp,n}} \\ &= N^{-2}\sum_{l,l'} \Delta\tilde{\sigma}^\dagger_{\boldsymbol{k}_{\perp,l}}\tilde{\sigma}_{\boldsymbol{k}_{\perp,l'}} N^{-2}\sum_n e^{-i(\boldsymbol{k}_{\perp,l}-\boldsymbol{k}_{\perp,l'})\cdot\boldsymbol{r}_{\perp,n}} \\ &= N^{-2}\sum_l \Delta\tilde{\sigma}^\dagger_{\boldsymbol{k}_{\perp,l}}\tilde{\sigma}_{\boldsymbol{k}_{\perp,l}} \quad . \end{aligned} \tag{1.70}$$

Likewise the driving term is found to be

$$\sum_n \left(\boldsymbol{d}^\dagger \boldsymbol{E}_{\text{in}}(\boldsymbol{r}_n)\sigma_n^\dagger + \text{H.c.}\right) = N^{-2}\sum_l \left(\boldsymbol{d}^\dagger \tilde{\boldsymbol{E}}_{\text{in}}(\boldsymbol{k}_{\perp,l}, z)\sigma^\dagger_{\boldsymbol{k}_{\perp,l}} + \text{H.c.}\right) \quad , \tag{1.71}$$

where $z$ is the $z$-coordinate of the lattice and

$$\tilde{\boldsymbol{E}}_{\text{in}}(\boldsymbol{k}_{\perp,l}, z) = \sum_n \boldsymbol{E}_{\text{in}}(\boldsymbol{r}_n)e^{-i\boldsymbol{k}_{\perp,l}\cdot\boldsymbol{r}_{\perp,n}} \quad . \tag{1.72}$$

We note that $\tilde{\boldsymbol{E}}_{\text{in}}(\boldsymbol{k}_\perp, z) = \frac{1}{a^2}\sum_m \boldsymbol{E}_{\text{in}}(\boldsymbol{k}_\perp + \boldsymbol{q}_m, z)$. Hence, the collective states are not only driven by the incoming light at their own momentum, but also at all momenta found by adding a reciprocal lattice vector, i.e. all corresponding Bragg-scattered momenta. Notice that since $\boldsymbol{E}_{\text{in}}(\boldsymbol{k}_\perp, z) = 0$ for $|\boldsymbol{k}_\perp| > k$, there will only be a finite number of contributions, corresponding to the finite number of "open" Bragg channels (diffraction orders). We will likewise see how the collective states emit into the different Bragg channels, resulting in the actual Bragg scattering of incoming light.

Finally, we calculate the interaction term

$$\begin{aligned}\sum_{\substack{n,m\\n\neq m}} J_{nm}\sigma_n^\dagger\sigma_m &= \sum_{\substack{n,m\\n\neq m}} J_{nm}N^{-4}\sum_l \tilde{\sigma}^\dagger_{\boldsymbol{k}_{\perp,l}} e^{-i\boldsymbol{k}_{\perp,l}\cdot\boldsymbol{r}_{\perp,n}} \sum_{l'} \tilde{\sigma}_{\boldsymbol{k}_{\perp,l'}} e^{i\boldsymbol{k}_{\perp,l'}\cdot\boldsymbol{r}_{\perp,m}} \\ &= N^{-4}\sum_{l,l'} \tilde{\sigma}^\dagger_{\boldsymbol{k}_{\perp,l}}\tilde{\sigma}_{\boldsymbol{k}_{\perp,l'}} \sum_{\substack{n,m\\n\neq m}} J_{nm} e^{-i(\boldsymbol{k}_{\perp,l}\cdot\boldsymbol{r}_{\perp,n}-\boldsymbol{k}_{\perp,l'}\cdot\boldsymbol{r}_{\perp,m})} \\ &= N^{-4}\sum_{l,l'} \tilde{\sigma}^\dagger_{\boldsymbol{k}_{\perp,l}}\tilde{\sigma}_{\boldsymbol{k}_{\perp,l'}} \sum_n e^{-i(\boldsymbol{k}_{\perp,l}-\boldsymbol{k}_{\perp,l'})\cdot\boldsymbol{r}_{\perp,n}} \\ &\qquad\times \sum_{\substack{m\\m\neq n}} J_{nm} e^{-i\boldsymbol{k}_{\perp,l'}\cdot(\boldsymbol{r}_{\perp,n}-\boldsymbol{r}_{\perp,m})} \quad .\end{aligned} \tag{1.73}$$

We can now use that $\boldsymbol{G}(\boldsymbol{r}_{\perp,n},\boldsymbol{r}_{\perp,m},k_a)$, and thus $J_{nm}$, only depends on $\boldsymbol{r}_{\perp,n}-\boldsymbol{r}_{\perp,m}$, and that (neglecting the lattice edges, which is exact for infinite lattices) a double sum over lattice sites is identical to a single sum paired with a sum over relative lattice vectors, i.e. $\sum_{\boldsymbol{r}_n,\boldsymbol{r}_m} = \sum_{\boldsymbol{r}_n}\sum_{\Delta\boldsymbol{r}_{nm}=\boldsymbol{r}_n-\boldsymbol{r}_m}$. Following the definition of $J_{nm}$ and $\Gamma_{nm}$ (given below Eq. (1.23)) we define

$$-\tilde{\Delta}_{\boldsymbol{k}_\perp}(z) + i\tilde{\Gamma}_{\boldsymbol{k}_\perp}(z) = \mu_0\omega_a^2\boldsymbol{d}^\dagger\tilde{\boldsymbol{G}}(\boldsymbol{k}_\perp,z,k_a)\boldsymbol{d} \quad , \tag{1.74}$$

and furthermore introduce the notation $\tilde{\Delta}^0_{\boldsymbol{k}_\perp} \equiv \tilde{\Delta}_{\boldsymbol{k}_\perp}(0)$ and $\tilde{\Gamma}^0_{\boldsymbol{k}_\perp} \equiv \tilde{\Gamma}_{\boldsymbol{k}_\perp}(0)$, as these quantities turn out to have a special role, as we will see. Using the discrete Fourier transformed Green's function from Eqs. (1.41) and (1.45), we can then write

$$\begin{aligned}\sum_{\substack{n,m\\n\neq m}} J_{nm}\sigma_{ij}^\dagger\sigma_m &= N^{-4}\sum_{l,l'} \tilde{\sigma}^\dagger_{\boldsymbol{k}_{\perp,l}}\tilde{\sigma}_{\boldsymbol{k}_{\perp,l'}} \sum_n e^{-i(\boldsymbol{k}_{\perp,l}-\boldsymbol{k}_{\perp,l'})\cdot\boldsymbol{r}_{\perp,n}} \\ &\qquad\times \sum_{\Delta\boldsymbol{r}_{nm}\neq 0} J_{nm} e^{-i\boldsymbol{k}_{\perp,l'}\cdot\Delta\boldsymbol{r}_{\perp,nm}} \\ &= -N^{-4}\sum_{l,l'} \tilde{\sigma}^\dagger_{\boldsymbol{k}_{\perp,l}}\tilde{\sigma}_{\boldsymbol{k}_{\perp,l'}} \sum_n e^{-i(\boldsymbol{k}_{\perp,l}-\boldsymbol{k}_{\perp,l'})\cdot\boldsymbol{r}_{\perp,n}} \tilde{\Delta}^0_{\boldsymbol{k}_{\perp,l'}} \\ &= -N^{-2}\sum_l \tilde{\Delta}^0_{\boldsymbol{k}_{\perp,l}} \tilde{\sigma}^\dagger_{\boldsymbol{k}_{\perp,l}}\tilde{\sigma}_{\boldsymbol{k}_{\perp,l}} \quad ,\end{aligned} \tag{1.75}$$

where $\tilde{\Delta}^0_{\boldsymbol{k}_{\perp,l}}$ is evaluated at $z=0$, because all the atoms have the same $z$-coordinate such that $z_n - z_m = 0$. The momentum-space Lindbladian is found in a similar way, and in total we have

$$H = -N^{-2}\sum_{l}\Bigg[(\Delta - \tilde{\Delta}^0_{\boldsymbol{k}_{\perp,l}})\tilde{\sigma}^\dagger_{\boldsymbol{k}_{\perp,l}}\tilde{\sigma}_{\boldsymbol{k}_{\perp,l}} + \left(\boldsymbol{d}^\dagger\tilde{\boldsymbol{E}}_{\text{in}}(\boldsymbol{k}_{\perp,l},z)\sigma^\dagger_{\boldsymbol{k}_{\perp,l}} + \text{H.c.}\right)\Bigg] , \tag{1.76a}$$

$$\mathcal{L}[\rho] = N^{-2}\sum_{l}\tilde{\Gamma}^0_{\boldsymbol{k}_{\perp,l}}\left(2\tilde{\sigma}_{\boldsymbol{k}_{\perp,l}}\rho\tilde{\sigma}^\dagger_{\boldsymbol{k}_{\perp,l}} - \left\{\tilde{\sigma}^\dagger_{\boldsymbol{k}_{\perp,l}}\tilde{\sigma}_{\boldsymbol{k}_{\perp,l}},\rho\right\}\right) . \tag{1.76b}$$

In the limit of an infinite array, $N^2 \to \infty$, the sums in the above expression become integrals according to $N^{-2}\sum_{\boldsymbol{k}_{\perp,l}} \to a^2\int_{\text{BZ}}\frac{\mathrm{d}^2k_\perp}{(2\pi)^2}$ (using that $2\pi/aN$ is the difference of two adjacent momenta in the Brillouin zone). At the same time we absorb a factor of $a$ in the definition of $\tilde{\sigma}_{\boldsymbol{k}_\perp}$ and $\tilde{\boldsymbol{E}}_{\text{in}}$, i.e. we replace $a\tilde{\sigma}_{\boldsymbol{k}_\perp} \to \tilde{\sigma}_{\boldsymbol{k}_\perp}$ and $a\tilde{\boldsymbol{E}}_{\text{in}} \to \tilde{\boldsymbol{E}}_{\text{in}}$. Hence, the above becomes

$$H = -\int_{\text{BZ}}\frac{\mathrm{d}^2k_\perp}{(2\pi)^2}\left[(\Delta - \tilde{\Delta}^0_{\boldsymbol{k}_\perp})\tilde{\sigma}^\dagger_{\boldsymbol{k}_\perp}\tilde{\sigma}_{\boldsymbol{k}_\perp} + \left(\boldsymbol{d}^\dagger\tilde{\boldsymbol{E}}_{\text{in}}(\boldsymbol{k}_\perp,z)\sigma^\dagger_{\boldsymbol{k}_\perp} + \text{H.c.}\right)\right] , \tag{1.77a}$$

$$\mathcal{L}[\rho] = \int_{\text{BZ}}\frac{\mathrm{d}^2k_\perp}{(2\pi)^2}\tilde{\Gamma}^0_{\boldsymbol{k}_\perp}\left(2\tilde{\sigma}_{\boldsymbol{k}_\perp}\rho\tilde{\sigma}^\dagger_{\boldsymbol{k}_\perp} - \left\{\tilde{\sigma}^\dagger_{\boldsymbol{k}_\perp}\tilde{\sigma}_{\boldsymbol{k}_\perp},\rho\right\}\right) . \tag{1.77b}$$

For the case of multiple arrays, as described above Eq. (1.55), we would arrive at the following momentum-space description

$$H = -\sum_j\int_{\text{BZ}}\frac{\mathrm{d}^2k_\perp}{(2\pi)^2}\Bigg[\Delta\tilde{\sigma}^\dagger_{\boldsymbol{k}_\perp,j}\tilde{\sigma}_{\boldsymbol{k}_\perp,j} + \left(\boldsymbol{d}^\dagger\tilde{\boldsymbol{E}}_{\text{in}}(\boldsymbol{k}_\perp,z_j)\sigma^\dagger_{\boldsymbol{k}_\perp,j} + \text{H.c.}\right) - \sum_{j'}\tilde{\Delta}_{\boldsymbol{k}_\perp}(z_j - z_{j'})\tilde{\sigma}^\dagger_{\boldsymbol{k}_\perp,j}\tilde{\sigma}_{\boldsymbol{k}_\perp,j'}\Bigg] , \tag{1.78a}$$

$$\mathcal{L}[\rho] = \sum_{j,j'}\int_{\text{BZ}}\frac{\mathrm{d}^2k_\perp}{(2\pi)^2}\tilde{\Gamma}_{\boldsymbol{k}_\perp}(z_j - z_{j'})\left(2\tilde{\sigma}_{\boldsymbol{k}_\perp,j}\rho\tilde{\sigma}^\dagger_{\boldsymbol{k}_\perp,j'} - \left\{\tilde{\sigma}^\dagger_{\boldsymbol{k}_\perp,j}\tilde{\sigma}_{\boldsymbol{k}_\perp,j'},\rho\right\}\right) , \tag{1.78b}$$

where $\tilde{\sigma}_{\boldsymbol{k}_\perp,j}$ is the Fourier transformed atomic operator for the $j$'th array, as defined in Eq. (1.55). Hence, we see how the single array Hamiltonian is repeated for each lattice and in addition there are interactions, both coherent and dissipative, between arrays. The collective energies $\tilde{\Delta}^0_{\boldsymbol{k}_\perp}$ and $\tilde{\Gamma}^0_{\boldsymbol{k}_\perp}$ pertain to excited state energies and widths within the individual lattices, while $\tilde{\Delta}_{\boldsymbol{k}_\perp}(z)$ and $\tilde{\Gamma}_{\boldsymbol{k}_\perp}(z)$ pertain to collective interactions between the lattices.

### 1.5.2 Understanding the momentum-space Hamiltonian

At first glance the Hamiltonian in Eq. (1.77a) might appear diagonal, except for the driving term. However, while the $\tilde{\sigma}^\dagger_{\boldsymbol{k}_\perp}\tilde{\sigma}_{\boldsymbol{k}_\perp}$-operator has the appearance of a counting operator, it is in fact not. Had the atomic operators been bosonic or fermionic it would have been the case. However, while Fourier transformed bosonic and fermionic operators are themselves bosonic or fermionic respectively, two-level operators to not transform "nicely". This can be seen by considering the commutators for these three types of operators. For bosonic, two-level, and fermionic operators defined on a lattice, the nontrivial commutators are

$$[a_n, a^\dagger_m] = \delta_{n,m} \ , \tag{1.79a}$$

$$[\sigma_n, \sigma^\dagger_m] = \delta_{n,m}(1 - 2\sigma^\dagger_n\sigma_n) \ , \tag{1.79b}$$

$$[\psi_n, \psi^\dagger_m] = \delta_{n,m} - 2\psi^\dagger_n\psi_m \ . \tag{1.79c}$$

As an aside, we can also see here how the two-level operators in some sense are "in between" being bosonic or fermionic, reflecting the fact that two-level systems commute, like bosons, but only allow for a single excitation, like fermions. Fourier transforming the above (for an infinite lattice and momenta within the first Brillouin zone, and scaling the operators by a factor of $a$ as before), would result in

$$[a_{\boldsymbol{k}_\perp}, a^\dagger_{\boldsymbol{q}_\perp}] = (2\pi)^2\delta(\boldsymbol{k}_\perp - \boldsymbol{q}_\perp) \ , \tag{1.80a}$$

$$[\tilde{\sigma}_{\boldsymbol{k}_\perp}, \tilde{\sigma}^\dagger_{\boldsymbol{q}_\perp}] = (2\pi)^2\delta(\boldsymbol{k}_\perp - \boldsymbol{q}_\perp) - 2\sum_n \tilde{\sigma}^\dagger_n\tilde{\sigma}_n e^{-i(\boldsymbol{k}_\perp - \boldsymbol{q}_\perp)\cdot\boldsymbol{r}_{\perp,n}} \ , \tag{1.80b}$$

$$[\psi_{\boldsymbol{k}_\perp}, \psi^\dagger_{\boldsymbol{q}_\perp}] = (2\pi)^2\delta(\boldsymbol{k}_\perp - \boldsymbol{q}_\perp) - 2\psi^\dagger_{\boldsymbol{k}_\perp}\psi_{\boldsymbol{q}_\perp} \ . \tag{1.80c}$$

Hence, the bosonic and fermionic commutation relations retain their form, such that the Fourier transformed operators are again bosonic or fermionic in nature respectively. The two-level commutation relation, however, involves the Fourier transform of the counting operator $\sigma^\dagger_n\sigma_n$, which can not be simply expressed in terms of $\tilde{\sigma}_{\boldsymbol{k}_\perp}$-operators. The nature of $\tilde{\sigma}_{\boldsymbol{k}_\perp}$ is thus *not* that of a two-level operator, and the eigenstates of $\tilde{\sigma}^\dagger_{\boldsymbol{k}_\perp}\tilde{\sigma}_{\boldsymbol{k}_\perp}$ are *not* in general given by products of $\tilde{\sigma}^\dagger_{\boldsymbol{k}_\perp}$-operators acting on the vacuum state, $|0\rangle$. However, in the specific case of only a single excitation, two-level systems are identical to bosons, and in fact $\tilde{\sigma}^\dagger_{\boldsymbol{k}_\perp}|0\rangle$ *is* an eigenstate of $\tilde{\sigma}^\dagger_{\boldsymbol{k}_\perp}\tilde{\sigma}_{\boldsymbol{k}_\perp}$. Hence, the *linear* sector of Eq. (1.77a) (i.e. the part of the operator pertaining to states of at most one excitation) is diagonalized, while the *nonlinear* (more than one excitation) is not.

We thus see how the nonlinearity of the system (which is what we will be interested in) comes about as a consequence of the nature of the saturable two-level atoms, rather than a "direct" interaction between quasimomentum modes. That is, there is no explicit interaction term between different $\boldsymbol{k}_\perp$-modes in Eq. (1.77a), and yet they are not eigenmodes of the system for more than one excitation, and must thus be interacting – this interaction is implicitly contained in the two-level operators. For the case of multiple lattices (Eq. (1.78)) the linear sector is also diagonalized with respect to the momentum degree of freedom (modes with different quasimomentum do not mix for a single excitation), but there is a further degree of freedom in the lattice index which is mixed through the interaction term. Due to the simple $z$-dependence of the discrete Fourier transformed Green's function, Eqs. (1.44) and (1.45), it is generally not difficult to diagonalize this remaining degree of freedom, in particular for evenly spaced lattices.

We will perform this diagonalization for the dual array in the next chapter. In order to use the more analytical approach of the next part of this thesis, we will find a way to write the inter-mode interaction explicitly. For the moment, we will proceed with analytics for the linear dynamics of this system, and numerics for the nonlinear dynamics, in the following chapters.

### 1.5.3 Collective energies

We now study the collective energies $\tilde{\Delta}^0_{\boldsymbol{k}_\perp}$, $\tilde{\Gamma}^0_{\boldsymbol{k}_\perp}$, $\tilde{\Delta}_{\boldsymbol{k}_\perp}(z)$, and $\tilde{\Gamma}_{\boldsymbol{k}_\perp}(z)$. To do so we consider the specific case of $\boldsymbol{d} = d\hat{\boldsymbol{e}}_+$, as we will do throughout the thesis. That is, from now on we consider the dipole moment of the atoms to be right-circularly polarized. For simplicity we will ignore the regularization of the real part of the discrete Fourier transformed Green's function, as discussed in Section 1.3.4, as it does not affect the general observations we will make. Using Eqs. (1.44) and (1.74) we can find

$$-\tilde{\Delta}_{\boldsymbol{k}_\perp}(z) + i\tilde{\Gamma}_{\boldsymbol{k}_\perp}(z) = \frac{i\tilde{\Gamma}}{k_a}\sum_m \frac{k_a^2 - (\boldsymbol{k}_\perp + \boldsymbol{q}_m)^2/2}{\sqrt{k_a^2 - (\boldsymbol{k}_\perp + \boldsymbol{q}_m)^2}} e^{i\sqrt{k_a^2 - (\boldsymbol{k}_\perp + \boldsymbol{q}_m)^2}|z|} \quad , \tag{1.81}$$

where $\tilde{\Gamma} = \mu_0\omega_a d^2/2a^2 = 3\pi\gamma/a^2k_a^2$. For $-\tilde{\Delta}^0_{\boldsymbol{k}_\perp} + i\tilde{\Gamma}^0_{\boldsymbol{k}_\perp}$ the expression is the same, except the exponential is unity, and we subtract $\mu_0\omega_a^2\boldsymbol{d}^\dagger\Re[\boldsymbol{G}(0,k_a)]\boldsymbol{d}$, see Eq. (1.45). There are several observations to be made here. We see how the contribution of the right hand side to either the collective energy shift or decay rate depends on whether $\sqrt{k_a^2 - (\boldsymbol{k}_\perp + \boldsymbol{q}_m)^2}$ is real or imaginary, i.e. whether $|\boldsymbol{k}_\perp + \boldsymbol{q}_m|$ is smaller or greater than $k_a$. Momenta for which this square root is real define the "light cone", i.e. the area of momenta where light can propagate. As the lattice spacing becomes larger, the Brillouin zones become smaller, and so more of them fit inside the light cone, corresponding to the square root remaining real for more reciprocal lattice vectors.

As mentioned, the discrete translation symmetry of the lattices we work with, result in transverse momentum being conserved only up to the addition of a reciprocal lattice vector $\boldsymbol{q}_m$ (Bragg scattering). Thus, we will see how light that is absorbed in a lattice, may be emitted with any $\boldsymbol{q}_m$ added to its transverse momentum, with each $\boldsymbol{q}_m$ corresponding to a Bragg channel. Barring other scattering effects, and assuming the light to be on resonance with the atomic transition, the longitudinal momentum of such a Bragg-scattered wave will be $\sqrt{k_a^2 - (\boldsymbol{k}_\perp + \boldsymbol{q}_m)^2}$. Thus there is finite number of small $\boldsymbol{q}_m$ where this is real (open channels), and the wave propagates along the shifted wave vector, and an infinite number of $\boldsymbol{q}_m$ where this is imaginary and the wave is evanescent on the surface of the lattice (closed channels). Whether a channel is open for a particular $\boldsymbol{k}_\perp$ depends on the lattice spacing $a$, as the $\boldsymbol{q}_m$ are inversely proportional to it. Hence, for small enough $a$, the $\boldsymbol{q}_m$ are so large that only $\boldsymbol{q}_0 = \boldsymbol{0}$ is open and only for momenta with $|\boldsymbol{k}_\perp| < k_a$. In particular, for $a < \lambda_a/2$ all $\boldsymbol{k}_\perp$ are either evanescent or have only the zeroth order Bragg channel open, while for $a > \lambda_a$ all $\boldsymbol{k}_\perp$ have at least two open Bragg channels available.

For this reason we will consider subwavelength lattices, i.e. $a < \lambda_a$, where scattering is reduced for small momenta, as they do not experience Bragg scattering. This intrinsic weak scattering, and thus low loss, is the basis for the popularity of ordered quantum emitters as optical platforms, as it allows for an approximate single-mode behaviour of the system [25, 26, 33]. Furthermore, the fact that modes outside the light cone only decay due to imperfections [51] allows for long-lived states that can be utilized for the implementation of quantum memories, as we will briefly touch on in the next chapter. Notice that $\tilde{\Gamma}$ is the contribution from the zeroth order Bragg channel to $\tilde{\Gamma}_0^0$ (i.e. for $a < \lambda_a$ it is decay rate of the zero momentum collective state).

If we let $\sum_m'$ and $\sum_m''$ denote sums over $\boldsymbol{q}_m$, where $\sqrt{k_a^2 - (\boldsymbol{k}_\perp + \boldsymbol{q}_m)^2}$ is real or imaginary respectively (i.e. where $|\boldsymbol{k}_\perp + \boldsymbol{q}_m|$ is smaller or greater than $k_a$), we can write the collective energies for $z \neq 0$, pertaining to interactions between the lattices,

$$\tilde{\Delta}_{\boldsymbol{k}_\perp}(z) = \frac{\tilde{\Gamma}}{k_a}\sum_m{}' \frac{k_a^2 - (\boldsymbol{k}_\perp + \boldsymbol{q}_m)^2/2}{\sqrt{k_a^2 - (\boldsymbol{k}_\perp + \boldsymbol{q}_m)^2}} \sin(\sqrt{k_a^2 - (\boldsymbol{k}_\perp + \boldsymbol{q}_m)^2}|z|) + \frac{\tilde{\Gamma}}{k_a}\sum_m{}'' \frac{(\boldsymbol{k}_\perp + \boldsymbol{q}_m)^2/2 - k_a^2}{\sqrt{(\boldsymbol{k}_\perp + \boldsymbol{q}_m)^2 - k_a^2}} e^{-\sqrt{(\boldsymbol{k}_\perp + \boldsymbol{q}_m)^2 - k_a^2}|z|} \ , \tag{1.82a}$$

$$\tilde{\Gamma}_{\boldsymbol{k}_\perp}(z) = \frac{\tilde{\Gamma}}{k_a}\sum_m{}' \frac{k_a^2 - (\boldsymbol{k}_\perp + \boldsymbol{q}_m)^2/2}{\sqrt{k_a^2 - (\boldsymbol{k}_\perp + \boldsymbol{q}_m)^2}} \cos(\sqrt{k_a^2 - (\boldsymbol{k}_\perp + \boldsymbol{q}_m)^2}|z|) \ , \tag{1.82b}$$

and for $z = 0$, pertaining to the individual lattices

$$\tilde{\Delta}^0_{\boldsymbol{k}_\perp} = \frac{\tilde{\Gamma}}{k_a}\sum_m{}'' \frac{(\boldsymbol{k}_\perp + \boldsymbol{q}_m)^2/2 - k_a^2}{\sqrt{(\boldsymbol{k}_\perp + \boldsymbol{q}_m)^2 - k_a^2}} + \mu_0\omega_a^2 \boldsymbol{d}^\dagger \Re[\boldsymbol{G}(0, k_a)]\boldsymbol{d} \ , \tag{1.83a}$$

$$\tilde{\Gamma}^0_{\boldsymbol{k}_\perp} = \frac{\tilde{\Gamma}}{k_a}\sum_m{}' \frac{k_a^2 - (\boldsymbol{k}_\perp + \boldsymbol{q}_m)^2/2}{\sqrt{k_a^2 - (\boldsymbol{k}_\perp + \boldsymbol{q}_m)^2}} \ . \tag{1.83b}$$

Per the above discussion the primed sum is finite, while the double primed sum is infinite. Thus, both the collective energy shift and decay rate, for $z \neq 0$, get a finite number of contributions that oscillate in $|z|$, while the energy shift also gets an infinite number of contribution that decay exponentially in $|z|$. The collective energies in Eq. (1.78) are evaluated at $z_j - z_{j'}$, and so we see how the two types of contributions correspond to interactions due to propagating photons exchanged between lattices, or due to the evanescent fields of the lattices. For sufficiently large distance between lattices the evanescent contribution is negligible, while the propagating contribution may become zero for sufficiently large $\boldsymbol{k}_\perp$ depending on $a$, if there are no open Bragg channels. For $z = 0$, the energy shift gets an infinite number of contributions from the reciprocal lattice vectors corresponding to closed Bragg channels (which needs regularization), while the decay rate gets only a finite number of contributions from the open Bragg channels. For Eq. (1.77) and for the fully diagonalized version of Eq. (1.78), we thus see how the decay of single excitations gets a contribution from each open Bragg channel, corresponding to emitting light that is correspondingly Bragg-scattered.

We also see how the collective energies diverge, when $k_a^2 - (\boldsymbol{k}_\perp + \boldsymbol{q}_m)^2 = 0$, see Fig. 1.3. This is exactly the border between a Bragg channel opening or closing, and corresponds to light propagating with zero longitudinal momentum, i.e. in the plane of the array. Such a photon would interact with infinitely many atoms, and this leads to the divergence. Quasimomentum modes near these lines are greatly detuned from others and have very short lifetimes. We will see how this becomes relevant for the lossy scattering of modes of light that are broad in momentum space.

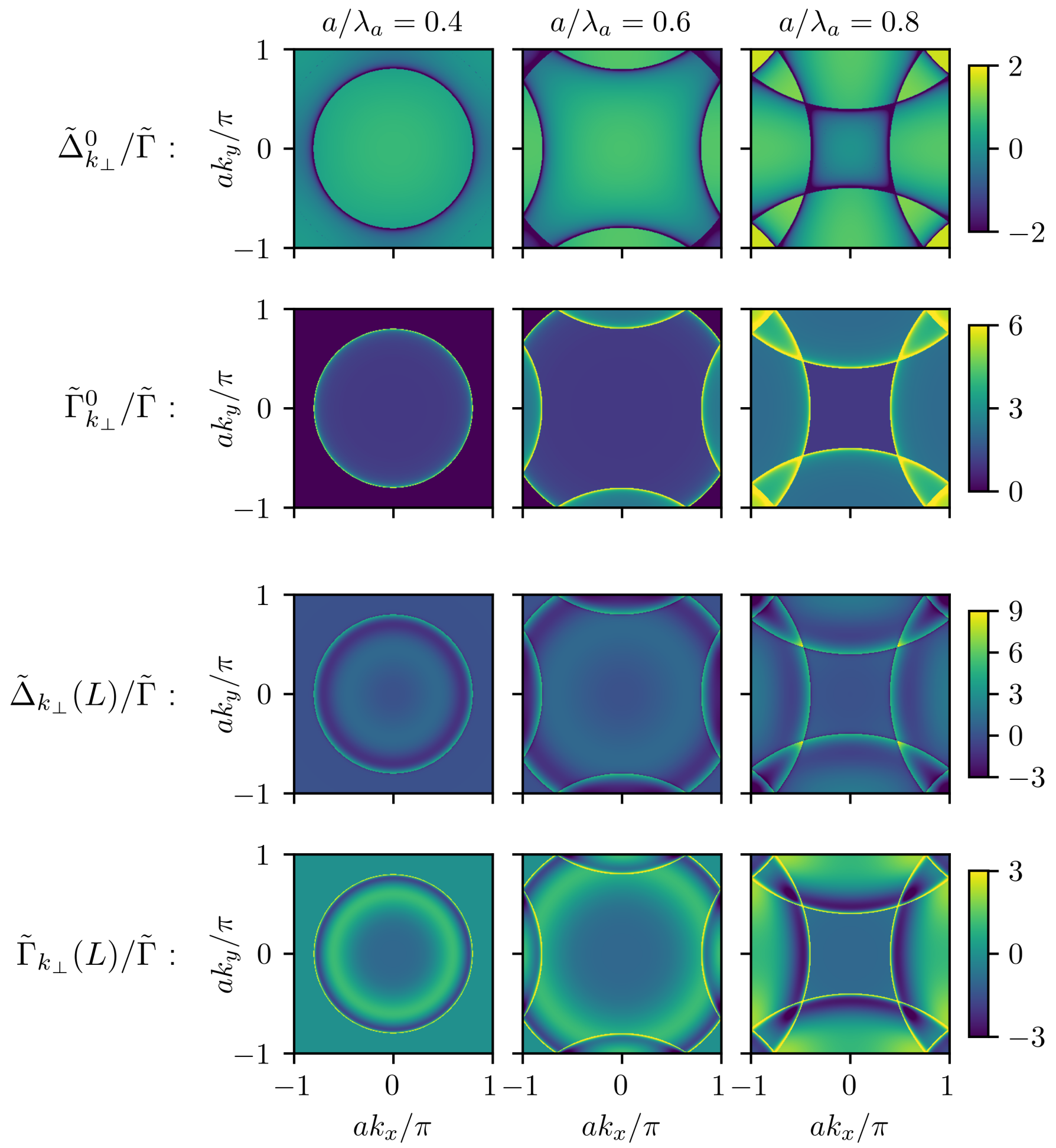

**Figure 1.3 | Intra- and inter-array collective energies.** $\tilde{\Delta}^0_{\boldsymbol{k}_\perp}$, $\tilde{\Gamma}^0_{\boldsymbol{k}_\perp}$, $\tilde{\Delta}_{\boldsymbol{k}_\perp}(z)$, and $\tilde{\Gamma}_{\boldsymbol{k}_\perp}(z)$ for $z = 1.5\lambda_a$, plotted in the first Brillouin zone for different values of the lattice spacing $a$. The divergences of the collective energies on the circular contours have been cut off (the energy shifts diverge towards $-\infty$, while the decay rates diverge towards $+\infty$). Note that each row has their own colour scale.

This circular contour, for $\boldsymbol{q}_0 = \mathbf{0}$, borders the light cone. We see in Eq. (1.81) how the collective energies are found by repeating the $\boldsymbol{q}_0 = \mathbf{0}$ contribution, but translated by a reciprocal lattice vectors. Thus, there is such a circular contour of divergence in each Brillouin zone, centred on each reciprocal lattice vector. For larger lattice spacing the Brillouin zones becomes smaller, and the contours start to extend into other Brillouin zones, in particular the first. Alternatively, this can be seen as the original circular contour around the light cone, folding in on itself at the periodic boundaries of the first Brillouin zone. Figure 1.3 shows nicely how the collective decay is greater in areas where multiple Bragg channels are open, and how for larger $a$ these circular contours of divergence move closer to the centre of the first Brillouin zone (as the shifts $\boldsymbol{q}_m$ become smaller). For small $a$, however, the collective energies have only a single contour of divergence, corresponding to only the zeroth order Bragg channel being available, such that the collective energies are close to flat for small momenta.

# 2

# Single-excitation physics

In this chapter, we study the linear physics of the atomic cavity, comparing with the atomic mirror and a single atom along the way. As the two-level atoms become equivalent to bosons when there is just a single excitation present, the Hamiltonian and Lindbladian in Eq. (1.78) are effectively diagonal in the linear sector, as described in Section 1.5.2. Hence, we can generally proceed analytically in our study of the properties of infinitely extended lattices. We focus on the steady state of the system as we consider continuous-wave driving. With only a single frequency in the incoming field, as assumed in Section 1.2.2, the strength of the driving is constant in time, eventually balancing the dissipation of the system, such that it reaches a steady state. We will not consider the transient dynamics of the system. For the infinite arrays we will be focusing on plane-wave driving and detection, using Eq. (1.56) to calculate the E-field. As mentioned, we will also be interested in driving and detecting with a Gaussian mode of light, and we do so at the end of this chapter, and also in the next when we consider the nonlinear behaviour of finite arrays.

As mentioned in the beginning of the previous chapter, the single array and other configurations of quantum emitters have already been studied extensively, in particular their linear response. We will review some of these studies to give an overview of the context of the present work. We have already mentioned studies pertaining to the basic transmission and reflection properties of the single array [17, 18, 28], where it is found that a single sheet of atoms can in theory implement a perfect mirror, constituting a strong collective interaction with light, despite the weak coupling of the underlying individual atoms.

The collective coupling to light means arrays can be employed as a medium for other atoms to affect how these interact or decay. Thus,

an ordered array of emitters can implement a waveguide which is an active, quantum component of the dynamics of other atoms coupled through it. This is in contrast to ordinary waveguides, which are passive elements simply setting the stage for the optical dynamics. Essentially an atomic array constitutes a tunable and active environment that can be used to shape the behaviour of the external atoms, in particular their resonance energies and emission [66–68], as well as their coupling to other external atoms. In Refs. [69, 70] external atoms are coupled to the subradiant states[1] of a one-dimensional atomic array, which then mediate interactions between the external atoms, conceptually identical to how guided modes in a waveguide work. Unlike conventional waveguides, the atomic waveguide takes an active part in the dynamics, and can host nonlinear [69] and non-Markovian effects [70], due to the saturability of the waveguide atoms and the slow group velocity of spin waves in the atomic chain. As we will see, the dual array that we analyse here, behaves like an optical cavity, and in relation to the ideas of the referred works, it might be interesting to consider how familiar experiments with atoms in cavities would change if the cavity was itself formed by atomic lattices. The mirrors would then be an active, quantum, nonlinear part of the dynamics, rather than passive, classical elements, potentially leading to interesting effects.

Likewise, it has been proposed that the cooperative behaviour of atomic arrays can be used to engineer the wavefront of light [30]. It has been shown that an array could be used to focus and steer light, as well as generate states of light with orbital angular momentum [71] or non-trivial topology [72]. By having a unit cell which contains multiple atoms, and including a more complicated level structure in each atom, corresponding to transitions with different polarizations, it is possible to achieve collective modes that correspond to each of these unit cells having a strong electric dipole moment or *magnetic* dipole moment (and even higher multipoles). This is opposed to natural materials, which usually only interact weakly with the magnetic component of light. Upon tuning atomic and lattice properties (transition frequencies, lattice geometry, etc.), these different modes can be used to shape light transmitted through the array to have non-trivial properties (angular momentum, topology) despite the input state being "ordinary" (e.g. a Gaussian beam).

The highly subradiant states of ordered arrays, originating in destructive interference effects or the mode quasimomentum being outside the

1: Specifically those modes which for an infinite array are evanescent (their quasimomenta lie outside the light cone, such that they become perfectly subradiant), while for finite arrays they can radiate at the edges.

light cone, are strong candidates for implementing quantum memories [41, 42, 51, 73, 74]. It is intrinsically challenging to store and retrieve information in a quantum memory, as a good memory is defined by being decoupled from its surrounding, but at the same time it is exactly this coupling that is needed in order to access the memory. The works referenced here all utilize additional excited states in the atoms and externally controlled light to temporarily couple or detune states in such a way that population can be transferred into a collective mode, which upon turning these external fields off is highly subradiant. The photon is thus stored in this long-lived state, and it can be retrieved by time-inverting the storage pulse scheme [75].

A big part of the literature, which we have excluded, is the work done to generate optical nonlinearities with arrays. This we will return to in the next chapter. The dual array, which we will consider, has also specifically been the focus of a few papers, concerning its linear properties. In particular, its subradiant states have been studied with the intention of implementing a quantum memory [76, 77], and it has also been proposed to use the dual array as a "quantum antenna", coupling distant external atoms to implement a quantum link between qubits [78], similar to the mentioned papers regarding the use of arrays as mediators of interaction. A system similar to the dual array, is that of a single sheet of atoms in front of a classical mirror. As the atomic array "sees" another copy of itself, its reflection, we would expect some similar physics. In Ref. [79] they consider the linear transmission of a generalized version of such a system, focussing on normal-incidence light, and find narrow transmission resonances, similar to what we find below.

Here we consider basic transmission and reflection properties of the dual array, as well as the time scale of photon confinement in the system. We will see how referring to the dual array as an atomic cavity is well justified. We will get an idea of the atomic cavity's potential for generating nonlinearities through its extremely narrow transmission resonances and the correspondingly long confinement time of photons.

In Section 2.1 we consider the single-excitation eigenmodes of the dual array, diagonalizing the corresponding sector of Eq. (1.78). Section 2.2 looks at the steady state expectation values of the atomic operators, which in Section 2.3 are used together with the input-output relation to study the transmission and reflection of light off the dual array. We determine two regimes of the dual array according to whether the individual lattices are close enough for evanescent fields to have an effect or far enough that these can be neglected. Section 2.4 introduces the delay time, which gives us a time scale for how long a photon is confined in the system. We find that around extremely narrow features of transmission or reflection the delay time diverges, meaning the photons, in this idealized setup, are confined arbitrarily long in the system. This should lead to very strong nonlinear effects, which we will study in the following chapter. Finally, in Section 2.5 we will consider the effects of having finite lattices rather than the infinite lattices otherwise considered in this chapter, as we will be dealing with finite arrays in the next chapter.

## 2.1 Single-excitation eigenmodes

To understand the linear physics of the infinite dual array, we will be looking at the transmission, reflection and delay time. To calculate these we need the expectation value of the E-field, which we find via the input-output relation, Eq. (1.78). Hence, we ultimately require the steady value of $\langle \tilde{\sigma}_{\boldsymbol{k}_\perp,j} \rangle$, i.e. the expectation value of atomic operators pertaining to the quasi-momentum modes. This operator evolves according to $i\partial_t \tilde{\sigma}_{\boldsymbol{k}_\perp,j} = [\tilde{\sigma}_{\boldsymbol{k}_\perp,j}, H] + i\mathcal{L}[\tilde{\sigma}_{\boldsymbol{k}_\perp,j}]$ (as described in Section 1.2.2) with

$$H = -\sum_{j=1,2} \left[ \Delta \tilde{\sigma}^\dagger_{\boldsymbol{k}_\perp,j} \tilde{\sigma}_{\boldsymbol{k}_\perp,j} + \left( \tilde{\Omega}_{\boldsymbol{k}_\perp} e^{ik_{z,k} z_j} \sigma^\dagger_{\boldsymbol{k}_\perp,j} + \text{H.c.} \right) \right.$$
$$\left. - \sum_{j'} \tilde{\Delta}_{\boldsymbol{k}_\perp}(z_j - z_{j'}) \tilde{\sigma}^\dagger_{\boldsymbol{k}_\perp,j} \tilde{\sigma}_{\boldsymbol{k}_\perp,j'} \right] , \tag{2.1a}$$

$$\mathcal{L}[A] = \sum_{j,j'} \tilde{\Gamma}_{\boldsymbol{k}_\perp}(z_j - z_{j'}) \left( 2\tilde{\sigma}^\dagger_{\boldsymbol{k}_\perp,j} A \tilde{\sigma}_{\boldsymbol{k}_\perp,j'} - \left\{ \tilde{\sigma}^\dagger_{\boldsymbol{k}_\perp,j} \tilde{\sigma}_{\boldsymbol{k}_\perp,j'}, A \right\} \right) , \tag{2.1b}$$

where, again, $k_{z,k} = \sqrt{k^2 - k_\perp^2}$ with $k$ the wave number of the continuous-wave drive, and we have defined $\tilde{\Omega}_{\boldsymbol{k}_\perp} e^{ik_{z,k} z} \equiv \boldsymbol{d}^\dagger \tilde{\boldsymbol{E}}_{\text{in}}(\boldsymbol{k}_\perp, z)$, such that $\tilde{\Omega}_{\boldsymbol{k}_\perp}$ is the discrete Fourier transformed Rabi frequency. The Hamiltonian and Lindbladian from Eq. (1.78) have been written without the integrals over momentum, as there is no mixing between the momenta for the linear physics, as described in Section 1.5.2. As long as we replace $(2\pi)^2 \delta(\boldsymbol{k}_\perp - \boldsymbol{k}'_\perp)$ with $\delta_{\boldsymbol{k}_\perp, \boldsymbol{k}'_\perp}$ in the commutation relations, Eq. (1.80b), we will get the correct results.

As the dual array is symmetric under the exchange of the two lattices (as seen by $\tilde{\Delta}_{\boldsymbol{k}_\perp}(z_1 - z_2) = \tilde{\Delta}_{\boldsymbol{k}_\perp}(z_2 - z_1)$, and likewise for $\tilde{\Gamma}_{\boldsymbol{k}_\perp}$), the Hamiltonian can be fully diagonalized (for a single excitation), by introducing the even and odd modes

$$\tilde{\sigma}_{\boldsymbol{k}_\perp,\pm} = \frac{\tilde{\sigma}_{\boldsymbol{k}_\perp,1} \pm \tilde{\sigma}_{\boldsymbol{k}_\perp,2}}{\sqrt{2}} . \tag{2.2}$$

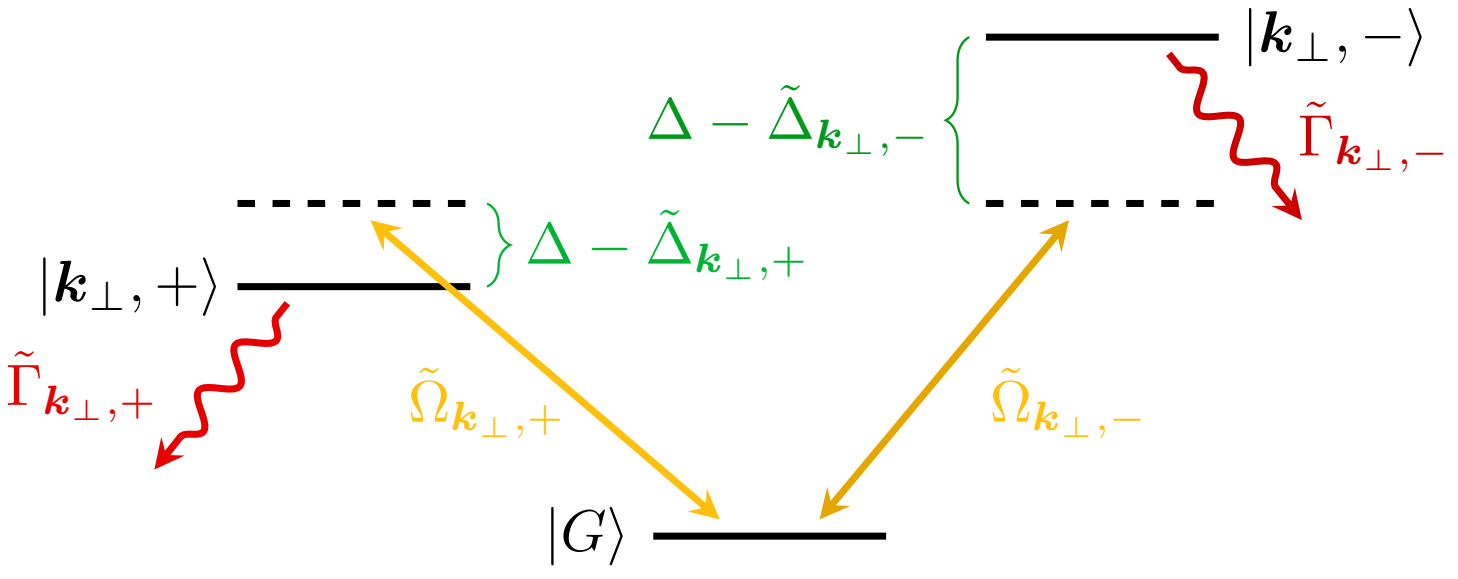

**Figure 2.1 | Dual array single-excitation level structure.** Designations of the parameters pertaining to the even and odd collective momentum states.

As this is a unitary transformation of the operators it conserves the commutators. If we furthermore let $z_1 = -L/2$ and $z_2 = L/2$, we can find

$$H = -\sum_{\alpha=\pm}\left[(\Delta - \tilde{\Delta}_{\boldsymbol{k}_\perp,\alpha})\tilde{\sigma}^\dagger_{\boldsymbol{k}_\perp,\alpha}\tilde{\sigma}_{\boldsymbol{k}_\perp,\alpha} + \left(\tilde{\Omega}_{\boldsymbol{k}_\perp,\alpha}\sigma^\dagger_{\boldsymbol{k}_\perp,\alpha} + \text{H.c.}\right)\right] , \tag{2.3a}$$

$$\mathcal{L}[A] = \sum_{\alpha=\pm}\tilde{\Gamma}_{\boldsymbol{k}_\perp,\alpha}\left(2\tilde{\sigma}^\dagger_{\boldsymbol{k}_\perp,\alpha}A\tilde{\sigma}_{\boldsymbol{k}_\perp,\alpha} - \left\{\tilde{\sigma}^\dagger_{\boldsymbol{k}_\perp,\alpha}\tilde{\sigma}_{\boldsymbol{k}_\perp,\alpha}, A\right\}\right) , \tag{2.3b}$$

where

$$\tilde{\Delta}_{\boldsymbol{k}_\perp,\pm} = \tilde{\Delta}^0_{\boldsymbol{k}_\perp} \pm \tilde{\Delta}_{\boldsymbol{k}_\perp}(L) , \tag{2.4a}$$

$$\tilde{\Gamma}_{\boldsymbol{k}_\perp,\pm} = \tilde{\Gamma}^0_{\boldsymbol{k}_\perp} \pm \tilde{\Gamma}_{\boldsymbol{k}_\perp}(L) , \tag{2.4b}$$

$$\tilde{\Omega}_{\boldsymbol{k}_\perp,+} = \sqrt{2}\tilde{\Omega}_{\boldsymbol{k}_\perp}\cos(k_{z,k}L/2) , \tag{2.4c}$$

$$\tilde{\Omega}_{\boldsymbol{k}_\perp,-} = -\sqrt{2}i\tilde{\Omega}_{\boldsymbol{k}_\perp}\sin(k_{z,k}L/2) . \tag{2.4d}$$

Hence, the modes are now decoupled (for a single excitation) and have different drives. The collective energies and drives depend on the distance $L$ between the lattices. A schematic of the level structure after this change of basis can be seen in Fig. 2.1.

## 2.2 Steady state

With this we can now write down the equation of motion of $\tilde{\sigma}_{\boldsymbol{k}_\perp,\pm}$. As discussed in Section 1.5.2, two-level systems are equivalent to bosons when there is only a single excitation present, and so we use bosonic commutation relations $[\tilde{\sigma}_{\boldsymbol{k}_\perp,\alpha}, \tilde{\sigma}^\dagger_{\boldsymbol{k}_\perp,\beta}] = \delta_{\alpha,\beta}$. We then find

$$\begin{aligned} i\partial_t \tilde{\sigma}_{\boldsymbol{k}_\perp,\pm} &= [\tilde{\sigma}_{\boldsymbol{k}_\perp,\pm}, H] + i\mathcal{L}[\tilde{\sigma}_{\boldsymbol{k}_\perp,\pm}] \\ &= -(\Delta - \tilde{\Delta}_{\boldsymbol{k}_\perp,\pm})\tilde{\sigma}_{\boldsymbol{k}_\perp,\pm} - \tilde{\Omega}_{\boldsymbol{k}_\perp,\pm} - i\tilde{\Gamma}_{\boldsymbol{k}_\perp,\pm}\tilde{\sigma}_{\boldsymbol{k}_\perp,\pm} \\ &= -(\Delta - \tilde{\Delta}_{\boldsymbol{k}_\perp,\pm} + i\tilde{\Gamma}_{\boldsymbol{k}_\perp,\pm})\tilde{\sigma}_{\boldsymbol{k}_\perp,\pm} - \tilde{\Omega}_{\boldsymbol{k}_\perp,\pm} \ , \end{aligned} \tag{2.5}$$

where we have used $\tilde{\sigma}_{\boldsymbol{k}_\perp,\pm}\tilde{\sigma}_{\boldsymbol{k}_\perp,\pm} = 0$ for the linear sector. Notice how the combination $-\tilde{\Delta}_{\boldsymbol{k}_\perp}(z) + i\tilde{\Gamma}_{\boldsymbol{k}_\perp}(z) = \mu_0\omega_a^2 \boldsymbol{d}^\dagger \tilde{\boldsymbol{G}}(\boldsymbol{k}_\perp, z, k_a)\boldsymbol{d}$ appears in the equation. That is, despite the real and imaginary part of the EM Green's function having different physical interpretation and having been previously split accordingly, the full complex Green's function is what appears in our equations in the end. Taking the expectation value of the above, and setting the time-derivative to zero, to impose the steady state, the equation is easily solved. We find

$$\langle \tilde{\sigma}_{\boldsymbol{k}_\perp,\pm} \rangle = -\frac{\tilde{\Omega}_{\boldsymbol{k}_\perp,\pm}}{\Delta - \tilde{\Delta}_{\boldsymbol{k}_\perp,\pm} + i\tilde{\Gamma}_{\boldsymbol{k}_\perp,\pm}} \ . \tag{2.6}$$

It can be noted here that the equations of motion for $\tilde{\sigma}_{\boldsymbol{k}_\perp,\pm}$ could also have been solved using the full two-level commutation relations while still assuming only one excitation, and the above expression would then appear as a lowest order expansion in $\tilde{\Omega}_{\boldsymbol{k}_\perp,\pm}$. In other words, assuming the two-level operators to behave bosonically corresponds to assuming both having only a single excitation *and* weak driving. This makes sense, as the driving must necessarily be weak to ensure only a single excitation.

For comparison, the single array and single atom steady values of the operators corresponding to $\tilde{\sigma}_{\boldsymbol{k}_\perp,\pm}$ can be found to be respectively

$$\left\langle \tilde{\sigma}^{(1)}_{\boldsymbol{k}_\perp} \right\rangle = -\frac{\tilde{\Omega}_{\boldsymbol{k}_\perp}}{\Delta - \tilde{\Delta}^0_{\boldsymbol{k}_\perp} + i\tilde{\Gamma}^0_{\boldsymbol{k}_\perp}} \ , \tag{2.7a}$$

$$\left\langle \sigma^{(0)} \right\rangle = -\frac{\Omega}{\Delta + i\gamma} \ , \tag{2.7b}$$

where for the single atom, $\sigma^{(0)}$ and $\Omega$ are the real-space atomic operator and driving pertaining to that single atom. Notice how the structure is the same in all three cases. Specifically, the linear physics of the dual or single array is qualitatively the same as for the single atom, but with shifted excited state energy and width given by the collective energies. This can also be seen directly from the Hamiltonian and Lindbladian Eq. (2.3), which for each mode are identical in form with the Hamiltonian and Lindbladian of a single atom. Thus we see how in the linear regime, for each transverse quasi-momentum, the array reduces to a single two-level system. This is the basis of the low-loss, single-mode behaviour of the atomic lattices. We will see how a single-photon plane wave experience no other scattering than Bragg scattering, and how a mode of light (consisting of many plane waves) is only additionally scattered due to the dispersion of the atomic modes (as mentioned in Section 1.5.3).

## 2.3 Transmission and reflection

To study the linear physics of the atomic arrays, we consider the transmission and reflection amplitudes of plane waves with polarization matching the atomic dipoles (i.e. right-circular). That is, we take the incoming light to have this polarization and we pick out that component of the full outgoing field. This polarization is not physical (i.e. orthogonal to the momentum) for other transverse momenta than $\boldsymbol{k}_\perp = 0$, and so we are effectively ignoring a component of the light (the component with polarization orthogonal to the atomic dipoles). It is only the matching component of the incoming light, which couples to the atoms, as can be seen from the dot product between $\boldsymbol{d}$ and $\tilde{\boldsymbol{E}}_{\text{in}}$ in our definition of the Rabi frequency below Eq. (2.1). However, as light emitted by the atoms would of course have a physical polarization (and not simply the polarization of the dipoles), such that it contributes to the orthogonally polarized light, this light will have a non-trivial transmission and reflection. We choose to consider the component of light matching the dipoles as it behaves nicely, as we will focus on the $\boldsymbol{k}_\perp = 0$ case, and as it could in principle be picked out in an experiment. Considering detection with the same polarization as the incoming light or its orthogonal partner may reveal interesting behaviour, and would show how much of the full light field is transmitted or reflected in the general case, but we will not perform this study here.

### 2.3.1 Transmitted and reflected light

We identify the transmitted light of the dual array as having $k_z > 0$ and $z > L/2$, while reflected light has $k_z < 0$ and $z < -L/2$. Identifying these two components from Eq. (1.56), and approximating $k_{z,k} \simeq k_{z,k_a}$ due to the small detuning of the incoming light, as we did in Section 1.4.3, the $z$-dependence of the input-output relation is simply an overall factor of $e^{\pm i k_{z,k_a} z}$ (as can be seen from Eq. (1.40)). This overall factors makes no difference for the quantities we will consider, as the transmitted or reflected plane-wave light is the same no matter how far from the array it is observed. Therefore, we do not include it in our expression for the transmitted or reflected light. We do note, however, that we are only interested in propagating light, and so require that $k_{z,k_a}$ is real. This means we only consider $|\boldsymbol{k}_\perp| < k_a$. In total we define the transmitted and reflected E-field operators as

$$E_T(\boldsymbol{k}_\perp) = \frac{\Omega_{\boldsymbol{k}_\perp}}{d} + \mu_0\omega_a^2 d\hat{\boldsymbol{e}}_+^\dagger \boldsymbol{G}(\boldsymbol{k}_\perp, 0, k_a)\hat{\boldsymbol{e}}_+ \sum_j e^{-ik_{z,k_a}z_j}\tilde{\sigma}_{\boldsymbol{k}_\perp,j} \quad , \tag{2.8a}$$

$$E_R(\boldsymbol{k}_\perp) = \mu_0\omega_a^2 d\hat{\boldsymbol{e}}_+^\dagger \boldsymbol{G}(\boldsymbol{k}_\perp, 0, k_a)\hat{\boldsymbol{e}}_+ \sum_j e^{ik_{z,k_a}z_j}\tilde{\sigma}_{\boldsymbol{k}_\perp,j} \quad , \tag{2.8b}$$

where we have defined the continuous Fourier transformed Rabi frequency, $\Omega_{\boldsymbol{k}_\perp}$, analogously to the discrete version as defined below Eq. (2.1). Similarly, in analogy to Eq. (1.74), we define

$$-\Delta_{\boldsymbol{k}_\perp}(z) + i\Gamma_{\boldsymbol{k}_\perp}(z) = \mu_0\omega_a^2 \boldsymbol{d}^\dagger \boldsymbol{G}(\boldsymbol{k}_\perp, z, k_a)\boldsymbol{d} \quad , \tag{2.9}$$

such that $\tilde{\Delta}_{\boldsymbol{k}_\perp}(z) = \frac{1}{a^2}\sum_m \Delta_{\boldsymbol{k}_\perp+\boldsymbol{q}_m}(z)$ and likewise for $\tilde{\Gamma}_{\boldsymbol{k}_\perp}(z)$. We also introduce the notation $\Delta^0_{\boldsymbol{k}_\perp} = \Delta_{\boldsymbol{k}_\perp}(0)$ and $\Gamma^0_{\boldsymbol{k}_\perp} = \Gamma_{\boldsymbol{k}_\perp}(0)$, and note that $\Gamma^0_0 = a^2\tilde{\Gamma}$.

As $\boldsymbol{G}(\boldsymbol{k}_\perp, z, k_a)$ is purely real for $|\boldsymbol{k}_\perp| > k_a$ (see Eq. (1.40)), $\Gamma_{\boldsymbol{k}_\perp+\boldsymbol{q}_m}(z)$ is zero for most $\boldsymbol{q}_m$. For future reference we note that

$$\Gamma_{\boldsymbol{k}_\perp}(z) = \begin{cases} a^2\tilde{\Gamma}\frac{k_a}{k_{z,k_a}}\left(1 - \frac{k_\perp^2}{2k_a^2}\right)\cos(k_{z,k_a}|z|), & \text{for } k_\perp < k_a \quad , \\ 0, & \text{for } k_\perp > k_a \quad , \end{cases} \tag{2.10}$$

which also holds for $z = 0$. For the momenta we consider, $|\boldsymbol{k}_\perp| < k_a$, we thus get the simple expressions

$$E_T(\boldsymbol{k}_\perp) = \frac{\Omega_{\boldsymbol{k}_\perp}}{d} + \frac{i\Gamma^0_{\boldsymbol{k}_\perp}}{d}\sum_j e^{-ik_{z,k_a}z_j}\tilde{\sigma}_{\boldsymbol{k}_\perp,j} \quad , \tag{2.11a}$$

$$E_R(\boldsymbol{k}_\perp) = \frac{i\Gamma^0_{\boldsymbol{k}_\perp}}{d}\sum_j e^{ik_{z,k_a}z_j}\tilde{\sigma}_{\boldsymbol{k}_\perp,j} \quad . \tag{2.11b}$$

Carrying out the sums and writing this in terms of the even and odd modes, we get

$$E_T(\boldsymbol{k}_\perp) = \frac{\Omega_{\boldsymbol{k}_\perp}}{d} + \frac{\sqrt{2}i\Gamma^0_{\boldsymbol{k}_\perp}}{d}(\cos(k_{z,k_a}L/2)\tilde{\sigma}_{\boldsymbol{k}_\perp,+} + i\sin(k_{z,k_a}L/2)\tilde{\sigma}_{\boldsymbol{k}_\perp,-}) \quad , \tag{2.12a}$$

$$E_R(\boldsymbol{k}_\perp) = \frac{\sqrt{2}i\Gamma^0_{\boldsymbol{k}_\perp}}{d}(\cos(k_{z,k_a}L/2)\tilde{\sigma}_{\boldsymbol{k}_\perp,+} - i\sin(k_{z,k_a}L/2)\tilde{\sigma}_{\boldsymbol{k}_\perp,-}) \quad . \tag{2.12b}$$

Before defining the transmission and reflection coefficients, we will study these fields a bit more. Using the result from the previous section, Eq. (2.6), we can find the expectation value of the transmitted field in the steady state to be

$$\begin{aligned}\langle E_T(\boldsymbol{k}_\perp)\rangle \\ = \frac{\Omega_{\boldsymbol{k}_\perp}}{d} - i\Gamma^0_{\boldsymbol{k}_\perp}\frac{\tilde{\Omega}_{\boldsymbol{k}_\perp}}{d}\left(\frac{1+\cos(k_{z,k_a}L)}{\Delta-\tilde{\Delta}_{\boldsymbol{k}_\perp,+}+i\tilde{\Gamma}_{\boldsymbol{k}_\perp,+}} + \frac{1-\cos(k_{z,k_a}L)}{\Delta-\tilde{\Delta}_{\boldsymbol{k}_\perp,-}+i\tilde{\Gamma}_{\boldsymbol{k}_\perp,-}}\right),\end{aligned} \tag{2.13}$$

where we have used the definitions in Eq. (2.4) after approximating $k_{z,k} \simeq k_{z,k_a}$. We now note that $\Gamma_{\boldsymbol{k}_\perp}(L) = \Gamma^0_{\boldsymbol{k}_\perp}\cos(k_{z,k_a}L)$, and defining $\Gamma_{\boldsymbol{k}_\perp,\pm} = \Gamma^0_{\boldsymbol{k}_\perp} \pm \Gamma_{\boldsymbol{k}_\perp}(L)$ (where we yet again have the relation $\tilde{\Gamma}_{\boldsymbol{k}_\perp,\pm} = \frac{1}{a^2}\sum_m \Gamma_{\boldsymbol{k}_\perp+\boldsymbol{q}_m,\pm}$), we can finally write

$$\begin{aligned}\langle E_T(\boldsymbol{k}_\perp)\rangle \\ = \frac{\Omega_{\boldsymbol{k}_\perp}}{d} - \frac{\tilde{\Omega}_{\boldsymbol{k}_\perp}}{d}\left(\frac{i\Gamma_{\boldsymbol{k}_\perp,+}}{\Delta-\tilde{\Delta}_{\boldsymbol{k}_\perp,+}+i\tilde{\Gamma}_{\boldsymbol{k}_\perp,+}} + \frac{i\Gamma_{\boldsymbol{k}_\perp,-}}{\Delta-\tilde{\Delta}_{\boldsymbol{k}_\perp,-}+i\tilde{\Gamma}_{\boldsymbol{k}_\perp,-}}\right).\end{aligned} \tag{2.14}$$

Likewise we can find

$$\langle E_R(\boldsymbol{k}_\perp)\rangle = -\frac{\tilde{\Omega}_{\boldsymbol{k}_\perp}}{d}\left(\frac{i\Gamma_{\boldsymbol{k}_\perp,+}}{\Delta-\tilde{\Delta}_{\boldsymbol{k}_\perp,+}+i\tilde{\Gamma}_{\boldsymbol{k}_\perp,+}} - \frac{i\Gamma_{\boldsymbol{k}_\perp,-}}{\Delta-\tilde{\Delta}_{\boldsymbol{k}_\perp,-}+i\tilde{\Gamma}_{\boldsymbol{k}_\perp,-}}\right). \tag{2.15}$$

Here we see the Bragg scattering of light directly. All quantities with a tilde have the periodicity of the reciprocal lattice, i.e. they are invariant under addition of $\boldsymbol{q}_m$ to $\boldsymbol{k}_\perp$. Hence, if the incoming light is a plane wave, such that $\Omega_{\boldsymbol{k}_\perp}$ is a delta-function, $\tilde{\Omega}_{\boldsymbol{k}_\perp}$ will be a sum of delta-functions, one for each Bragg channel, such that the emitted light gets a contribution for each of these. Whether a channel is open or not (i.e. whether the contribution is non-zero or not) is then determined by the factor $\Gamma_{\boldsymbol{k}_\perp,\pm}$, which is zero for evanescent waves. Furthermore, the sum of $\Gamma_{\boldsymbol{k}_\perp,\pm}$ over all open Bragg channels, exactly yields $\tilde{\Gamma}_{\boldsymbol{k}_\perp,\pm}$, which leads to no more light being emitted than was incoming, ensuring conservation of energy.

### 2.3.2 Transmission and reflection coefficients

We define the intensity transmission and reflection coefficients as the ratio of the transmitted or reflected intensity to the intensity of the (Bragg scattered) incoming field

$$T(\boldsymbol{k}_\perp) = \frac{\left\langle E_T^\dagger(\boldsymbol{k}_\perp) E_T(\boldsymbol{k}_\perp)\right\rangle}{(a^2\tilde{\Omega}_{\boldsymbol{k}_\perp}/d)^2} \quad , \tag{2.16a}$$

$$R(\boldsymbol{k}_\perp) = \frac{\left\langle E_R^\dagger(\boldsymbol{k}_\perp) E_T(\boldsymbol{k}_\perp)\right\rangle}{(a^2\tilde{\Omega}_{\boldsymbol{k}_\perp}/d)^2} \quad , \tag{2.16b}$$

for $\boldsymbol{k}_\perp$ where $\tilde{\Omega}_{\boldsymbol{k}_\perp} \neq 0$. We use $\tilde{\Omega}_{\boldsymbol{k}_\perp}$ as this takes Bragg scattering into account. That is, if light is transmitted or reflected and at the same time Bragg-scattered (such that its outgoing momentum is different from its initial momentum), the above coefficients are still always well-defined. We can thus quantify how much of light is transmitted or reflected into every Bragg channel. The additional factors in the denominator are to make the transmission and reflection coefficients unitless. We furthermore define the complex transmission and reflection amplitudes

$$t(\boldsymbol{k}_\perp) = \frac{\langle E_T(\boldsymbol{k}_\perp)\rangle}{a^2\tilde{\Omega}_{\boldsymbol{k}_\perp}/d} \quad , \tag{2.17a}$$

$$r(\boldsymbol{k}_\perp) = \frac{\langle E_R(\boldsymbol{k}_\perp)\rangle}{a^2\tilde{\Omega}_{\boldsymbol{k}_\perp}/d} \quad , \tag{2.17b}$$

for $\boldsymbol{k}_\perp$ where $\tilde{\Omega}_{\boldsymbol{k}_\perp} \neq 0$. As we are working in the linear regime of only a single excitation in the array, it can be shown that expectation values of products of atomic operators, splits into products of expectation values of the individual operators (i.e. there are no correlations). In other words, we have $T = |t|^2$ and $R = |r|^2$. Using the calculation of the transmitted and reflected field of the previous section (Eqs. (2.14) and (2.15)), we can immediately find the transmission and reflection amplitudes. We get

$$t(\boldsymbol{k}_\perp) = \frac{\Omega_{\boldsymbol{k}_\perp}}{a^2\tilde{\Omega}_{\boldsymbol{k}_\perp}} - \frac{1}{a^2}\left(\frac{i\Gamma_{\boldsymbol{k}_\perp,+}}{\Delta - \tilde{\Delta}_{\boldsymbol{k}_\perp,+} + i\tilde{\Gamma}_{\boldsymbol{k}_\perp,+}} + \frac{i\Gamma_{\boldsymbol{k}_\perp,-}}{\Delta - \tilde{\Delta}_{\boldsymbol{k}_\perp,-} + i\tilde{\Gamma}_{\boldsymbol{k}_\perp,-}}\right) \quad , \tag{2.18a}$$

$$r(\boldsymbol{k}_\perp) = -\frac{1}{a^2}\left(\frac{i\Gamma_{\boldsymbol{k}_\perp,+}}{\Delta - \tilde{\Delta}_{\boldsymbol{k}_\perp,+} + i\tilde{\Gamma}_{\boldsymbol{k}_\perp,+}} - \frac{i\Gamma_{\boldsymbol{k}_\perp,-}}{\Delta - \tilde{\Delta}_{\boldsymbol{k}_\perp,-} + i\tilde{\Gamma}_{\boldsymbol{k}_\perp,-}}\right) \quad . \tag{2.18b}$$

For the case of no Bragg scattering (sufficiently small $a$ and $\boldsymbol{k}_\perp$) the continuous and discrete Fourier transformed quantities are proportional to each other by a factor of $a^2$, and we get

$$t(\boldsymbol{k}_\perp) = 1 - \left( \frac{i\tilde{\Gamma}_{\boldsymbol{k}_\perp,+}}{\Delta - \tilde{\Delta}_{\boldsymbol{k}_\perp,+} + i\tilde{\Gamma}_{\boldsymbol{k}_\perp,+}} + \frac{i\tilde{\Gamma}_{\boldsymbol{k}_\perp,-}}{\Delta - \tilde{\Delta}_{\boldsymbol{k}_\perp,-} + i\tilde{\Gamma}_{\boldsymbol{k}_\perp,-}} \right) , \tag{2.19a}$$

$$r(\boldsymbol{k}_\perp) = - \left( \frac{i\tilde{\Gamma}_{\boldsymbol{k}_\perp,+}}{\Delta - \tilde{\Delta}_{\boldsymbol{k}_\perp,+} + i\tilde{\Gamma}_{\boldsymbol{k}_\perp,+}} - \frac{i\tilde{\Gamma}_{\boldsymbol{k}_\perp,-}}{\Delta - \tilde{\Delta}_{\boldsymbol{k}_\perp,-} + i\tilde{\Gamma}_{\boldsymbol{k}_\perp,-}} \right) , \tag{2.19b}$$

which is still only defined for $\boldsymbol{k}_\perp$ where $\tilde{\Omega}_{\boldsymbol{k}_\perp} \neq 0$. For this case it can be calculated that $T + R = 1$, which is a consequence of there being no loss for plane waves in the linear regime without Bragg scattering. While a plane may be reflected, its transverse momentum is not changed, and so is not scattered into any other mode. With Bragg scattering present we get "loss" into other Bragg channels, and so to get unity we need to sum the transmission and reflection amplitudes over all Bragg channels, before adding the squared norms of these. Let us briefly compare with the corresponding results of the single array and a single atom. For the single array, we can find

$$t^{(1)}(\boldsymbol{k}_\perp) = 1 - \frac{i\tilde{\Gamma}_{\boldsymbol{k}_\perp}}{\Delta - \tilde{\Delta}_{\boldsymbol{k}_\perp} + i\tilde{\Gamma}_{\boldsymbol{k}_\perp}} , \tag{2.20a}$$

$$r^{(1)}(\boldsymbol{k}_\perp) = - \frac{i\tilde{\Gamma}_{\boldsymbol{k}_\perp}}{\Delta - \tilde{\Delta}_{\boldsymbol{k}_\perp} + i\tilde{\Gamma}_{\boldsymbol{k}_\perp}} , \tag{2.20b}$$

and for the single atom

$$t^{(0)}(\boldsymbol{k}_\perp) = 1 - \frac{\Omega}{\Omega_{\boldsymbol{k}_\perp}} \frac{i\tilde{\Gamma}_{\boldsymbol{k}_\perp}}{\Delta + i\gamma} , \tag{2.21a}$$

$$r^{(0)}(\boldsymbol{k}_\perp) = - \frac{\Omega}{\Omega_{\boldsymbol{k}_\perp}} \frac{i\tilde{\Gamma}_{\boldsymbol{k}_\perp}}{\Delta + i\gamma} , \tag{2.21b}$$

where, as before, $\Omega$ is the real-space driving at the position of the atom.

For the single atom, a spatially broad mode (e.g. a plane wave), would have vanishing $\frac{\Omega}{\Omega_{\boldsymbol{k}_\perp}}$, such that the atom has only a very small effect on the light, which is intuitively consistent. For a spatially focused mode, the atom might have a greater effect, but such a mode is ultimately diffraction-limited [7] in how much it can be focused. Thus, it turns out that a beam cannot be sufficiently focused compared to the cross section of the photon-absorption process, which goes as the square of the resonant wavelength [60].

For the single array there is a Lorentzian resonance of perfect reflection, i.e. $R^{(1)}(\boldsymbol{k}_\perp)$ is Lorentzian function of $\Delta$ centred at $\Delta = \tilde{\Delta}_{\boldsymbol{k}_\perp}$ with a peak value of 1, and with a width given by the collective state width $\tilde{\Gamma}_{\boldsymbol{k}_\perp}$. The fact that a single sheet of two-level atoms can ideally behave as a perfect mirror, and indeed as a single two-level emitter (as we found in Section 2.2) is the basis for the study of this thesis, and for the attention is has gotten in the literature. Despite consisting of a large number of two-level emitters in free space, which on their own would scatter light in all directions, the structure of an ordered lattice results in the potential for no scattering at all at the single-photon level. The system is essentially reduced to that of a single two-level emitter in a one-dimensional space, similar to an atom in a waveguide. As the single array constitutes a perfect mirror, we expect the dual array will behave as a high-quality cavity. Such a cavity confines photons for a long time, allowing them to interact indirectly with each other through the nonlinear medium that is the lattices themselves, a nonlinearity inherited from the saturable atoms. Thus, despite the individual lattices being largely linear, as the nonlinearity of the atoms is washed away exactly by the collective behaviour of the system, involving many atoms, making their saturability less significant, the dual array revives the nonlinearity of the atoms by exposing the photons to it for a long time. That is indeed what we will find in the following sections and chapter.

### 2.3.3 Normal incidence transmission

As we are interested in achieving perfect reflection or transmission with the dual array, specifically narrow resonances, we will only consider the regime of no Bragg scattering from here on. As the physics is qualitatively the same for all momenta that have no Bragg scattering (i.e. only the zeroth order channel is open), we focus on $\boldsymbol{k}_\perp = 0$ and $a < \lambda_a$. Furthermore, we only look at the transmission, as the behaviour of the reflection coefficient follows from $R = 1 - T$. Dropping the momentum subscript for a compact notation, i.e. $\tilde{\Gamma}_\pm = \tilde{\Gamma}_{0,\pm}$ and likewise for $\tilde{\Delta}_\pm$, the object we will be studying is

$$t = 1 - \left( \frac{i\tilde{\Gamma}_+}{\Delta - \tilde{\Delta}_+ + i\tilde{\Gamma}_+} + \frac{i\tilde{\Gamma}_-}{\Delta - \tilde{\Delta}_- + i\tilde{\Gamma}_-} \right) . \tag{2.22}$$

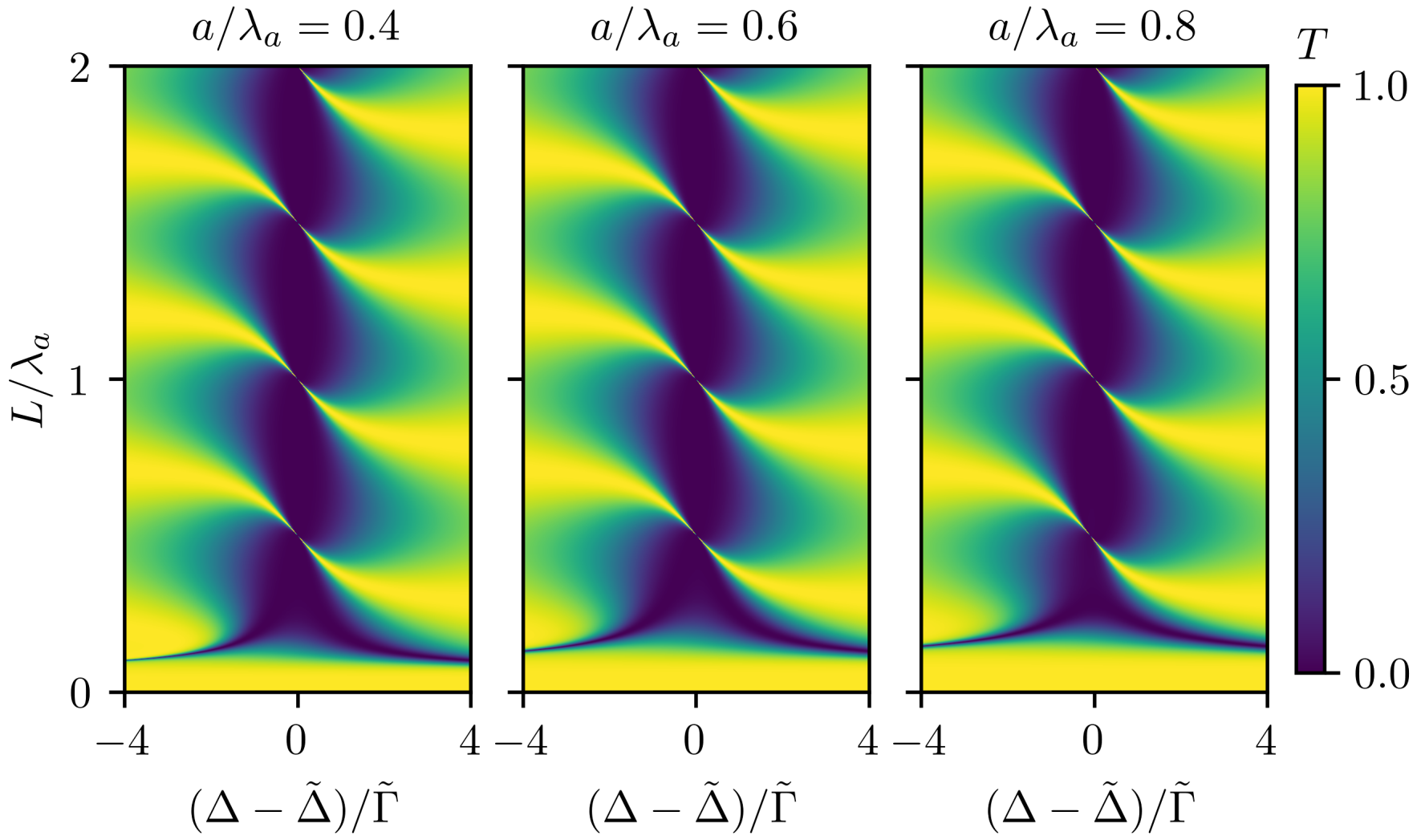

**Figure 2.2** | **Transmission coefficient** $T = |t|^2$. Plotted as a function of detuning $\Delta$ and inter-array distance $L$ for different lattice spacings $a$. $\Delta$ has been shifted and scaled according to the single-array collective energy shift and width, $\tilde{\Delta}$ and $\tilde{\Gamma}$, respectively.

Let us first write down the collective energies $\tilde{\Delta}_\pm$ and $\tilde{\Gamma}_\pm$ for later reference. Using Eqs. (1.82) and (1.83) we have

$$\tilde{\Delta}_\pm = \tilde{\Delta} \pm \tilde{\Gamma}\left(\sin(k_a L) + \frac{1}{k_a}\sum_m{}'' \frac{q_m^2/2 - k_a^2}{\sqrt{q_m^2 - k_a^2}} e^{-\sqrt{q_m^2 - k_a^2}L}\right) \quad , \tag{2.23a}$$

$$\tilde{\Gamma}_\pm = \tilde{\Gamma}(1 \pm \cos(k_a L)) \quad , \tag{2.23b}$$

where $\tilde{\Delta} \equiv \tilde{\Delta}_0^0$ is the zero-momentum single array collective energy shift. We can clearly see the contribution from the single array itself, from photons propagating between the lattices, and, in the case of energy shift, also a contribution from the evanescent fields. From $\tilde{\Gamma}_\pm$ we can see how the even and odd states each oscillate between being superradiant and subradiant as a function of $L$. For $L = n\lambda_a/2$ (with $n$ an integer) the subradiant state is perfectly subradiant, i.e. it has zero decay rate.

Plotting $T = |t|^2$ as a function of the detuning, shifted and scaled by the collective energies, $(\Delta - \tilde{\Delta})/\tilde{\Gamma}$, and of the inter-lattice spacing $L$ for a few different values of $a$, Fig. 2.2, we can a make some observations. Firstly, as the shifted and scaled plots are almost identical, the value of $a$ simply changes the position and width of the features (the collective energies depend on $a$), as long as no new Bragg channels are opened. We will therefore focus on one value of $a$, namely $a = 0.6\lambda_a$ (as this turns out to be a good choice for later numerical treatment of the nonlinear physics), unless explicitly stated otherwise. Secondly, we see the emergence of two regimes with respect to $L$. For small inter-lattice spacing there appear two resonances of low transmission, and thus high reflection, moving to greater and smaller $\Delta$ respectively, as $L$ goes to zero. For larger spacing there appears a repeating pattern of ridges of extremely narrow transmission resonances every time $L$ increases by $\lambda_a/2$. We will now analyse $t$ in these regimes to understand this behaviour.

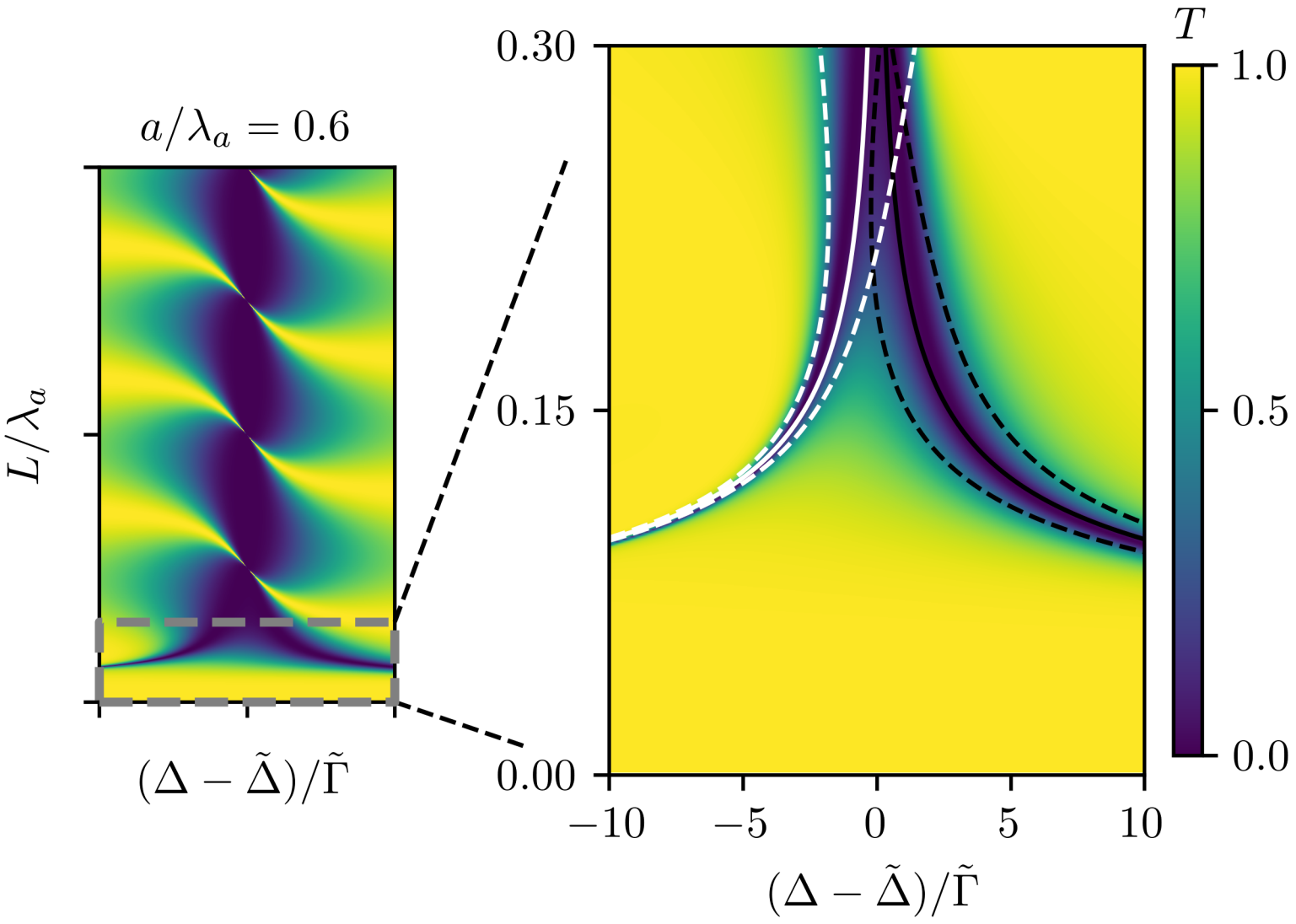

**Figure 2.3 | Transmission coefficient for small $L$.** The right panel shows the evanescent field dominated regime of $L$ of the middle panel of Fig. 2.2 (here reprinted as the left panel). The approximate collective energies are plotted as solid black and white lines for the even and odd state respectively, and these energies plus/minus their collective widths are plotted as dashed lines.

### 2.3.4 Small inter-lattice spacing

For small $L$ the evanescent term in Eq. (2.23a) becomes very large (it diverges for $L \to 0$, where we recover the un-regularized $\tilde{\Delta}$), such that $\tilde{\Delta}_\pm$ are also very large. Hence, if we consider tuning the incoming light to either the even or odd collective energy, $\Delta \simeq \tilde{\Delta}_+$ or $\Delta \simeq \tilde{\Delta}_-$, the other

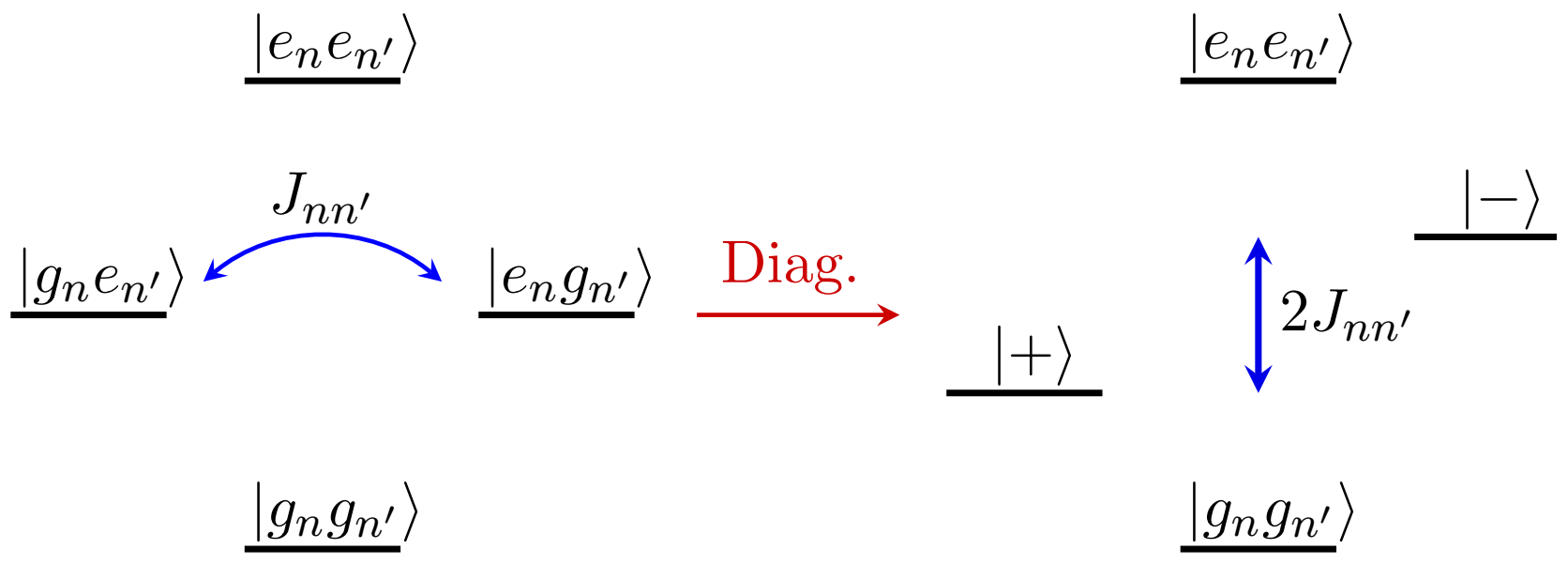

**Figure 2.4 | Level structure of atomic dimer.** The level structure of two atoms, labelled $n$ and $n'$, forming a dimer. Before diagonalization the single-excitation levels interact with a coupling strength $J_{nn'}$, and afterwards the states $|\pm\rangle = (|g_n e_{n'}\rangle \pm |e_n g_{n'}\rangle)/\sqrt{2}$ are detuned by $2J_{nn'}$.

of the two fractions in Eq. (2.22) will be very small. In other words, we can approximate

$$t \simeq 1 - \frac{i\tilde{\Gamma}_\pm}{\Delta - \tilde{\Delta}_\pm + i\tilde{\Gamma}_\pm} \, , \tag{2.24}$$

where the $\pm$ is for $\Delta \simeq \tilde{\Delta}_+$ or $\Delta \simeq \tilde{\Delta}_-$ respectively. The transmission thus becomes identical in form to the single array transmission, Eq. (2.20a). We would therefore find the same perfect-reflection Lorentzian resonance, but now we have one for each of the even and odd states, corresponding to the pairs of resonances seen at small $L$ plotted in Fig. 2.3.

We can also understand this single array-like behaviour by considering the level structure of two closely spaced atoms. As $L$ becomes smaller than $a$, the interaction between atoms at the same lattice position, but opposite lattices, will be stronger than the interaction between atoms in the same lattice. These pairs of atoms can be diagonalized and considered as four-level dimers, see Fig. 2.4. We can therefore think of the dual array as a single lattice of dimers. The interaction between two such atoms will be

$$J_{nn'} = \frac{6\pi\gamma}{k_a} \Re\left[\frac{e^{ik_aL}}{4\pi L}\left(1 + \frac{ik_aL - 1}{k_a^2 L^2}\right)\right] \simeq -\frac{3\gamma}{2(k_aL)^3} \, , \tag{2.25}$$

for $L \ll a < \lambda_a$, where $n$ and $n'$ label two atoms at the same lattice position but opposite lattices, and we have used Eq. (1.38). For small $L$ the two single-excitation states of the dimers will thus be detuned by $2|J_{nn'}| = 3\gamma/(k_aL)^3 \gg \gamma$, such that any incoming light tuned near resonance with one of these states will not affect the other and so the dimer effectively behaves like a two-level system (in the regime of at most one excitation). Hence, the array is reduced to a single lattice of effective two-level systems, and we recover the physics of the single array. Specifically, we see Lorentzian reflection resonances at the energies of the collective dimer states, which correspond to the dual array even and odd collective states. For small $L$ these energies are approximately $\tilde{\Delta}_\pm \simeq \omega_a + \tilde{\Delta} \mp J_{nn'}$, as contributions to the shift in energy due to the interaction between atoms of different lattice sites is negligible in comparison to the interaction between same-site atoms. Indeed, in Fig. 2.3 we again have the transmission coefficient, now only for small $L$, with the energies $\tilde{\Delta}_\pm$ plotted as solid lines (using the approximate expression). We see how the reflection resonances lie along these lines. The widths of these resonances are $\tilde{\Gamma}_\pm = \tilde{\Gamma}(1 \pm \cos(k_a L)) \simeq \tilde{\Gamma}(1 \pm [1 - (k_a L)^2/2])$, such that for vanishing $L$, the width of the even resonance goes to $2\tilde{\Gamma}$, while that of the odd resonance goes to zero. These widths, using the approximate expression, are plotted as dashed lines in Fig. 2.3, and again we see that it captures the width of the resonances perfectly. Hence, the dual array produces a atomic mirror-like reflection resonance at small $L$, but of arbitrary narrowness in contrast to the actual atomic mirror, whose resonances have finite width.

### 2.3.5 Large inter-lattice spacing

For large $L$ the evanescent term in Eq. (2.23a) is negligible, and we can rewrite the transmission amplitude, Eq. (2.22), in a more explicit form

$$\begin{aligned} t = 1 - \Bigg( & \frac{i\tilde{\Gamma}(1+\cos(k_a L))}{\Delta - \tilde{\Delta} - \tilde{\Gamma}\sin(k_a L) + i\tilde{\Gamma}(1+\cos(k_a L))} \\ & + \frac{i\tilde{\Gamma}(1-\cos(k_a L))}{\Delta - \tilde{\Delta} + \tilde{\Gamma}\sin(k_a L) + i\tilde{\Gamma}(1-\cos(k_a L))} \Bigg) \\ = & \frac{(\Delta - \tilde{\Delta})^2}{(\Delta - \tilde{\Delta} + i\tilde{\Gamma})^2 + \tilde{\Gamma}^2 e^{2ik_a L}} \ , \end{aligned} \tag{2.26}$$

where the final expression is found after some algebra. Comparing with Eq. (2.20), we see that this can be written as

$$t = \frac{(t^{(1)})^2}{1 - (r^{(1)})^2 e^{2ik_a L}} \ , \tag{2.27}$$

where the single array amplitudes are evaluated at $\boldsymbol{k}_\perp = 0$. Up to a phase, this is exactly the transmission amplitude of a Fabry-Pérot cavity (or resonator) consisting of two mirrors with transmission and reflection amplitudes $t^{(1)}$ and $r^{(1)}$ [80]. Given two parallel mirrors with identical transmission and reflection amplitudes $t_0$ and $r_0$, the total transmission amplitude for normal incidence light can be found by summing up the amplitudes of each possible trajectory of light that is eventually transmitted. These are individually found by multiplying $t_0$, $r_0$, and $e^{ik_aL}$ according to how many times light is transmitted, reflected, or propagates during the trajectory, see Fig. 2.5. The transmission amplitude of the Fabry-Pérot cavity is thus

$$\begin{aligned} t_{FB} &= t_0 e^{ik_aL} t_0 + t_0 e^{ik_aL} r_0 e^{ik_aL} r_0 e^{ik_aL} t_0 + \dots \\ &= t_0 e^{ik_aL} \left( \sum_{n=0}^{\infty} (r_0^2 e^{2ik_aL})^n \right) t_0 \\ &= \frac{t_0^2}{1 - r_0^2 e^{2ik_aL}} e^{ik_aL} \quad , \end{aligned} \tag{2.28}$$

where we have recognized the geometric series, and used the fact that $\lim_{n\to\infty} r_0^{2n} = 0$. With this, we reach the very intuitively reasonable result that the dual array, for large $L$, behaves exactly like a Fabry-Pérot cavity consisting of two individual single arrays, which themselves behave like mirrors. Hence, referring to the dual array as the atomic cavity.

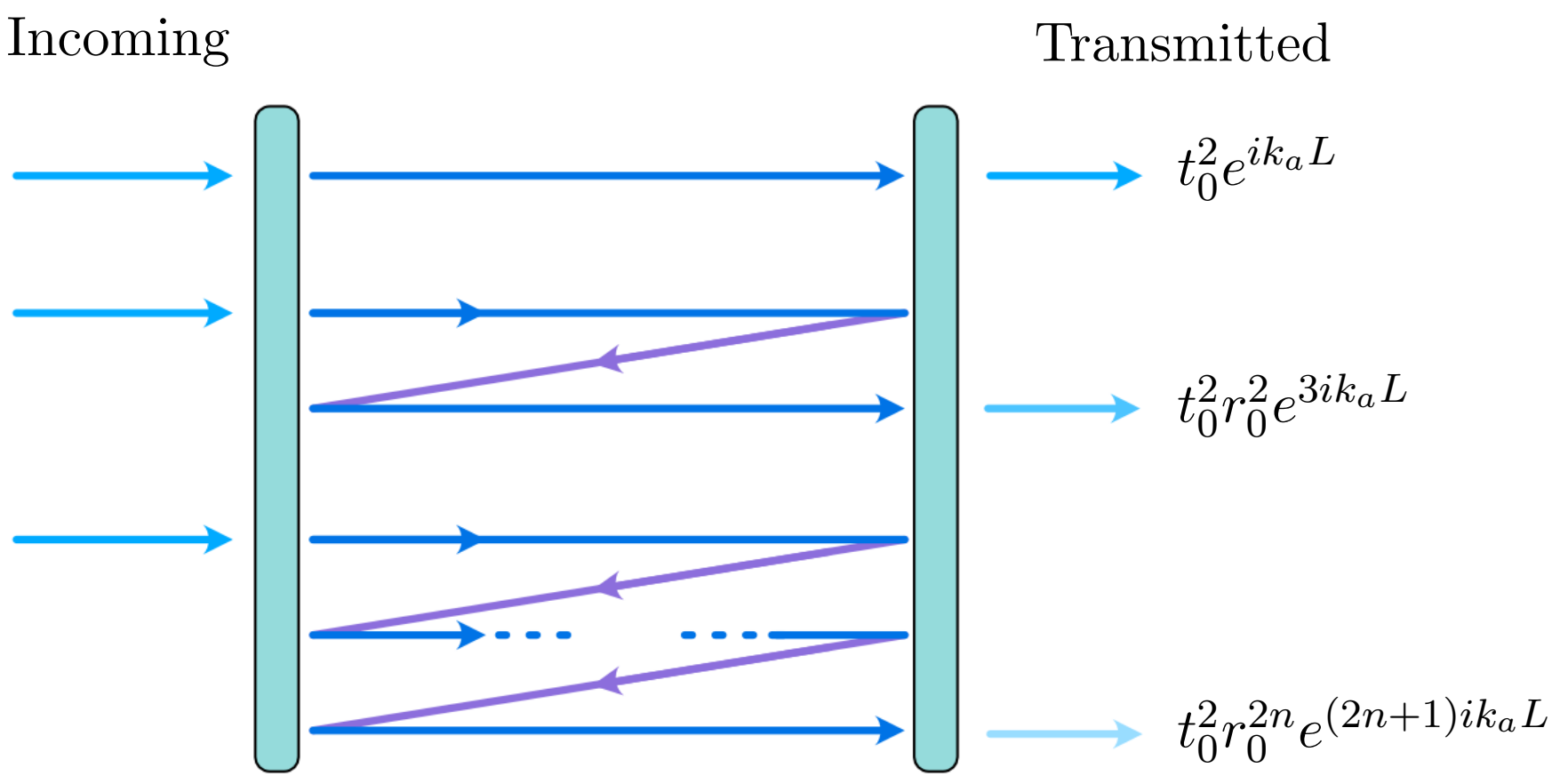

**Figure 2.5 | Possible trajectories of a Fabry-Pérot cavity ending in transmission.** Light that is transmitted into the cavity can be reflected any number of times before being transmitted out again. Each time light is transmitted, reflected, or propagates the length of the cavity, the corresponding contribution to the total transmission accumulates a factor of $t_0$, $r_0$, or $e^{ik_aL}$.

At certain values of $L$, a Fabry-Pérot cavity of identical, zero-loss (i.e. $|t_0|^2 + |r_0|^2 = 1$) mirrors becomes completely transparent for any $|r_0| < 1$. That is, no matter how reflective the individual mirrors are, short of being perfectly reflective, the pair of mirrors can be completely transparent. In particular, it can be found, with some algebra, that the equation $|t_{FB}|^2 = 1$ is solved by

$$\tan(k_a L) = \frac{\Im[r_0]}{\Re[r_0]} \quad . \tag{2.29}$$

Hence, in the case of the dual array, where the reflection amplitude of the individual lattices is $r_0 = r^{(1)}$, there will be a perfect transmission resonance when

$$\tan(k_a L) = \frac{\Im[r^{(1)}]}{\Re[r^{(1)}]} = -\frac{\Delta - \tilde{\Delta}}{\tilde{\Gamma}} \quad . \tag{2.30}$$

Note that the left hand side is periodic in $L$ with a period of $\lambda_a/2$, such that this condition of perfect transmission resonances is satisfied repeatedly, as we indeed saw in Fig. 2.2. This periodicity makes sense, as the physics is dominated by the propagating photons, and the periodicity corresponds to the condition for a standing wave between the two lattices. Calculating the width of this resonance is messy, as it is asymmetric with respect to the line of perfect transmission. We therefore extract the width from a calculation in the next section, where we consider the time scale of the confinement of photons in the atomic cavity. We can find that the transmission amplitude along the line defined by Eq. (2.30) explicitly becomes

$$t = 1 - \left(\frac{1}{2}(1 + e^{2ik_aL}) + \frac{1}{2}(1 + e^{2ik_aL})\right) = -e^{2ik_aL} \quad , \tag{2.31}$$

where the first expression is written to underline what the contribution from each term in Eq. (2.22) is. In other words, we have perfect transmission with a phase that is tunable through $L$, and each of the super- and subradiant states makes an identical contribution to the field. When $L$ is an integer multiple of $\lambda_a/2$, the transmission amplitude is $t = -1$, such that single photons are perfectly transmitted and acquire a phase of $\pi$.

In Fig. 2.6 we again plot the transmission, now for larger $L$, and mark the transmission ridge given by $k_a L = -\arctan((\Delta - \tilde{\Delta})/\tilde{\Gamma}) + n\pi$ for $n = 3$ (we can choose the positive integer $n$ freely, corresponding to the periodicity of Eq. (2.30)). Furthermore, we plot the energies of the even and odd collective states and their widths given by Eq. (2.23)

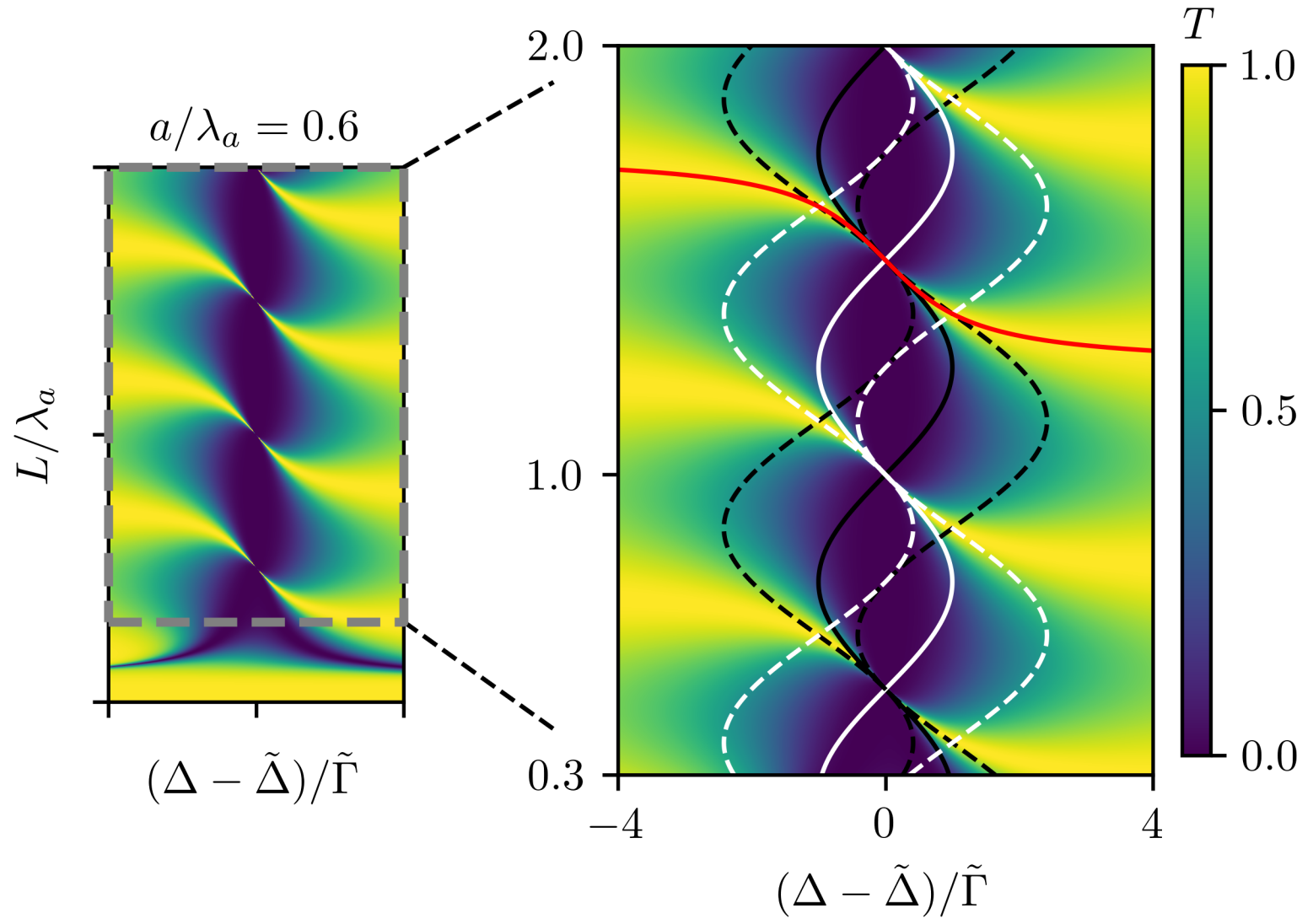

**Figure 2.6 | Transmission coefficient for larger $L$.** The right panel shows the non-evanescent regime of $L$ of the middle panel of Fig. 2.2 (here reprinted as the left panel). The line defined by the perfect-transmission condition of Eq. (2.30) is plotted in red for a certain interval of $L$. The collective energies and widths are again plotted as solid and dashed lines, as they were in Fig. 2.3.

(neglecting the evanescent term of Eq. (2.23a)). We see how the shape of the overall reflective region is given by the interplay of the two states as they oscillate between being super- or subradiant, respectively, as a function of $L$, with the narrow transmission resonances appearing when one state becomes perfectly subradiant. This origin of the transmission resonance is similar to the physics of Fano resonances [79, 81], including the asymmetric lineshape of the resonances with respect to detuning [82].

The atomic cavity thus also shows repeated, extremely narrow transmission resonances. Contrary to the reflection resonances, these exist at finite values of the inter-lattice spacing and for detuning near the single array collective frequency, making them experimentally more accessible. As discussed, it is these narrow resonances we are interested in, as they, in some intuitive sense, should be very sensitive to nonlinear effects. That is, if the presence of a second photon disturbs the behaviour of the first, a narrow feature, where a tiny change causes the photon to go from perfect transmission to high reflection, should result in a stronger nonlinear response. In the next section we will see how the narrow frequency width of these resonances corresponds to a long time scale, which pertains to the confinement of photons in the atomic cavity. Hence, we may also understand the potential of the atomic cavity for high nonlinearity via the picture that the photons simply spend more time in a nonlinear environment and so accumulate a strong effect. Indeed, we will see in the next chapter that the atomic mirror is only weakly nonlinear, but the accumulated effect of the atomic cavity is very strong.

## 2.4 Delay time

We can define a time which gives the scale of how long a single photon is absorbed in a system before emission. This is the delay or confinement time (also known as the Wigner delay time [83–85]).

### 2.4.1 General derivation

Consider a field passing through a medium of length $L$ along the direction of propagation. The field just after the medium is related to the incident field via

$$\boldsymbol{E}_{\text{out}}(\omega) = t(\omega)\boldsymbol{E}_{\text{in}}(\omega) \ . \tag{2.32}$$

These are the fields Fourier transformed in time, and $t(\omega) = |t(\omega)|e^{i\phi(\omega)}$ is the transmission amplitude at the frequency $\omega$. Let us assume that the incoming wave is concentrated around a carrier frequency $\omega_0$ (e.g. it could be a Gaussian mode at this frequency). Let us furthermore assume that $|t(\omega)|$ varies slowly around $\omega_0$, such that we can approximate it with the simple expansion

$$t(\omega) \simeq |t(\omega_0)|e^{i(\phi(\omega_0)+\phi'(\omega_0)(\omega-\omega_0))} \ , \tag{2.33}$$

where the prime indicates a derivative. Using this expansion, we can evaluate the time-dependence of the outgoing field

$$\begin{aligned}
\boldsymbol{E}_{\text{out}}(t) &= \int t(\omega)\boldsymbol{E}_{\text{in}}(\omega)e^{-i\omega t}\mathrm{d}t \\
&\simeq \int |t(\omega_0)|e^{i(\phi(\omega_0)+\phi'(\omega_0)(\omega-\omega_0))}\boldsymbol{E}_{\text{in}}(\omega)e^{-i\omega t}\mathrm{d}\omega \\
&= |t(\omega_0)|e^{i\phi(\omega_0)-\phi'(\omega_0)\omega_0}\int \boldsymbol{E}_{\text{in}}(\omega)e^{-i\omega(t-\phi'(\omega_0))}\mathrm{d}\omega \\
&= |t(\omega_0)|e^{i\phi(\omega_0)-\phi'(\omega_0)\omega_0}\boldsymbol{E}_{\text{in}}(t-\phi'(\omega_0)) \ .
\end{aligned} \tag{2.34}$$

Thus we see that the outgoing field has the same shape as the incoming wave, but it has changed its overall amplitude, acquired a phase, and it has been shifted by a time $\phi'(\omega_0)$. Had the medium been vacuum, we would simply have $\phi'(\omega_0) = L$ (where $c = 1$), i.e. the shift in time is simply the travel time through the medium. We can therefore define a delay time as

$$\tau = \left.\frac{\mathrm{d}\phi}{\mathrm{d}\omega}\right|_{\omega_0} - L \ . \tag{2.35}$$

From $\log t = \log|t| + i\phi$ and $\frac{\mathrm{d}\log(t)}{\mathrm{d}\omega} = \frac{1}{t}\frac{\mathrm{d}t}{\mathrm{d}\omega}$, we find

$$\tau = \Im\left[\frac{1}{t}\frac{\mathrm{d}t}{\mathrm{d}\omega}\right]\Bigg|_{\omega_0} - L \ . \tag{2.36}$$

In the cases we are considering, the term $-L$ will be neglected as we are working in a regime where the speed of light is very large. We have not included retardation effects in our derivations so far, and this amounts to letting the speed of light go to infinity, such that $L = L/c$ indeed goes to zero. Furthermore, we are not working with pulses but rather continuous beams, and therefore the carrier frequency $\omega_0$ must effectively be replaced by the single frequency of the beam, which we will simply call $\omega$. In terms of the detuning $\Delta = \omega - \omega_a$, we can thus write

$$\tau = \Im\left[\frac{1}{t}\frac{\mathrm{d}t}{\mathrm{d}\Delta}\right] \ , \tag{2.37}$$

which is implicitly evaluated at $\omega$.

### 2.4.2 Atomic cavity delay time

Using the normal incidence transmission amplitude found for the dual array, Eq. (2.22), we can find

$$\begin{aligned} \tau &= \Im\left[\frac{1}{t}\frac{\mathrm{d}t}{\mathrm{d}\Delta}\right] \\ &= \Im\left[\frac{1}{t}\left(\frac{i\tilde{\Gamma}_+}{(\Delta - \tilde{\Delta}_+ + i\tilde{\Gamma}_+)^2} + \frac{i\tilde{\Gamma}_-}{(\Delta - \tilde{\Delta}_- + i\tilde{\Gamma}_-)^2}\right)\right] \ . \end{aligned} \tag{2.38}$$

This expression could be reduced further, but is not illuminating in the general case.

For comparison the single array delay time can be found to have the simple form

$$\tau^{(1)} = \frac{R^{(1)}}{\tilde{\Gamma}} \ . \tag{2.39}$$

In other words, the delay time pertaining to zero transverse momentum light is given by the reflection coefficient and the lifetime of the corresponding collective state, $1/\tilde{\Gamma}$. Indeed, on the reflection resonance, $R^{(1)} = 1$, the delay time is equal to the lifetime of the collective state. In the limit of small $L$, where the evanescent fields dominate, and the transmission of the dual array has the same form as for the single array, $\tau$ can likewise be found to have the simple form

$$\tau \simeq \frac{R}{\tilde{\Gamma}_{\pm}} \ , \tag{2.40}$$

near the even or odd reflection resonance. This is consistent with the picture that the dual array behaves like an effective single array composed of subradiant or superradiant atomic dimers.

In the large $L$ regime, neglecting evanescent fields, we can find with some algebra that

$$\tau = \frac{2}{\tilde{\Gamma}\Delta'} \Im\left[\frac{i\Delta' - 1 + e^{2ikL}}{(\Delta' + i)^2 + e^{2ikL}}\right] \ , \tag{2.41}$$

where $\Delta' = (\Delta - \tilde{\Delta})/\tilde{\Gamma}$. Again, the expression could be rewritten in a different form, but is not illuminating. Along the line of perfect transmission, defined by $\Delta' = -\tan(k_a L)$, however, we can find the simple expression

$$\tau = \frac{2\tilde{\Gamma}}{(\Delta - \tilde{\Delta})^2} \ . \tag{2.42}$$

Thus the delay time diverges as $(\Delta - \tilde{\Delta})^{-2}$ when the transmission resonances becomes narrow. If we consider this near the divergence, where $\Delta - \tilde{\Delta} \simeq 0$ and thus $\tan(k_a L) \simeq 0$. such that $k_a L - n\pi \simeq 0$ for some integer $n$, we can rewrite the delay time

$$\tau = \frac{2}{\tilde{\Gamma}\tan^2(k_a L)} \simeq \frac{2}{\tilde{\Gamma}(k_a L - n\pi)^2} \simeq \frac{1}{\tilde{\Gamma}(1 - (-1)^n \cos(k_a L))} \ , \tag{2.43}$$

where we have made a lowest order Taylor expansion of the tangent, and rewritten that in terms of an expression involving a cosine, which shares the same lowest order Taylor expansion. Comparing with Eq. (2.23b), we see that we have recovered the decay rate of the subradiant zero transverse momentum collective state. Hence, the delay time is given by the lifetime of that state, similar to the reflection resonance case. In other words, on the transmission resonance when $\Delta - \tilde{\Delta} \simeq 0$, we can write

$$\tau \simeq \frac{1}{\tilde{\Gamma}_\pm} \ , \tag{2.44}$$

where the $\pm$ is now determined by which state is subradiant, $\tilde{\Gamma}_\pm \simeq 0$.

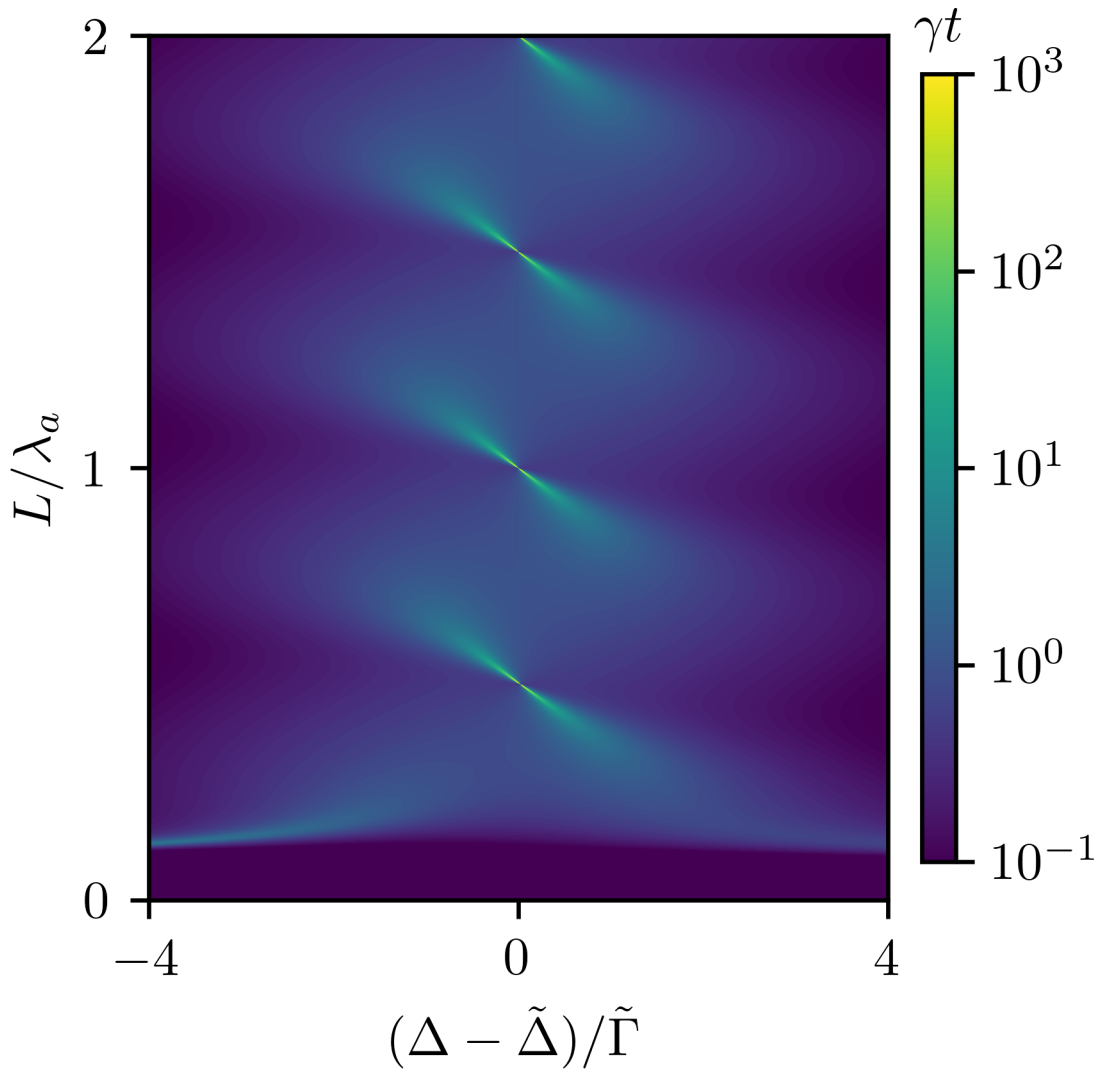

**Figure 2.7 | Delay time $\gamma\tau$ for the dual array.** Plotted for the same parameters as the transmission is in middle panel of Fig. 2.2.

In total, we have shown the expected relationship between the resonance widths, the collective state decay rates, and the delay time, namely that widths are equal to rates, and time scales are given by the reciprocal of these. The delay time is plotted in Fig. 2.7, and we see how it has a structure similar to the transmission coefficient in Fig. 2.2, with large delay times when the transmission (or reflection) has a narrow resonance. Indeed, following the above result, we take the width of the transmission resonances to vanish as $\sim (\Delta - \tilde{\Delta})^2$, for $\Delta - \tilde{\Delta} \simeq 0$ on the lines of perfect transmission. For the atomic mirror, the collective state has finite lifetime, resulting in a finite width resonance and finite delay time. The atomic cavity shows, on resonance, a perfectly subradiant state, resulting in an infinitely narrow resonance and diverging delay time. Hence, while we expect a weak nonlinearity of the atomic mirror, and the photons being delayed only a relatively short time to appreciate the nonlinear environment, the atomic cavity ideally confines photons arbitrarily long and so the accumulated nonlinear effect should be very strong.

## 2.5 Finite arrays

Before moving on to study the nonlinear behaviour of the atomic cavity, we will consider and study how the linear physics described so far in this chapter for infinite lattices carries over to finite lattices. The study of nonlinearity in the following chapter will proceed by numerically time-evolving the wave function of a finite dual array, and so we need to understand at least the most immediate finite size effects. In particular we will be considering two identical square lattices, parallel to each other, of $N \times N$ atoms illuminated with a Gaussian beam of light. We must use a finite beam that is smaller than the array, as any light going around the lattice or being scattered on its edges, would of course not behave as in the ideal situation of infinite arrays. We choose a Gaussian mode as this corresponds to a standard laser beam [7].

### 2.5.1 Solving the linear dynamics numerically

While the infinite lattices allow for an analytical solution of the linear dynamics, as the Fourier modes are eigenmodes of the system, in the finite case we proceed numerically. As in Section 2.2 we take the expectation value of the equations of motion, but now for real space $\sigma_n$, using bosonic commutation relations, and set time-derivatives to zero. In particular, with $i\partial_t \sigma_n = [\sigma_n, H] + i\mathcal{L}[\sigma_n]$ using Eqs. (1.26) and (1.30), we get the equation

$$0 = -\Delta \langle \sigma_n \rangle - \Omega_n - \sum_{m \neq n} J_{nm} \langle \sigma_m \rangle - \sum_m i\Gamma_{nm} \langle \sigma_m \rangle \tag{2.45}$$

$$\Rightarrow \sum_m [\Delta \delta_{n,m} + (1 - \delta_{n,m}) J_{nm} + i\Gamma_{nm}] \langle \sigma_m \rangle = -\Omega_n \quad , \tag{2.46}$$

where we have defined the individual Rabi frequencies of the atoms $\Omega_n = \boldsymbol{d}^\dagger \boldsymbol{E}_{\text{in}}(\boldsymbol{r}_n)$ (the discrete or continuous Fourier transform of which are the previously employed $\tilde{\Omega}_{\boldsymbol{k}_\perp}$ and $\Omega_{\boldsymbol{k}_\perp}$). This is a system of linear equations and inverting the matrix, whose diagonal entries are $\Delta + i\Gamma_{nn} = \Delta + i\gamma$, and whose off-diagonal entries are $J_{nm} + i\Gamma_{nm} = \mu_0 \omega_a^2 \boldsymbol{d}^\dagger \boldsymbol{G}(\boldsymbol{r}_n, \boldsymbol{r}_m, k_a) \boldsymbol{d}$, we can find $\langle\sigma_n\rangle$ in the steady state. Using Eq. (1.65), we then find the E-field component in the mode $\boldsymbol{f}$, which we take to be a Gaussian beam in the next section, and with it calculate transmission and reflection amplitudes and coefficients defined analogously to Eqs. (2.16) and (2.17).

### 2.5.2 Gaussian driving mode

The paraxial Gaussian mode we consider is described by [7]

$$\boldsymbol{f}_G(\boldsymbol{r},k) = \hat{\boldsymbol{e}}_+\sqrt{\frac{2}{\pi}}\frac{1}{w(z)}\exp\left[\frac{-r_\perp^2}{w(z)^2}\right]\exp\left[i\left(kz+k\frac{r_\perp^2}{2R(z)}-\phi(z)\right)\right] , \tag{2.47}$$

where $w(z) = w_0\sqrt{1+\left(\frac{z}{z_R}\right)^2}$ is the beam waist at $z$, with $w_0$ the waist at $z=0$, $z_R = \pi w_0^2/\lambda = kw_0^2/2$ is the Rayleigh range, and $R(z) = z\left[1+(z_R/z)^2\right]$ is the $z$-dependent radius of the wavefront curvature of the beam. Finally, $\phi(z) = \arctan(z/z_R)$ denotes the Gouy phase. The mode is normalized such that $\int \mathrm{d}^2r_\perp \boldsymbol{f}_G^\dagger(\boldsymbol{r},k)\boldsymbol{f}_G(\boldsymbol{r},k) = 1$. We have taken the polarization to be right-circular to match that of the atomic dipole moments. The modulus and phase of the real space mode has been plotted in Fig. 2.8 for a waist of $w_0 = 1.5\lambda_a$ with the different quantities marked. This is the value of the beam waist we will use, unless otherwise stated. We see how the Rayleigh range gives the region with the highest intensity of light, and how the beam broadens linearly beyond this range. Performing a transverse Fourier transformation of $\boldsymbol{f}_G(\boldsymbol{r},k)$ yields

$$\boldsymbol{f}_G(\boldsymbol{k}_\perp,z,k) = \hat{\boldsymbol{e}}_+\sqrt{2\pi}w_0 e^{-k_\perp^2 w_0^2/4} e^{ik\left(1-\frac{k_\perp^2}{2k^2}\right)z} . \tag{2.48}$$

Hence, in terms of Eq. (1.57), we can write the paraxial mode as $\boldsymbol{f}_G(\boldsymbol{k}_\perp) = \hat{\boldsymbol{e}}_+\sqrt{2\pi}w_0e^{-k_\perp^2w_0^2/4}$, and the $z$-dependence of the above then follows from making the paraxial expansion $k_z \simeq k - k_\perp^2/2k$ in $e^{ik_zz}$.

### 2.5.3 Transmission and loss

Numerically solving the dynamics of the finite system as described in Section 2.5.1, and using Eq. (1.65) to calculate the E-field pertaining to the Gaussian mode, we get the corresponding transmission, reflection and loss. The loss is defined as $1-T-R$, and represents the fraction of light, which is scattered into a different mode than the chosen detection mode, i.e. the Gaussian mode. The loss is due to scattering at the edges of the lattices, and due to different transmission and reflection of the different momentum components of the chosen mode of light. As the collective energies depend on momentum, and the transmission and reflection amplitudes depend on these, there would be some scattering into other modes even for the infinite arrays.

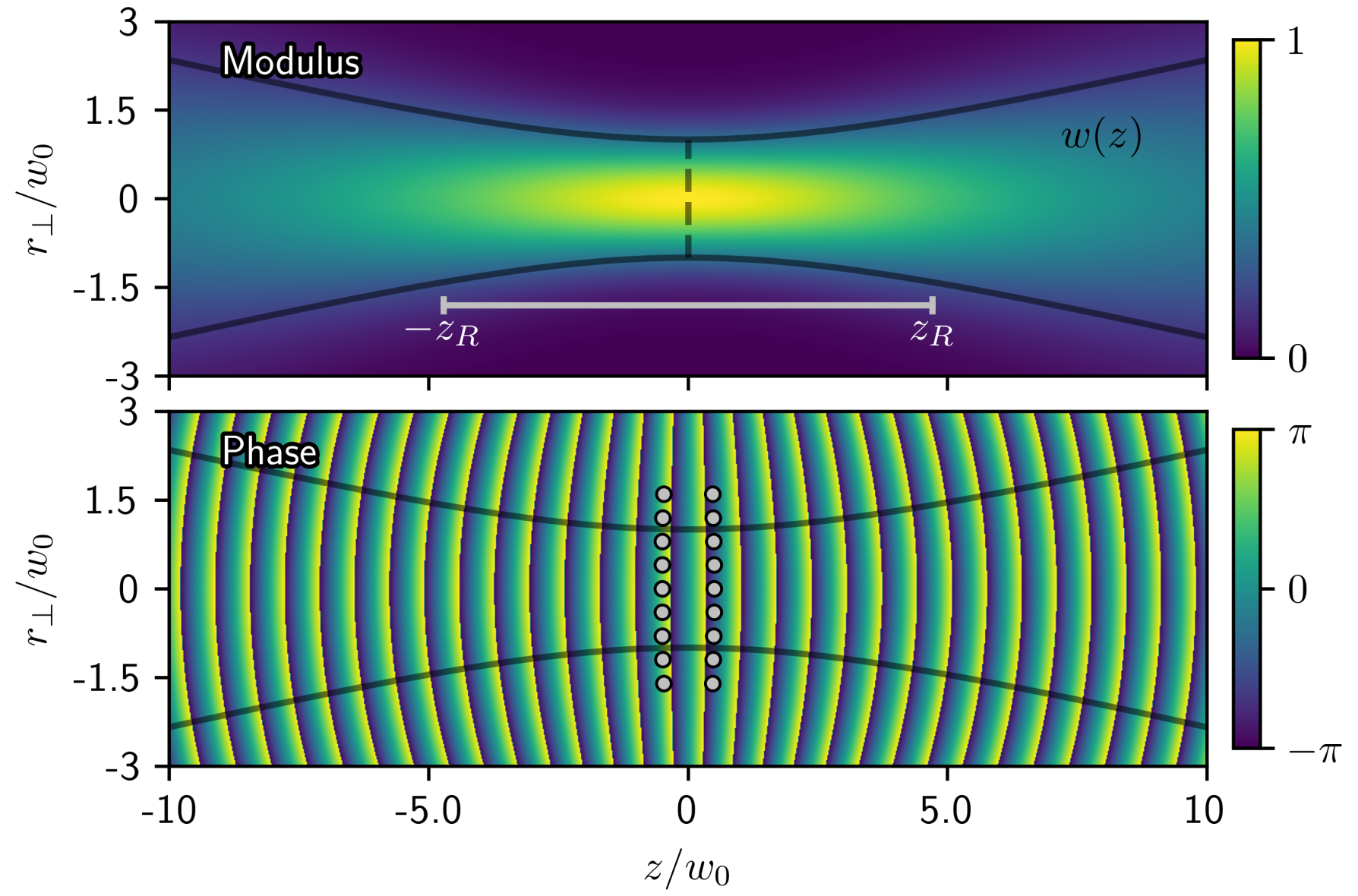

**Figure 2.8 | Modulus and phase of the Gaussian mode.** The modulus (top) normalized to a peak value of unity and the phase (bottom) of Eq. (2.47), for a waist of $w_0 = 1.5\lambda_a$. The width of the beam, $\pm w(z)$, is plotted as black lines with the waist, $\pm w_0$, marked with a dashed line, and the Rayleigh range, $\pm z_R$, is indicated by a white scale bar. The atom positions of a curved dual array of $9 \times 9$ atoms in each lattice with $a = 0.6\lambda_a$ and $L = 1.5\lambda_a$ are marked with grey dots.

The transmission and loss coefficients of a dual array of $9 \times 9$ atoms in each lattice, illuminated by a Gaussian beam, can be seen in Fig. 2.9. We see how the transmission is nearly identical to the infinite array and plane-wave driving case, Fig. 2.2, but the narrowest features are gone and replaced by losses. Furthermore, we see how the loss increases as the lattice are farther separated, as light more easily escapes through the side of the cavity-like system. As discussed, the most narrow features are associated with long delay times, and therefore it is exactly around these that the light has a long time to explore the edges of the array and to be scattered into other modes. In this way, loss will indeed be greatest near the areas we are most interested in. We will therefore consider some ways to mitigate these effects. We will be using lattices of $9 \times 9$ atoms throughout unless otherwise stated.

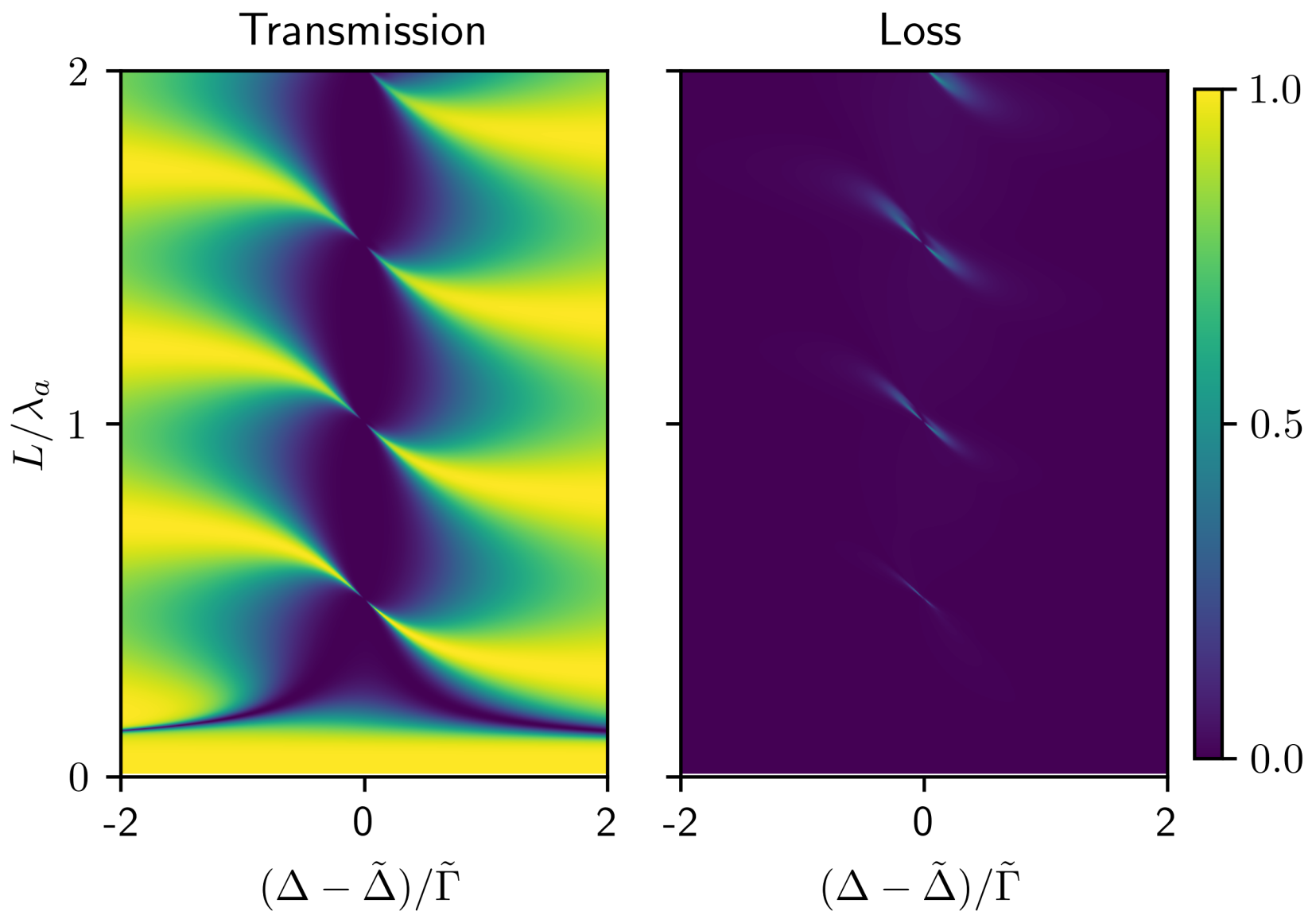

**Figure 2.9 | Transmission and loss,** $1 - T - R$**, for a flat dual array.** Each lattice has $9 \times 9$ atoms with lattice spacing $a = 0.6\lambda_a$, illuminated by a Gaussian beam with waist $w_0 = 1.5\lambda_a$.

### 2.5.4 Curved lattices

Firstly, to mitigate the effects of having a finite array, namely the scattering at the edges of the lattices, we curve these slightly to match the phase front of the Gaussian beam, as is done in Ref. [76]. That is, rather than consider planar lattices, we will shift the atoms relative to each other along the axis of incidence, $z$, while keeping the $xy$-coordinates the same. We will shift the atoms such that they lie along a surface of constant phase of the Gaussian beam, see bottom panel of Fig. 2.8. Intuitively, one can think that with planar arrays, the components of light, which are not at perfect normal incidence will be reflected back and forth towards the edges, and eventually be emitted out of the side of the array. With the curved lattices this light will be reflected back towards the centre of the array and thus not be lost. Alternatively, by matching the atoms to the Gaussian wave front, the Gaussian mode becomes an (at least approximate) spatial eigenmode of the system, and so is less likely to scatter into other modes.

Looking at Eq. (2.47), we see that choosing the $z$-coordinates of the atoms in the first array (i.e. the one with negative $z$-coordinates), such that

$$kz_n + k\frac{r_{n\perp}^2}{2R(z_n)} - \phi(z_n) = -kL/2 + k\frac{r_\perp^2}{2R(-L/2)} - \phi(-L/2) \ , \quad (2.49)$$

then the atom at $r_\perp = 0$ (assuming the number of atoms in each lattice is odd) will have $z_n = -L/2$, and all other atoms will be positioned such that the phase of the beam is the same. The second lattice is then determined by mirroring the first one. This is similar to how regular dielectric mirrors are curved when used to form optical cavities. The middle column of Fig. 2.10 shows the transmission and loss for the same parameters as in Fig. 2.9, but now using the curved lattices, and we see how the loss is significantly reduced. The curvature of the lattices is quite small, with the relative shift in the $z$-direction of the corner atoms being $\Delta z/(L/2) \sim 0.1$ for a lattice with the same parameters as in Fig. 2.9 at $L = 1.5\lambda_a$, and only $\sim 0.01$ if the beam waist is increased to $w_0 = 2.5\lambda_a$. The curved lattice used to calculate the data in the middle panel of Fig. 2.10 at $L = 1.5\lambda_a$ is shown in the bottom panel of Fig. 2.8 as grey dots, showing the small curvature. For a larger array, a broader beam could be used, further reducing the curvature. Similarly, if the lattices are large, but the beam size is kept relatively narrow, the effect of the edges is reduced and it should not be necessary to use curved lattices. This would be more easily implemented in an experimental context, and we use the curved arrays here only due to numerical restrictions to the size of arrays we can simulate.

### 2.5.5 Diverging collective energy shift

Looking at Fig. 2.10 we see that for $a$ approaching $\lambda_a$, the loss starts to increase dramatically. This is due to a particularly critical instance of the different momentum components of the mode being transmitted and reflected differently due to the momentum dependent collective energies. As discussed in Section 1.5.3, the collective energies diverge at momenta where a Bragg channel opens or closes. As plotted in Fig. 1.3, the circular contours where this takes place approach the centre of Brillouin zone for larger $a$. If we plot the contours of the Gaussian mode on top of these, Fig. 2.11, we see how it avoids the divergences for $a/\lambda_a = 0.4, 0.6,$ while for $a = 0.8\lambda_a$ a significant part of the Gaussian overlaps with them. This reflects the large increase in loss at this lattice spacing as seen in Fig. 2.10, and we conclude that the loss is indeed due to the vastly different transmission and reflection of the different momentum components of the Gaussian. Likewise, we see an increase of loss at $a = 0.4\lambda_a$, which is simply due to the array becoming so small that the beam starts to be significantly scattered by the edges.

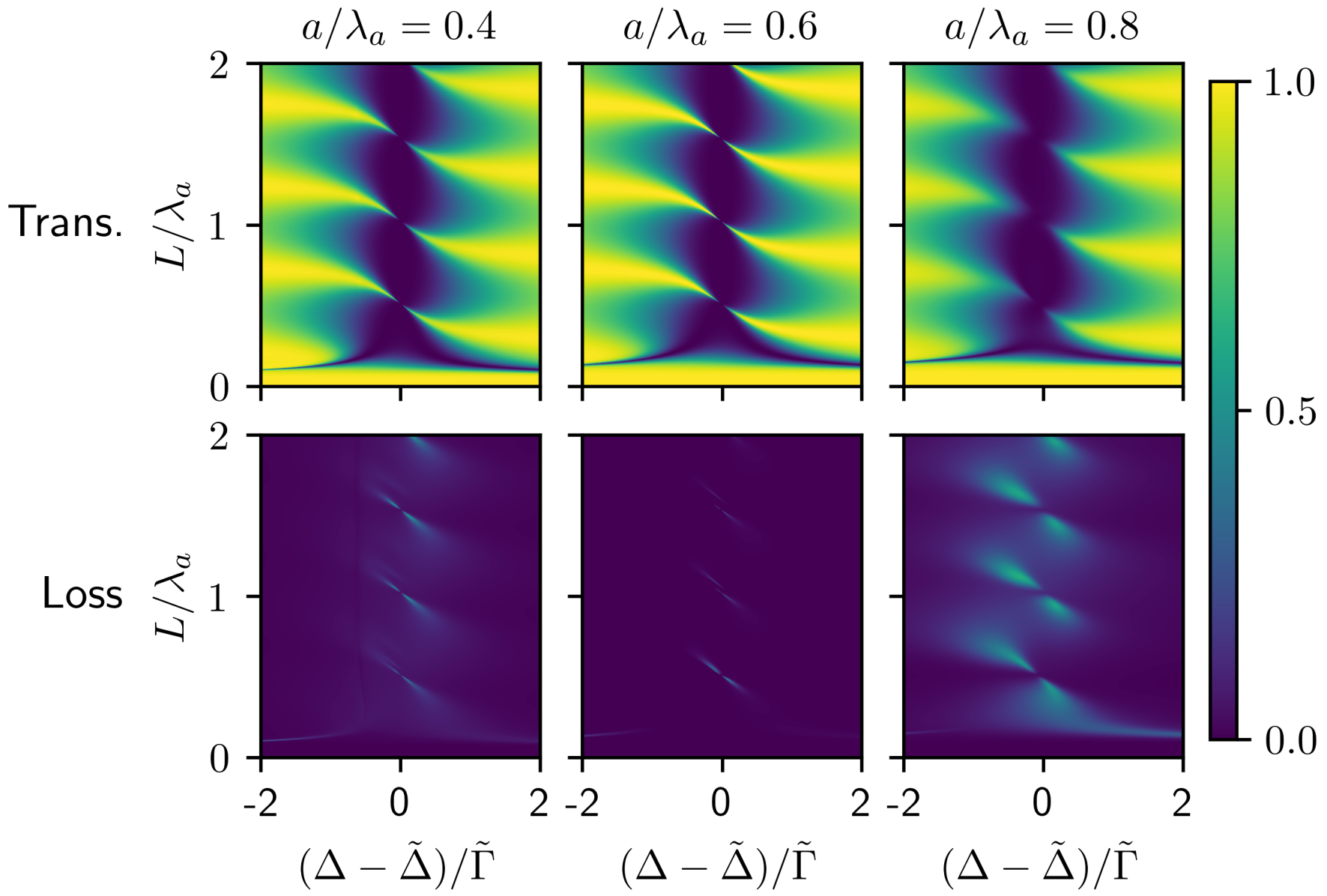

**Figure 2.10 | Transmission and loss for a curved dual array.** Same as Fig. 2.9, but for curved arrays, and different values of the lattice spacing $a$, as indicated.

Thus, when working with a finite beam of light, as we must with the finite arrays, we must choose the width of the beam and lattice spacing such that the mode does not spill over the edges and does not overlap with the divergences of the collective energies. A spatially broad beam corresponds to a narrow distribution in momentum space, and so we will generally want a broad beam, and a small lattice spacing. However, these two conditions push towards the beam becoming larger than the array, such that the number of atoms must also be large. Hence, an appropriate compromise between these two parameters must be found, as we are limited in the number of atoms we can simulate. With the code used to generate data for this thesis, it became impractical to simulate more than $9 \times 9$ atoms in each lattice (a total of 162 atoms), when calculating the nonlinear behaviour, and so we generally will be working with $a = 0.6\lambda_a$, as mentioned, where loss is low. This does mean, however, that there is Bragg scattering for large momenta, as we will see in the next chapter when we consider momentum correlations.

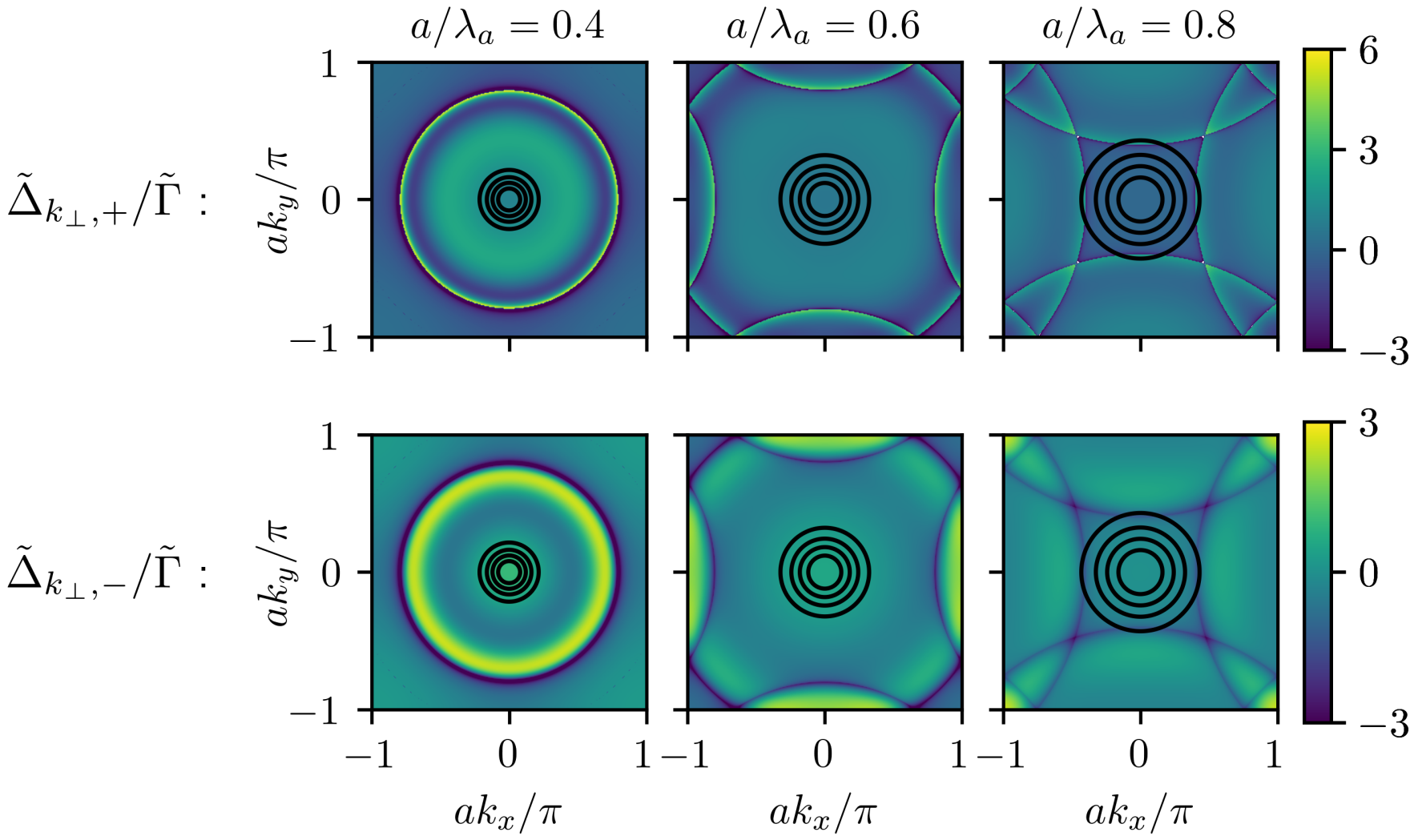

**Figure 2.11 | Even and odd collective energy shifts, $\tilde{\Delta}_{\boldsymbol{k}_\perp,\pm}$.** Same parameters as in Fig. 1.3. The contours of a Gaussian driving beam with $w_0 = 1.5\lambda_a$ are plotted in black.

# 3

# Double-excitation physics

In the final chapter of this part of the thesis we come to the analysis of the nonlinear behaviour of the atomic cavity. We will look at momentum and temporal correlations between two photons emitted from a finite dual array that is driven by a Gaussian beam of light. As we study the nonlinear physics using a numerical approach, we must use a finite array and beam due to computational limitations, as discussed in Section 2.5. We focus on driving the system at the narrow transmission resonances, where the photon delay time is extremely long.

Before starting the study at hand we will again briefly review some of the existing literature regarding nonlinear optics, focussing on the use of regular arrays of emitters. As mentioned, and as we will see in this chapter, the atomic mirror is mostly linear. A popular approach to reintroduce a nonlinear element to the atomic arrays is to include Rydberg levels (i.e. highly excited levels) in the atoms, in particular in conjunction with electromagnetically induced transparency (EIT) [14]. EIT, briefly put, is a phenomenon where an otherwise non-transparent medium, consisting of three-level systems, can be turned fully transparent by an external driving that causes a destructive interference between different excitation pathways (at certain conditions) resulting in no photon being absorbed on resonance, but rather being fully transmitted. The strong and long-range interaction among Rydberg atoms [86] means that if one atom gets excited to the Rydberg level, the Rydberg level of all other atoms within a certain radius gets detuned, such that no additional Rydberg excitations occur, the famous Rydberg blockade [87, 88]. At the same time this breaks the EIT condition, turning the medium within the blockade radius from transmissive to opaque. This constitutes a strong nonlinearity as it only requires the absorption of a single photon to drastically change the response of the medium (very similar to the saturation of a single two-level system) and has been used in the context of arrays [49, 89–93], and to a great extent also in atomic gasses [94, 95]. Hence, one can imagine a situation where a single photon is always transmitted, but any additional photons are not, resulting in the transmitted light consisting of well-separated individual photons [89–91, 96, 97].

Another direction of studies regarding the nonlinear behaviour of ordered atomic arrays concerns the phenomenon of superradiant bursts [98]. Here, the array is fully inverted (i.e. all atoms are excited), and the subsequent decay of the system shows an initial "burst" of radiation. That is, the emitted light starts at a certain intensity, but then increases strongly for a short time, before decreasing again as the atoms shed their energy. This is counter to the more intuitive behaviour of individual emitters where the intensity decays exponentially from the start. This so-called Dicke superradiance (from Dicke's original work concerning a gas of atoms in a point [99]) has also been observed in many other optical systems (see Ref. [98] and references therein). The phenomenon comes about as a collective effect, as the emitters synchronize their phase (becoming correlated) during the initial decay and subsequently a cascade of stimulated emission results in the concentrated burst of radiation. The phenomenon has been found in arrays of different dimensionalities and geometries [98, 100, 101], described using different methods of circumventing the exponential complexity of such a many-body effect [31, 98, 102], and using atoms with a more realistic level structure [103].

Some work has also been done to generate correlations among photons using just the single array [104–107]. While extended arrays of many atoms are generally mostly linear (due to the loss of saturability), these papers consider arrays of only a few atoms at small lattice spacings with an overall size of the arrays comparable to or smaller than a single wavelength of the resonant light. Furthermore, when calculating photon correlation functions, they only consider the light directly emitted from the atoms, rather than the full field consisting of the driving light and the atomic contribution. This results in a response similar to a single atom, which is also capable of producing highly anticorrelated light (see Fig. 3.6), but with only a very small intensity in the nonlinear signal due to its weak coupling to light. Hence, these proposals indeed generate nonlinearities with the single array, but the generated nonlinear signal might be very weak, and the small size of the arrays might make it experimentally challenging to achieve the exact configuration of atoms. The low number of atoms also means that the system does not show the properties of collective quasimomentum modes of a translationally invariant array, but rather simply has some interesting or useful states in a very finite Hilbert space. In contrast to this, the work at hand considers extended arrays of many atoms, exploiting the collective quasimomentum modes, and we will see how transverse momentum is conserved for interacting photons.

In Ref. [43], however, they do succeed in exploiting the nonlinearity of extended one-dimensional arrays. The nonlinear shift of doubly excited states, as well as an increased decay rate of doubly excited state, allows for the coherent preparation of single-excitation subradiant states. This is used to implement a two-qubit gate in a system consisting of two copies of such one-dimensional arrays. Their results work best for a very tight lattice, but are shown to also extend to a lattice spacing of a quarter wavelength.

Finally, let us mention some works that deal with systems similar to the dual array. In Ref. [108] they consider a toy model of the nonlinearity of the single array in front of a classical mirror, finding correlations similar to the ones we will study below. However, they do not derive these directly from the full system model, and do not discuss whether the used parameters are consistent with the capabilities of the full system model. Here, we study the nonlinearity of the system at hand directly. Refs. [39, 40] consider two or three two-level systems in a one-dimensional waveguide, finding a transmission spectrum qualitatively very similar to the one we saw in the last chapter, and correlations, which are again similar to the ones we will study below. Likewise, Ref. [109] shows experimental results regarding a superconducting circuit, which implements artificial atoms in a waveguide, with pairs of atoms constituting atomic mirrors, forming a cavity in which there is another (probe) atom. They mostly consider linear properties, finding similar physics to what we have studied, and briefly consider nonlinear transmission and decay of the system.

Here we will see how the long delay time of the atomic cavity indeed results in very strong and long-lasting correlations. Studying these correlations we will be able to get an intuitive picture of photon pairs scattering off of each other, exiting the system highly correlated and in a different transverse mode than the original Gaussian. Thus, the transmitted Gaussian mode consists mainly of transmitted single photons, while bunched pairs of photons can be found in a higher mode.

In Section 3.1 we define the momentum and temporal correlators that we will employ. Section 3.2 describes briefly the overall numerical approach we have taken to calculate the correlators based on performing time-evolution of the system via non-unitary quantum trajectory wave function simulation. We then go on to study the momentum correlations in Section 3.3, finding photon-photon scattering under total momentum-conservation, resulting in strong correlations which, however, vanish for a very short time-delay between photon detections. Section 3.4 further studies the temporal correlation, finding that photons exit the system in pairs or as well-separated singles. We relate the shape of the temporal correlation function to the populations of the two dominant single-excitation states after the detection of the first photon, and again relate these states to the super- and subradiant collective modes of the system.

## 3.1 Photon correlations

Let us first define the correlators we will be considering. We will look both at correlations between the quasimomentum modes, revealing the photon-photon scattering and the shape of the correlated light, and correlations in time, revealing the statistics of the emitted photons.

### 3.1.1 Two-photon momentum density

As mentioned we will consider correlations among different momentum components of the emitted light, and temporal correlations between photons in the Gaussian mode that is also used to drive the system. The momentum correlations are studied using the two-photon momentum density[1]

$$\rho(\boldsymbol{k}_{1\perp}, \boldsymbol{k}_{2\perp}, t) = \left\langle E^\dagger(\boldsymbol{k}_{1\perp}, t')E^\dagger(\boldsymbol{k}_{2\perp}, t'+t)E(\boldsymbol{k}_{2\perp}, t'+t)E(\boldsymbol{k}_{1\perp}, t')\right\rangle \ . \tag{3.1}$$

Here, the expectation value is with respect to the system ground state, and $t'$ is any time large enough for the system to be in the steady state, such that $t$ is the difference in detection time of the two photons. The density can be calculated for both transmitted and reflected photons, by using either Eq. (2.11a) or Eq. (2.11b) for $E(\boldsymbol{k}_\perp, t)$. This expectation value can be understood as the squared norm of the state, which is found by time-evolving the ground state (the state with no excitations in the array) to the steady state, detecting an emitted photon with transverse momentum $\boldsymbol{k}_{1\perp}$, and then detecting another photon a time $t$ later with transverse momentum $\boldsymbol{k}_{2\perp}$. Thus, the correlator tells us the likelihood of detecting two photons with certain transverse momenta separated by

1: The variable $t$ now represents time, rather than the atomic cavity transmission amplitude.

a certain amount of time. We will mostly be considering the same-time momentum correlations, i.e. $t = 0$.

### 3.1.2 Two-time correlation function

To study the temporal correlations we will use the two-photon two-time correlation function

$$g^{(2)}(t) = \frac{\left\langle E^{\dagger}_{\boldsymbol{f}_G}(t')E^{\dagger}_{\boldsymbol{f}_G}(t'+t)E_{\boldsymbol{f}_G}(t'+t)E_{\boldsymbol{f}_G}(t')\right\rangle}{\left\langle E^{\dagger}_{\boldsymbol{f}_G}(t')E_{\boldsymbol{f}_G}(t')\right\rangle \left\langle E^{\dagger}_{\boldsymbol{f}_G}(t'+t)E_{\boldsymbol{f}_G}(t'+t)\right\rangle} \ , \tag{3.2}$$

where again the expectation value is with respect to the ground state, and $t'$ is a steady state time. Here $E_{\boldsymbol{f}_G}(t)$ is given by Eq. (1.65) with $\boldsymbol{f}_G$ being the Gaussian mode described in Section 2.5.2. As with the momentum density the correlation function tells us a likelihood of a series of detection events, this time detecting two photons in the Gaussian mode separated by a time $t$. The correlation function is normalized such that no correlation corresponds to unity, as the expectation values in the numerator splits into a product, when the two detection events do not affect each other. A value smaller than unity at some time then indicates that it is less likely to observe the photons separated by that time, while a value greater than unity indicates it is more likely. In particular, we can speak of bunched or antibunched photons depending on whether $g^{(2)}(0)$ is greater or smaller than its later values, giving the intuitive picture of photons preferring to be emitted from the system in groups or separately.

## 3.2 Numerical approach

Before moving on to the results of calculating the correlators, we will briefly go through the numerical approach used to perform these calculations [110]. The time-evolution is based on the Monte-Carlo wave function approach [111]. Here, a non-Hermitian Hamiltonian is defined from the original Hermitian Hamiltonian and the Lindbladian, by including in the Hamiltonian the anticommutator terms from the Lindbladian. In our case we get from Eqs. (1.26) and (1.30), the non-Hermitian Hamiltonian

$$\begin{aligned} H = &-\sum_n \Delta\sigma_n^{\dagger}\sigma_n - \sum_n \left(\Omega_n\sigma_n^{\dagger} + \Omega_n^{*}\sigma_n\right) \\ &-\sum_{\substack{n,m \\ n\neq m}} J_{nm}\sigma_n^{\dagger}\sigma_m - i\sum_{n,m}\Gamma_{nm}\sigma_n^{\dagger}\sigma_m \ , \end{aligned} \tag{3.3}$$

where again $\Omega_n = \boldsymbol{d}^\dagger \boldsymbol{E}_{\text{in}}(\boldsymbol{r}_n)$. This describes the coherent dynamics and the continuous loss of population to the surrounding environment (photons emitted away from the array in our case). However, it lacks the first term of Eq. (1.30), which describes quantum jumps between states. These jumps are included via the Monte-Carlo algorithm.

A state $|\psi\rangle$ is then time-evolved in small time steps $\delta t$ by a series of operations. First, we write

$$\left|\psi'(t+\delta t)\right\rangle = (1 - iH\delta t)\left|\psi(t)\right\rangle \quad , \tag{3.4}$$

which is acting on the state with the time-evolution operator $e^{-iH\delta t}$, expanded to first order in $\delta t$. As $H$ is not Hermitian, this state is not normalized, and we write

$$\langle\psi'(t+\delta t)|\psi'(t+\delta t)\rangle = 1 - \delta p \quad . \tag{3.5}$$

A quick calculation reveals that

$$\delta p = 2\sum_{n,m} \Gamma_{nm} \langle\psi(t)|\sigma_n^\dagger \sigma_m|\psi(t)\rangle \delta t \quad . \tag{3.6}$$

The next operation involves picking a random number $\epsilon$ uniformly distributed between 0 and 1, and if $\delta p \leq \epsilon$, we simply take the time-evolved state to be

$$|\psi(t+\delta t)\rangle = \frac{|\psi'(t+\delta t)\rangle}{\sqrt{1-\delta p}} \quad , \tag{3.7}$$

but if $\delta p > \epsilon$ a quantum jump is performed. However, as discussed below, we will be performing the time-evolution without employing jumps, so we will not detail how they are performed.

A central part of the approach is to use very weak driving. While the analytical calculations of the linear physics of the previous chapter did not need to specify the driving strength,, for the numerics we need to specify some specific value. By picking a very small driving amplitude, $E_{\text{in}}$, in fact considering the limit of zero driving strength, we achieve three important simplifications.

Firstly, we can indeed neglect quantum jumps. Generally we can say that the amplitude of single-excited state will be proportional to $E_{\text{in}}$, double-excited state amplitudes will be proportional to $E_{\text{in}}^2$, etc. As described above a jump takes place when the deficit in the norm of the time-evolved state, $\delta p$, is larger than some random number. However, looking at Eq. (3.6), we see that $\delta p$ must be proportional to $E_{\text{in}}^2$, as only excited state components of $|\psi(t)\rangle$ contribute to $\sigma_m |\psi(t)\rangle$. Therefore the likelihood of a quantum jump vanishes with vanishing $E_{\text{in}}$, and we do not include them in the numerics. It would then be more appropriate to call our approach to time-evolution quantum trajectory wave function simulation, as we no longer include the defining random element of Monte Carlo simulation.

Secondly, we can truncate the Hilbert space of the system at some total number of excitations. If the amplitude of states with $n$ excitations scales as $E_{\text{in}}^n$, the effect of highly-excited states is negligible. Indeed, for vanishing $E_{\text{in}}$, we can neglect all excited states that do not contribute at the lowest order to the quantities we wish to calculate. We therefore truncate the Hilbert space at two excitations, meaning we only include states with zero, one, or two excitations in our numerical simulation. Thus, we reduce the Hilbert space from being exponentially large in the number of atoms involved to being polynomially large (specifically going as the square of the number of atoms).

Thirdly, for vanishing $E_{\text{in}}$, the value of $g^{(2)}$ converges to its lowest order contribution, which is independent of $E_{\text{in}}$ (such that the actual value of $E_{\text{in}}$ becomes unimportant). The reason we truncate at two excitations is that the lowest order contribution to two-photon correlators like the momentum density $\rho$ and the two-time correlation function $g^{(2)}$ come from the two-excitation states. Hence, the contribution from any more highly excited state would be smaller by at least a factor of $E_{\text{in}}$. In particular, since $g^{(2)}$ is normalized the way it is, it will in fact converge for vanishing $E_{\text{in}}$, and that is the value we will be interested in. The momentum density $\rho$ is not normalized and so, of course, would vanish in that limit. However, its structure and its value in comparison to other calculations of $\rho$ for different choices of system parameters, would converge. We will therefore not be concerned with the specific value of $\rho$, but only its relative value.

The numerical simulation proceeds with a simple step-by-step approach to calculating expectation values of interest. For example, the numerator of $g^{(2)}$ can be written as

$$\left\langle E^\dagger(t')E^\dagger(t'+t)E(t'+t)E(t')\right\rangle = \langle 0|U^\dagger(t')E^\dagger U^\dagger(t)E^\dagger E U(t) E U(t')|0\rangle \ , \tag{3.8}$$

where $U(t) = e^{-iHt}$ is the time-evolution operator. We thus see how calculating this expectation value can be done by first time-evolving $|0\rangle$ for a time $t'$, then acting on it with the E-field operator, time-evolve the result for another time $t$, and finally acting with the E-field operator

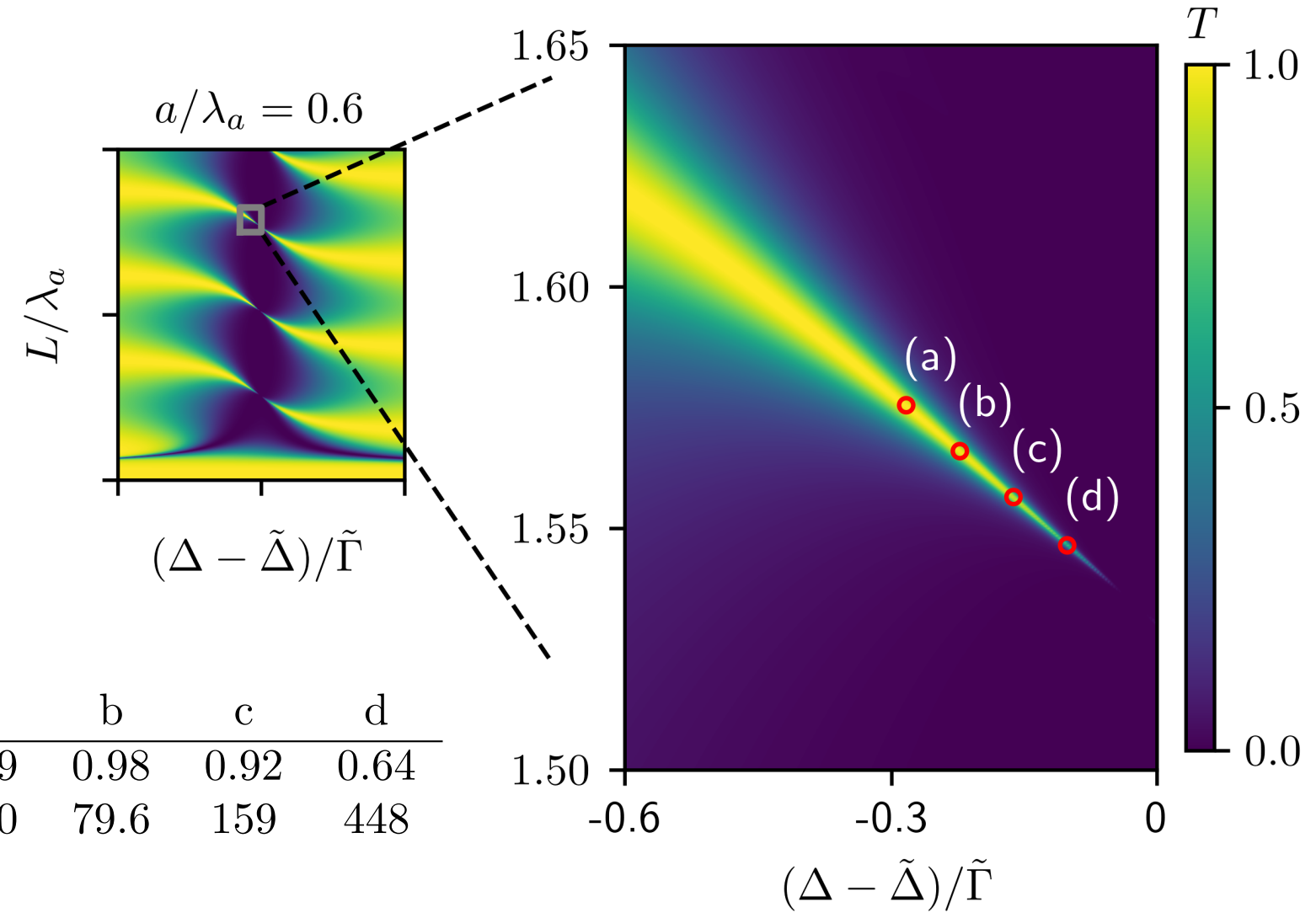

| | a | b | c | d |
|---|---|---|---|---|
| $T$ | 0.99 | 0.98 | 0.92 | 0.64 |
| $\gamma\tau$ | 47.0 | 79.6 | 159 | 448 |

**Figure 3.1 | Selected parameter points for correlation calculations.** The right panel shows a transmission ridge from the middle panel of Fig. 2.10 (here reprinted as the left panel), and we mark four peaks (red circles). We will be focussing on the corresponding four sets of parameter choices when calculating correlators. The values of the transmission, and delay time for the four points are shown in the table. The delay time corresponds to the reciprocal of the width of the resonances.

again. The squared norm of the resulting state would be equal to the desired expectation value.

## 3.3 Momentum density

We consider a dual array with $9 \times 9$ atoms in each lattice with a lattice spacing of $a = 0.6\lambda_a$ illuminated by a Gaussian beam with waist $w_0 = 1.5\lambda_a$. Figure 3.1 shows a specific ridge of transmission resonances marked with a few choices of $(\Delta, L)$, which we will use to calculate $\rho$ (and also $g^{(2)}$ in the next section). The table shows the transmission, and delay time (calculated numerically via Eq. (2.37)), the latter of which gives the reciprocal of the width of the transmission resonances. We can see how the transmission starts to drop rapidly as the resonances become narrow, where the rest of the light is mostly lost to other modes than the Gaussian (the reflection coefficient also increases, but is smaller than the loss). Even with the relatively small $9 \times 9$ lattices we still achieve good delay times before the loss becomes large.

### 3.3.1 Momentum-dependent interaction

Using the four chosen sets of system parameters, Fig. 3.2 shows the same-time momentum density $\rho(\boldsymbol{k}_{1\perp}, \boldsymbol{k}_{2\perp}, 0)$ as function of the $y$-components of $\boldsymbol{k}_{1\perp}$ and $\boldsymbol{k}_{2\perp}$ with the $x$-components fixed at zero. That is, we consider light propagating parallel to the $x$-axis. The density is high in two separate parts: firstly a bar along the antidiagonal with some variation along it, and secondly in the diagonal corners. The latter are due to Bragg scattering. As we consider $a = 0.6\lambda_a > \lambda_a/2$ there is Bragg scattering for the largest momenta, and so we see a signal at these high values of the momenta, as low momentum photons are Bragg-scattered into them. The bar along the antidiagonal corresponds to photons scattering and exchanging momentum, but with the total momentum conserved. Thus, as the driving Gaussian beam is centred on zero momentum with a width in momentum space that goes as $1/w_0$, we have high density along $k_{1,y} + k_{2,y} \sim 0$, and this bar in the momentum density indeed has a width of $\sim 1/w_0$. The variation along the bar corresponds to the scattering of photons having a momentum dependence, i.e. some momenta are preferred over others in the scattering processes. In other words, the photons of the system have a spatially dependent effective interaction.

We see how the narrower transmission resonances results in a higher density along this bar, i.e. the interaction among photons is stronger, and how the central Gaussian, which corresponds to the uncorrelated (linearly) transmitted light, diminishes at the same time. Thus, the more narrow resonances indeed result in stronger correlations and interactions among photons as anticipated, while also scattering photons out of the incoming Gaussian mode.

Finally, we note that the high value of the density at the edges of the plots are due to the Green's function in Eq. (1.56) diverging as $1/k_{z,k}$, as can be seen in Eq. (1.40). This divergence seems unphysical and may be due to the choice of polarization for the detected light, right-circular, which is only a physically correct polarization for light with $\boldsymbol{k}_\perp = 0$ (as the polarization must be orthogonal). This issue has not been solved here, but the results should still give the correct physical picture, though the divergence for large $k_\perp$ is unphysical. Understanding the Green's function $1/k_z$-dependence, and how to avoid the resulting divergence in the calculated fields is an obvious direction for future work. In the next part of the thesis, where we derive an analytical expression for the momentum density, we will have a better understanding of the origin of the exact structure of it.

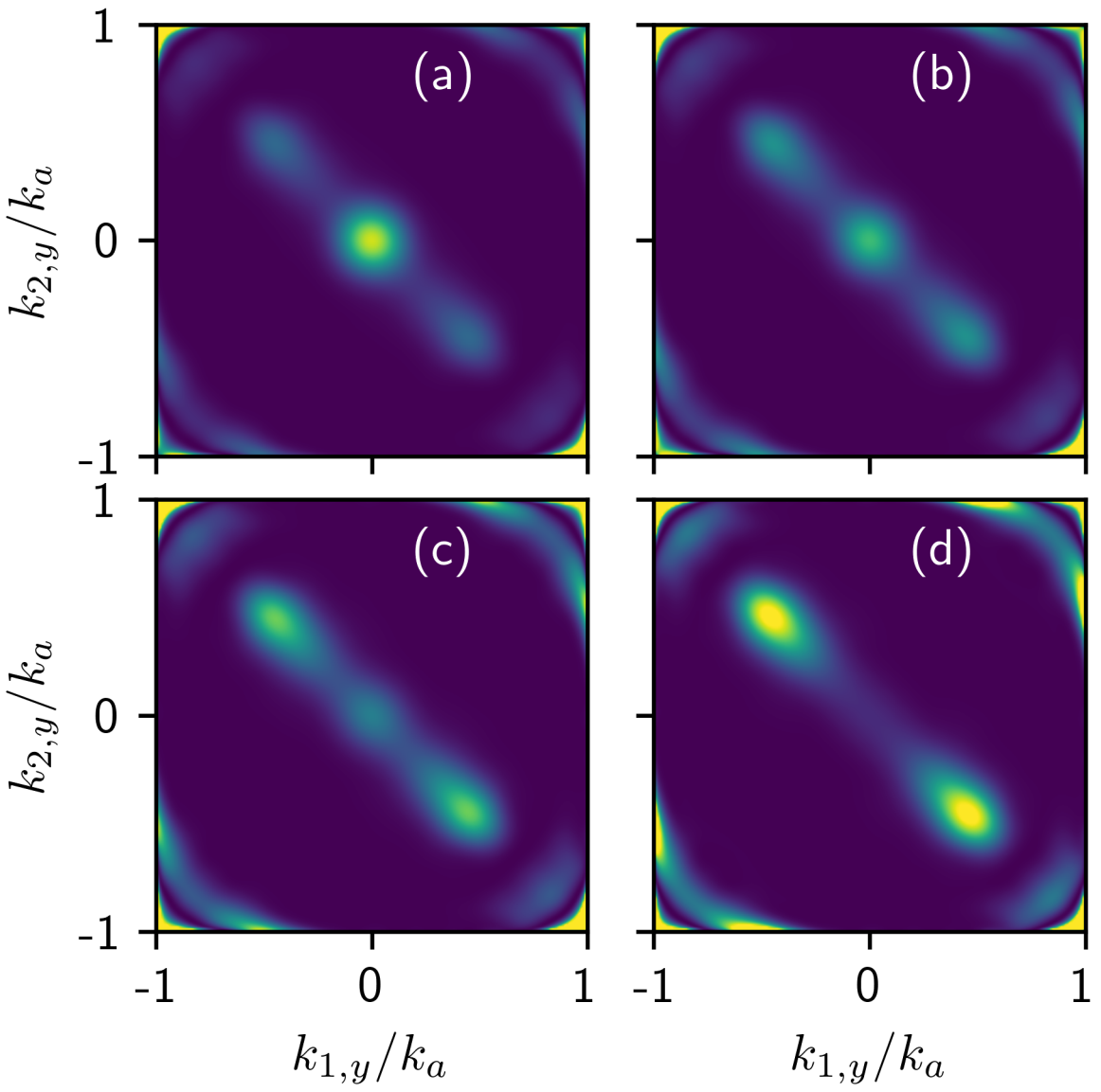

**Figure 3.2 | Momentum density $\rho$ at the four points indicated in Fig. 3.1.** The $x$-components of the two photon momenta have been fixed at zero, $k_{1,x} = k_{2,x} = 0$, and the density is plotted as function of the $y$-components.

### 3.3.2 The outgoing mode

To get a better understanding of the outgoing mode of the atomic cavity, we plot first $\rho(\boldsymbol{k}_{1\perp}, \boldsymbol{k}_{2\perp}, 0)$ as a function of $\boldsymbol{k}_{2\perp}$ with $\boldsymbol{k}_{1\perp}$ fixed at two arbitrary values in the first two panels of Fig. 3.3. We see how the density is peaked at and concentrated around $\boldsymbol{k}_{2\perp} = -\boldsymbol{k}_{1\perp}$ with a distribution that has the same width as the incoming Gaussian beam. There are some peaks near the border, $k_{2\perp} = k_a$, that are due to a mixture of Bragg scattering and the mentioned $1/k_{z,k_a}$ divergence. We then plot $\rho(\boldsymbol{k}_\perp, -\boldsymbol{k}_\perp, 0)$ in the third panel, i.e. the momentum density for $\boldsymbol{k}_{2\perp} = -\boldsymbol{k}_{1\perp} = \boldsymbol{k}_\perp$, to see the relative peak density for different values of the fixed momentum. The density is zero outside of $k_{1\perp}^2 + k_{2\perp}^2 = k_a^2$, as we only consider propagating momenta. The plot shows that the emitted light is structured into two rings with some fourfold-symmetric structure on top, originating in the square atomic lattice. The outermost ring of high density is due to the $1/k_{z,k_a}$ divergence and should be ignored. It thus appears that the emitted light is mainly found in some higher Laguerre-Gaussian modes, with noticeably little light in the central Gaussian mode. Calculating the overlap of the emitted light with such higher modes, or calculating correlations between the light components in these modes, could reveal in greater detail the exact mode structure of the emitted light.

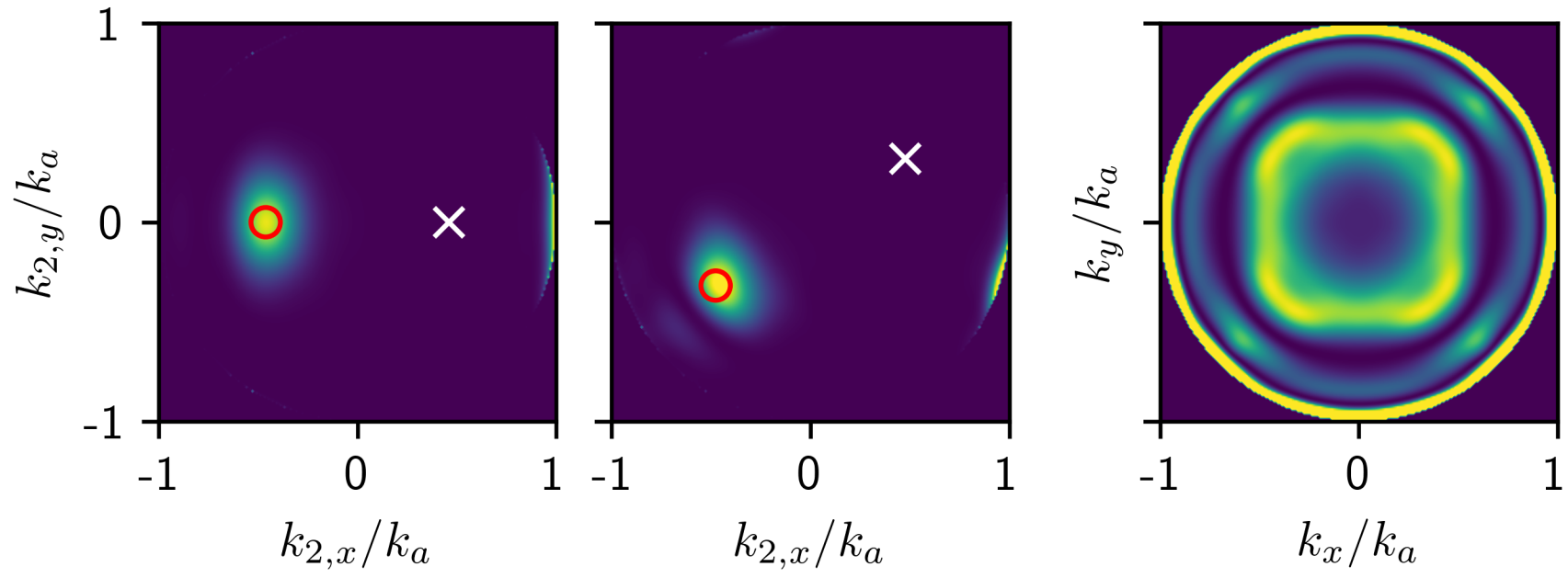

**Figure 3.3 | Momentum density for fixed $\boldsymbol{k}_{1\perp}$, and for $\boldsymbol{k}_{2\perp} = -\boldsymbol{k}_{1\perp} = \boldsymbol{k}_{\perp}$.** We plot two arbitrary examples of $\rho$ with $\boldsymbol{k}_{1\perp}$ fixed at the values indicated with a white cross, while at the same time marking $-\boldsymbol{k}_{1\perp}$ with a red circle. The final panel shows $\rho(\boldsymbol{k}_\perp, -\boldsymbol{k}_\perp, 0)$. The system parameters are as in Fig. 3.2d.

### 3.3.3 Comparison with the atomic mirror

The upper row of Fig. 3.4 shows $\rho$ for the same system parameters as Fig. 3.2d, but now also for the reflected light, and the lower row of Fig. 3.4 likewise shows $\rho$ for the transmitted and reflected light at a reflection resonance of the single array. We see that the single array also has similar and strong momentum correlations for its transmitted light, while its reflected light is completely dominated by the uncorrelated light corresponding to the incoming beam. It makes sense that the transmitted light is highly correlated, as $\rho$ has been calculated on a reflection resonance, such that photons are only transmitted if they have interacted with one another, resulting in an exchange of momentum. However, to make this correlation visible in comparison with the uncorrelated reflected light, we have multiplied the transmitted momentum density by a factor of 500. As light is emitted symmetrically from the array, the reflected light in fact also has a highly correlated structure that is not visible in the plot. Thus, the correlated signal is very small. In other words, while the transmitted light does show clear momentum correlations (and the reflected in fact also does), it is only a very small fraction of the total light, which is actually correlated. The atomic cavity, on the other hand, shows correlations in both the transmitted and reflected light, and with equal signal strength. Hence, we conclude that while the atomic mirror of course can generate correlations, the atomic cavity does so much more effectively. It is also interesting to note that the correlated light from the atomic mirror shows much less structure in the antidiagonal bar, such that the scattering of photons is closer to being isotropic, while the atomic cavity shows signs of a preferred direction of scattering. We will discuss this further when we derive exact results in the next part of the thesis.

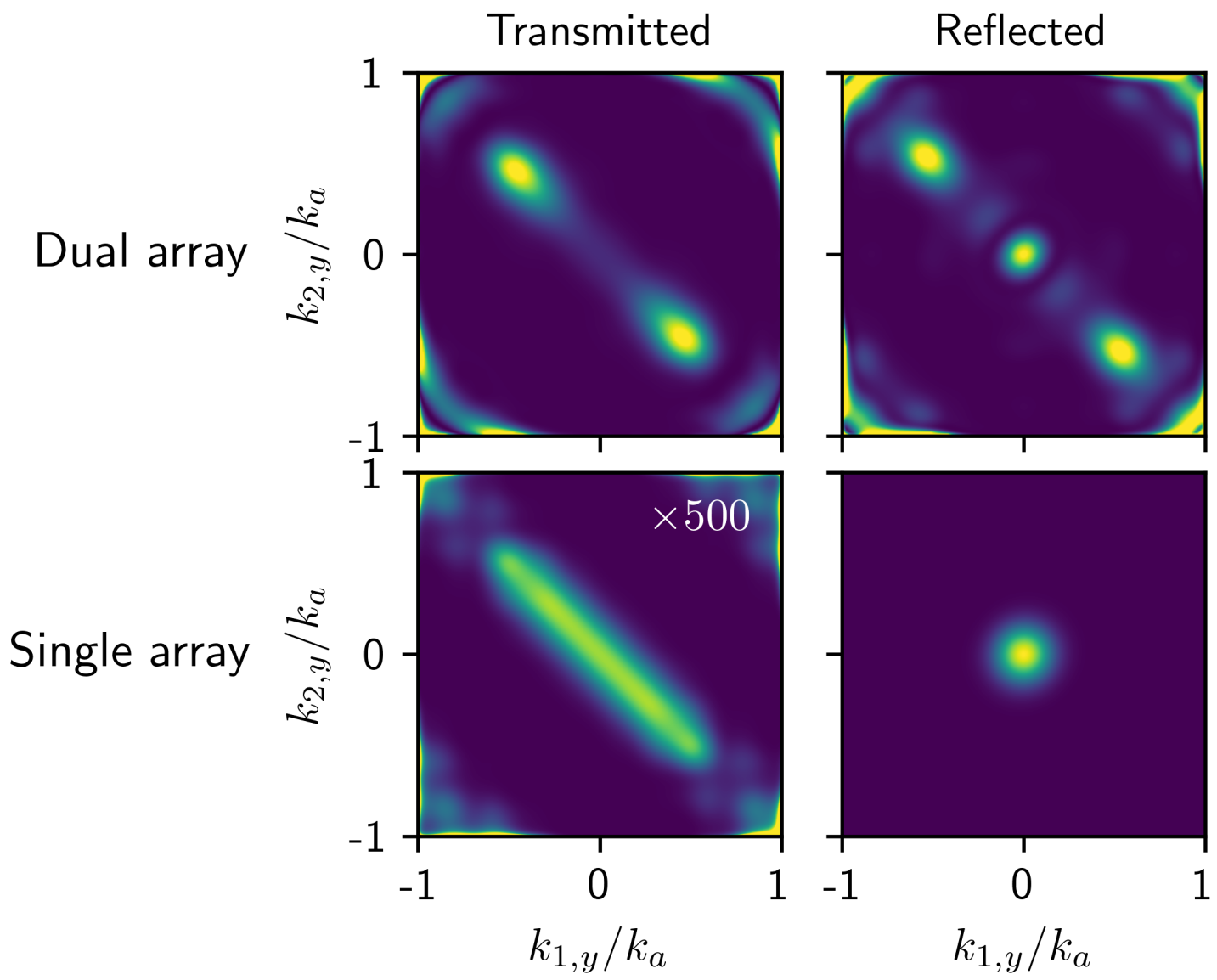

**Figure 3.4 | Comparison of the momentum correlation of the single and dual arrays.** The upper row shows $\rho(\boldsymbol{k}_{1\perp}, \boldsymbol{k}_{2\perp}, 0)$ with $k_{1,x} = k_{2,x} = 0$ for both the transmitted (first column) and reflected (second column) light with the same parameters as in Fig. 3.2d. The lower row shows the same, but for the single array, driven at a reflection resonance (with $a = 0.6\lambda_a$ and $15 \times 15$ atoms). The single array transmitted light density has been multiplied with a factor of $500$ to make it visible in comparison with the reflected light density.

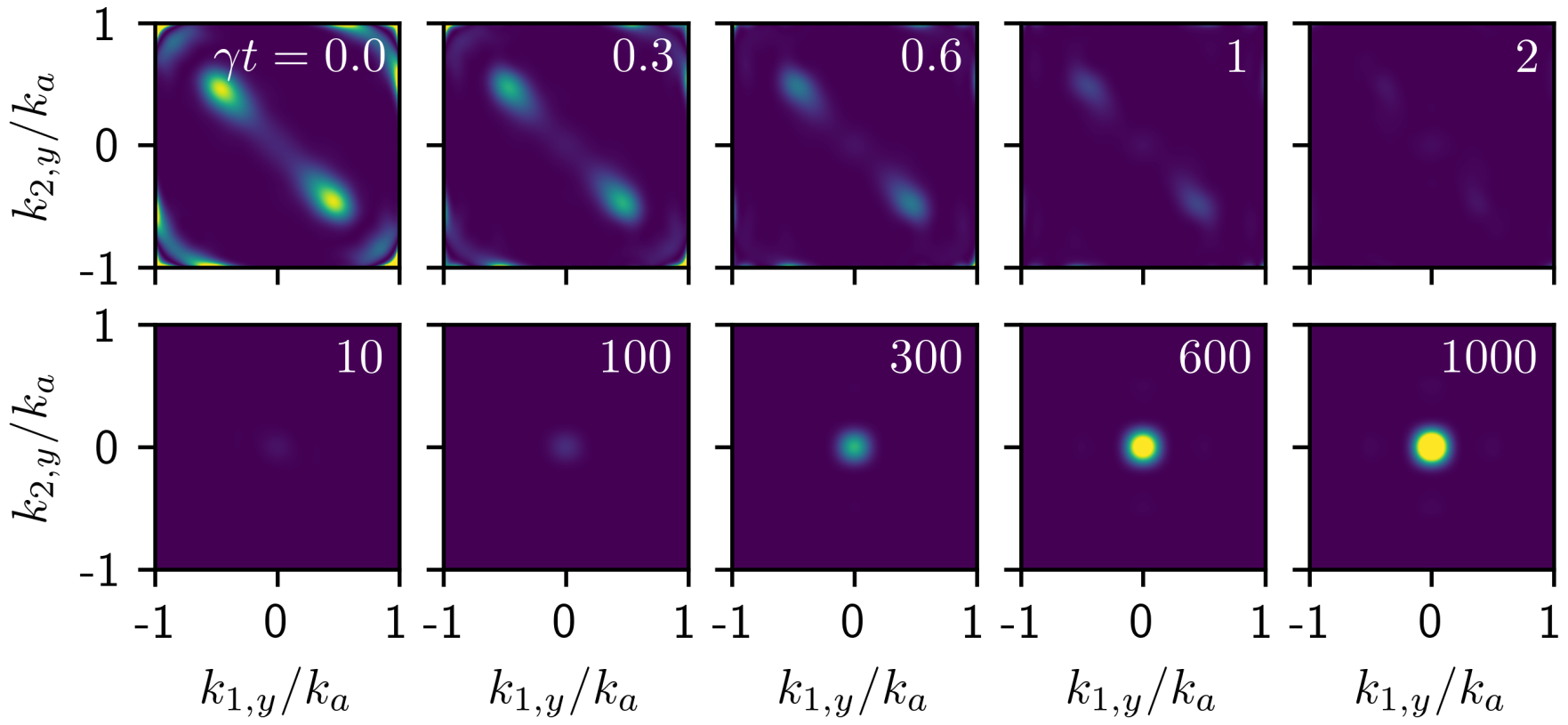

**Figure 3.5 | Momentum density at different times.** The panels show $\rho(\boldsymbol{k}_{1\perp}, \boldsymbol{k}_{2\perp}, t)$, with the same parameters as in Fig. 3.2d, but at different times, $t$, as indicated in upper right corner of each panel.

### 3.3.4 Momentum density with time delay

Figure 3.5 again shows $\rho(\boldsymbol{k}_{1\perp}, \boldsymbol{k}_{2\perp}, t)$ for the system parameters pertaining to Fig. 3.2d, but now for different times as indicated in the upper right corners of each panel. We can see how, within a very short timescale $\sim \gamma^{-1}$, the overall value of the density decreases and the correlations vanish, being replaced by the uncorrelated Gaussian signal, which then increases its strength over a very long time scale $\sim 100\gamma^{-1}$. To understand this, we move on to consider the temporal correlations of the system.

## 3.4 Temporal correlator

### 3.4.1 Atomic mirror and single atom

single

To use as a basis of comparison we start by considering the $g^{(2)}$ correlation function for the light emitted by a weakly driven single atom and for the light reflected off the single array, with the driving tuned to a reflection resonance. This is plotted in Fig. 3.6 in dark blue for the single atom and green for the single array. For a free two-level atom, the light emitted by the atom is perfectly antibunched ($g^{(2)}(0) = 0$), as the atom is saturated by a single photon, and cannot emit more than one photon at a time. This anticorrelation decays exponentially with the lifetime of the atom. Specifically, it can be found [80, 112] that $g^{(2)}(t) = (1 - e^{-\gamma t})^2$ for a weakly driven atom. For the single array, there is weak antibunching, and it decays within approximately the lifetime of a free atom. We can find numerically that the decay rate of $g^{(2)}$ is approximately the decay rate of the zero-momentum collective state. For the single array the collective decay rates, $\tilde{\Gamma}^0_{\boldsymbol{k}_\perp}$, are indeed on the order of $\gamma$ (see Fig. 1.3 and confer $\tilde{\Gamma} = 3\pi\gamma/a^2k_a^2 \simeq 0.66\gamma$ for $a = 0.6\lambda_a$). Later, we will similarly find that the time scales of the correlations of the dual array originate in the lifetimes of the near-zero momentum collective states and elaborate on the physics behind this there. Thus, the correlation generated by the atomic mirror is indeed weaker than that of the single atom and does not last for much longer.

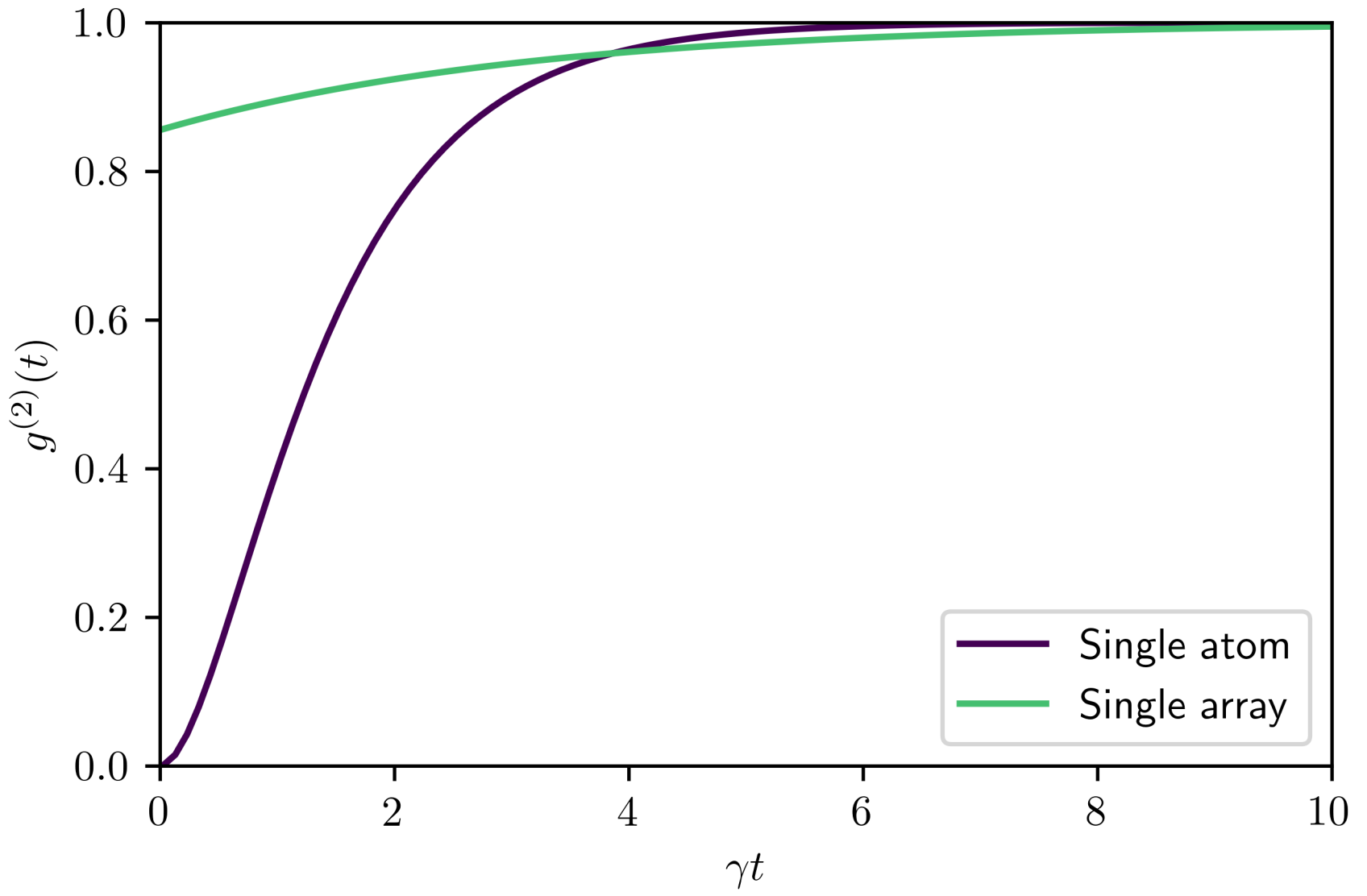

**Figure 3.6 | Temporal correlation of the atomic mirror and a single atom.** $g^{(2)}(t)$ is calculated and plotted for a single array of $9 \times 9$ atoms with lattice spacing $a = 0.6\lambda_a$, driven by a Gaussian beam with waist $w_0 = 1.5\lambda_a$ (green curve), and for a single weakly driven atom (dark blue curve).

As discussed previously, we can attribute this to the collective behaviour of the atomic mirror. Namely that while the individual atoms are saturated by a single photon, the array can absorb multiple photons and so becomes more linear, although the lifetime of the collective states can be slightly longer than that of the single atom. What the normalized correlator does not reveal of the single atom, of course, is that the highly correlated light from it would be very weak. The atom only interacts weakly with light and so the signal, though strongly correlated, would be weak. The atomic mirror, on the other hand, has a strong interaction (we have seen it is capable of perfect reflection for example), but this comes at the price of losing the nonlinearity (the antibunching is weak, and we likewise saw in Fig. 3.4 that the correlated signal is weak). This is rectified with the atomic cavity, which has both a strong interaction with light and shows strong correlations.

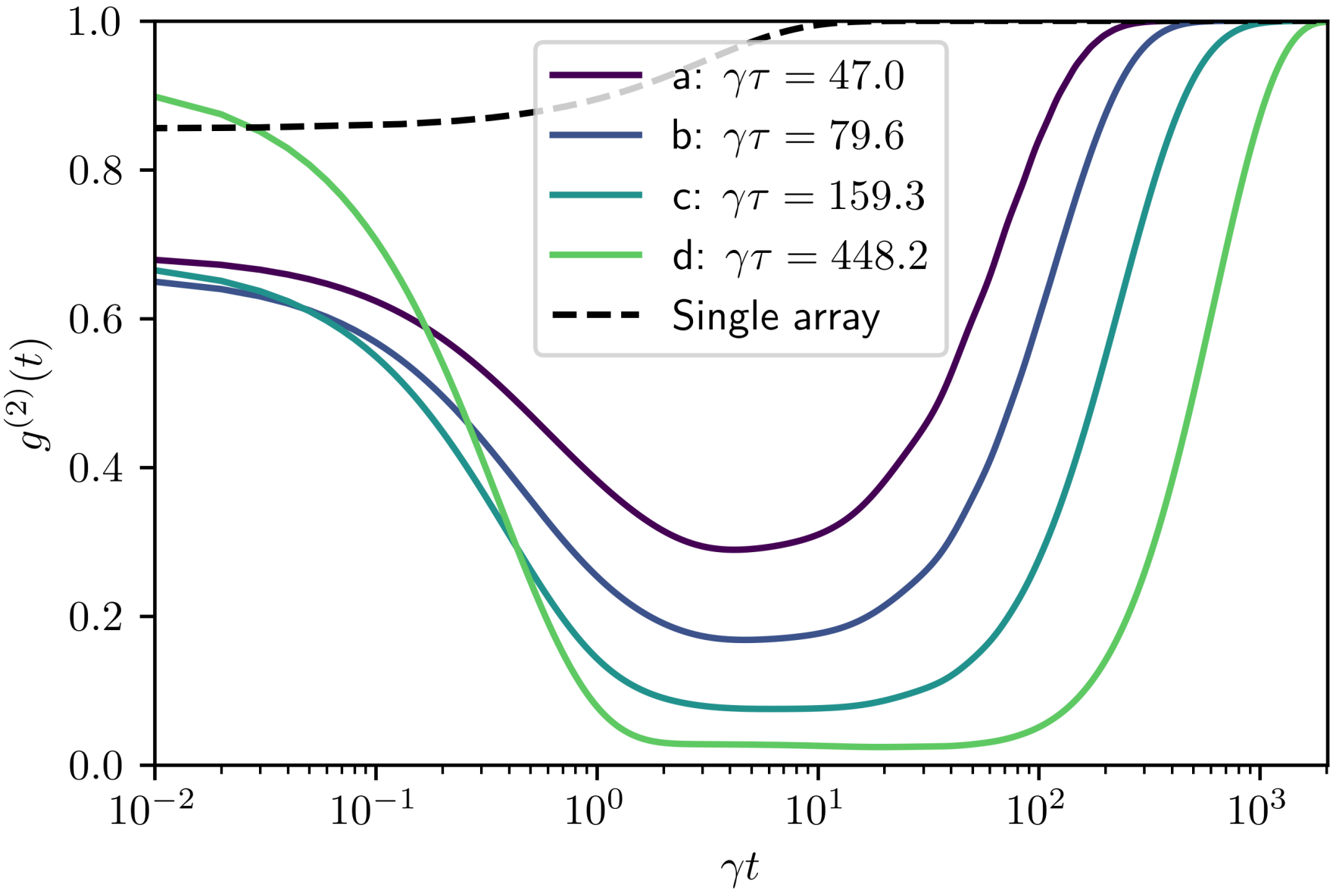

**Figure 3.7 | Two-photon temporal correlation function calculated at the points indicated in Fig. 3.1.** $g^{(2)}(t)$ is calculated and plotted for the same points (see Fig. 3.1) as were used to study $\rho$ in Section 3.3.1. The legend shows the delay times pertaining to these points. For comparison we plot the same single array $g^{(2)}(t)$ (black dashed line) as plotted in Fig. 3.6.

### 3.4.2 Atomic cavity

Figure 3.7 shows $g^{(2)}(t)$ calculated for the dual array with the same sets of system parameters as were used for $\rho$, marked in Fig. 3.1, and for comparison repeats the plot of $g^{(2)}$ for the light reflected off of the single array. The time-axis has a logarithmic scale to make the early behaviour clearer. We can see how $g^{(2)}$ for the dual array generally starts out at a moderate value, smaller than one, very quickly drops to a deep minimum, and then slowly returns to unity. We see how narrower resonances result in stronger and longer-lasting anticorrelations. Indeed, the minimal value even extends and remains flat for a while for the narrowest resonances. The timescales are the same as we found when considering $\rho(\boldsymbol{k}_{1\perp}, \boldsymbol{k}_{2\perp}, t)$ in Fig. 3.5. Clearly, the atomic cavity shows much stronger and long-lasting correlations than the atomic mirror, and below we will expand on how this is related to its long delay time.

Comparing Figs. 3.5 and 3.7, the array appears to emit highly bunched photons, followed by a long period of anticorrelation. The bunched pairs appear to have been scattered to a different transverse mode than the Gaussian, which is used both as driving and detection mode when calculating $g^{(2)}$. While the initial value of $g^{(2)}$ is significantly higher than its minimum value, the narrowness of this peaks nonetheless implies a very weak signal in the Gaussian mode (note the logarithmic scale on the first axis of Fig. 3.7). This is also seen in Figs. 3.2, 3.3 and 3.5, where we see only a small density near zero momentum, and so only a small contribution from the Gaussian mode. Indeed, the fact that $g^{(2)}$ is less than unity at all times, indicates that the overall likelihood of detecting two photons at all in the Gaussian mode is less than in the steady state (as we will detail on below). The return to unity of $g^{(2)}$ is easily understood in terms of the continuous driving of the system, as it is thus continuously repopulated, and emitted photons eventually become uncorrelated with those emitted at earlier times. In this way, if two photons are detected (before the repopulation of the array), they will be detected within a very short time of each other.

All this implies the atomic cavity produces a highly non-classical state of light. To understand the initial dip and the time scales involved, we will describe $g^{(2)}$ in terms of single-excitation states in the coming section.

### 3.4.3 Dominant state time scales

Looking Eq. (3.2), we first note that the two factors in the denominator are identical, as $t'$ is a steady state time, such that adding an additional $t$ to it does not change the expectation value, by definition of the steady state. If we denote the steady state as $|\psi\rangle$, and define the normalized state of the system after detection of the first photon as

$$|\overline{\psi}\rangle = \frac{E\,|\psi\rangle}{\sqrt{\langle\psi|E^\dagger E|\psi\rangle}} \quad , \tag{3.9}$$

we can write $g^{(2)}$ as

$$g^{(2)}(t) = \frac{\langle\overline{\psi}(t)|E^\dagger E|\overline{\psi}(t)\rangle}{\langle\psi|E^\dagger E|\psi\rangle} \quad . \tag{3.10}$$

Thus, $g^{(2)}$ becomes the ratio of the expectation values of the intensity in the post-detection state and in the steady state. In other words, the value of $g^{(2)}$ a time $t$ after detecting the first photon is greater or smaller than unity depending on whether the intensity of emitted light is greater or smaller than in the steady state. Due to the weak driving, the intensity is dominated by the ground state and the single-excitation manifold. Hence, we are able to understand the behaviour of $g^{(2)}$ in terms of the linear physics of the array. However, actually understanding the origin of the correlation, namely why $|\overline{\psi}\rangle$ takes on the specific superposition it does, can of course not be done with the linear physics, as the correlation is specifically a nonlinear effect. We come closer to understanding this in the next part of the thesis, where we derive an analytical description of the correlations.

In performing the numerical simulation of the system we notice that the free decay of $|\overline{\psi}\rangle$ is dominated by two decay rates, one superradiant and one subradiant. We find that the dynamics of $|\overline{\psi}\rangle$ is generally determined by two corresponding non-interacting single-excitation states. These states, $|\pm\rangle$, are the even or odd superposition of having a single excitation in the first or second lattice distributed approximately according to the driving beam profile, i.e. a Gaussian distribution on the lattice. Which of the two states is super- or subradiant depends on which branch of transmission resonance is considered (i.e. the value of $L$), and so these states have a behaviour identical to that of the quasimomentum modes we considered in the previous chapter, in particular the zero-momentum mode. In the limit of an infinitely broad Gaussian beam (and a correspondingly large array), these states would indeed become identical to the even and odd zero-momentum collective states. Evidently even for a finite beam and array we already see the corresponding behaviour.

We find that the time-evolution of $|\overline{\psi}\rangle$ can be described in terms of the ground state $|0\rangle$, and the two states $|\pm\rangle$, evolving according to a non-Hermitian Hamiltonian of the form

$$\begin{aligned} H = &-(\Delta_+ + i\gamma_+)|+\rangle\langle+| - (\Delta_- + i\gamma_-)|-\rangle\langle-| \\ &- g_+(|+\rangle\langle 0| + |0\rangle\langle+|) - g_-(|-\rangle\langle 0| + |0\rangle\langle-|) \ , \end{aligned} \tag{3.11}$$

where we numerically fit the parameters $\Delta_\pm$, $\gamma_\pm$, and $g_\pm$. This is identical in form to Eq. (2.3), if quantum jumps are ignored (as we have in our numerics). Hence, despite the many single-excitation states included in the simulation (one for each atom), we can understand the qualitative behaviour of the array with just these two. We find that the two time-scales of $g^{(2)}$ are given by the lifetimes of these two states (see paragraph below). In the next part of the thesis we will analytically derive that the zero-momentum collective energies indeed have a major role in determining the behaviour of $g^{(2)}$, when driving with a Gaussian. In particular, we will find that the time scales of $g^{(2)}$ are given by the zero-momentum collective decay rates.

As we have written $g^{(2)}$ as the ratio of intensities, we expect the populations of $|\pm\rangle$ to be dominant contributors to the behaviour of the correlation function. Had there been only a single state, the intensity would be proportional to its population, but with two states there will additionally be some interference between their contributions. We therefore plot the population of the two states, $|\langle\pm|\overline{\psi}\rangle|^2$, as a function of time after the detection of the first photon. This is done in Fig. 3.8, which also plots $g^{(2)}(t)$, with the parameters corresponding to the "d" point of Fig. 3.1. We can see the super- and subradiant nature of the two states from the respectively fast and slow return to the steady state values of the populations[2]. We see how the timing and behaviour of the two populations is mimicked by $g^{(2)}$ as expected. In other words, the short and long time scales of the correlation function come from the lifetimes of the single-excitation super- and subradiant states that are predominantly excited by the Gaussian drive.

We understand, then, the nonlinear behaviour of the atomic cavity, in the following way. After the detection of the first photon, the superradiant state has $> 3$ times its steady state population, while the subradiant state is reduced to $< 1/100$ of its steady state population. Thus the superradiant state, due to its strong coupling to the light field, contributes greatly to $g^{(2)}$, but only at very short times as it rapidly decays. This is the initial peak and rapid drop of $g^{(2)}(t)$, corresponding to the initial high and correlated momentum density. The subsequent long-lasting small value of $g^{(2)}(t)$ is due to the array being largely depleted of excitations. Only as the subradiant state is slowly repopulated, returning the array to its steady state, does the array emit further photons, with the corresponding rise in intensity contributing to $g^{(2)}(t)$ until it reaches unity. We can thus understand the bunching of photons in terms of the second photon inhabiting the superradiant state, such that it is emitted very quickly after the first. This leaves the array depleted, such that any subsequent photons first come much later, when the subradiant state is repopulated. Since these latter photons never "met" the initially detected photon, their momenta are uncorrelated as we saw in Fig. 3.5.

2: The steady state values must necessarily be the final values in these plots, as the system is driven back to its steady state after detection of the first photon.

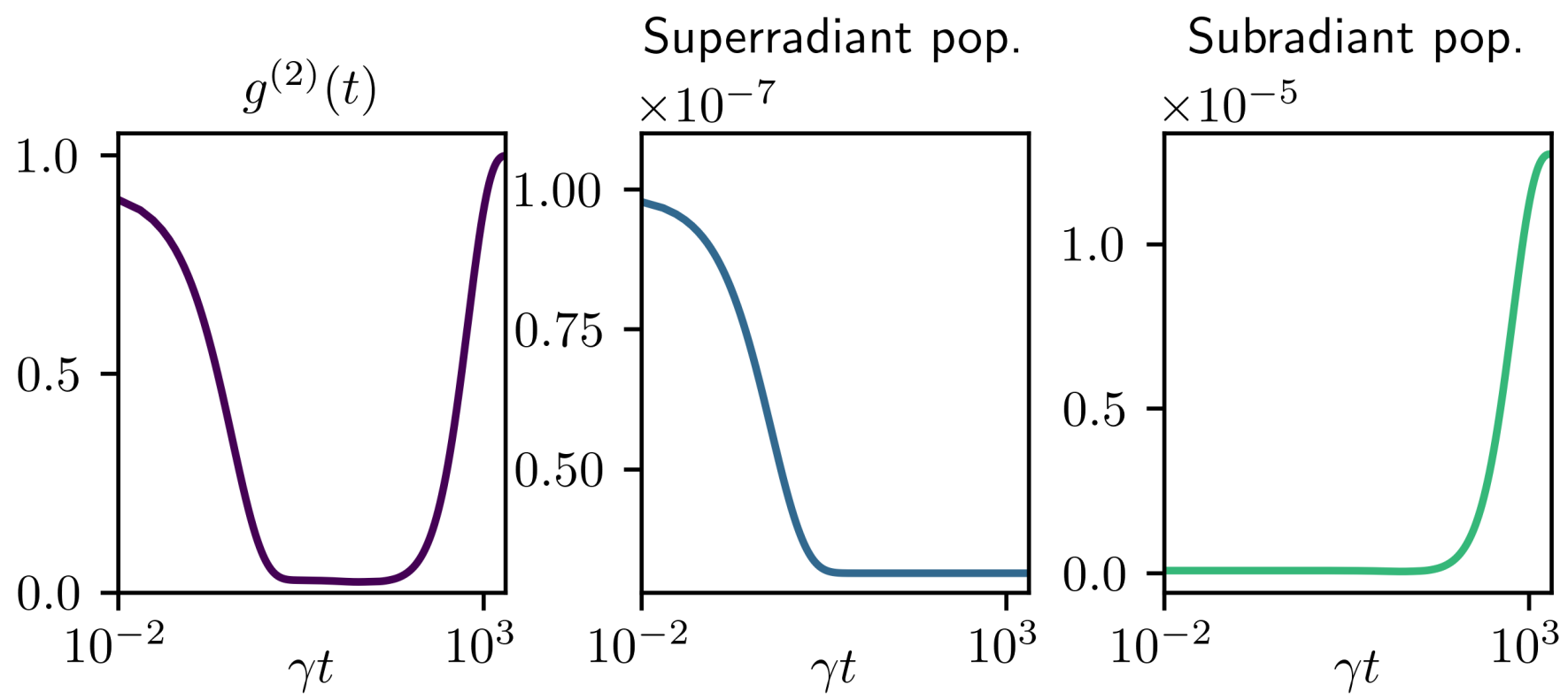

**Figure 3.8 | Temporal correlator compared to single-excitation state populations.** The data is calculated at the "d" point of Fig. 3.1. Hence, the $g^{(2)}$ plotted here is the same as the "d" curve of Fig. 3.7, and the populations are calculated as $|\langle \pm|\overline{\psi}\rangle|^2$.

Comparing the single-photon transmission for the Gaussian mode, Fig. 2.10, the lack of a Gaussian mode component in the two-photon correlated light, Figs. 3.2 and 3.3, and the bunching of photons and subsequent strong anticorrelation, Fig. 3.7, we conclude that the transmitted light from the atomic cavity consists of single photons in the Gaussian mode, with only a very small two-photon contribution, and of highly bunched pairs of photons in higher momentum modes. In other words, the atomic cavity generates both single photons in the driving mode, and at the same time, bunched pairs of momentum-correlated photons.

# Part II

# *Green's Function Approach*

# 4

# Illustration using the Atomic Mirror

In the second part of this thesis we will turn our attention to an analytical approach for studying atomic-optical systems, endeavouring to describe the scattering between photons mediated by atoms using the full Hamiltonian of the system, i.e. without integrating out the photonic degrees of freedom. We use a diagrammatic approach to find the Green's functions, the propagators, of the system, which in turn are used to perform time-evolution via an integration. Due to the simple nature of the interactions in the system, exact expressions for the propagators can be found with a minimum of approximations, and considering only the light propagating away from the atoms in the steady state, the integral of time-evolution can be done with complex contour integration. In fact, we will find exact expressions for both the linear and nonlinear quantities studied in the previous analysis without any use of additional approximations. We illustrate this method by example, calculating the exact nonlinear response of the infinite single array and of the infinite dual array. The exact analytical results makes physical interpretation and understanding easier, and could lead to an analytical study of the many-body physics taking place in the atomic cavity.

Some of the results presented in this part, particularly the ones pertaining to the atomic mirror, can be found in Ref. [113].

In this chapter we will develop and present the method at hand, first laying out some generalities and then applying it to the atomic mirror. In the next chapter we will then apply it to the atomic cavity. To describe the coupling between atomic excitations due to saturation of the two-level systems, we describe these as bosons with a hard-core repulsion. Analytically taking the repulsion to infinity then allows us to describe the interaction due to saturation with no approximation involved. In the previous part of the thesis we considered the atomic response in the limit of vanishing driving, such that the different quantities calculated were dominated by their lowest order contributions, i.e. single- or two-photon contributions. We do the same here by only including the lowest number of contributing photons in the wave function when calculating expectation values. Hence, we find the exact steady state of the propagating light in the one- and two-photon sector.

We will find, both for the atomic mirror and for the atomic cavity, that the photons, which have interacted, are emitted in a manner of bound state, similar to what has been found in one-dimensional waveguides with two-level atoms [114, 115]. However, in contrast to these one-dimensional systems, the "bound" photons in our system can be propagating in completely different directions, even on different sides of the array. Their bounding is in terms of an exponential localization to each other with respect to the distance travelled away from the array, and so it may be more appropriate to think of them as being in a *timed* state. It is the ability to exchange momentum due to the three-dimensional nature of the system at hand, which sets it apart from these existing proposals.

The method developed here is based on the well-known theory of Green's functions from quantum field theory that can be found in text-books [116–118]. We will therefore only make sparse references to these books, and otherwise take it to be implicit that almost everything in the following sections is derived from them. Indeed, similar applications of Green's functions already exist in the literature [39, 119–124]. Especially relevant is the work by Poddubny and co-workers [107, 125], which also uses propagators to explicitly time-evolve wave functions for (small) arrays of atoms and extracts correlations from these. Here we apply this method to extended arrays in free space. The single atomic array is also studied in Ref. [126] to see the effects of imperfect atomic lattices and motion of the atoms using a diagrammatic approach to find the scattering matrix.

We start the chapter by quickly deriving the basic formula for time-evolution using the time-ordered Green's function in Section 4.1. The Feynman rules and the basic diagrams of our theory are then presented in Section 4.2. With these we derive the dressed single-particle propagators in Section 4.3, and the two-photon propagator, written in terms of the atomic $\mathcal{T}$-matrix, in Section 4.4. In Section 4.5, we then time-evolve the propagating part of an initial continuous-wave but otherwise arbitrary two-photon wave function to its steady state. From this, the same momentum and temporal correlators as considered in the previous part of the thesis are extracted and analysed in Section 4.6.

## 4.1 Time-evolution via the time-ordered Green's function

As the very first step, we will derive the equation that is the basis of the method employed in this part of the thesis. We will derive how a wave function can be time-evolved via a convolution of the time-ordered Green's function and an initial state [117]. Consider the following initial state of $n$ identical particles with real-space creation operator $\psi^\dagger(\boldsymbol{r}, t)$

$$\left|i(t')\right\rangle = \int \mathrm{d}^{3n}r' \psi^\dagger(\boldsymbol{r}'_1, t') \cdots \psi^\dagger(\boldsymbol{r}'_n, t') \left|0\right\rangle \Psi_{\text{in}}^{(n)}(\boldsymbol{r}', t') \quad , \tag{4.1}$$

where $\boldsymbol{r}' = (\boldsymbol{r}'_1, \ldots, \boldsymbol{r}'_n)$ is the position vector of all the particles, $\Psi_{\text{in}}^{(n)}(\boldsymbol{r}', t')$ is the initial $n$-particle wave function, and $|0\rangle$ is the vacuum state. Then define a final state as

$$\left|f(\boldsymbol{r}, t)\right\rangle = \psi^\dagger(\boldsymbol{r}_1, t) \cdots \psi^\dagger(\boldsymbol{r}_n, t) \left|0\right\rangle \quad , \tag{4.2}$$

for $t \geq t'$. Hence, in the initial state the $n$ particles are distributed according to $\Psi_{\text{in}}^{(n)}(\boldsymbol{r}', t')$ at time $t'$, while in the final state they are exactly localized at $\boldsymbol{r}$ at time $t$ (corresponding to a delta-function wave function). The overlap between these two states would be the probability amplitude for finding the particles at exactly $\boldsymbol{r}$ at time $t$ given that they were distributed according to $\Psi_{\text{in}}^{(n)}(\boldsymbol{r}', t')$ at the earlier time $t'$. That quantity is itself the wave function $\Psi^{(n)}(\boldsymbol{r}, t)$ at time $t$. In other words, taking the initial time to be $t' = 0$, we can write

$$\begin{aligned}
\Psi^{(n)}(\boldsymbol{r}, t) &= \langle f(\boldsymbol{r}, t) | i(0) \rangle \\
&= \int \mathrm{d}^{3n}r' \left\langle \psi(\boldsymbol{r}_1, t) \cdots \psi(\boldsymbol{r}_n, t) \psi^\dagger(\boldsymbol{r}'_1 0) \cdots \psi^\dagger(\boldsymbol{r}'_n 0) \right\rangle \Psi_{\text{in}}^{(n)}(\boldsymbol{r}', 0) \\
&= -(-i)^n \int \mathrm{d}^{3n}r' \left( -i^n \left\langle T[\psi(\boldsymbol{r}_1, t) \cdots \psi^\dagger(\boldsymbol{r}'_n, 0)] \right\rangle \right) \Psi_{\text{in}}^{(n)}(\boldsymbol{r}', 0) \\
&= -(-i)^n \int \mathrm{d}^{3n}r' G^{(n)}(\boldsymbol{r}, \boldsymbol{r}', t) \Psi_{\text{in}}^{(n)}(\boldsymbol{r}', 0) \quad ,
\end{aligned} \tag{4.3}$$

where we have used $t \geq t' = 0$ to introduce the time-ordering operator, $T[\cdot]$, the expectation value is with respect to $|0\rangle$, and we have defined the $n$-particle time-ordered Green's function

$$G^{(n)}(\boldsymbol{r}, \boldsymbol{r}', t) = -i^n \left\langle T[\psi(\boldsymbol{r}_1, t) \cdots \psi(\boldsymbol{r}_n, t)\psi^\dagger(\boldsymbol{r}'_1 0) \cdots \psi^\dagger(\boldsymbol{r}'_n 0)] \right\rangle \ , \quad (4.4)$$

with all creation operator times set to 0 and all annihilation operator times set to $t$. Thus, if we can determine the Green's function, or propagator, and we can evaluate the integral in Eq. (4.3), we will be able to perform time-evolution. We now proceed to first determine the appropriate propagators for the time-evolution of one- and two-photon wave functions.

## 4.2 Feynman rules and basic diagrams

We write the basic Hamiltonian, Eq. (1.1), for interacting two-level atoms and photons again, this time writing the photon-atom coupling explicitly and neglecting terms that vanish in a RWA

$$\begin{aligned} H = & \sum_{\boldsymbol{k}\nu} \omega_{\boldsymbol{k}} b^\dagger_{\boldsymbol{k}\nu} b_{\boldsymbol{k}\nu} + \omega_a \sum_n \sigma^\dagger_n \sigma_n \\ & - \frac{i}{\sqrt{2\epsilon_0 V}} \sum_{n,\boldsymbol{k}\nu} \sqrt{\omega_{\boldsymbol{k}}} \left( \boldsymbol{d}^\dagger \hat{\boldsymbol{e}}_{\boldsymbol{k}\nu} e^{i\boldsymbol{k}\cdot\boldsymbol{r}_n} \sigma^\dagger_n b_{\boldsymbol{k}\nu} - \text{H.c.} \right) \ . \end{aligned} \quad (4.5)$$

As discussed in Section 1.5.2, there is no explicit interaction between atomic excitations in this Hamiltonian, but rather it is the nature of the two-level operators that defines the interaction. In order to make this interaction explicit, and to facilitate the diagrammatic approach we will take to determine propagators, we now bosonize the two-level atoms. That is, we replace the two-level operators with bosonic operators, $\sigma_n \to a_n$, and put in an on-site repulsive interaction between the atomic excitations, which in the limit of infinite repulsion recovers the two-level structure of the atoms[1]. Thus, we have

$$\begin{aligned} H = & \sum_{\boldsymbol{k}\nu} \omega_{\boldsymbol{k}} b^\dagger_{\boldsymbol{k}\nu} b_{\boldsymbol{k}\nu} + \omega_a \sum_n a^\dagger_n a_n \\ & - \frac{i}{\sqrt{2\epsilon_0 V}} \sum_{n,\boldsymbol{k}\nu} \sqrt{\omega_{\boldsymbol{k}}} \left( \boldsymbol{d}^\dagger \hat{\boldsymbol{e}}_{\boldsymbol{k}\nu} e^{i\boldsymbol{k}\cdot\boldsymbol{r}_n} a^\dagger_n b_{\boldsymbol{k}\nu} - \text{H.c.} \right) + \frac{U}{2} \sum_n a^\dagger_n a^\dagger_n a_n a_n \ . \end{aligned} \quad (4.6)$$

1: This is a known trick, see for example Refs. [39, 121, 122, 127, 128]

We now assume that the atoms are arranged in the single array configuration of an infinite, square lattice in the $xy$-plane. We then Fourier transform the Hamiltonian (as we did in Section 1.5) to arrive at

$$\begin{aligned} H = &\sum_{\nu} \int \frac{\mathrm{d}^3 k}{(2\pi)^3} \omega_{\boldsymbol{k}} b^{\dagger}_{\boldsymbol{k}\nu} b_{\boldsymbol{k}\nu} + \int_{\mathrm{BZ}} \frac{\mathrm{d}^2 k_{\perp}}{(2\pi)^2} \omega_a a^{\dagger}_{\boldsymbol{k}_{\perp}} a_{\boldsymbol{k}_{\perp}} \\ &- \sum_{\nu} \int \frac{\mathrm{d}^3 k}{(2\pi)^3} \left( g_{\boldsymbol{k}\nu} b^{\dagger}_{\boldsymbol{k}\nu} a_{\boldsymbol{k}_{\perp}} + \mathrm{H.c.} \right) \\ &+ \frac{U a^2}{2} \int_{\mathrm{BZ}} \frac{\mathrm{d}^2 k_{\perp} \mathrm{d}^2 k'_{\perp} \mathrm{d}^2 q_{\perp}}{(2\pi)^6} a^{\dagger}_{\boldsymbol{k}_{\perp}+\boldsymbol{q}_{\perp}} a^{\dagger}_{\boldsymbol{k}'_{\perp}-\boldsymbol{q}_{\perp}} a_{\boldsymbol{k}_{\perp}} a_{\boldsymbol{k}'_{\perp}} \ . \end{aligned} \tag{4.7}$$

The quasi-momentum of the atomic modes takes on values within the first BZ, and $a_{\boldsymbol{k}_{\perp}+\boldsymbol{q}_m} = a_{\boldsymbol{k}_{\perp}}$ for any reciprocal lattice vector $\boldsymbol{q}_m$. Hence, we see in the third term how the same atomic mode couples to photons, whose momenta differ by a reciprocal lattice vector. This is Bragg scattering. Here $g_{\boldsymbol{k}\nu} = i\sqrt{\omega_{\boldsymbol{k}}/2\epsilon_0 a^2}\hat{\boldsymbol{e}}^{\dagger}_{\boldsymbol{k}\nu}\boldsymbol{d} = i\sqrt{\omega_{\boldsymbol{k}}\tilde{\Gamma}/\omega_a}\hat{\boldsymbol{e}}^{\dagger}_{\boldsymbol{k}\nu}\hat{\boldsymbol{e}}_{+}$. Finally, the last term is the Fourier transformed on-site repulsion of the atomic excitations. As this interaction is local and uniform it results in a momentum-independent interaction that allows scattering modes to exchange momentum, while conserving the total momentum.

We will eventually consider an initial state of light corresponding to driving with a Gaussian beam, illuminating a finite number of atoms, such that the behaviour of infinite arrays should be identical to that of simply very large arrays (the localized beam is not affected by atoms very far away in the array). Hence, taking the arrays to be infinite eases the theoretical analysis, but the results should still be relevant for realistic experiments, and identical to those of our previous analysis.

With the momentum-space Hamiltonian in hand, we can derive the Feynman rules for our system. These tell us how to evaluate complicated, time-dependent expectation value, like the one involved in Eq. (4.4), by drawing a series of diagrams, each representing a contribution to the expectation value in a certain systematic expansion of it, and each having a simple physical interpretation. For the expectation value in Eq. (4.4), we would say there are $n$ particles "incoming" at $\boldsymbol{r}'$, represented by the creation operators at the early time, and then a corresponding $n$ particles "outgoing" at $\boldsymbol{r}$, represented by the annihilation operators at later times. We will soon consider the propagator in momentum space, where we deal with incoming and outgoing particles with certain momenta, rather than position. The diagrams then represent processes that result in such outgoing particles from an initial set of such incoming particles.

The diagrams can be used to perform mathematical manipulations of otherwise notationally heavy quantities in a straightforward manner. We can thus translate a complicated expectation value into a sum of a series of diagrams, which we manipulate until we find a simple diagrammatic expression for this sum, which can then be translated back to an explicit mathematical expression. In this way, the diagrams allow us replace long and complicated calculations with merely drawing and manipulating these diagrams, which is far easier and more physically intuitive, as we will see. Deriving the Feynman rules is a complex but illuminating exercise that is nonetheless "textbook". We include it for completeness, but place it in Appendix A, and simply quote the Feynman rules here and show the basic diagrams.

Let us start by considering the basic diagrams, the building blocks of more advanced processes. First, the bare propagators, which represent time-evolution according to the first two terms of Eq. (4.7), i.e. the non-interacting part of the Hamiltonian. The diagrams and corresponding mathematical quantities are

$$= G^0_{\mathrm{at}}(\boldsymbol{k}_\perp, \boldsymbol{k}'_\perp, \omega) \quad , \tag{4.8a}$$

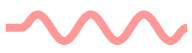

$$= G^0_{\mathrm{ph}}(\boldsymbol{k}\nu, \boldsymbol{k}'\nu', \omega) \quad , \tag{4.8b}$$

where the primed (un-primed) momenta and polarizations are incoming (outgoing). For single-particle propagators the total energy is always conserved (while momentum will not be, once we consider the full propagators), so we write the propagators as functions of a single frequency $\omega$. For two-particle propagators we will write a frequency for each particle, as they can exchange energy as well as momentum. The atom-photon interaction vertices (i.e. emission and absorption) are

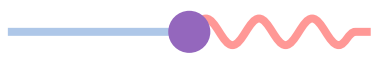

$$= g_{k\nu} \quad , \tag{4.9a}$$

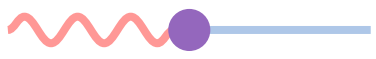

$$= g^*_{k\nu} \quad . \tag{4.9b}$$

This vertex conserves transverse momentum up to Bragg scattering, and does not conserve longitudinal momentum. Specifically, for emission the transverse momentum of the emitted photon is equal to that of the decayed atomic excitation plus a reciprocal lattice vector, and for absorption the momentum of the atomic excitation is given by that of the photon, projected into the first BZ. Notice, that for diagrams like these, where a single vertex is in focus, the propagator lines are only drawn to indicate which type of propagators can be connected with the vertex. When translating the vertex to an explicit mathematical expression we therefore do not include these propagators, but only the factor due to the vertex itself. Finally, the hardcore scattering between atomic excitations is

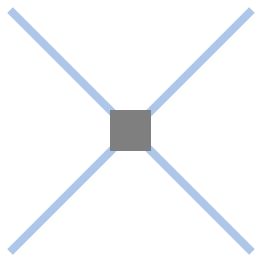

$$= \frac{Ua^2}{2} \quad . \tag{4.10}$$

This vertex conserves total (transverse) momentum.

The bare propagators in Fourier space have the usual form[2], $G^0_{\rm at}(\boldsymbol{k}_\perp, \boldsymbol{k}'_\perp, \omega) = (2\pi)^2\delta(\boldsymbol{k}_\perp - \boldsymbol{k}'_\perp)G^0_{\rm at}(\omega)$ and $G^0_{\rm ph}(\boldsymbol{k}\nu, \boldsymbol{k}'\nu', \omega) = (2\pi)^3\delta(\boldsymbol{k}-\boldsymbol{k}')\delta_{\nu,\nu'}G^0_{\rm ph}(\boldsymbol{k}, \omega)$, with

$$G^0_{\rm at}(\omega) = \frac{1}{\omega + i\eta - \omega_a} \quad , \tag{4.11a}$$

$$G^0_{\rm ph}(\boldsymbol{k}, \omega) = \frac{1}{\omega + i\eta - \omega_{\boldsymbol{k}}} \quad , \tag{4.11b}$$

where $\eta = 0^+$ is a positive infinitesimal. Usually one does not write the momentum conservation of bare propagation explicitly in terms of delta-functions like this, but simply include it implicitly by writing the propagators as functions of only a single momentum (as we do for energy). However, since momentum is not conserved by the on-site repulsion and only conserved up to Bragg scattering by emission, the full propagators we will derive, do not conserve momentum, and so we already write it explicitly here to have a consistent and explicit notation. As mentioned, we will do the same for energy once we consider two-particle propagators.

With the basic diagrams in hand, we can write the Feynman rules for drawing the contributions to a certain expectation value. They are

1. Draw all topologically distinct, fully connected diagrams consisting of $n$ vertices of the types defined by the interactions in the system, connected with lines according to which types of particles are involved in the interaction, with appropriate loose-ended lines representing the incoming and outgoing particles.

2. Assign to each line an outgoing and incoming momentum (and polarization if needed) and an energy ($\boldsymbol{k}\nu, \boldsymbol{k}'\nu', \omega$ for photons, or $\boldsymbol{k}_\perp, \boldsymbol{k}'_\perp, \omega$ for atomic excitations).

2: We derive this form in a generalized setting in Appendix A, specifically Eq. (A.24).

3. Reduce these momenta and energies by conserving whatever is conserved at each vertex.
4. Write for each line in the resulting diagram the appropriate propagator, evaluated at the appropriate momentum and energy.
5. Write for each vertex the appropriate coupling strength, evaluated at the appropriate momentum and energy.
6. Multiply with the number of permutations of the incoming and outgoing lines which leave the topology of the diagram unchanged.
7. Integrate and sum over all remaining internal variables (i.e. remaining variables other than those pertaining to the incoming and outgoing particles).
8. Multiply with $-i$ for each interaction vertex, and $i$ for each propagator. Furthermore, multiply with constants from the quantity being calculated (e.g. $(-i)^n$ in the case of Eq. (4.4)).

We can now proceed to derive the full single- and two-particle propagators. The single-particle propagators we will refer to as "dressed propagators", as they refer to single-particle propagation "dressed" by the presence of the other type of particle. That is, atomic excitations are dressed by photons, and vice-versa, as these repeatedly change into another.

## 4.3 Single-particle dressed propagators

We draw and represent the atomic and photonic dressed propagators, i.e. $G^{(n)}$ from Eq. (4.4) with $n = 1$ for atomic excitations and photon respectively, as

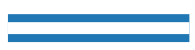 $$= G^{(1)}_{\text{at}}(\boldsymbol{k}_\perp, \boldsymbol{k}'_\perp, \omega) \ , \tag{4.12a}$$

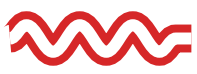 $$= G^{(1)}_{\text{ph}}(\boldsymbol{k}\nu, \boldsymbol{k}'\nu', \omega) \ . \tag{4.12b}$$

In both cases, the propagator is found by adding all diagrams starting and ending with a single of the corresponding particle type's bare propagator, and using only the vertices of emission and absorption. These are the Dyson's equations [117]. Diagrammatically, we have

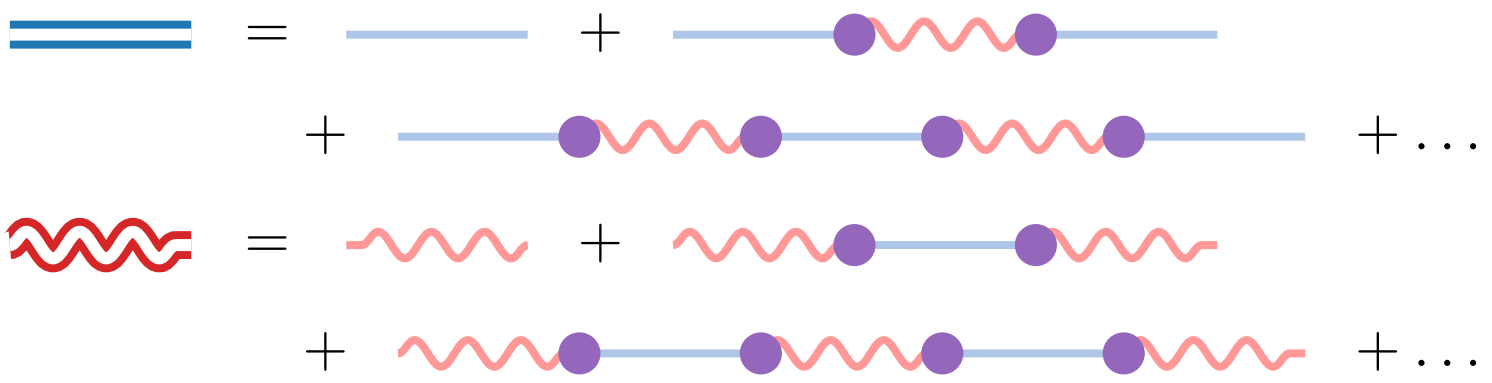

(4.13a)

(4.13b)

In both cases we can note an iterative pattern, that allows us to draw the above as

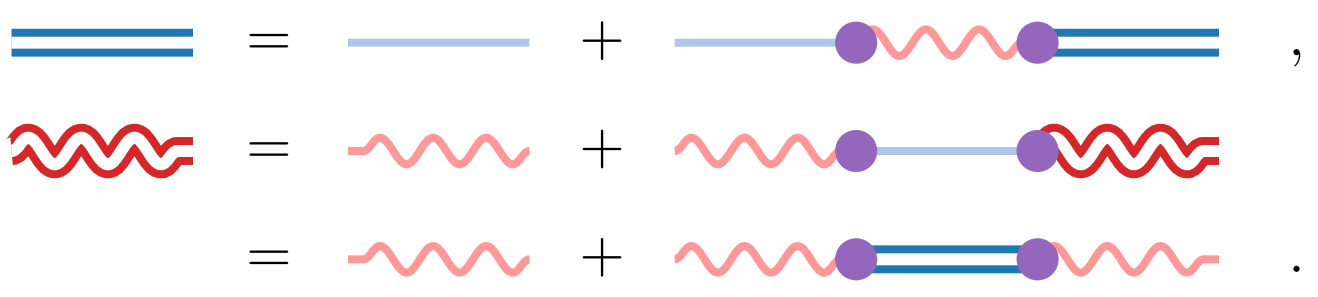

(4.14a)
(4.14b)
(4.14c)

That is, the infinite Dyson series can be recovered by iteratively inserting the dressed propagators into the expression of themselves.

Due to the conservation of transverse momentum (within the first BZ) in absorption, the Dyson equation for the dressed atomic propagator, Eq. (4.14a), factorizes to a simple algebraic equation. That is, the terms in the equation are simply products of propagators or integrals, such that we can solve for $G^{(1)}_{\text{at}}$. In particular, the explicit mathematical equation corresponding to Eq. (4.14a) is

$$G^{(1)}_{\text{at}}(\boldsymbol{k}_\perp, \boldsymbol{k}'_\perp, \omega) = (2\pi)^2\delta(\boldsymbol{k}_\perp - \boldsymbol{k}'_\perp)G^0_{\text{at}}(\omega) \\ + (2\pi)^2\delta_{\text{BZ}}(\boldsymbol{k}_\perp - \boldsymbol{k}'_\perp)G^{(1)}_{\text{at}}(\boldsymbol{k}_\perp, \boldsymbol{k}'_\perp, \omega)\Sigma(\boldsymbol{k}'_\perp, \omega)G^0_{\text{at}}(\omega) \ , \tag{4.15}$$

where we have introduced $\delta_{\text{BZ}}(\boldsymbol{k}_\perp - \boldsymbol{k}'_\perp) = \sum_{\boldsymbol{q}_m} \delta(\boldsymbol{k}_\perp + \boldsymbol{q}_m - \boldsymbol{k}'_\perp)$, which acts as a delta-function up to Bragg scattering (we will refer to it as a *Bragg delta-function*), and we have defined the (proper) self-energy

$\Sigma(\boldsymbol{k}_\perp, \omega) =$ 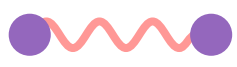

$$= \sum_{\boldsymbol{q}_m, \nu} \int \frac{\mathrm{d}k_z}{2\pi} g^*_{\boldsymbol{k}\nu} G^0_{\text{ph}}(\boldsymbol{k}, \omega) g_{\boldsymbol{k}\nu} \ , \tag{4.16}$$

where $\boldsymbol{k} = (\boldsymbol{k}_\perp + \boldsymbol{q}_m, k_z)$ on the right hand side. Using the fact that $G^{(1)}_{\text{at}}(\boldsymbol{k}_\perp, \boldsymbol{k}'_\perp, \omega)$ only takes on momenta within the first BZ, we have $\delta_{\text{BZ}}(\boldsymbol{k}_\perp - \boldsymbol{k}'_\perp) = \delta(\boldsymbol{k}_\perp - \boldsymbol{k}'_\perp)$ in Eq. (4.15), and we can define $G^{(1)}_{\text{at}}(\boldsymbol{k}_\perp, \boldsymbol{k}'_\perp, \omega) = (2\pi)^2\delta(\boldsymbol{k}_\perp - \boldsymbol{k}'_\perp)G^{(1)}_{\text{at}}(\boldsymbol{k}_\perp, \omega)$. We can then write

$$G^{(1)}_{\text{at}} = G^0_{\text{at}} + G^0_{\text{at}}\Sigma G^{(1)}_{\text{at}} \ , \tag{4.17}$$

where we have suppressed the $(\boldsymbol{k}_\perp, \omega)$-dependence, which is identical for all factors. Here, we can easily solve for $G^{(1)}_{\text{at}}$

$$(G^{(1)}_{\text{at}})^{-1} = (G^0_{\text{at}})^{-1} - \Sigma \ . \tag{4.18}$$

We have thus found the dressed atomic propagator in terms of the self-energy, which we calculate in a moment.

The same trick does not work for the dressed photonic propagator. Due to Bragg scattering in emission, the $G^{(1)}_{\mathrm{ph}}$-factor on the right hand side of Eq. (4.14b) is summed over reciprocal lattice vectors, such that it can not be easily isolated. We therefore instead use the expression in Eq. (4.14c), expressing $G^{(1)}_{\mathrm{ph}}$ in terms of $G^{(1)}_{\mathrm{at}}$. Specifically, we have

$$\begin{aligned} G^{(1)}_{\mathrm{ph}}(\boldsymbol{k}\nu, &\boldsymbol{k}'\nu', \omega) \\ &= (2\pi)^3 \delta(\boldsymbol{k} - \boldsymbol{k}')\delta_{\nu,\nu'} G^0_{\mathrm{ph}}(\boldsymbol{k}', \omega) \\ &\quad + (2\pi)^2 \delta_{\mathrm{BZ}}(\boldsymbol{k}_\perp - \boldsymbol{k}'_\perp) G^0_{\mathrm{ph}}(\boldsymbol{k}, \omega) g_{\boldsymbol{k}\nu} G^{(1)}_{\mathrm{at}}(\boldsymbol{k}'_\perp, \omega) g^*_{\boldsymbol{k}'\nu'} G^0_{\mathrm{ph}}(\boldsymbol{k}', \omega) \ . \end{aligned} \tag{4.19}$$

Note that here, the Bragg delta-function can not be reduced to a simple delta-function, and so we see how momentum conservation is different for the two parts of the dressed propagator (which is why we write the delta-functions explicitly). Also, it is noteworthy that the dressed photonic propagator has a form comparable to the input-output relation used in the previous part of the thesis, Eq. (1.19). That is, the full propagation of the photons in the presence of the atoms is given by their free propagation plus a term that can be simply interpreted as photons being absorbed, propagating as excitations before being emitted again. This is very similar to the input-output relation, where the full field is given by the freely evolving field plus the contribution emitted by the atoms.

We now only need to evaluate the expression for the self-energy to fully have the single-particle dressed propagators. We have

$$\begin{aligned} \Sigma(\boldsymbol{k}_\perp, \omega) &= \sum_{\boldsymbol{q}_m, \nu} \int \frac{\mathrm{d}k_z}{2\pi} \frac{g^*_{\boldsymbol{k}\nu} g_{\boldsymbol{k}\nu}}{\omega + i\eta - \omega_{\boldsymbol{k}}} \\ &= \frac{1}{4\pi\epsilon_0 a^2} \sum_{\boldsymbol{q}_m} \int \mathrm{d}k_z \frac{\omega_{\boldsymbol{k}} \boldsymbol{d}^\dagger \boldsymbol{Q} \boldsymbol{d}}{\omega + i\eta - \omega_{\boldsymbol{k}}} \ , \end{aligned} \tag{4.20}$$

where $\boldsymbol{Q} = \sum_\nu \hat{\boldsymbol{e}}_{\boldsymbol{k}\nu} \hat{\boldsymbol{e}}^\dagger_{\boldsymbol{k}\nu} = 1 - \hat{\boldsymbol{k}}\hat{\boldsymbol{k}}^\dagger$ is the same projection matrix, as we used in the first part of the thesis. Indeed, introducing the zeta-function in the above, changing the variable of integration to $\omega_{\boldsymbol{k}}$, and comparing with the expression in Eq. (1.40) for the continuous transverse Fourier transformed EM dyadic Green's function $\boldsymbol{G}(\boldsymbol{k}_\perp, z, \omega)$, we can write the above as

$$\Sigma(\boldsymbol{k}_\perp, \omega) = \frac{1}{\pi\epsilon_0 a^2} \boldsymbol{d}^\dagger \sum_{\boldsymbol{q}_m} \int_0^\infty \mathrm{d}\omega_{\boldsymbol{k}} \omega^2_{\boldsymbol{k}} \Im\left[\boldsymbol{G}(\boldsymbol{k}_\perp + \boldsymbol{q}_m, z = 0, \omega_{\boldsymbol{k}})\right] \boldsymbol{d} \zeta(\omega - \omega_{\boldsymbol{k}}) \ . \tag{4.21}$$

This integral is reminiscent of the ones we dealt with in Section 1.2. A similar line of calculations leads us to

$$\Sigma(\boldsymbol{k}_\perp,\omega) = -\mu_0\omega^2\boldsymbol{d}^\dagger\tilde{\boldsymbol{G}}(\boldsymbol{k}_\perp,z=0,\omega)\boldsymbol{d} = -\frac{i\tilde{\Gamma}}{\omega_a}\sum_{\boldsymbol{q}_m}\frac{\omega^2-(\boldsymbol{k}_\perp+\boldsymbol{q}_m)^2/2}{\sqrt{\omega^2-(\boldsymbol{k}_\perp+\boldsymbol{q}_m)^2}} \ , \tag{4.22}$$

where the final expression is found using Eqs. (1.44) and (1.45). We thus have an explicit expression for

$$(G_{\text{at}}^{(1)})^{-1}(\boldsymbol{k}_\perp,\omega) = \omega + i\eta - \omega_a - \Sigma(\boldsymbol{k}_\perp,\omega) \ . \tag{4.23}$$

To finish the connection with our previous analysis, we note here that the poles of the dressed propagator, i.e. the zeros of $(G_{\text{at}}^{(1)})^{-1}$, reveal the energies of dressed single-particle excitations [116, 117]. That is, the complex energy $\omega$ at which $(G_{\text{at}}^{(1)})^{-1}(\boldsymbol{k}_\perp,\omega) = 0$, gives the energy and decay rate (real and imaginary part respectively) of the single-particle excitations taking emission and absorption into account. For comparison we can see that the bare propagators in Eq. (4.11) indeed have simple poles at the single-particle energies of the bare atomic excitations and bare photons. As mentioned previously, we take $\gamma/\omega_a$ to be extremely small, such that $\tilde{\Gamma}/\omega_a = 3\pi\gamma/a^2\omega_a^3$ is also small. Iteratively solving $(G_{\text{at}}^{(1)})^{-1}(\boldsymbol{k}_\perp,\omega) = 0$ by substituting $\omega = \omega_a - i\eta + \Sigma(\boldsymbol{k}_\perp,\omega)$ into the expression for $\Sigma(\boldsymbol{k}_\perp,\omega) \propto \gamma/\omega_a$, we can therefore stop the iteration at first order in $\gamma/\omega_a$, to arrive at the equation

$$(G_{\text{at}}^{(1)})^{-1}(\boldsymbol{k}_\perp,\omega) = \omega + i\eta - \omega_a - \Sigma(\boldsymbol{k}_\perp,\omega_a) \ . \tag{4.24}$$

That is, the self-energy is evaluated at the atomic transition frequency, and the dressed atomic propagator takes on the same simple form as the bare propagators (one explicit simple pole). This corresponds to the Markov approximation we made in Section 1.2. With this, we see $G_{\text{at}}^{(1)}$ has a simple pole at

$$\omega = \omega_a - i\eta + \Sigma(\boldsymbol{k}_\perp,\omega_a) = \omega_a - i\eta + \tilde{\Delta}^0_{\boldsymbol{k}_\perp} - i\tilde{\Gamma}^0_{\boldsymbol{k}_\perp} \ , \tag{4.25}$$

where we have recovered the single array collective energies (Eq. (1.74)). In other words, we find that the energies of the single-particle excitations in the single array are given by the previously found collective energies, making the present analysis consistent with the former. The underlying reason why it is possible to fully determine the dressed propagators is exactly that within the approximations we have made, the linear sector of Eq. (4.7) is diagonalizable, as we also discussed in Section 1.5.2.

We have thus found explicit expressions for the dressed single-particle propagators, with which we could time-evolve single-particle wave functions, determining the linear response of the atomic mirror. We will return to this later and see how it recovers the familiar results, as can already be glimpsed in the fact that the photonic dressed propagator has a form similar to the input-output relation and the fact that we have recovered the collective energies. We now move on to find the two-photon propagator, which can tell us about the nonlinear response of the array at the level of two photons.

The interaction between photons and matter excitations leads to the true excitations of a system to be mixtures of these two types of bare excitations, a type of quasiparticles called polaritons [4, 6]. This would also be the case for our system, and it should therefore be possible to write the dressed propagators for the atomic excitations and the photons in terms of polariton propagators, which should each have just a single pole, corresponding to a true excitation of the system. The fact that the atomic dressed propagator takes on the same form as a bare propagator with only a single pole, implies that this is in fact the polaritonic propagator of a true excitation. We conclude that in performing the approximation $\Sigma(\boldsymbol{k}_\perp, \omega) \simeq \Sigma(\boldsymbol{k}_\perp, \omega_a)$, we neglected the contribution to the dressed atomic propagator from other polaritonic excitations than this central one. Indeed, if one does not make this approximation the propagator shows a seemingly infinite range of additional poles at complex energies with extremely large imaginary parts, corresponding to extremely short lifetimes. The large imaginary parts seemingly comes about as a consequence of the smallness of $\gamma/\omega_a$. It is these short-lived (polaritonic) excitations that we have neglected. This, in addition to the fact that the polaritons of this system are mixtures of intrinsically two-dimensional atomic excitations and three-dimensional photons, makes the study of the polaritonic excitations of the system an exciting prospect for future work.

## 4.4 Two-photon propagator

We again use the Feynman rules from Section 4.2 to construct the diagrammatic equation for the two-photon propagator, by drawing and summing all diagrams that have two incoming photons and two outgoing. Here it is necessary to keep track of the individual energies, as the photons can exchange energy, though the total energy is conserved. We introduce the photon multi-index $x = (\boldsymbol{k}, \nu, \omega)$ and correspondingly for the atomic excitations $y = (\boldsymbol{k}_\perp, \omega)$ for a more compact notation. We then find the following diagrammatic expression for the two-photon propagator

$$G^{(2)}_{\mathrm{ph}}(x_1, x_2; x'_1, x'_2) =$$

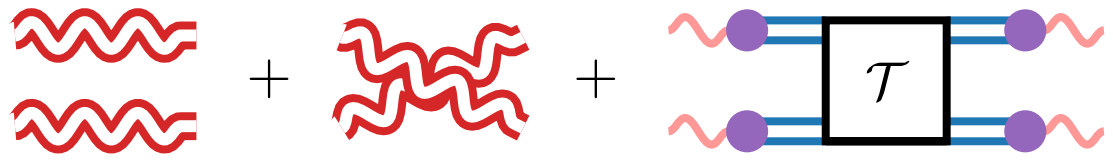

$$. \qquad (4.26)$$

The first two terms represent photon propagating without scattering off of each other, but still interacting with the array (propagating according to the dressed propagators). There are two such terms for the simple combinatorial reason that there are two ways of pairing two incoming and two outgoing photons. The third term represents photon-photon scattering. Here, two photons propagate, are absorbed in the array, propagate as atomic excitations and then interact via the full atomic interaction vertex (the atomic $\mathcal{T}$-matrix, see below), propagate again, and are finally emitted as photons. There is only one interaction term in Eq. (4.26) (though there are two free terms), as the interaction term with crossed propagators yields an identical contribution to the one without. This is due to the fact that the bare scattering is momentum-independent (it only conserves the total momentum), and so the contribution is independent of how the outgoing photons have been matched. We therefore write only one interaction term, and absorb this factor in the definition of the $\mathcal{T}$-matrix. As all other factors in Eq. (4.26) are single-particle quantities known from the previous section, we need only determine the atomic $\mathcal{T}$-matrix to fully know the two-photon propagator.

### 4.4.1 The atomic $\mathcal{T}$-matrix

The atomic $\mathcal{T}$-matrix is represented by a large square vertex

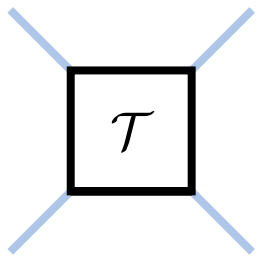

$$= \mathcal{T}(y_1, y_2; y'_1, y'_2) \ . \qquad (4.27)$$

This quantity can be considered as the dressed version of the bare excitation-excitation scattering of Eq. (4.10). Within the so-called ladder approximation, we can write the $\mathcal{T}$-matrix diagrammatically as

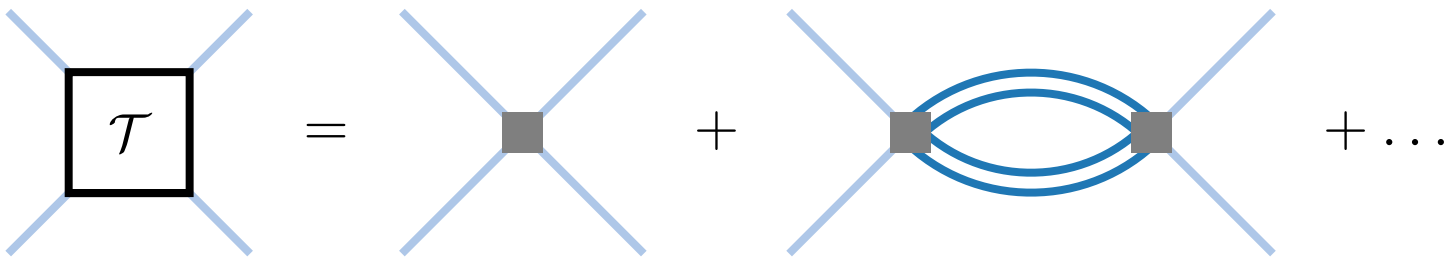

$$(4.28)$$

It is the interaction vertex which takes multiple scattering and intermediate propagation into account. The ladder approximation means we include only diagrams with "one interaction at a time", i.e. there are never more than two propagation lines and these repeatedly meet in a bare scattering vertex[3]. This approximation becomes exact when there are only two particles involved [116], which is precisely the case we consider. Thus, for the case at hand there is no approximation performed, but the calculation we do here would hold for the two-body interactions of a larger number of photons within the ladder approximation.

Similar to the Dyson equation for the dressed propagators, the iterative pattern of the above, can be summed by expression the $\mathcal{T}$-matrix in terms of itself. This is done in the Bethe-Salpeter equation [116, 118]

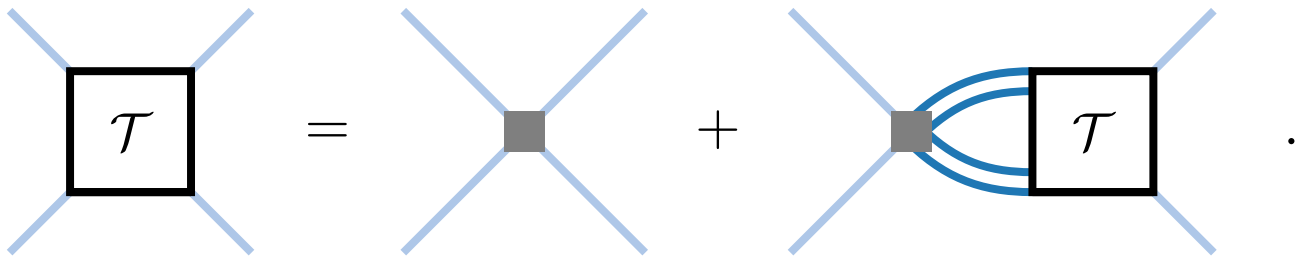

(4.29)

Using the Feynman rules (noting that the $\mathcal{T}$-matrix is a vertex, and absorbing a factor of 2 in the $\mathcal{T}$-matrix for later ease), these diagrams can be translated to

$$\mathcal{T}(y_1, y_2; y_1', y_2') = U + iUa^2 \int \frac{\mathrm{d}^3 y}{(2\pi)^3} G_{\mathrm{at}}^{(1)}(y) G_{\mathrm{at}}^{(1)}(Y' - y) \times \mathcal{T}(y, Y' - y; y_1', y_2') \ , \tag{4.30}$$

where the integration is over (transverse) momentum in the first Brillouin zone and energy, $y = (\boldsymbol{k}_\perp, \omega)$. We have introduced the total momentum and energy $Y' = y_1' + y_2' = (\boldsymbol{K}'_\perp, \Omega')$ of the incoming particles, with a corresponding $Y = (\boldsymbol{K}_\perp, \Omega)$ for the outgoing. As the right hand side of Eq. (4.30) is independent of the outgoing $y_1, y_2$ (as a consequence of the bare interaction being independent of momentum and energy), we conclude that $\mathcal{T}(y_1, y_2; y_1', y_2')$ is independent of its first two arguments. The $\mathcal{T}$-matrix then decouples from the integral in the above, and we can isolate it to yield

$$\mathcal{T}^{-1}(y_1, y_2; y_1', y_2') = U^{-1} - a^2 \int_{\mathrm{BZ}} \frac{\mathrm{d}^2 q_\perp}{(2\pi)^2} \Pi(Y', \boldsymbol{q}_\perp) \ . \tag{4.31}$$

Here, we have defined the pair propagator

$$\Pi(Y, \boldsymbol{q}_\perp) = i \int \frac{\mathrm{d}\omega}{2\pi} G_{\mathrm{at}}^{(1)}(\boldsymbol{q}_\perp, \omega) G_{\mathrm{at}}^{(1)}(\boldsymbol{K}_\perp - \boldsymbol{q}_\perp, \Omega - \omega) \ , \tag{4.32}$$

3: Each vertex thus forms a rung on a ladder of propagators.

which includes a factor of $i$ for later convenience. We see that the right hand side of Eq. (4.31) is independent of the individual incoming $y_1', y_2'$, and conclude that $\mathcal{T}$ only depends on the total incoming momentum and energy. This simple dependence comes from the underlying bare interaction being constant and simply conserving total momentum and energy. We can now easily take the limit $U \to \infty$, returning to the situation of two-level atoms. In this limit we find the simple expression

$$\mathcal{T}^{-1}(Y) = -a^2 \int_{\mathrm{BZ}} \frac{\mathrm{d}^2 q_\perp}{(2\pi)^2} \Pi(Y, \boldsymbol{q}_\perp) \ . \tag{4.33}$$

The pair propagator describes the free propagation of two atomic excitations after exchanging energy. The frequency integral is quickly calculated. We have

$$\begin{aligned} \Pi(\boldsymbol{K}_\perp, \Omega, \boldsymbol{q}_\perp) = i \int \frac{\mathrm{d}\omega}{2\pi} & \frac{1}{\omega + i\eta - \omega_a - \Sigma(\boldsymbol{q}_\perp)} \\ & \times \frac{1}{\Omega - \omega + i\eta - \omega_a - \Sigma(\boldsymbol{K}_\perp - \boldsymbol{q}_\perp)} \ , \end{aligned} \tag{4.34}$$

where from now on we suppress the frequency dependence of the self-energy as we will always take it to be evaluated at $\omega_a$. This integral can be calculated using complex contour integration (expanding the integration to a contour closed by an infinite semicircle in the upper or lower complex plane), or using the convolution theorem for Fourier transformations (as the integral is indeed a convolution and the Fourier transform of each fraction is an exponential that is easily multiplied and Fourier transformed back again). We find that the pair propagator takes on the form of a free propagator

$$\Pi(\boldsymbol{K}_\perp, \Omega, \boldsymbol{q}_\perp) = \frac{1}{\Omega + i\eta - 2\omega_a - \Sigma(\boldsymbol{q}_\perp) - \Sigma(\boldsymbol{K}_\perp - \boldsymbol{q}_\perp)} \ , \tag{4.35}$$

where we have absorbed a factor of 2 in $\eta$ as this is a positive infinitesimal.

Unfortunately the momentum integral in Eq. (4.33) can not be done easily, as we do not have a closed analytical expression for the self-energy. The self-energy (or rather, the discrete Fourier transformed EM Green's function) must be regularized and calculated numerically (as described in Section 1.3), and so we must also calculate the $\mathcal{T}$-matrix numerically. These are the only two quantities in the present approach that must be handled numerically. However, their physical interpretation is clear, and we will see that they enter only as algebraic factors in the results we derive, such that these will still be clear to understand and use.

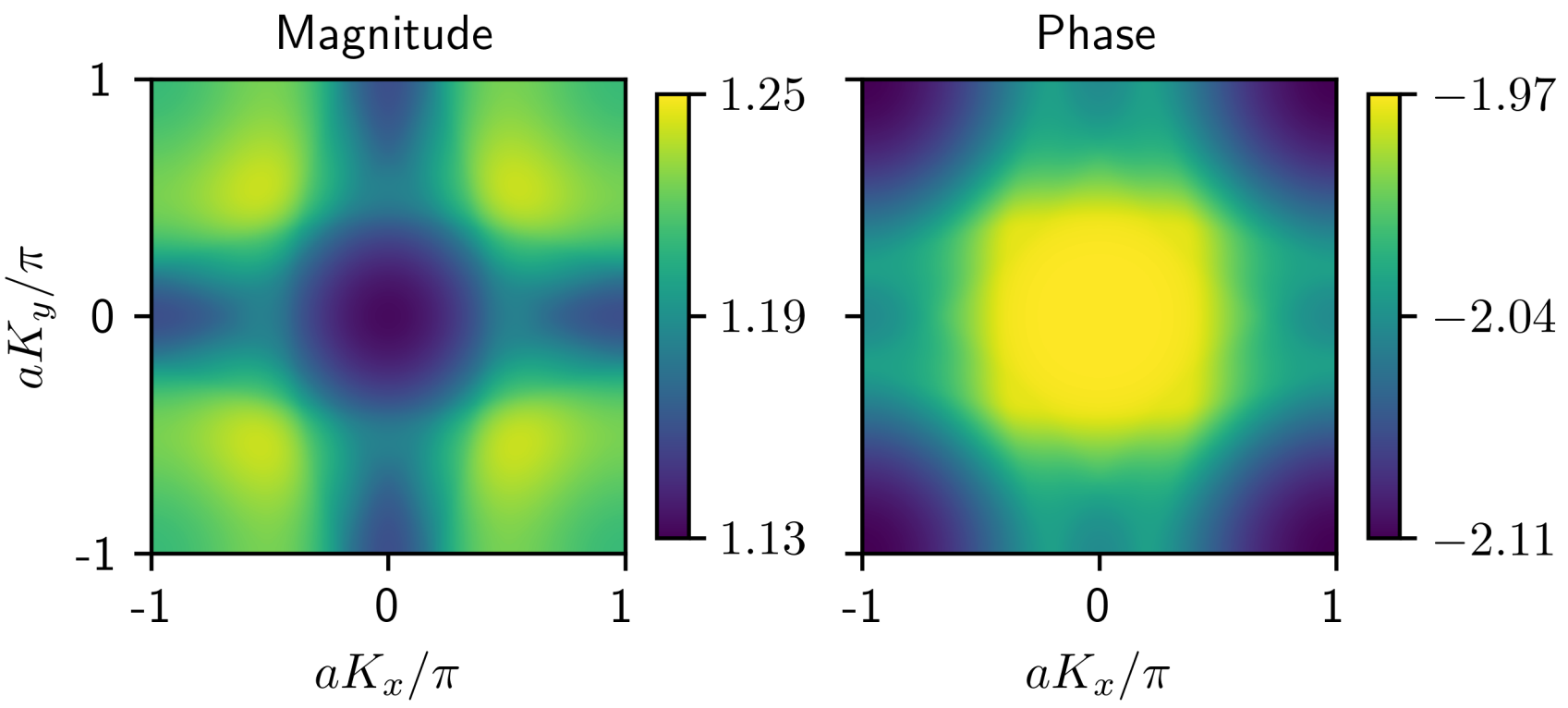

**Figure 4.1 | Example of atomic $\mathcal{T}$-matrix for the single array.** Equation (4.33) is numerically evaluated at the reflection resonance of the single array with $a = 0.6\lambda_a$, and the magnitude and phase of $\mathcal{T}/\tilde{\Gamma}$ is plotted as a function of $\boldsymbol{K}_\perp$.

Fig. 4.1 shows an example of the $\mathcal{T}$-matrix. We have numerically evaluated Eq. (4.33) at the reflection resonance of the single array with lattice spacing $a = 0.6\lambda_a$, and plotted $\mathcal{T}/\tilde{\Gamma}$ as a function of the transverse momentum $\boldsymbol{K}_\perp$. We see that neither its magnitude nor its phase vary much across the Brillouin zone. There is some structure similar to that of the collective energies. We will later on perform approximations due to the near-flat value of $\mathcal{T}$ at small momenta. The near-flatness of the $\mathcal{T}$-matrix originates in the fact that the bare interaction is truly flat, and the energies of the dressed atomic excitations, which might suppress scattering between certain momenta, do not vary much. Hence, scattering from one momentum to another is nearly independent of the involved momenta.

### 4.4.2 Explicit two-photon propagator

We can now write the explicit mathematical expression corresponding to Eq. (4.26)

$$\begin{aligned} &G^{(2)}_{\text{ph}}(x_1, x_2; x'_1, x'_2) \\ &\quad = - G^{(1)}_{\text{ph}}(x_1, x'_1) G^{(1)}_{\text{ph}}(x_2, x'_2) - G^{(1)}_{\text{ph}}(x_1, x'_2) G^{(1)}_{\text{ph}}(x_2, x'_1) \\ &\qquad - 2ia^2 (2\pi)^3 \delta(\Omega - \Omega') \delta_{\text{BZ}}(\boldsymbol{K}_\perp - \boldsymbol{K}'_\perp) \\ &\qquad\quad \times G^0_{\text{ph}}(\boldsymbol{k}_1, \omega_1) g_{\boldsymbol{k}_1 \nu_1} G^{(1)}_{\text{at}}(y_1) G^0_{\text{ph}}(\boldsymbol{k}_2, \omega_2) g_{\boldsymbol{k}_2 \nu_2} G^{(1)}_{\text{at}}(y_2) \\ &\qquad\quad \times \mathcal{T}(Y') \\ &\qquad\quad \times G^{(1)}_{\text{at}}(y'_1) g^*_{\boldsymbol{k}'_1 \nu'_1} G^0_{\text{ph}}(\boldsymbol{k}'_1, \omega'_1) G^{(1)}_{\text{at}}(y'_2) g^*_{\boldsymbol{k}'_2 \nu'_2} G^0_{\text{ph}}(\boldsymbol{k}'_2, \omega'_2) \quad , \end{aligned} \tag{4.36}$$

where $G^{(1)}_{\text{ph}}(\boldsymbol{k}\nu\omega, \boldsymbol{k}'\nu'\omega') = 2\pi\delta(\omega - \omega') G^{(1)}_{\text{ph}}(\boldsymbol{k}\nu, \boldsymbol{k}'_1\nu', \omega)$, and we have again used the multi-indices $x_i = (\boldsymbol{k}_i, \nu_i, \omega_i)$, $y_i = (\boldsymbol{k}_{i\perp}, \omega_i)$, and the total momentum-energy $Y = y_1 + y_2 = (\boldsymbol{K}_\perp, \Omega)$ (with a prime indicating it pertains to the incoming particles). The first line corresponds to no interaction between the photons, simply propagating (using the dressed propagator) from the initial momentum to the final, with two terms corresponding to the two possible combinations of initial and final photons. The last term is the actual interaction and has been ordered to make it most easily read backwards from the last factor. It corresponds to two photons propagating freely, getting absorbed into the atomic array, propagating as excitations (with the dressed propagator), interacting via their $\mathcal{T}$-matrix, and then the reverse process, yielding photons in the end. The phases of the terms and the factor of $2a^2$ come from the Feynman rules and our definition of the $\mathcal{T}$-matrix.

With this large expression in hand, and all factors made explicit (up to the numerical calculation of the atomic self-energies and $\mathcal{T}$-matrix), we can now proceed to employ Eq. (4.3) to find the two-photon steady state of the system given a single-frequency, but otherwise arbitrary, initial state.

## 4.5 Exact two-photon steady state

We now commence with the challenging task of evaluating the integral in Eq. (4.3). First we will consider what to actually calculate. The propagators we have found are fully in momentum-energy space, while Eq. (4.3) is in real space and time. Furthermore, since we are interested in studying the transmitted and reflected light of the system as separate quantities in the steady state, and as transverse momentum is conserved, up to Bragg scattering, while longitudinal momentum is not, it turns out to be convenient for the derivations to in the space of transverse momentum, longitudinal spatial coordinate, and real time, i.e. $(\boldsymbol{k}_\perp, z, t)$ (as we also did in the first part of the thesis). We introduce the new multi-index $\tilde{x} = (\boldsymbol{k}_\perp, z, \nu)$, keeping time separately. With this in mind, the two-photon version of Eq. (4.3) we will use is

$$\Psi^{(2)}(\tilde{x}_1, \tilde{x}_2, t) = \int \frac{\mathrm{d}k_{1z}\mathrm{d}k_{2z}\mathrm{d}\omega_1\mathrm{d}\omega_2}{(2\pi)^4} e^{i(k_{1z}z_1 + k_{2z}z_2)} e^{-i(\omega_1+\omega_2)t} \times \sum\!\!\!\!\!\!\int \frac{\mathrm{d}^4x_1'\mathrm{d}^4x_2'}{(2\pi)^8} G^{(2)}_{\text{ph}}(x_1, x_2; x_1', x_2') \Psi^{(2)}_{\text{in}}(\boldsymbol{k}_1'\nu_1', \boldsymbol{k}_2'\nu_2') \ . \tag{4.37}$$

The first line is the Fourier transformation from momentum-energy space to longitudinal real space coordinate and time, and the second line is the overlap of the propagator and the initial state. Here, the integral-sum over $x_i'$ implies a summation over the polarization indices, $\nu_i'$, and an integration over the momentum-energy, $\boldsymbol{k}_i', \omega_i'$. The fact that we take the initial time to be 0 means there is no Fourier exponential for the $\omega_i'$-integrations (i.e. no factor of $e^{-i(\omega_1'+\omega_2')t'}$). As it is the regime of interest and to make the integration feasible, we will find the wave function in the steady state (i.e. taking $t \to \infty$) and we will neglect evanescent fields (i.e. take both $z_1$ and $z_2$ to be large and neglect terms that exponentially decrease with these). We will be able to evaluate the parts of this expression that are independent of the initial state for an almost arbitrary initial state. We will only be imposing that the initial state is continuous-wave, i.e. that it has only one frequency component, as discussed below. The integrals directly pertaining to the initial state can of course only be done after choosing the initial state.

We will handle the linear and nonlinear parts of Eq. (4.36) separately. In particular, assuming the initial state to be uncorrelated such that it can be written as a product of identical initial wave functions, $\Psi^{(2)}_{\text{in}}(\boldsymbol{k}_1'\nu_1', \boldsymbol{k}_2'\nu_2') = \Psi^{(1)}_{\text{in}}(\boldsymbol{k}_1'\nu_1')\Psi^{(1)}_{\text{in}}(\boldsymbol{k}_2'\nu_2')$, the integration according to Eq. (4.37) of the first two terms of Eq. (4.36) both separate into identical integrals. These correspond to the time-evolution of just a single-photon wave function according to the single-photon propagator $G^{(1)}_{\text{ph}}(x, x')$. From Eq. (4.19) we see that this integral will itself split into a sum of two terms, corresponding to free evolution and to scattering off the atomic array. In

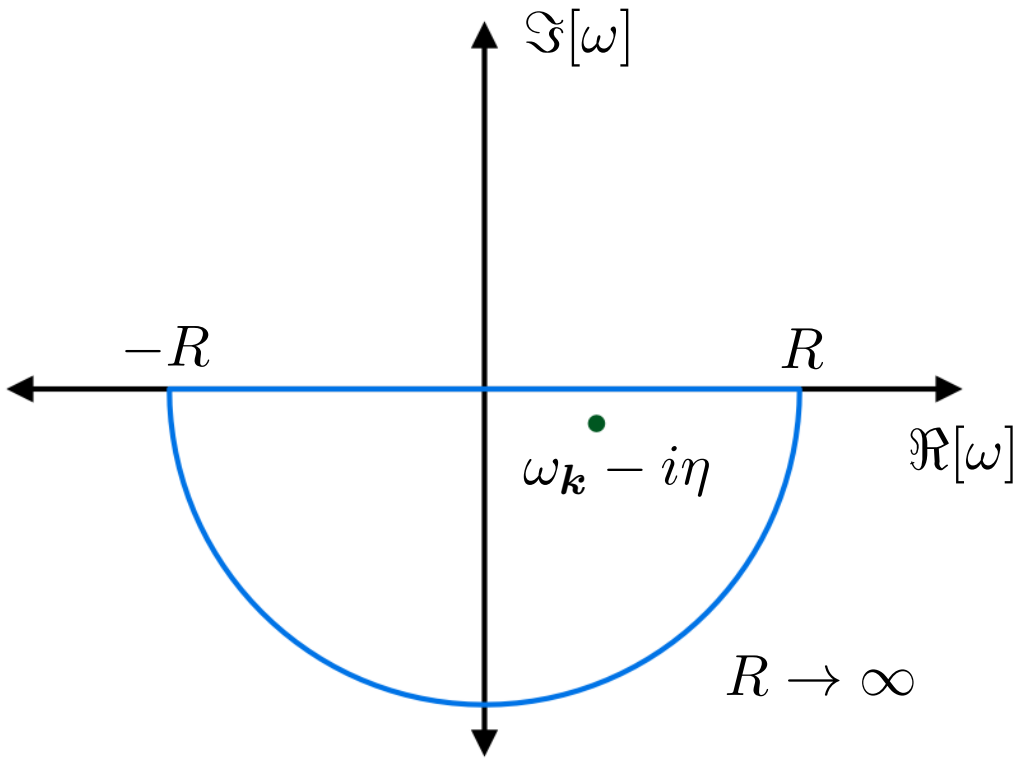

**Figure 4.2 | Contour for $\omega$-integral in Eq. (4.39).** The contour is shown in blue in the complex $\omega$-plane, with the single pole marked by a green dot.

other words, we will write

$$\begin{aligned}\Psi^{(2)}(\tilde{x}_1, \tilde{x}_2, t) = 2[\Psi_0^{(1)}(\tilde{x}_1, t) + \Psi_{\text{sc}}^{(1)}(\tilde{x}_1, t)] \\ \times [\Psi_0^{(1)}(\tilde{x}_2, t) + \Psi_{\text{sc}}^{(1)}(\tilde{x}_2, t)] \\ + \Psi_{\text{int}}^{(2)}(\tilde{x}_1, \tilde{x}_2, t) \quad .\end{aligned} \tag{4.38}$$

The factor of 2 comes from the fact that the two linear terms in Eq. (4.36) contribute identically due to the initial state being a product of identical functions. We will now proceed to calculate these partial wave functions individually.

### 4.5.1 Free propagation

We start by finding the freely propagating single-photon wave function, $\Psi_0^{(1)}(\tilde{x}, t)$, using Eqs. (4.8b) and (4.11b). We calculate

$$\begin{aligned}\Psi_0^{(1)}(\tilde{x}, t) &= i \int \frac{\mathrm{d}\omega}{2\pi} \int \frac{\mathrm{d}k_z}{2\pi} e^{-i\omega t} e^{ik_z z} \sum\!\!\!\!\!\!\int \frac{\mathrm{d}^4 x'}{(2\pi)^4} \\ &\qquad (2\pi)^4 \delta(\omega - \omega') \delta(\boldsymbol{k} - \boldsymbol{k}') \delta_{\nu,\nu'} \frac{1}{\omega + i\eta - \omega_{\boldsymbol{k}}} \Psi_{\text{in}}^{(1)}(\boldsymbol{k}'\nu') \\ &= i \int \frac{\mathrm{d}\omega}{2\pi} \int \frac{\mathrm{d}k_z}{2\pi} e^{-i\omega t} e^{ik_z z} \frac{1}{\omega + i\eta - \omega_{\boldsymbol{k}}} \Psi_{\text{in}}^{(1)}(\boldsymbol{k}\nu) \\ &= i \frac{1}{2\pi} \int \frac{\mathrm{d}k_z}{2\pi} (-2\pi i) e^{-i\omega_{\boldsymbol{k}} t} e^{ik_z z} \Psi_{\text{in}}^{(1)}(\boldsymbol{k}\nu) \\ &= \int \frac{\mathrm{d}k_z}{2\pi} e^{-i(\omega_{\boldsymbol{k}} t - k_z z)} \Psi_{\text{in}}^{(1)}(\boldsymbol{k}\nu) \ ,\end{aligned} \tag{4.39}$$

where we did the $\omega$-integral using complex contour integration, closing the contour with a semicircle at infinity in the lower half of the complex $\omega$-plane (yielding a minus due to the negative orientation of the contour), see Fig. 4.2. We choose the lower half-plane in order for the factor $e^{-i\omega t}$ to decrease exponentially, such that the contribution from the semicircle is zero. Generally, we will perform contour integrations by identifying which semi- or quarter circle can close the contour with a vanishing contribution. The time-dependence for a freely propagating photon is indeed $e^{-i\omega_{\boldsymbol{k}} t}$, such that the above result is the expected longitudinal Fourier transform of the state $\Psi_{\text{in}}^{(1)}(\boldsymbol{k}\nu)$ evolved for a time $t$.

We can already spot the condition we must make on the initial state in order to be able to go to the steady state. The limit $t \to \infty$ would imply the different frequency components of the above expression have highly different phases, irregardless of how close their frequencies are. In other words, a convergent result does not seem possible. If we simply consider physically the idea of a steady state, we can also see that we require the initial wave function to be constant. We therefore impose that the initial wave function be continuous-wave. Furthermore, we will also impose that it has only positive $k_z$-components, as both of these conditions correspond to the situation we considered in the first part of the thesis. Namely, we consider driving with light of a single frequency that is incident on the system from the left, propagating in the positive $z$-direction. On a technical level this means we take the initial wave function to be proportional to a delta-function in the frequency or in the longitudinal momentum. For simplicity we choose to assume $\Psi^{(1)}_{\text{in}}(\boldsymbol{k}\nu) = 2\pi\delta(k_z - k_{z,\omega})\Psi^{(1)}_{\text{in}}(\boldsymbol{k}_\perp\nu)$, where $\omega$ is the frequency of the incoming light, and $k_{z,\omega} = \sqrt{\omega^2 - k_\perp^2}$ is the positive longitudinal momentum pertaining to the frequency $\omega$ given the transverse momentum $\boldsymbol{k}_\perp$. The factor $e^{-i\omega t}$, of course, is not convergent for $t \to \infty$. However, if such a phase is global in the wave function it makes no physical difference, and we can artificially remove it. A stringent approach would perhaps be to overlap the wave function with the mode that we imagine is used for detection, and include this phase in the detection mode, such that it cancels, before taking the limit. For the sake of ease, we will simply remove the factor.

For the present case of a freely propagating photon we employ these considerations in Eq. (4.39) to write the steady state wave function as the simple expression

$$\Psi^{(1)}_0(\tilde{x}) = e^{ik_{z,\omega}z}\Psi^{(1)}_{\text{in}}(\boldsymbol{k}_\perp\nu) \ . \tag{4.40}$$

We will proceed like this for the other components of the wave function as well.

### 4.5.2 Atomic array scattering

We now proceed to calculate the steady state of $\Psi^{(1)}_{\text{sc}}(\tilde{x}, t)$, using the scattering part of $G^{(1)}_{\text{ph}}(x, x')$ given by second term of Eq. (4.19). The calculation at hand is

$$\begin{aligned}
&\Psi^{(1)}_{\text{sc}}(\tilde{x}, t) \\
&\quad = i\int \frac{\mathrm{d}\omega}{2\pi}\int \frac{\mathrm{d}k_z}{2\pi} e^{-i\omega t} e^{ik_z z} \sum\!\!\!\!\!\!\int \frac{\mathrm{d}^4 x'}{(2\pi)^4}(2\pi)^3\delta(\omega-\omega')\delta_{\text{BZ}}(\boldsymbol{k}_\perp - \boldsymbol{k}'_\perp) \\
&\qquad \times \frac{\tilde{\Gamma}}{\omega_a}\frac{1}{\omega + i\eta - \omega_{\boldsymbol{k}}}\frac{1}{\omega + i\eta - \omega_{\boldsymbol{k}'}}\frac{\sqrt{\omega_{\boldsymbol{k}}\omega_{\boldsymbol{k}'}}\hat{\boldsymbol{e}}^\dagger_{\boldsymbol{k}\nu}\hat{\boldsymbol{e}}_+\hat{\boldsymbol{e}}^\dagger_+\hat{\boldsymbol{e}}_{\boldsymbol{k}'\nu'}}{\omega + i\eta - \omega_a - \Sigma(\boldsymbol{k}'_\perp)}\Psi^{(1)}_{\text{in}}(\boldsymbol{k}'\nu') \\
&\quad = i\frac{\tilde{\Gamma}}{\omega_a}\int \frac{\mathrm{d}\omega}{2\pi} e^{-i\omega t}\frac{1}{\omega + i\eta - \omega_a - \Sigma(\boldsymbol{k}_\perp)}\int \frac{\mathrm{d}k_z}{2\pi} e^{ik_z z}\frac{\sqrt{\omega_{\boldsymbol{k}}}}{\omega + i\eta - \omega_{\boldsymbol{k}}} p(\boldsymbol{k}\nu) \\
&\qquad \times \int \frac{\mathrm{d}^3\boldsymbol{k}'}{2\pi}\delta_{\text{BZ}}(\boldsymbol{k}_\perp - \boldsymbol{k}'_\perp)\frac{\sqrt{\omega_{\boldsymbol{k}'}}}{\omega + i\eta - \omega_{\boldsymbol{k}'}}\sum_{\nu'} p^*(\boldsymbol{k}'\nu')\Psi^{(1)}_{\text{in}}(\boldsymbol{k}'\nu') \ ,
\end{aligned} \tag{4.41}$$

where we have ordered the factors according to their dependencies, and we have used the fact that the self-energy has the periodicity of the reciprocal lattice, i.e. $\Sigma(\boldsymbol{k}_\perp + \boldsymbol{q}_m) = \Sigma(\boldsymbol{k}_\perp)$ for any reciprocal lattice vector $\boldsymbol{q}_m$. Furthermore, we have introduced the overlap between the photonic polarization vector and the atomic dipole polarization $p(\boldsymbol{k}\nu) = \hat{\boldsymbol{e}}^\dagger_{\boldsymbol{k}\nu}\hat{\boldsymbol{e}}_+$. We could evaluate the primed transverse momentum integration using the Bragg delta-function, but wait to do so until a bit later.

We then proceed to calculate the integrals via complex contour integration. We perform the $k_z$-integral first. It is

$$\begin{aligned} I_{k_z} &= \int \frac{\mathrm{d}k_z}{2\pi} e^{ik_z z} \frac{\sqrt{\omega_{\boldsymbol{k}}}}{\omega + i\eta - \omega_{\boldsymbol{k}}} p(\boldsymbol{k}\nu) \\ &= \int \frac{\mathrm{d}k_z}{2\pi} e^{ik_z |z|} \frac{\sqrt{\omega_{\boldsymbol{k}}}}{\omega + i\eta - \omega_{\boldsymbol{k}}} p_{\mathrm{sgn}(z)}(\boldsymbol{k}\nu) \ , \end{aligned} \tag{4.42}$$

where $p_\pm(\boldsymbol{k}\nu)$ is the same as $p(\boldsymbol{k}\nu)$, but with $k_z \to \pm k_z$, and we have used the fact that the integrand is almost even in $k_z$ to take the absolute value of $z$, moving the sign of $z$ to the polarization overlap. The integrand has poles at $k_z = \pm k_{z,\omega} = \pm\sqrt{(\omega + i\eta)^2 - k_\perp^2}$ in the upper and lower half-plane respectively, where retaining $i\eta$ in the definition of $k_{z,\omega}$ ensures the poles are appropriately shifted into the complex plane. However, this is only for positive $\omega$, as $\omega_{\boldsymbol{k}} = \sqrt{k_\perp^2 + k_z^2}$ has a positive real part[4] for any complex $k_z$. It is also only for $k_\perp^2 \leq \omega^2$, as $k_{z,\omega}$ becomes purely imaginary and corresponds to an evanescent wave for $k_\perp^2 > \omega^2$. As we neglect evanescent contributions, we will take $I_{k_z}$ to be equal to zero for $k_\perp^2 > \omega^2$. Furthermore, the integrand has branch cuts from $k_z = \pm ik_\perp$ to $\pm i\infty$ respectively. These clearly have to do with evanescent fields, so we will neglect their contribution.

We therefore close the contour with a semicircle at infinite (with a dip to circumvent the branch cut) in the upper complex half-plane, see Fig. 4.3. The factor $e^{ik_z|z|}$ ensures the contribution from the semicircle vanishes. We thus only retain the contribution from the positive pole. The residues of the poles can be found in the following way. If a function $f(z) = g(z)/h(z)$ has a simple pole at $z_0$, where $h(z_0) = 0$, but its derivative $h'(z_0) \neq 0$, then the residue at $z_0$ is $g(z_0)/h'(z_0)$. Thus the residue of $1/(\omega + i\eta - \omega_{\boldsymbol{k}})$ at $\pm k_{z,\omega}$ is $\mp(\omega + i\eta)/k_{z,\omega}$ (the sign is a combination of the sign of $\pm k_{z,\omega}$, and a sign from the fact that $\omega_{\boldsymbol{k}}$ has a sign in the denominator). The rest of the integrand in Eq. (4.42) is simply evaluated at the point of the pole (corresponding to $g(z_0)$). In total, we find the following

$$I_{k_z} = -i\theta(\omega - k_\perp) e^{ik_{z,\omega}|z|} \frac{(\omega + i\eta)^{3/2}}{k_{z,\omega}} p_{\mathrm{sgn}(z),\omega}(\boldsymbol{k}_\perp\nu) \ , \tag{4.43}$$

4: The conventional principal branch of the square root is defined by having a positive real part for any argument.

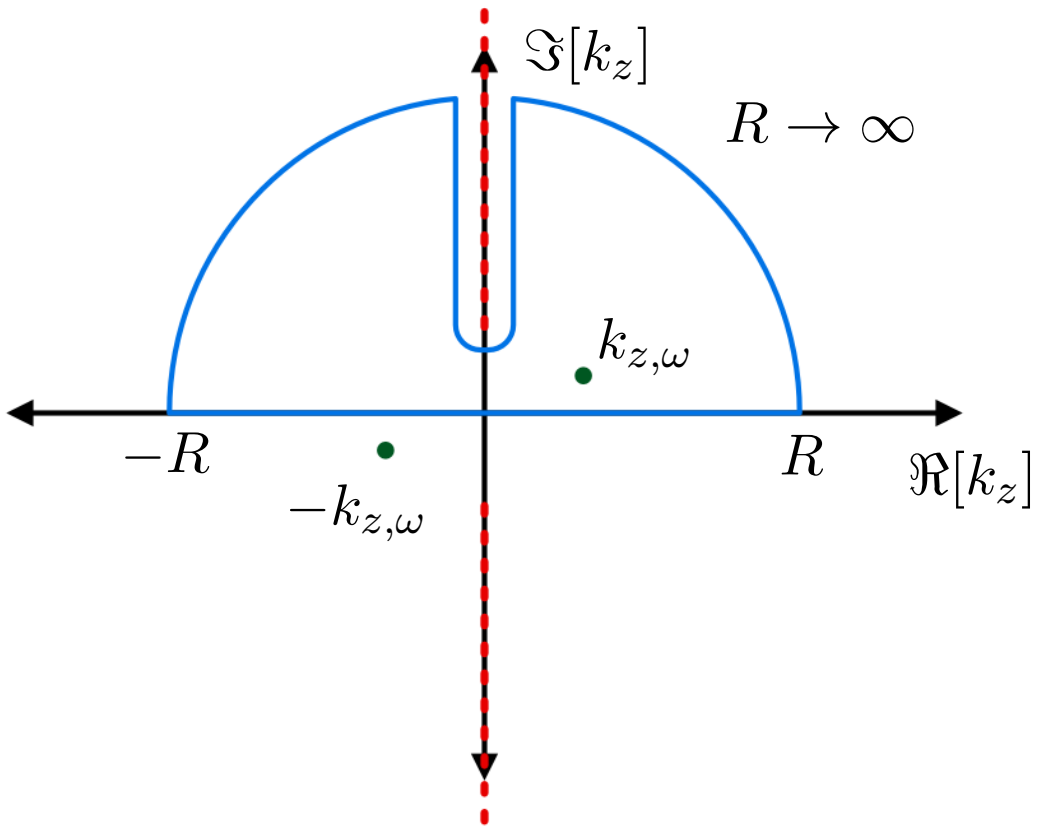

**Figure 4.3 | Contour for $k_z$-integral in Eq. (4.42).** The contour is shown in blue in the complex $k_z$-plane, with the poles marked by green dots and the branch cuts by red dashed lines.

where $p_{\mathrm{sgn}(z),\omega}(\boldsymbol{k}_\perp\nu)$ is $p_{\mathrm{sgn}(z)}(\boldsymbol{k}\nu)$, but with $k_z$ given by $k_{z,\omega}$.

We now do the $\omega$-integral of Eq. (4.41), writing only the factors that depend on $\omega$. It is

$$I_\omega = \int_{k_\perp}^{\infty} \frac{\mathrm{d}\omega}{2\pi} e^{-i\omega t} e^{ik_{z,\omega}|z|} \frac{(\omega+i\eta)^{3/2}}{k_{z,\omega}} p_{\mathrm{sgn}(z),\omega}(\boldsymbol{k}_\perp\nu) \times \frac{1}{\omega+i\eta-\tilde{\omega}_{\boldsymbol{k}_\perp}} \frac{1}{\omega+i\eta-\omega_{\boldsymbol{k}'}}, \tag{4.44}$$

where $\tilde{\omega}_{\boldsymbol{k}_\perp} \equiv \omega_a + \Sigma(\boldsymbol{k}_\perp)$. The integrand has poles at $\omega = \tilde{\omega}_{\boldsymbol{k}_\perp} - i\eta, \omega_{\boldsymbol{k}'} - i\eta$, and has branch cuts at $\omega = k_\perp - i\eta$ to $-k_\perp - i\eta$, and along all of the imaginary axis (both due to $k_{z,\omega}$). The pole at $\omega = \tilde{\omega}_{\boldsymbol{k}_\perp} - i\eta$ results in a complex frequency and thus exponential damping in time (due to the imaginary part of the self-energy for transverse momenta within the light cone, i.e. momenta corresponding to propagating light). Hence, we neglect any contribution from this pole. By closing the contour with a quarter circle at infinite in the lower right complex quadrant (the fourth quadrant) and a straight line parallel to the imaginary axis at real part $k_\perp$, we avoid the branch cuts, see Fig. 4.4. The contribution from this straight line we again neglect, as it has finite imaginary part, resulting in exponential damping in time. We take $\omega_{\boldsymbol{k}'} > k_\perp$ such that this pole is within the contour. We will eventually assume the incoming has only one frequency component, as we did before, and that the detuning of this frequency with respect to the atomic transition frequency is on the scale of the collective energies, i.e. $\gamma/\omega_a$. As $k_\perp$ ranges from 0 to $k_a$ (at the edge of the light cone), the condition $\omega_{\boldsymbol{k}'} > k_\perp$ is thus only broken by $\boldsymbol{k}_\perp$ right at the edge of the light cone. In other words, our calculation is valid for all momenta except those that are almost parallel to the atomic array, which we deem acceptable. Ultimately, we again only retain the contribution from a single pole to reach

$$I_\omega = -\, ie^{-i\omega_{\boldsymbol{k}'}t}e^{ik_{z,\omega_{\boldsymbol{k}'}}|z|}\frac{\omega_{\boldsymbol{k}'}^{3/2}}{k_{z,\omega_{\boldsymbol{k}'}}}p_{\mathrm{sgn}(z),\omega_{\boldsymbol{k}'}}(\boldsymbol{k}_\perp\nu)\frac{1}{\omega_{\boldsymbol{k}'} - \tilde{\omega}_{\boldsymbol{k}_\perp}} \ , \tag{4.45}$$

where we have finally let $\eta \to 0$.

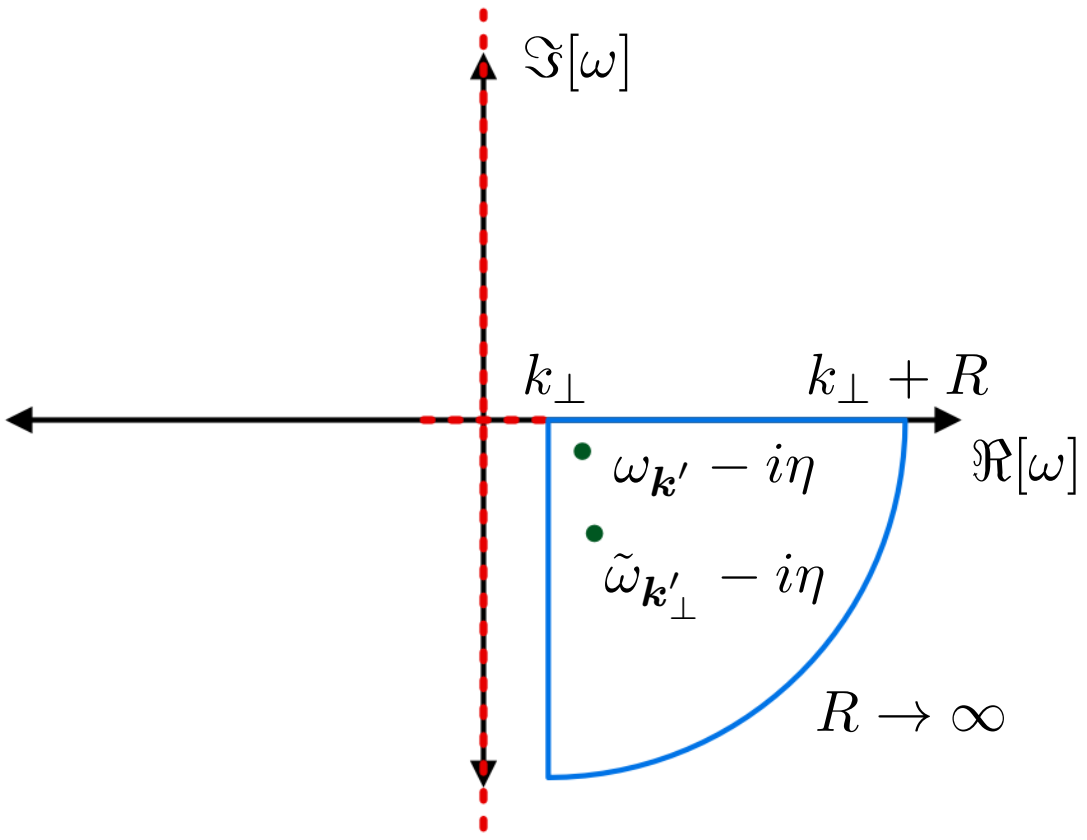

**Figure 4.4 | Contour for $\omega$-integral in Eq. (4.44).** The contour is shown in blue in the complex $\omega$-plane, with the poles marked by green dots and the branch cuts by red dashed lines.

Taking the initial wave function to have only a single frequency, $\omega$, specifically assuming $\Psi_{\mathrm{in}}^{(1)}(\boldsymbol{k}\nu) = 2\pi\delta(k_z - k_{z,\omega})\Psi_{\mathrm{in}}^{(1)}(\boldsymbol{k}_\perp\nu)$, as discussed in the previous section, and removing an overall factor of $e^{-i\omega t}$, we can thus write the steady state scattered single-photon wave function as

$$\begin{aligned}\Psi_{\mathrm{sc}}^{(1)}(\tilde{x}) = &-\, i\frac{\tilde{\Gamma}}{\omega_a}\frac{\omega^2}{k_{z,\omega}}\frac{1}{\Delta - \tilde{\Delta}_{\boldsymbol{k}_\perp}^0 + i\tilde{\Gamma}_{\boldsymbol{k}_\perp}^0}e^{ik_{z,\omega}|z|} \\ &\times p_{\mathrm{sgn}(z),\omega}(\boldsymbol{k}_\perp\nu)\sum_{\boldsymbol{q}_m,\nu'} p^*(\boldsymbol{k}'\nu')\Psi_{\mathrm{in}}^{(1)}(\boldsymbol{k}'_\perp\nu') \ ,\end{aligned} \tag{4.46}$$

where we have introduced the detuning $\Delta = \omega - \omega_a$, explicitly written the collective energies, evaluated the primed transverse momentum integral using the Bragg delta-function, and set $\boldsymbol{k}' = (\boldsymbol{k}_\perp + \boldsymbol{q}_m, k_{z,\omega})$ in the final expression. We have thus found the steady state wave function of the single-photon scattered light.

We see that Eq. (4.46) consists of a sum over different Bragg-scattered components, propagating symmetrically away from the array. As the propagation factor $e^{ik_{z,\omega}|z|}$ depends on $\boldsymbol{k}_\perp$, while the initial wave function $\Psi_{\text{in}}^{(1)}$ is evaluated at the Bragg-scattered $\boldsymbol{k}_\perp + \boldsymbol{q}_m$, means the wave function becomes evanescent for large $k_\perp$. If we imagine $\Psi_{\text{in}}^{(1)}$ is only non-zero for a certain region of transverse momentum, then for a large $\boldsymbol{q}_m$, $\boldsymbol{k}_\perp$ also needs to be large in order for $\boldsymbol{k}_\perp + \boldsymbol{q}_m$ to be in the region where $\Psi_{\text{in}}^{(1)}$ is non-zero. But then $k_{z,\omega}$ becomes imaginary, and $e^{ik_{z,\omega}|z|}$ becomes exponentially small (this expression is derived under the assumption that $|z|$ is large). This corresponds to the closed Bragg channels.

The factor $[\Delta - \tilde{\Delta}^0_{\boldsymbol{k}_\perp} + i\tilde{\Gamma}^0_{\boldsymbol{k}_\perp}]^{-1}$ can be understood as a complex detuning punishment. That is, the more detuned the incoming light is from the complex energy of the atomic collective mode off of which it is scattering, the less scattering we will see. This kind of factor, whose imaginary part is a Lorentzian, is ubiquitous in optics. We will refer to it as a *complex Lorentzian*. From the analysis in the previous part of the thesis, we also know this factor can be understood as the reflection amplitude of the array.

The polarization overlap factors can be understood as punishing mismatch between the polarization of the incoming light and the atomic dipoles, as well as between the outgoing light and the dipoles. The sum $\sum_{\nu'} p^*(\boldsymbol{k}'\nu')\Psi_{\text{in}}^{(1)}(\boldsymbol{k}'\nu')$ picks out the $\hat{\boldsymbol{e}}_+$ component of the initial *vector* wave function $\boldsymbol{\Psi}_{\text{in}}^{(1)}(\boldsymbol{k}'_\perp) = \sum_{\nu'} \hat{\boldsymbol{e}}_{\boldsymbol{k}'\nu'}\Psi_{\text{in}}^{(1)}(\boldsymbol{k}'_\perp\nu'))$, and $p_{\text{sgn}(z),\omega}(\boldsymbol{k}_\perp\nu)$ is the amplitude of the projection of $\hat{\boldsymbol{e}}_+$ onto $\hat{\boldsymbol{e}}_{(\boldsymbol{k}_\perp,\text{sgn}(z)k_{z,\omega})\nu}$.

The remaining factor can together with the polarization factors be seen as the building blocks of a factor of $\Gamma^0_{\boldsymbol{k}_\perp}$ from Eqs. (2.9) and (2.10), i.e. the continuous Fourier transformed decay rate. To get this factor explicitly, we consider the $\hat{\boldsymbol{e}}_+$-component of $\Psi_{\text{sc}}^{(1)}(\tilde{x})$, which we find via

$$\begin{aligned}\Psi_{\text{sc},+}(\boldsymbol{k}_\perp z) &= \hat{\boldsymbol{e}}_+^\dagger \sum_\nu \hat{\boldsymbol{e}}_{(\boldsymbol{k}_\perp,\text{sgn}(z)k_{z,\omega})\nu}\Psi_{\text{sc}}^{(1)}(\tilde{x}) \\ &= \sum_\nu p^*_{\text{sgn}(z),\omega}(\boldsymbol{k}_\perp\nu)\Psi_{\text{sc}}^{(1)}(\tilde{x}) \quad .\end{aligned} \tag{4.47}$$

As we take $\Delta$ to be on the scale of on the scale of $\gamma/\omega_a$, we can then approximate $\omega \simeq \omega_a$ everywhere except in the exponential, which may be sensitive to small changes of $k_{z,\omega}$ due to the largeness of $|z|$. If we furthermore say that $\Psi_{\text{sc}}^{(1)}(\tilde{x})$ is zero for evanescent $\boldsymbol{k}_\perp$, i.e. where $k_{z,\omega}$ is imaginary and $e^{ik_{z,\omega}|z|}$ causes exponential suppression, we can freely replace the factor $1/k_{z,\omega}$ with its real part $\Re[1/k_{z,\omega}]$. Using Eqs. (1.40) and (2.9), we can then find with some manipulations

$$\Psi_{\mathrm{sc},+}(\boldsymbol{k}_\perp z) = -\frac{i\Gamma^0_{\boldsymbol{k}_\perp}/a^2}{\Delta - \tilde{\Delta}^0_{\boldsymbol{k}_\perp} + i\tilde{\Gamma}^0_{\boldsymbol{k}_\perp}} e^{ik_{z,\omega}|z|} \sum_{\boldsymbol{q}_m} \Psi^{(1)}_{\mathrm{in},+}(\boldsymbol{k}_\perp + \boldsymbol{q}_m) \quad , \tag{4.48}$$

where $\Psi^{(1)}_{\mathrm{in},+}$ is the $\hat{\boldsymbol{e}}_+$-component of $\Psi^{(1)}_{\mathrm{in}}$. We remind ourselves that $\Gamma^0_{\boldsymbol{k}_\perp}$ is zero for evanescent $\boldsymbol{k}_\perp$, see Eq. (2.10). Considering the full single-photon wave function

$$\begin{aligned}
&\Psi^{(1)}_+(\boldsymbol{k}_\perp z) \\
&\quad = \Psi_{0,+}(\boldsymbol{k}_\perp z) + \Psi_{\mathrm{sc},+}(\boldsymbol{k}_\perp z) \\
&\quad = e^{ik_{z,\omega}z}\Psi^{(1)}_{\mathrm{in},+}(\boldsymbol{k}_\perp) - \frac{i\Gamma^0_{\boldsymbol{k}_\perp}/a^2}{\Delta - \tilde{\Delta}^0_{\boldsymbol{k}_\perp} + i\tilde{\Gamma}^0_{\boldsymbol{k}_\perp}} e^{ik_{z,\omega}|z|} \sum_{\boldsymbol{q}_m} \Psi^{(1)}_{\mathrm{in},+}(\boldsymbol{k}_\perp + \boldsymbol{q}_m) \quad ,
\end{aligned} \tag{4.49}$$

we could now easily find the transmission and reflection amplitudes of Eq. (2.20).

### 4.5.3 Photon-photon scattering

Finally, we consider the interacting part of the two-photon wave function. The procedure to calculate this quantity is very similar to that of the previous section, with one additional subtlety due to having two photons, which will lead to a manner of bound state between them. The integral to be evaluated is

$$\begin{aligned}
&\Psi^{(2)}_{\mathrm{int}}(\tilde{x}_1, \tilde{x}_2, t) \\
&= -2ia^2 \int \frac{\mathrm{d}\omega_1 \mathrm{d}\omega_2}{(2\pi)^2} e^{-i(\omega_1+\omega_2)t} \\
&\quad \times \int \frac{\mathrm{d}k_{1z}}{2\pi} e^{ik_{1z}z_1} G^0_{\mathrm{ph}}(\boldsymbol{k}_1, \omega_1) g_{\boldsymbol{k}_1\nu_1} \int \frac{\mathrm{d}k_{2z}}{2\pi} e^{ik_{2z}z_2} G^0_{\mathrm{ph}}(\boldsymbol{k}_2, \omega_2) g_{\boldsymbol{k}_2\nu_1} \\
&\quad \times G^{(1)}_{\mathrm{at}}(\boldsymbol{k}_{1\perp}, \omega_1) G^{(1)}_{\mathrm{at}}(\boldsymbol{k}_{2\perp}, \omega_2) \mathcal{T}(\boldsymbol{K}_\perp, \Omega) \\
&\quad \times \sum_{\nu_1'\nu_2'} \int \frac{\mathrm{d}^3k_1' \mathrm{d}^3k_2'}{(2\pi)^4} \delta_{\mathrm{BZ}}(\boldsymbol{K}_\perp - \boldsymbol{K}'_\perp) g^*_{\boldsymbol{k}_1'\nu_1'} g^*_{\boldsymbol{k}_2'\nu_2'} \Psi^{(2)}_{\mathrm{in}}(\boldsymbol{k}_1'\nu_1', \boldsymbol{k}_2'\nu_2') \\
&\quad \times \int \frac{\mathrm{d}\omega_1'}{2\pi} G^{(1)}_{\mathrm{at}}(\boldsymbol{k}'_{1\perp}, \omega_1') G^0_{\mathrm{ph}}(\boldsymbol{k}_1', \omega_1') \\
&\qquad\quad \times G^{(1)}_{\mathrm{at}}(\boldsymbol{k}'_{2\perp}, \Omega - \omega_1') G^0_{\mathrm{ph}}(\boldsymbol{k}_2', \Omega - \omega_1') \quad ,
\end{aligned} \tag{4.50}$$

where we have again arranged the factors according to their dependencies, and we have already carried out the $\omega_2'$-integration using the energy-conserving delta-function, $2\pi\delta(\Omega-\Omega')$. We have also used the fact that the atomic $\mathcal{T}$-matrix has the periodicity of the reciprocal lattice, just as the self-energy does. We perform the integrations one at a time.

First the $\omega_2'$-integral. It is

$$\begin{aligned} I_{\omega_2'} = \int \frac{\mathrm{d}\omega_1'}{2\pi} & \frac{1}{\omega_1' + i\eta - \tilde{\omega}_{\boldsymbol{k}_{1\perp}'}} \frac{1}{\omega_1' + i\eta - \omega_{\boldsymbol{k}_1'}} \\ & \times \frac{1}{\omega_1' - i\eta - \Omega + \tilde{\omega}_{\boldsymbol{k}_{2\perp}'}} \frac{1}{\omega_1' - i\eta - \Omega + \omega_{\boldsymbol{k}_2'}} \quad . \end{aligned} \tag{4.51}$$

This can be done with complex contour integration. There are no branch cuts and we can close the contour with a semicircle at infinity in either the upper or lower half-plane, which each contain two poles. Choosing either yields the same result (as it must be). Performing this integral, and then some algebra, it can be shown that

$$\begin{aligned} I_{\omega_2'} = & -i \left( 1 + \frac{\Omega - \omega_{\boldsymbol{k}_1'} - \omega_{\boldsymbol{k}_2'}}{\Omega - \tilde{\omega}_{\boldsymbol{k}_{1\perp}'} - \tilde{\omega}_{\boldsymbol{k}_{2\perp}'}} \right) \\ & \times \frac{1}{(\Omega - \omega_{\boldsymbol{k}_1'} - \omega_{\boldsymbol{k}_2'})(\Omega - \omega_{\boldsymbol{k}_2'} - \tilde{\omega}_{\boldsymbol{k}_{1\perp}'})(\Omega - \omega_{\boldsymbol{k}_1'} - \tilde{\omega}_{\boldsymbol{k}_{2\perp}'})} \quad . \end{aligned} \tag{4.52}$$

We see that the two $k_z$-integrals are identical in form, and they are also identical to the $k_z$-integral performed in the single-photon case, Eq. (4.42). Hence, we simply translate the result of Eq. (4.43) to find

$$\begin{aligned} \int & \frac{\mathrm{d}k_{1z}}{2\pi} e^{ik_{1z}z_1} G^0_{\mathrm{ph}}(\boldsymbol{k}_1, \omega_1) g_{\boldsymbol{k}_1\nu_1} \int \frac{\mathrm{d}k_{2z}}{2\pi} e^{ik_{2z}z_2} G^0_{\mathrm{ph}}(\boldsymbol{k}_2, \omega_2) g_{\boldsymbol{k}_2\nu_1} \\ & = \tilde{\Gamma}/\omega_a \theta(\omega_1 - k_{1\perp}) e^{ik_{1z,\omega_1}|z_1|} \frac{\omega_1^{3/2}}{k_{1z,\omega_1}} p_{\mathrm{sgn}(z_1),\omega_1}(k_{1\perp}\nu_1) \\ & \quad \times \theta(\omega_2 - k_{2\perp}) e^{ik_{2z,\omega_2}|z_2|} \frac{\omega_2^{3/2}}{k_{2z,\omega_2}} p_{\mathrm{sgn}(z_2),\omega_2}(k_{2\perp}\nu_2) \quad , \end{aligned} \tag{4.53}$$

where $k_{1z,\omega} = \sqrt{\omega^2 - k_{1\perp}^2}$ and likewise for $k_{2z,\omega}$.

Let us put together the results so far

$$\begin{aligned}
&\Psi^{(2)}_{\text{int}}(\tilde{x}_1, \tilde{x}_2, t) \\
&=2a^2 \left(\frac{\tilde{\Gamma}}{\omega_a}\right)^2 \int_{k_{1\perp}}^{\infty} \frac{\mathrm{d}\omega_1}{2\pi} \int_{k_{2\perp}}^{\infty} \frac{\mathrm{d}\omega_2}{2\pi} e^{-i(\omega_1+\omega_2)t} e^{ik_{1z,\omega_1}|z_1|} e^{ik_{2z,\omega_2}|z_2|} \\
&\quad \times \frac{(\omega_1\omega_2)^{3/2}}{k_{1z,\omega_1} k_{2z,\omega_2}} p_{\text{sgn}(z_1),\omega_1}(k_{1\perp}\nu_1) p_{\text{sgn}(z_2),\omega_2}(k_{2\perp}\nu_2) \\
&\quad \times G^{(1)}_{\text{at}}(\boldsymbol{k}_{1\perp}, \omega_1) G^{(1)}_{\text{at}}(\boldsymbol{k}_{2\perp}, \omega_2) \mathcal{T}(\boldsymbol{K}_\perp, \Omega) \\
&\quad \times \sum_{\nu_1'\nu_2'} \int \frac{\mathrm{d}^3k_1' \mathrm{d}^3k_2'}{(2\pi)^4} \delta_{\text{BZ}}(\boldsymbol{K}_\perp - \boldsymbol{K}'_\perp) \sqrt{\omega_{\boldsymbol{k}_1'} \omega_{\boldsymbol{k}_2'}} \\
&\quad \times p^*(\boldsymbol{k}_1', \nu_1') p^*(\boldsymbol{k}_2', \nu_2') \Psi^{(2)}_{\text{in}}(\boldsymbol{k}_1'\nu_1', \boldsymbol{k}_2'\nu_2') \\
&\quad \times \left(1 + \frac{\Omega - \omega_{\boldsymbol{k}_1'} - \omega_{\boldsymbol{k}_2'}}{\Omega - \tilde{\omega}_{\boldsymbol{k}_{1\perp}'} - \tilde{\omega}_{\boldsymbol{k}_{2\perp}'}}\right) \\
&\quad \times \frac{1}{(\Omega - \omega_{\boldsymbol{k}_1'} - \omega_{\boldsymbol{k}_2'})(\Omega - \omega_{\boldsymbol{k}_2'} - \tilde{\omega}_{\boldsymbol{k}_{1\perp}'})(\Omega - \omega_{\boldsymbol{k}_1'} - \tilde{\omega}_{\boldsymbol{k}_{2\perp}'})} \ .
\end{aligned} \tag{4.54}$$

Due to the complexity of this expression, we will perform the next few steps by considering a generalized expression, keeping the properties of our specific case in mind. Let us therefore look at

$$I = \int_{k_{1\perp}}^{\infty} \frac{\mathrm{d}\omega_1}{2\pi} \int_{k_{2\perp}}^{\infty} \frac{\mathrm{d}\omega_2}{2\pi} F_1(\omega_1) F_2(\omega_2) H(\Omega) e^{ik_{1z,\omega_1}|z_1|} e^{ik_{2z,\omega_2}|z_2|} e^{-i\Omega t} \ , \tag{4.55}$$

where $F_1$ and $F_2$ have only poles with positive real part and negative imaginary part, and there are no branch cuts (or we neglect their contribution). We start by performing the $\omega_1$-integral, closing the contour in the same way as when we calculated Eq. (4.44) (see Fig. 4.4). Thus, we also have the same restriction here, namely that our result is invalid for transverse momenta that are almost parallel with the array. This yields

$$\begin{aligned}
I = &\int_{k_{2\perp}}^{\infty} \frac{\mathrm{d}\omega_2}{2\pi} F_2(\omega_2) e^{ik_{2z,\omega_2}|z_2|} \\
&\times (-i) \Bigg[ \sum_{z_{F_1}} \mathrm{R}[F_1(z_{F_1})] H(z_{F_1} + \omega_2) e^{ik_{1z,z_{F_1}}|z_1|} e^{-i(z_{F_1}+\omega_2)t} \\
&\quad + \sum_{\substack{z_H \\ \Re[z_H - \omega_2] > k_{1\perp}}} F_1(z_H - \omega_2) \mathrm{R}[H(z_H)] e^{ik_{1z,z_H-\omega_2}|z_1|} e^{-iz_H t} \Bigg] \ ,
\end{aligned} \tag{4.56}$$

where $z_f$ indicates the poles of the function $f$ in the fourth quadrant with real part greater than $k_{1\perp}$, and $\mathrm{R}[f(z)]$ indicates the residue of $f$ at $z$. The two lines in this expression are quite different to integrate, as the first has a factor of $e^{-i\omega_2 t}$ such that it can be integrated in exactly the same way as the $\omega_1$-integral, while the second line must be handled differently. Let us therefore split the integral into two parts, and start with the first line, $I_1$. Using the same contour as above we find

$$\begin{aligned} I_1 = &- \sum_{z_{F_1}, z_{F_2}} \mathrm{R}[F_1(z_{F_1})]\mathrm{R}[F_2(z_{F_2})]H(z_{F_1}+z_{F_2}) \\ &\qquad \times e^{ik_{1z,z_{F_1}}|z_1|}e^{ik_{2z,z_{F_2}}|z_2|}e^{-i(z_{F_1}+z_{F_2})t} \\ &- \sum_{\substack{z_{F_1}, z_H \\ \Re[z_H - z_{F_1}]>0 \\ \Im[z_H - z_{F_1}]<0}} \mathrm{R}[F_1(z_{F_1})]F_2(z_H - z_{F_1})\mathrm{R}[H(z_H)] \\ &\qquad \times e^{ik_{1z,z_{F_1}}|z_1|}e^{ik_{2z,z_H - z_{F_1}}|z_2|}e^{-iz_H t} \quad . \end{aligned} \tag{4.57}$$

We will now use the fact that we want to consider the steady state, $t \to \infty$. This means only the exponents $e^{-izt}$ where $z$ is real will survive. If we look at back at the integrals we must do in Eq. (4.54), we see that only the pole at $z_H = \omega_{\boldsymbol{k}'_1} + \omega_{\boldsymbol{k}'_2}$ is real (after letting $\eta \to 0$, and assuming the $\mathcal{T}$-matrix does not have any real poles, which we can check numerically). Hence, the first sum in Eq. (4.57) is negligible. Furthermore, the functions $F_1$ and $F_2$ only have poles at $z_{F_i} = \tilde{\omega}_{\boldsymbol{k}_{i\perp}}$, and we know that $\Im[z_H - z_{F_1}] = \Im[\omega_{\boldsymbol{k}'_1} + \omega_{\boldsymbol{k}'_2} - \tilde{\omega}_{\boldsymbol{k}_{1\perp}}] = \tilde{\Gamma}^0_{\boldsymbol{k}_{1\perp}} > 0$ (for all momenta in the light cone, which is what we consider). Hence, the second sum in Eq. (4.57) is empty. Thus, $I_1 = 0$. Let us now consider the second line of Eq. (4.56), $I_2$, without the $\omega_2$-independent factors

$$I_2 = \int_{k_{2\perp}}^{z_H - k_{1\perp}} \frac{\mathrm{d}\omega_2}{2\pi} F_1(z_H - \omega_2)F_2(\omega_2)e^{ik_{1z,z_H-\omega_2}|z_1|}e^{ik_{2z,\omega_2}|z_2|} \quad . \tag{4.58}$$

Naively this integral can be calculated using a rectangular contour in the complex $\omega_2$-plane, by adding two straight lines parallel to the imaginary axis, and connecting them at infinity. However, the lines parallel to the imaginary axis, which we would normally expect to pertain to evanescent fields and thus be negligible, would yield important contributions. This is due to the fact that while the $|z_i|$ are large enough to justify neglecting evanescent fields, the integrals would involve $|z_2| - |z_1|$, which does take on small values. In other words, since we have two quantities with equally large scales, we can not immediately neglect contributions, as the two scales can cancel each other.

To calculate this integral then, we will return to the specific expressions of Eq. (4.54), and we will invoke approximations pertaining to the assumption that the detuning of the incoming light is small. Inserting the expressions of our case, and changing integration variable to $u = \omega_2 - \omega_{\boldsymbol{k}_2'}$, yields

$$
\begin{aligned}
I_2 = - \int_{-\omega_{\boldsymbol{k}_2'}+k_{2\perp}}^{\omega_{\boldsymbol{k}_1'}-k_{1\perp}} \frac{\mathrm{d}u}{2\pi} & \frac{(\omega_{\boldsymbol{k}_1'} - u)^{3/2}}{k_{1z,\omega_{\boldsymbol{k}_1'}-u}} p_{\mathrm{sgn}(z_1),\omega_{\boldsymbol{k}_1'}-u}(k_{1\perp}\nu_1) \\
& \times \frac{(u+\omega_{\boldsymbol{k}_2'})^{3/2}}{k_{2z,u+\omega_{\boldsymbol{k}_2'}}} p_{\mathrm{sgn}(z_2),u+\omega_{\boldsymbol{k}_2'}}(k_{2\perp}\nu_2) \\
& \times \frac{1}{u - \Delta_{\boldsymbol{k}_1'} + \tilde{\Delta}^0_{\boldsymbol{k}_{1\perp}} - i\tilde{\Gamma}^0_{\boldsymbol{k}_{1\perp}}} \frac{1}{u + \Delta_{\boldsymbol{k}_2'} - \tilde{\Delta}^0_{\boldsymbol{k}_{2\perp}} + i\tilde{\Gamma}^0_{\boldsymbol{k}_{2\perp}}} \\
& \times e^{ik_{1z,\omega_{\boldsymbol{k}_1'}-u}|z_1|} e^{ik_{2z,u+\omega_{\boldsymbol{k}_2'}}|z_2|} \quad .
\end{aligned}
\tag{4.59}
$$

We have introduced the detunings of the incoming light $\Delta_{\boldsymbol{k}_i'} = \omega_{\boldsymbol{k}_i'} - \omega_a$. We will now assume that these are on the same scale as the collective energies. The complex Lorentzians in the integrand will be strongly peaked around $\pm(\Delta_{\boldsymbol{k}_i'} - \tilde{\Delta}^{(0)}_{\boldsymbol{k}_{i\perp}})$, which thus are also small. We therefore perform expansions in $u$, assuming that contributions from the integrand, where $u$ is not on the scale of collective energies, are suppressed. We expand the $k_z$-factors in the exponentials to first order, but will in fact approximate all remaining terms, except the Lorentzians, to zeroth order, i.e. we simply evaluate them at $u = 0$. This is justified by the fact that the other terms are all polynomial in nature, and therefore change much more slowly than the exponentials, and by the extreme smallness of $\gamma/\omega_a$ (as mentioned, it may be as small as $\sim 10^{-7}$ [28]). Continuing this line of approximation, we will even approximate $\omega_{\boldsymbol{k}_i'} \simeq \omega_a$ for the polynomial factors, i.e. fully neglect the detuning. Using $k_{z,\omega} = k_{z,-\omega}$, the relevant expansion is

$$
\begin{aligned}
k_{iz,\omega_{\boldsymbol{k}_i'}\pm u} &= \sqrt{(\omega_{\boldsymbol{k}_i'} \pm u)^2 - k_{i\perp}^2} \\
&\simeq \sqrt{\omega_{\boldsymbol{k}_i'}^2 - k_{i\perp}^2} \pm \frac{\omega_{\boldsymbol{k}_i'}}{\sqrt{\omega_{\boldsymbol{k}_i'}^2 - k_{i\perp}^2}} u \quad .
\end{aligned}
\tag{4.60}
$$

With this, and performing zeroth order approximation for the remaining terms as explained, we continue the calculation

$$
\begin{aligned}
I_2 = &- \frac{\omega_a^3}{k_{1z,\omega_a} k_{2z,\omega_a}} p_{\mathrm{sgn}(z_1),\omega_a}(k_{1\perp}\nu_1) p_{\mathrm{sgn}(z_2),\omega_a}(k_{2\perp}\nu_2) \\
&\times e^{ik_{1z,\omega_{\boldsymbol{k}_1'}}|z_1|} e^{ik_{2z,\omega_{\boldsymbol{k}_2'}}|z_2|} \\
&\times \int_{-\omega_{\boldsymbol{k}_2'}+k_{2\perp}}^{\omega_{\boldsymbol{k}_1'}-k_{1\perp}} \frac{\mathrm{d}u}{2\pi} \frac{1}{u - \Delta_{\boldsymbol{k}_1'} + \tilde{\Delta}^0_{\boldsymbol{k}_{1\perp}} - i\tilde{\Gamma}^0_{\boldsymbol{k}_{1\perp}}} \frac{1}{u + \Delta_{\boldsymbol{k}_2'} - \tilde{\Delta}^0_{\boldsymbol{k}_{2\perp}} + i\tilde{\Gamma}^0_{\boldsymbol{k}_{2\perp}}} \\
&\qquad\qquad \times e^{i\left(\frac{\omega_{\boldsymbol{k}_2'}}{k_{2z,\omega_{\boldsymbol{k}_2'}}}|z_2| - \frac{\omega_{\boldsymbol{k}_1'}}{k_{1z,\omega_{\boldsymbol{k}_1'}}}|z_1|\right)u} .
\end{aligned}
\tag{4.61}
$$

This integral now has no branch cuts, and just two simple poles. Again because of the extremely narrow Lorentzians, we expand the integral to be over the full real axis. Introducing $\Delta r_{\boldsymbol{k}_1'\boldsymbol{k}_2'} = \frac{\omega_{\boldsymbol{k}_2'}}{k_{2z,\omega_{\boldsymbol{k}_2'}}}|z_2| - \frac{\omega_{\boldsymbol{k}_1'}}{k_{1z,\omega_{\boldsymbol{k}_1'}}}|z_1|$, we then calculate the integral via complex contour integration, closing the contour with a half-circle at infinite either in the upper or lower half-plane according to whether $\Delta r_{\boldsymbol{k}_1'\boldsymbol{k}_2'}$ is positive or negative. Performing the contour integration yields

$$
\begin{aligned}
I_2 = &- i\frac{\omega_a^3}{k_{1z,\omega_a} k_{2z,\omega_a}} p_{\mathrm{sgn}(z_1),\omega_a}(k_{1\perp}\nu_1) p_{\mathrm{sgn}(z_2),\omega_a}(k_{2\perp}\nu_2) \\
&\times e^{ik_{1z,\omega_{\boldsymbol{k}_1'}}|z_1|} e^{ik_{2z,\omega_{\boldsymbol{k}_2'}}|z_2|} \\
&\times \frac{1}{\Delta_{\boldsymbol{k}_1'} + \Delta_{\boldsymbol{k}_2'} - \tilde{\Delta}^0_{\boldsymbol{k}_{1\perp}} - \tilde{\Delta}^0_{\boldsymbol{k}_{2\perp}} + i(\tilde{\Gamma}^0_{\boldsymbol{k}_{1\perp}} + \tilde{\Gamma}^0_{\boldsymbol{k}_{2\perp}})} \\
&\times \Big[\theta(\Delta r_{\boldsymbol{k}_1'\boldsymbol{k}_2'}) e^{\left(i(\Delta_{\boldsymbol{k}_1'} - \tilde{\Delta}^0_{\boldsymbol{k}_{1\perp}}) - \tilde{\Gamma}^0_{\boldsymbol{k}_{1\perp}}\right)\Delta r_{\boldsymbol{k}_1'\boldsymbol{k}_2'}} \\
&\quad + \theta(-\Delta r_{\boldsymbol{k}_1'\boldsymbol{k}_2'}) e^{\left(i(\Delta_{\boldsymbol{k}_2'} - \tilde{\Delta}^0_{\boldsymbol{k}_{2\perp}}) - \tilde{\Gamma}^0_{\boldsymbol{k}_{2\perp}}\right)(-\Delta r_{\boldsymbol{k}_1'\boldsymbol{k}_2'})}\Big] .
\end{aligned}
\tag{4.62}
$$

We note that $\frac{\omega}{k_{z,\omega}}|z|$ is the distance travelled away from the array, when following a straight line defined by the direction of $\boldsymbol{k} = (\boldsymbol{k}_\perp, k_{z,\omega})$ and whose endpoint has $z$-coordinate $z$. Hence, $\Delta r_{\boldsymbol{k}_1'\boldsymbol{k}_2'}$ is the difference of how far the two photons have travelled away from the array.

We can now write a final expression for the interacting part of the two-photon wave function, assuming as in the single-photon case $\Psi^{(2)}_{\mathrm{in}}(\boldsymbol{k}_1'\nu_1', \boldsymbol{k}_2'\nu_2') = (2\pi)^2\delta(k_{1z} - k_{1z,\omega})\delta(k_{2z} - k_{2z,\omega})\Psi^{(2)}_{\mathrm{in}}(\boldsymbol{k}_{1\perp}'\nu_1', \boldsymbol{k}_{2\perp}'\nu_2')$ (consistent with the assumptions utilized above, and assuming both initial photons to have the same frequency $\omega$), and performing the same small-detuning approximations as above

$$
\begin{aligned}
&\Psi^{(2)}_{\text{int}}(\tilde{x}_1, \tilde{x}_2) \\
&= -2\tilde{\Gamma}^2 \frac{\omega_a^2}{k_{1z,\omega_a} k_{2z,\omega_a}} p_{\text{sgn}(z_1),\omega_a}(k_{1\perp}\nu_1) p_{\text{sgn}(z_2),\omega_a}(k_{2\perp}\nu_2) \\
&\quad \times e^{i(k_{1z,\omega}|z_1| + k_{2z,\omega}|z_2|)} \\
&\quad \times \frac{\mathcal{T}(\boldsymbol{k}_{1\perp} + \boldsymbol{k}_{2\perp}, 2\omega)}{2\Delta - \tilde{\Delta}^0_{\boldsymbol{k}_{1\perp}} - \tilde{\Delta}^0_{\boldsymbol{k}_{2\perp}} + i(\tilde{\Gamma}^0_{\boldsymbol{k}_{1\perp}} + \tilde{\Gamma}^0_{\boldsymbol{k}_{2\perp}})} \\
&\quad \times \Bigg[\theta(\Delta r) e^{\left(i(\Delta - \tilde{\Delta}^0_{\boldsymbol{k}_{1\perp}}) - \tilde{\Gamma}^0_{\boldsymbol{k}_{1\perp}}\right)\Delta r} \\
&\qquad + \theta(-\Delta r) e^{\left(i(\Delta - \tilde{\Delta}^0_{\boldsymbol{k}_{2\perp}}) - \tilde{\Gamma}^0_{\boldsymbol{k}_{2\perp}}\right)(-\Delta r)}\Bigg] \\
&\quad \times a^2 \int \frac{\mathrm{d}^2 k'_{1\perp} \mathrm{d}^2 k'_{2\perp}}{(2\pi)^2} \delta_{\text{BZ}}(\boldsymbol{k}_{1\perp} + \boldsymbol{k}_{2\perp} - \boldsymbol{k}'_{1\perp} - \boldsymbol{k}'_{2\perp}) \\
&\qquad \times p^*(\boldsymbol{k}'_1, \nu'_1) p^*(\boldsymbol{k}'_2, \nu'_2) \Psi^{(2)}_{\text{in}}(\boldsymbol{k}'_{1\perp}\nu'_1, \boldsymbol{k}'_{2\perp}\nu'_2) \\
&\qquad \times \frac{1}{\Delta - \tilde{\Delta}^0_{\boldsymbol{k}'_{1\perp}} + i\tilde{\Gamma}^0_{\boldsymbol{k}'_{1\perp}}} \frac{1}{\Delta - \tilde{\Delta}^0_{\boldsymbol{k}'_{2\perp}} + i\tilde{\Gamma}^0_{\boldsymbol{k}'_{2\perp}}} \,,
\end{aligned}
\tag{4.63}
$$

where $\Delta r = \frac{\omega_a}{k_{2z,\omega_a}}|z_2| - \frac{\omega_a}{k_{1z,\omega_a}}|z_1|$, and $\boldsymbol{k}'_1 = (\boldsymbol{k}_{1\perp}, k_{1z,\omega})$ and likewise for $\boldsymbol{k}'_2$. Let us understand this expression.

The first line has factors that stem from the EM vacuum Green's function, and factors pertaining to overlap of polarization, which are understood in the same way as in the single-photon case. The usual plane wave propagation factors follow with propagation symmetrically away from the array. Then a factor that peaks when the *total* photonic energy matches the total energy of the collective atomic excitations that decay to the outgoing photons. The $\mathcal{T}$-matrix evaluated at the total transverse momentum and the total energy, sets an overall scale of the interacting part of the wave function stemming from the strength of the excitation-excitation coupling. The square parenthesis shows exponential localization of the two photons relative to each other. The localization is with respect to the relative distance travelled by the two photons. The relevant decay rate is given by the photon that has travelled the least distance. This expression is for the steady state and no evanescent fields (large $|z_i|$), so the exponential localization can be explained by imagining the following. After the first photon has been emitted from the array, the second simply experiences the dynamics of a single atomic excitation, accumulating a phase and decaying according to the complex energy of the atomic mode it inhabits. The exponential decay of that mode yields this apparent localization, or rather timing, of the photons. We then integrate over the initial state with polarization overlaps and a delta-function that conserves total transverse momentum up to Bragg scattering. The final line has factors that peak strongly when the *individual* incoming photons' energies match the energies of the corresponding collective atomic excitations that they excite. In other words, as the photons are absorbed individually it is their individual energies that must match the modes they excite, but after interacting it only the total energy that must match the modes that emit the photons. The three detuning terms could also be written as two dressed atomic propagators, $G^{(1)}_{\text{at}}$, and a pair propagator, $\Pi$, for the sake of a more compact notation.

We have thus derived an analytical expression for the full two-photon wave function in the steady state, neglecting evanescent waves, and assuming the initial wave function to have only a single frequency. There remains an integral over the initial wave function, and since we do not have explicit expression for the self-energies and the $\mathcal{T}$-matrix, not all initial wave functions would allow for analytical evaluation of these integrals. However, we shall see that for a Gaussian beam (which is what we considered when calculating correlations in the previous part of the thesis) the integral can be done. From this wave function we can thus extract any correlations between photons travelling away from the single atomic array.

In this way, we have not and will not, in fact, be invoking any additional restrictions compared to the previous numerical analysis. There we also considered the propagating part of the emitted field in the steady state for a continuous-wave drive. The basic Hamiltonian is the same and we have performed the Markov approximation in both analyses. We shall below be invoking approximations concerning the flatness of the collective energies within the area of transverse momentum where the Gaussian beam is located. However, while these approximations were not directly used in the previous analysis, they would have been valid, and this is in fact the reason why we saw that the Gaussian beam only experiences minimal scattering into other modes (loss). Since the collective energies have similar values for all the components of the Gaussian beam, these each experienced the same response from the atomic array, and so the mode was conserved. Likewise, we shall be invoking the paraxial approximation, but this was also implicitly included in the previous analysis by assuming the driving to be a Gaussian beam. In other words, the exact results of the presented approach have come at no additional cost and have the same regime of validity.

Before turning to the extraction of correlators, we will as a last detail write the right-circular component of $\Psi^{(2)}_{\text{int}}$, just as we did in the single-photon case, Eq. (4.48). Following the same approach, we find here

$$\begin{aligned}\Psi^{(2)}_{\text{int},+}&(\boldsymbol{k}_{1\perp}z_1,\boldsymbol{k}_{2\perp}z_2)\\ =&-2e^{i(k_{1z,\omega}|z_1|+k_{2z,\omega}|z_2|)}\\ &\times\frac{\mathcal{T}(\boldsymbol{k}_{1\perp}+\boldsymbol{k}_{2\perp},2\omega)}{2\Delta-\tilde{\Delta}^0_{\boldsymbol{k}_{1\perp}}-\tilde{\Delta}^0_{\boldsymbol{k}_{2\perp}}+i(\tilde{\Gamma}^0_{\boldsymbol{k}_{1\perp}}+\tilde{\Gamma}^0_{\boldsymbol{k}_{2\perp}})}\\ &\times\Big[\theta(\Delta r)e^{\left(i(\Delta-\tilde{\Delta}^0_{\boldsymbol{k}_{1\perp}})-\tilde{\Gamma}^0_{\boldsymbol{k}_{1\perp}}\right)\Delta r}\\ &\quad+\theta(-\Delta r)e^{\left(i(\Delta-\tilde{\Delta}^0_{\boldsymbol{k}_{2\perp}})-\tilde{\Gamma}^0_{\boldsymbol{k}_{2\perp}}\right)(-\Delta r)}\Big]\\ &\times a^2\int\frac{\mathrm{d}^2k'_{1\perp}\mathrm{d}^2k'_{2\perp}}{(2\pi)^2}\delta_{\text{BZ}}(\boldsymbol{k}_{1\perp}+\boldsymbol{k}_{2\perp}-\boldsymbol{k}'_{1\perp}-\boldsymbol{k}'_{2\perp})\\ &\quad\times p^*(\boldsymbol{k}'_1,\nu'_1)p^*(\boldsymbol{k}'_2,\nu'_2)\Psi^{(2)}_{\text{in}}(\boldsymbol{k}'_{1\perp}\nu'_1,\boldsymbol{k}'_{2\perp}\nu'_2)\\ &\quad\times\frac{\Gamma^0_{\boldsymbol{k}_{1\perp}}/a^2}{\Delta-\tilde{\Delta}^0_{\boldsymbol{k}'_{1\perp}}+i\tilde{\Gamma}^0_{\boldsymbol{k}'_{1\perp}}}\frac{\Gamma^0_{\boldsymbol{k}_{2\perp}}/a^2}{\Delta-\tilde{\Delta}^0_{\boldsymbol{k}'_{2\perp}}+i\tilde{\Gamma}^0_{\boldsymbol{k}'_{2\perp}}}\ .\end{aligned}\tag{4.64}$$

Using Eq. (4.38) and the derived components, Eqs. (4.40), (4.46) and (4.63), we now have the full steady state two-photon wave function.

## 4.6 Correlation functions

In the previous part of the thesis we considered the two-photon correlations in the limit of vanishing driving strength, meaning we saw only the lowest order contribution, namely that from the two-or-fewer photon sector of the system. To do the same here we must only include up to two photons in the total wave function, when calculating the correlators, i.e. we take the total wave function to be the two-photon wave function we calculated in the previous section. In accordance with the interpretation of the exponential localization of the photons as being due to one photon leaving the atomic array, propagating freely, while the other evolves as a single excitation and eventually is emitted, we will interpret the relative distance travelled by the photons, $\Delta r$, as the relative time between the emission of the two photons. That is, when calculating the correlation between two photons at different times, the time difference will be equal to the relative travelled distance. We can do this since the photons propagate freely at the speed of light away from the atomic array, such that a distance can be immediately translated into a time.

Calculating the momentum density,

$$\rho(\boldsymbol{k}_{1\perp}, \boldsymbol{k}_{2\perp}, t) = \left\langle E^\dagger(\boldsymbol{k}_{1\perp}, t')E^\dagger(\boldsymbol{k}_{2\perp}, t'+t)E(\boldsymbol{k}_{2\perp}, t'+t)E(\boldsymbol{k}_{1\perp}, t')\right\rangle \quad , \tag{4.65}$$

will thus be done by taking the squared norm of $\Psi^{(2)}$ at the appropriate momenta and replacing $\Delta r \to t$. Any further $z_i$-dependencies cancel. Likewise calculating the two-time correlation function

$$g^{(2)}(t) = \frac{\left\langle E^\dagger_{\boldsymbol{f}_G}(t')E^\dagger_{\boldsymbol{f}_G}(t'+t)E_{\boldsymbol{f}_G}(t'+t)E_{\boldsymbol{f}_G}(t')\right\rangle}{\left\langle E^\dagger_{\boldsymbol{f}_G}(t')E_{\boldsymbol{f}_G}(t')\right\rangle \left\langle E^\dagger_{\boldsymbol{f}_G}(t'+t)E_{\boldsymbol{f}_G}(t'+t)\right\rangle} \tag{4.66}$$

is done by overlapping $\Psi^{(2)}$ with a Gaussian mode (corresponding to detection in that mode), and taking the squared norm of the result, again performing the replacement $\Delta r \to t$, and then dividing by the square of same quantity but for $\Psi^{(1)} = \Psi^{(1)}_0 + \Psi^{(1)}_{\text{sc}}$. We start by finding the wave functions for Gaussian initial wave function, and taking the right-circular component of the wave function, as that is what we have considered previously.

## 4.6.1 Incoming Gaussian beam

With the initial wave function being a product $\Psi^{(2)}_{\text{in}}(\boldsymbol{k}'_{1\perp}\nu'_1, \boldsymbol{k}'_{2\perp}\nu'_2) = \Psi^{(1)}_{\text{in}}(\boldsymbol{k}'_{1\perp}\nu'_1)\Psi^{(1)}_{\text{in}}(\boldsymbol{k}'_{2\perp}\nu'_2)$, we now take the single-photon initial wave function to be given by the Gaussian mode of Eq. (2.48)

$$\Psi^{(1)}_{\text{in}}(\boldsymbol{k}_\perp\nu) = \sqrt{2\pi}w_0 p(\boldsymbol{k}\nu)e^{-k_\perp^2 w_0^2/4} \quad , \tag{4.67}$$

where $\boldsymbol{k} = (\boldsymbol{k}_\perp, k_{z,\omega})$, and the factor of $p(\boldsymbol{k}\nu)$ implies the mode is right-circularly polarized. We will now calculate each of $\Psi^{(1)}_{0,+}$, $\Psi^{(1)}_{\text{sc},+}$, and $\Psi^{(2)}_{\text{int},+}$ for this choice of initial state.

### Free single-photon wave function

For the free single-photon wave function, Eq. (4.40), we simply have

$$\begin{aligned} \Psi^{(1)}_{0,+}(\boldsymbol{k}_\perp z) &= \sum_\nu p^*(\boldsymbol{k}\nu)\sqrt{2\pi}w_0 p(\boldsymbol{k}\nu)e^{ik_{z,\omega}z}e^{-k_\perp^2 w_0^2/4} \\ &= \sqrt{2\pi}w_0\left(1 - \frac{k_\perp^2}{2k_a^2}\right)e^{ik_{z,\omega}z}e^{-k_\perp^2 w_0^2/4} \quad , \end{aligned} \tag{4.68}$$

where we have used $\sum_\nu p^*(\boldsymbol{k}\nu)p(\boldsymbol{k}\nu) = \hat{\boldsymbol{e}}_+(1-\hat{\boldsymbol{k}}\hat{\boldsymbol{k}}^\dagger)\hat{\boldsymbol{e}}_+ = 1-k_\perp^2/2k^2$, and we have furthermore performed the small detuning approximations. The factor $\left(1-\frac{k_\perp^2}{2k_a^2}\right)$ essentially quantifies the error due to the paraxial approximation, i.e. the error in assuming the Gaussian beam has a single polarization. For a sufficiently broad Gaussian beam we can approximate this factor to unity (as the paraxial approximation is valid). This, incidentally, is a good approximation already at $w_0 \gtrsim \lambda_a$, as the typical scale of the second term in the parenthesis would then be $k_\perp^2/2k_a^2 \simeq 1/2w_0^2k_a^2 = 1/2(2\pi)^2 \sim 10^{-2}$. Within this approximation we have

$$\Psi_{0,+}^{(1)}(\boldsymbol{k}_\perp z) = \sqrt{2\pi}w_0 e^{ik_{z,\omega}z}e^{-k_\perp^2 w_0^2/4} \quad . \tag{4.69}$$

### Scattered single-photon wave function

For the scattered single-photon wave function we use the expression in Eq. (4.48) to find

$$\begin{aligned}\Psi_{\text{sc},+}^{(1)}(\boldsymbol{k}_\perp z) =& -\sqrt{2\pi}w_0\frac{i\Gamma_{\boldsymbol{k}_\perp}^0/a^2}{\Delta-\tilde{\Delta}_{\boldsymbol{k}_\perp}^0+i\tilde{\Gamma}_{\boldsymbol{k}_\perp}^0}e^{ik_{z,\omega}|z|}\\ &\times\sum_{\boldsymbol{q}_m}\left(1-\frac{(\boldsymbol{k}_\perp+\boldsymbol{q}_m)^2}{2k_a^2}\right)e^{-(\boldsymbol{k}_\perp+\boldsymbol{q}_m)^2w_0^2/4} \quad .\end{aligned} \tag{4.70}$$

Here, we can also perform an approximation due to the narrowness of the Gaussian mode in transverse momentum space and due to the flatness of the collective energies for small momenta (see Section 1.5.3). In particular, due to the Gaussian factor in the Bragg sum, $\Psi_{\text{sc},+}^{(1)}(\boldsymbol{k}_\perp z)$ vanishes away from $\boldsymbol{k}_\perp = 0$, up to Bragg scattering. Therefore, we can approximate $\tilde{\Delta}_{\boldsymbol{k}_\perp}^0 \simeq \tilde{\Delta}_0^0 = \tilde{\Delta}$, $\tilde{\Gamma}_{\boldsymbol{k}_\perp}^0 \simeq \tilde{\Gamma}_0^0 = \tilde{\Gamma}$, and also $\left(1-\frac{(\boldsymbol{k}_\perp+\boldsymbol{q}_m)^2}{2k_a^2}\right) \simeq 1$ as for the free wave function. Within this approximation we have

$$\Psi_{\text{sc},+}^{(1)}(\boldsymbol{k}_\perp z) = -\sqrt{2\pi}w_0\frac{i\Gamma_{\boldsymbol{k}_\perp}^0/a^2}{\Delta-\tilde{\Delta}+i\tilde{\Gamma}}e^{ik_{z,\omega}|z|}\sum_{\boldsymbol{q}_m}e^{-(\boldsymbol{k}_\perp+\boldsymbol{q}_m)^2w_0^2/4} \quad . \tag{4.71}$$

Note here that the sum is over *all* reciprocal lattice vectors, including those pertaining to evanescent waves, but in order for $e^{-(\boldsymbol{k}_\perp+\boldsymbol{q}_m)^2w_0^2/4}$ to not be vanishing, $\boldsymbol{k}_\perp$ must be of the same size as $\boldsymbol{q}_m$, and $\Gamma_{\boldsymbol{k}_\perp}^0$ becomes zero outside the light cone, and thus evanescent contributions are indeed excluded from the above.

### Interacting two-photon wave function

For the interacting two-photon wave function we use the expression in Eq. (4.64) to find

$$\begin{aligned}
&\Psi^{(2)}_{\text{int},+}(\boldsymbol{k}_{1\perp}z_1, \boldsymbol{k}_{2\perp}z_2) \\
&\quad = -4\pi w_0^2 e^{i(k_{1z,\omega}|z_1|+k_{2z,\omega}|z_2|)} \\
&\qquad \times \frac{\mathcal{T}(\boldsymbol{k}_{1\perp}+\boldsymbol{k}_{2\perp}, 2\omega)}{2\Delta - \tilde{\Delta}^0_{\boldsymbol{k}_{1\perp}} - \tilde{\Delta}^0_{\boldsymbol{k}_{2\perp}} + i(\tilde{\Gamma}^0_{\boldsymbol{k}_{1\perp}} + \tilde{\Gamma}^0_{\boldsymbol{k}_{2\perp}})} \\
&\qquad \times \Big[\theta(\Delta r) e^{\left(i(\Delta - \tilde{\Delta}^0_{\boldsymbol{k}_{1\perp}}) - \tilde{\Gamma}^0_{\boldsymbol{k}_{1\perp}}\right)\Delta r} \\
&\qquad\quad + \theta(-\Delta r) e^{\left(i(\Delta - \tilde{\Delta}^0_{\boldsymbol{k}_{2\perp}}) - \tilde{\Gamma}^0_{\boldsymbol{k}_{2\perp}}\right)(-\Delta r)}\Big] \qquad (4.72) \\
&\qquad \times a^2 \int \frac{\mathrm{d}^2 k'_{1\perp} \mathrm{d}^2 k'_{2\perp}}{(2\pi)^2} \delta_{\mathrm{BZ}}(\boldsymbol{k}_{1\perp} + \boldsymbol{k}_{2\perp} - \boldsymbol{k}'_{1\perp} - \boldsymbol{k}'_{2\perp}) \\
&\qquad \times \left(1 - \frac{k'^2_{1\perp}}{2k_a^2}\right)\left(1 - \frac{k'^2_{2\perp}}{2k_a^2}\right) e^{-(k'^2_{1\perp}+k'^2_{2\perp})w_0^2/4} \\
&\qquad \times \frac{\Gamma^0_{\boldsymbol{k}_{1\perp}}/a^2}{\Delta - \tilde{\Delta}^0_{\boldsymbol{k}'_{1\perp}} + i\tilde{\Gamma}^0_{\boldsymbol{k}'_{1\perp}}} \frac{\Gamma^0_{\boldsymbol{k}_{2\perp}}/a^2}{\Delta - \tilde{\Delta}^0_{\boldsymbol{k}'_{2\perp}} + i\tilde{\Gamma}^0_{\boldsymbol{k}'_{2\perp}}} \quad .
\end{aligned}$$

Other than using the delta-function, the integral can not be reduced. However, we can employ the same approximations as we did above, taking the polarization factors to be unity, and the self-energies to be flat for small momenta. The remaining integrand is simply a Gaussian, which is easily integrated, yielding

$$\begin{aligned}
&\Psi^{(2)}_{\text{int},+}(\boldsymbol{k}_{1\perp}z_1, \boldsymbol{k}_{2\perp}z_2) \\
&\quad = -2a^2 e^{i(k_{1z,\omega}|z_1|+k_{2z,\omega}|z_2|)} \sum_{\boldsymbol{q}_m} e^{-(\boldsymbol{k}_{1\perp}+\boldsymbol{k}_{2\perp}+\boldsymbol{q}_m)^2 w_0^2/8} \\
&\qquad \times \left[\theta(\Delta r) e^{\left(i(\Delta - \tilde{\Delta}^0_{\boldsymbol{k}_{1\perp}}) - \tilde{\Gamma}^0_{\boldsymbol{k}_{1\perp}}\right)\Delta r} + \theta(-\Delta r) e^{\left(i(\Delta - \tilde{\Delta}^0_{\boldsymbol{k}_{2\perp}}) - \tilde{\Gamma}^0_{\boldsymbol{k}_{2\perp}}\right)(-\Delta r)}\right] \\
&\qquad \times \frac{\Gamma^0_{\boldsymbol{k}_{1\perp}}\Gamma^0_{\boldsymbol{k}_{2\perp}}/a^4}{(\Delta - \tilde{\Delta} + i\tilde{\Gamma})^2} \frac{\mathcal{T}(\boldsymbol{k}_{1\perp}+\boldsymbol{k}_{2\perp}, 2\omega)}{2\Delta - \tilde{\Delta}^0_{\boldsymbol{k}_{1\perp}} - \tilde{\Delta}^0_{\boldsymbol{k}_{2\perp}} + i(\tilde{\Gamma}^0_{\boldsymbol{k}_{1\perp}} + \tilde{\Gamma}^0_{\boldsymbol{k}_{2\perp}})} \quad .
\end{aligned} \qquad (4.73)$$

We can now extract the momentum density.

## 4.6.2 Momentum density

Following the discussed procedure, we can immediately write down the momentum density

$$\begin{aligned}
&\rho(\boldsymbol{k}_{1\perp}, \boldsymbol{k}_{2\perp}, t) \\
&\quad = \Big| 2[\Psi^{(1)}_{0,+}(\boldsymbol{k}_{1\perp}) + \Psi^{(1)}_{\mathrm{sc},+}(\boldsymbol{k}_{1\perp})][\Psi^{(1)}_{0,+}(\boldsymbol{k}_{2\perp}) + \Psi^{(1)}_{\mathrm{sc},+}(\boldsymbol{k}_{2\perp})] \\
&\qquad\quad + \Psi^{(2)}_{\mathrm{int},+}(\boldsymbol{k}_{1\perp}, \boldsymbol{k}_{2\perp}, \Delta r \to t) \Big|^2 \\
&\quad = (4\pi w_0^2)^2 \Bigg| \left( e^{-k_{1\perp}^2 w_0^2/4} - \frac{i\Gamma^0_{\boldsymbol{k}_{1\perp}}/a^2}{\Delta - \tilde{\Delta} + i\tilde{\Gamma}} \sum_{\boldsymbol{q}_m} e^{-(\boldsymbol{k}_{1\perp}+\boldsymbol{q}_m)^2 w_0^2/4} \right) \\
&\qquad\qquad \times \left( e^{-k_{2\perp}^2 w_0^2/4} - \frac{i\Gamma^0_{\boldsymbol{k}_{2\perp}}/a^2}{\Delta - \tilde{\Delta} + i\tilde{\Gamma}} \sum_{\boldsymbol{q}'_m} e^{-(\boldsymbol{k}_{2\perp}+\boldsymbol{q}'_m)^2 w_0^2/4} \right) \\
&\qquad\quad - \frac{a^2}{2\pi w_0^2} e^{\left(i(\Delta - \tilde{\Delta}^0_{\boldsymbol{k}_{2\perp}}) - \tilde{\Gamma}^0_{\boldsymbol{k}_{2\perp}}\right)t} \sum_{\boldsymbol{q}_m} e^{-(\boldsymbol{k}_{1\perp}+\boldsymbol{k}_{2\perp}+\boldsymbol{q}_m)^2 w_0^2/8} \\
&\qquad\qquad \times \frac{\Gamma^0_{\boldsymbol{k}_{1\perp}}\Gamma^0_{\boldsymbol{k}_{2\perp}}/a^4}{(\Delta - \tilde{\Delta} + i\tilde{\Gamma})^2} \frac{\mathcal{T}(\boldsymbol{k}_{1\perp}+\boldsymbol{k}_{2\perp}, 2\omega)}{2\Delta - \tilde{\Delta}^0_{\boldsymbol{k}_{1\perp}} - \tilde{\Delta}^0_{\boldsymbol{k}_{2\perp}} + i(\tilde{\Gamma}^0_{\boldsymbol{k}_{1\perp}} + \tilde{\Gamma}^0_{\boldsymbol{k}_{2\perp}})} \Bigg|^2 .
\end{aligned} \tag{4.74}$$

To calculate the density for other combinations of transmitted or reflected photons, it is merely the contribution from freely propagating photons that must be included or omitted, as light is emitted symmetrically from the array. Figure 4.5 shows the momentum density calculated on the $a = 0.6\lambda_a$ reflection resonance, as calculated by the numerical approach of the last chapter (see Fig. 3.4), and using Eq. (4.74). We see that the numerical and analytical calculations match very well.

To get a simpler expression for the momentum density we will specialize to driving on a reflection resonance, $\Delta = \tilde{\Delta}$, and same-time correlations, $t = 0$. Furthermore, we ignore Bragg scattering (that is, include only $\boldsymbol{q}_0 = \boldsymbol{0}$ in the sums), which only contributes to the peaks at the diagonal corners of Fig. 4.5. Thus, the linear transmitted contribution vanishes, and we can reduce Eq. (4.74) to the following

$$\begin{aligned}
\rho(\boldsymbol{k}_{1\perp}, \boldsymbol{k}_{2\perp}, 0) = &\frac{4(\Gamma^0_{\boldsymbol{k}_{1\perp}}\Gamma^0_{\boldsymbol{k}_{2\perp}})^2}{a^4\tilde{\Gamma}^4} e^{-(\boldsymbol{k}_{1\perp}+\boldsymbol{k}_{2\perp})^2 w_0^2/8} \\
&\times \frac{|\mathcal{T}(\boldsymbol{k}_{1\perp}+\boldsymbol{k}_{2\perp}, 2\omega_a + 2\tilde{\Delta})|^2}{(2\tilde{\Delta} - \tilde{\Delta}^0_{\boldsymbol{k}_{1\perp}} - \tilde{\Delta}^0_{\boldsymbol{k}_{2\perp}})^2 + (\tilde{\Gamma}^0_{\boldsymbol{k}_{1\perp}} + \tilde{\Gamma}^0_{\boldsymbol{k}_{2\perp}})^2} .
\end{aligned} \tag{4.75}$$

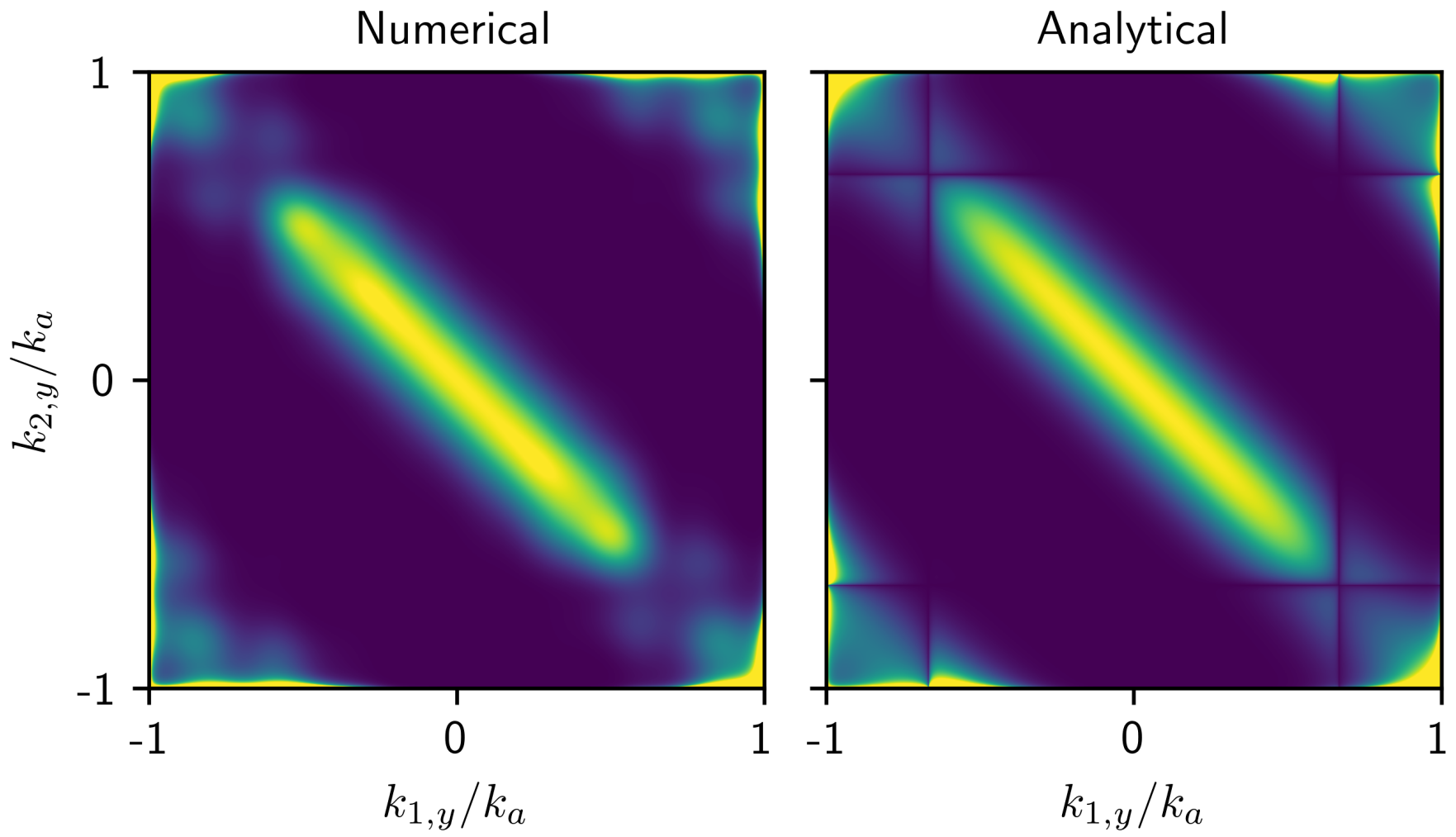

**Figure 4.5 | Comparison of numerical and analytical calculation of $\rho$.** The panels show $\rho(\boldsymbol{k}_{1\perp}, \boldsymbol{k}_{2\perp}, 0)$ with $k_{1,x} = k_{2,x} = 0$, calculated on a reflection resonance (same parameters as the lower left panel of Fig. 3.4), as calculated numerically in the previous chapter (left) and using Eq. (4.74) (right).

We can now easily determine that the straight lines of suppressed density that can be seen near $k_{i,y} \simeq \pm 0.7k_a$ in Fig. 4.5, for both $i = 1$ and 2, are due to divergence of the collective energies at $k_x = 0$ and $k_y = \pm(2\pi/a - k_a) = \pm 2k_a/3$ (for $a = 0.6\lambda_a$), where a new Bragg channel opens.

As noted earlier, the variation along the antidiagonal of $\rho(\boldsymbol{k}_{1\perp}, \boldsymbol{k}_{2\perp}, 0)$ is much more prominent for the atomic cavity than the atomic mirror. Plotting the momentum density for $a = 0.4\lambda_a$, where there is no Bragg scattering, in Fig. 4.6, it is even clearer that the high density corresponding to conservation of total momentum is nearly featureless and flat (except the $1/k_{z,k_a}$ divergence). This flat exchange of momentum among photons is what one would expect from the bare excitation-excitation scattering vertex, Eq. (4.10), the final term of Eq. (4.7), as it is momentum independent. The dressed excitation scattering, given by the atomic $\mathcal{T}$-matrix, is momentum-dependent due to multiple scattering and propagation in the collective modes, whose energies are momentum-dependent. Due to the absence of momentum dependence in the photon-photon interaction, as revealed by $\rho$ in Fig. 4.6, we conclude that photons do not propagate significantly in the collective modes, and do not scatter multiple times.

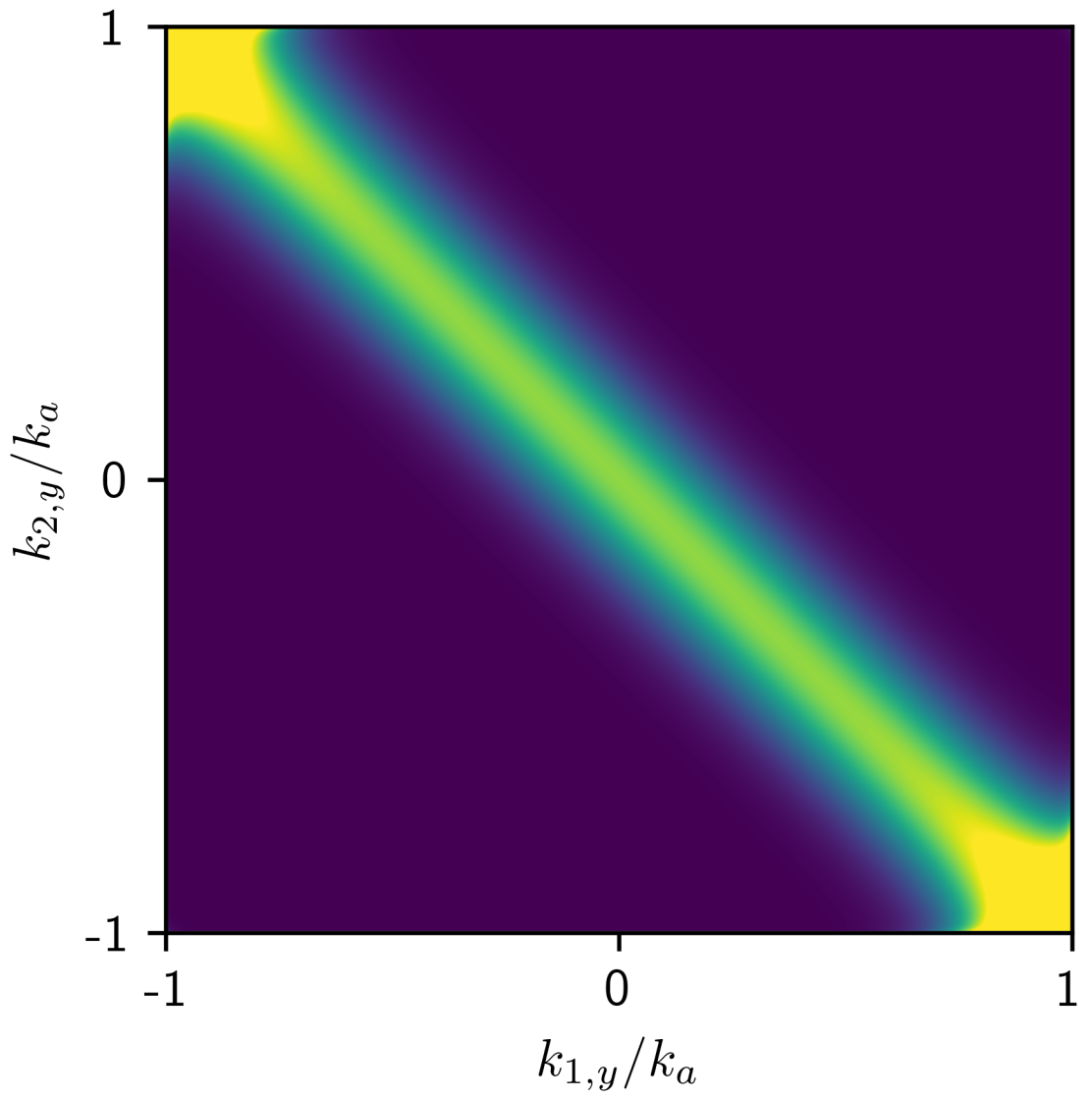

**Figure 4.6 | Momentum density with no Bragg scattering.** The plot shows $\rho(\boldsymbol{k}_{1\perp}, \boldsymbol{k}_{2\perp}, 0)$ with $k_{1,x} = k_{2,x} = 0$, as calculated on the reflection resonance of a single array with $a = 0.4\lambda_a$, driven by a Gaussian with waist $w_0 = 1.5\lambda_a$.

This is in contrast to the case of atomic cavity, see for example Fig. 3.2, where we do significant momentum dependence along the antidiagonal. This is consistent with our picture of a long confinement time for the atomic cavity, compared to the atomic mirror, where the delay time is close lifetime of the single atom excited state. We conclude the photon-photon scattering in the atomic mirror essentially takes place off of a single atom. Thus, the long confinement time of the atomic cavity not only causes the strength of the correlations, but is also the origin of the momentum dependence of the photon-photon scattering.

### 4.6.3 Gaussian detection mode

The temporal correlator is calculated for detection in a Gaussian mode, propagating in the negative $z$-direction. We therefore calculate the overlap between a Gaussian mode (Eq. (2.48)) and the wave functions found in Section 4.6.1. Since we consider the correlation of the reflected light, the mode is propagating in opposite direction than the driving beam, and thus the overlap with the free single-photon wave function is just zero.

### Scattered single-photon wave function

For the scattered single-photon wave function we can find

$$\begin{aligned}\Psi^{(1)}_{\mathrm{sc},G} &= \int \frac{\mathrm{d}^2 k_\perp}{(2\pi)^2}\sqrt{2\pi} w_0 e^{-ik_{z,\omega}z} e^{-k_\perp^2 w_0^2/4}\Psi^{(1)}_{\mathrm{sc},+}(\boldsymbol{k}_\perp z)\\ &= -\frac{i\tilde{\Gamma}}{\Delta-\tilde{\Delta}+i\tilde{\Gamma}}\ ,\end{aligned} \tag{4.76}$$

where we have used the flatness of $\Gamma^0_{\boldsymbol{k}_\perp}$ at small momenta, we approximated the overlap of Gaussians centred at different reciprocal lattice vectors to zero, and we have made paraxial approximation $\omega - k_\perp^2/2\omega \simeq k_{z,\omega}$ (see Section 2.5.2).

### Interacting two-photon wave function

Finally, we find the Gaussian mode component of the interacting two-photon wave function

$$\Psi^{(2)}_{\mathrm{int},G} = -\frac{a^2}{\pi w_0^2} e^{\left(i(\Delta-\tilde{\Delta})-\tilde{\Gamma}\right)\Delta r}\frac{\tilde{\Gamma}^2\mathcal{T}(0,2\omega)}{(\Delta-\tilde{\Delta}+i\tilde{\Gamma})^3}\ , \tag{4.77}$$

where we again performed the approximations we used above, as well as approximating the $\mathcal{T}$-matrix to be flat for small momenta. This is numerically justified, see Fig. 4.1. We can now extract the temporal correlator.

## 4.6.4 Temporal correlation

We can now find the two-time correlator as

$$g^{(2)}(t) = \left|1 + \frac{\Psi^{(2)}_{\mathrm{int},G}}{2(\Psi^{(1)}_{\mathrm{sc},G})^2}\right|^2\ , \tag{4.78}$$

with $\Delta r \to t$. This yields

$$g^{(2)}(t) = \left|1 + \frac{a^2}{2\pi w_0^2} e^{\left(i(\Delta-\tilde{\Delta})-\tilde{\Gamma}\right)t}\frac{\mathcal{T}(0,2\omega)}{\Delta-\tilde{\Delta}+i\tilde{\Gamma}}\right|^2\ . \tag{4.79}$$

Again, we could calculate the correlator for other combinations of transmitted or reflected light, simply by including or omitting the contribution from the freely photons in the single-photon wave function. We see how on resonance, $\Delta = \tilde{\Delta}$, $g^{(2)}(t)$ would show a simple exponential behaviour, with a time scale given by the collective decay $\tilde{\Gamma}$, and an initial value given by the $\mathcal{T}$-matrix and the beam width. This expression is similar in form to the correlation of the light from a weakly driven single atom, $g^{(2)}(t) = (1 - e^{-\gamma t})^2$ (which we discussed in Section 3.4.1 and plotted in Fig. 3.6), except for a front factor on the exponentially diminishing term. This factor result in having imperfect antibunching, but is also related to the atomic mirror's strong coupling to light.

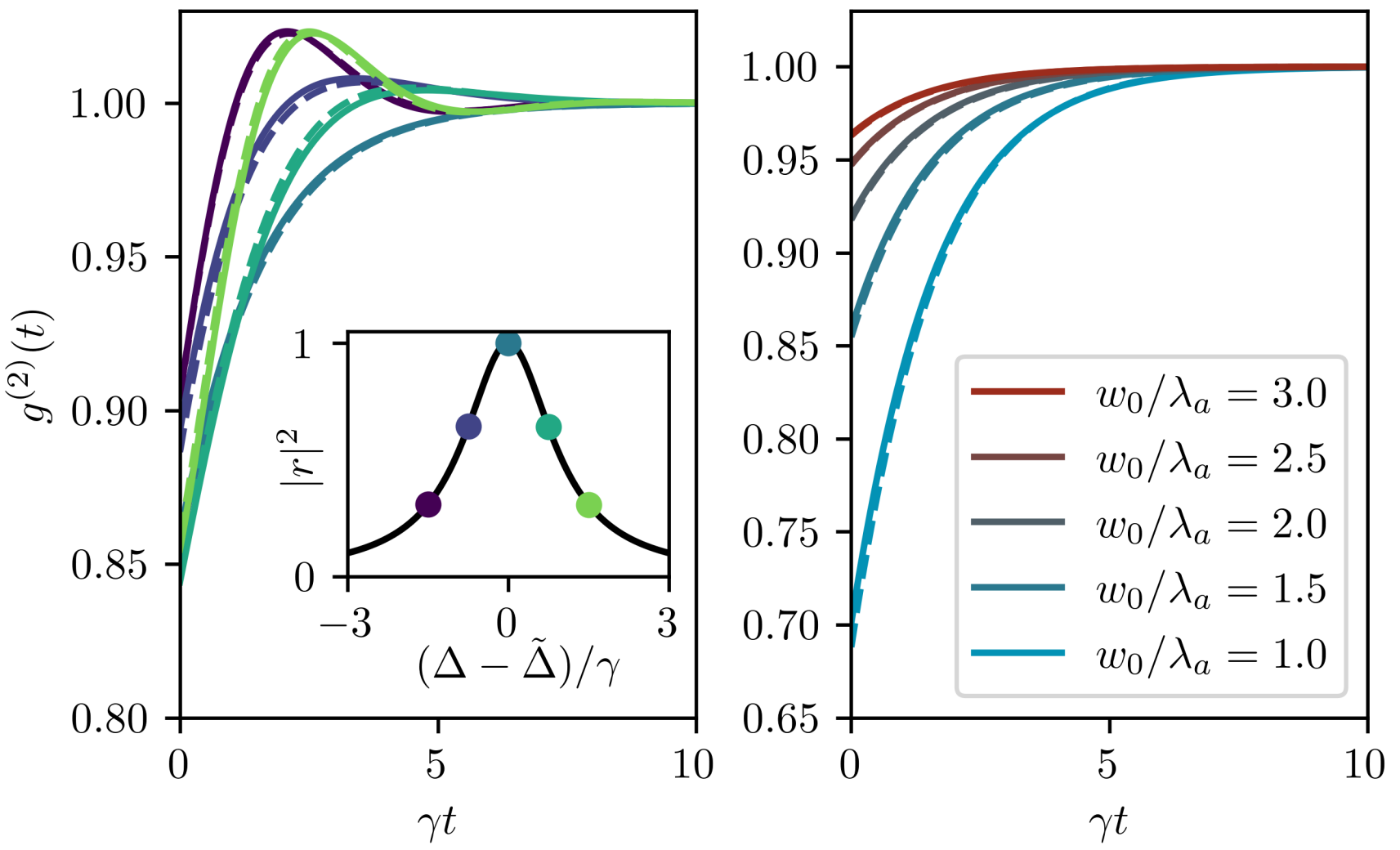

**Figure 4.7 | Comparison of numerical and analytical calculation of $g^{(2)}$.** Solid lines show $g^{(2)}(t)$ as calculated by Eq. (4.79), while dashed lines are calculated numerically as in the previous chapter. In the left panel $\Delta$ is varied across a reflection resonance (shown in inset), with $a = 0.6\lambda_a$ and $w_0 = 1.5\lambda_a$, while the right panel fixes $\Delta = \tilde{\Delta}$ at resonance and varies $w_0$. Numerical calculations are for a $15 \times 15$ lattice of atoms.

Fig. 4.7 shows a comparison of $g^{(2)}(t)$ as calculated by the numerical approach of the previous chapter (dashed lines) and using Eq. (4.79) (solid lines). In the first panel $\Delta$ is varied across the reflection resonance of the single array with $a = 0.6\lambda_a$ (shown in the inset, see also Eq. (2.20)), while the second panel fixes the detuning at resonance and varies the beam width instead. We see that the numerical and analytical results match nearly perfectly. Discrepancies might be explained by finite-size effects of the numerical approach. In particular, the finite size slightly shifts and diminishes the resonances, compared to the infinite case. Here, the numerical and analytical calculation are performed at the same values of $\Delta$, which are chosen according to the resonance of an infinite array. Thus, the finite array is in fact driven slightly off-resonance. We choose to use the same parameters for both calculations to show how close to the behaviour of the infinite case the finite array is, even without compensating for this shift. This kind of discrepancy will be much more prominent in the atomic cavity, due to its much more narrow resonances, and the fact that we consider curved finite arrays, such that the inter-lattice spacing $L$ takes on a slightly different meaning for the finite and the infinite arrays. We return to this in Section 5.2.2.

# 5

# Exact results of the Atomic Cavity

In this chapter we apply the method presented in the previous chapter to the atomic cavity. The extra lattice of atoms results in the atomic propagators, self-energy, and $\mathcal{T}$-matrix becoming tensors in an index pertaining to the two lattices. Hence, some matrix inversions will be necessary to solve the equations, and the results will generally involve more terms, stemming from contractions among these tensor quantities. Otherwise the derivations and results are nearly identical in form to those of the atomic mirror. We will therefore only highlight the differences in the derivations, and otherwise simply summarize the results before studying the correlators.

In Section 5.1 we thus summarize and discuss the results for the atomic cavity upon applying the Green's function approach. Section 5.2 then shows the exact correlators extracted from the two-photon wave function and discusses these.

## 5.1 Results

We will quickly go through the results pertaining to the atomic cavity, showing only the parts of the derivations which differ from the atomic mirror.

### 5.1.1 Hamiltonian and bare propagators

As in the first part of the thesis, we consider two identical planar square lattices of atoms parallel to the $xy$-plane at $z = \pm L/2$. Writing the

Hamiltonian in momentum space after bosonization as in Eq. (4.7), we have

$$
\begin{aligned}
H =&\sum_{\nu} \int \frac{\mathrm{d}^3 k}{(2\pi)^3} \omega_{\boldsymbol{k}} b^{\dagger}_{\boldsymbol{k}\nu} b_{\boldsymbol{k}\nu} + \sum_{j} \int_{\mathrm{BZ}} \frac{\mathrm{d}^2 k_{\perp}}{(2\pi)^2} \omega_a a^{\dagger}_{\boldsymbol{k}_{\perp},j} a_{\boldsymbol{k}_{\perp},j} \\
&- \sum_{\nu,j} \int \frac{\mathrm{d}^3 k}{(2\pi)^3} \left( g_{\boldsymbol{k}\nu j} b^{\dagger}_{\boldsymbol{k}\nu} a_{\boldsymbol{k}_{\perp},j} + \mathrm{H.c.} \right) \\
&+ \frac{U a^2}{2} \sum_{j} \int_{\mathrm{BZ}} \frac{\mathrm{d}^2 k_{\perp} \mathrm{d}^2 k'_{\perp} \mathrm{d}^2 q_{\perp}}{(2\pi)^6} a^{\dagger}_{\boldsymbol{k}_{\perp}+\boldsymbol{q}_{\perp},j} a^{\dagger}_{\boldsymbol{k}'_{\perp}-\boldsymbol{q}_{\perp},j} a_{\boldsymbol{k}_{\perp},j} a_{\boldsymbol{k}'_{\perp},j} \quad ,
\end{aligned}
\tag{5.1}
$$

where $g_{\boldsymbol{k}\nu j} = i\sqrt{\omega_{\boldsymbol{k}}\tilde{\Gamma}/\omega_a} p(\boldsymbol{k}\nu) e^{-(-1)^j i k_z L/2}$ where $j = 0, 1$ is the lattice index. The additional factor of $e^{-(-1)^j i k_z L/2}$ pertains to the acquired phase of propagating photons due to the shift of the lattices away from $z = 0$.

We use the same Feynman diagrams and rules as in the previous chapter (see Section 4.2) but now there is an additional degree of freedom in the form of the lattice index. We therefore write the bare atomic propagator as $G^0_{\mathrm{at},jj'}(\boldsymbol{k}_{\perp}, \boldsymbol{k}'_{\perp}, \omega) = (2\pi)^2 \delta(\boldsymbol{k}_{\perp} - \boldsymbol{k}'_{\perp}) \delta_{jj'} G^0_{\mathrm{at}}(\omega)$, where $G^0_{\mathrm{at}}(\omega)$ is given by Eq. (4.11a) again, and the factor due to emission and absorption is now given by $g_{\boldsymbol{k}\nu j}$ and $g^*_{\boldsymbol{k}\nu j}$ respectively. We write the lattice indices as subscripts, as we want to think of $G^0_{\mathrm{at},jj'}$, and other quantities, as matrices and vectors, and manipulate them as such. In particular, we will from now on use the Einstein summation convention and take a repeated index to imply summation, $A_{jj'} b_{j'} \equiv \sum_{j'} A_{jj'} b_{j'}$.

### 5.1.2 Dressed single-particle propagators

The dressed single-particle propagators are found via the same diagrammatic equations as for the atomic mirror, Eqs. (4.14a) and (4.14c). Following the Feynman rules and again defining the (proper) self-energy $\Sigma$, we arrive at the following summed Dyson's equation for the atomic propagator

$$
\begin{aligned}
&G^{(1)}_{\mathrm{at},jj'}(\boldsymbol{k}_{\perp}, \boldsymbol{k}'_{\perp}, \omega) \\
&\quad = (2\pi)^2 \delta(\boldsymbol{k}_{\perp} - \boldsymbol{k}'_{\perp}) \delta_{jj'} G^0_{\mathrm{at}}(\omega) \\
&\qquad + (2\pi)^2 \delta_{\mathrm{BZ}}(\boldsymbol{k}_{\perp} - \boldsymbol{k}'_{\perp}) G^{(1)}_{\mathrm{at},jj''}(\boldsymbol{k}_{\perp}, \boldsymbol{k}'_{\perp}, \omega) \Sigma_{j''j'}(\boldsymbol{k}'_{\perp}, \omega) G^0_{\mathrm{at}}(\omega) \quad ,
\end{aligned}
\tag{5.2}
$$

where

$$\Sigma_{jj'}(\boldsymbol{k}_\perp,\omega) = \sum_{\boldsymbol{q}_m,\nu} \int \frac{\mathrm{d}k_z}{2\pi} g^*_{\boldsymbol{k}\nu j} G^0_{\mathrm{ph}}(\boldsymbol{k},\omega) g_{\boldsymbol{k}\nu j'} \quad . \tag{5.3}$$

From $g_{\boldsymbol{k}\nu j} \propto e^{-(-1)^j i k_z L/2}$, we see that $\Sigma$ has the matrix-form

$$\Sigma \sim \begin{pmatrix} A & B \\ B & A \end{pmatrix} \quad , \tag{5.4}$$

such that its eigenvectors are simply $\boldsymbol{v}_\pm = (1, \pm 1)^T/\sqrt{2}$, with eigenvalues $\Sigma_\pm = A \pm B$. This is due to the parity symmetry of the two lattices, and is in this analysis the origin of the even and odd modes, previously considered (see Section 2.1). We will now see how these are also eigenvectors of the dressed propagator.

Using again the fact that the atomic propagator takes on momenta within the first BZ, we can replace $\delta_{\mathrm{BZ}}(\boldsymbol{k}_\perp - \boldsymbol{k}'_\perp)$ with $\delta(\boldsymbol{k}_\perp - \boldsymbol{k}'_\perp)$ in Eq. (5.2), and we can define $G^{(1)}_{\mathrm{at},jj'}(\boldsymbol{k}_\perp, \boldsymbol{k}'_\perp, \omega) = (2\pi)^2 \delta(\boldsymbol{k}_\perp - \boldsymbol{k}'_\perp) G^{(1)}_{\mathrm{at},jj'}(\boldsymbol{k}_\perp, \omega)$. We can then write the equation

$$G^{(1)}_{\mathrm{at},jj'} = \delta_{jj'} G^0_{\mathrm{at}} + G^{(1)}_{\mathrm{at},jj''} \Sigma_{j''j'} G^0_{\mathrm{at}} \quad , \tag{5.5}$$

where we have suppressed the common $(\boldsymbol{k}_\perp, \omega)$-dependence. $G^{(1)}_{\mathrm{at},jj'}$ can be isolated in terms of an inverse matrix

$$G^{(1)}_{\mathrm{at},jj'} = \left[\delta_{jj'} - \Sigma_{jj'} G^0_{\mathrm{at}}\right]^{-1} G^0_{\mathrm{at}} \quad . \tag{5.6}$$

Writing $\Sigma_{jj'} = U_{ji} \Lambda_{ii'} U^{-1}_{i'j'}$, where the columns of $U_{ij}$ are the eigenvectors of $\Sigma_{jj'}$, and $\Lambda_{ij}$ is a diagonal matrix of the eigenvalues, we can find after some matrix algebra and using the explicit expression for $G^0_{\mathrm{at}}$

$$G^{(1)}_{\mathrm{at},jj'} = U_{ji} \left[(\omega + i\eta - \omega_a)\delta_{ii'} - \Lambda_{ii'}\right]^{-1} U^{-1}_{i'j'} \quad . \tag{5.7}$$

In other words, in the basis of the eigenvectors of the self-energy, the dressed atomic propagator is also diagonal and each non-zero entry takes the same simple form of a bare propagator. We will therefore define, analogously to Eq. (4.24),

$$(G^{(1)}_{\mathrm{at},\alpha\alpha'})^{-1} = (\omega + i\eta - \omega_a - \Sigma_\alpha)\delta_{\alpha\alpha'} \quad , \tag{5.8}$$

where Greek indices run over $+, -$. We then only need to find the eigenvalues $\Sigma_\pm$ of the self-energy. Following the same derivation as we did for the atomic mirror (see Section 4.3), we find

$$\begin{aligned} \Sigma_\pm(\boldsymbol{k}_\perp,\omega) &= -\mu_0 \omega^2 \boldsymbol{d}^\dagger \left(\tilde{\boldsymbol{G}}(\boldsymbol{k}_\perp, z=0, \omega) \pm \tilde{\boldsymbol{G}}(\boldsymbol{k}_\perp, L, \omega)\right) \boldsymbol{d} \\ &= \tilde{\Delta}_{\boldsymbol{k}_\perp,\pm} - i\tilde{\Gamma}_{\boldsymbol{k}_\perp,\pm} \quad . \end{aligned} \tag{5.9}$$

We have thus again recovered the collective energies we found in the previous analysis (see Section 2.1), and we have found that the dressed propagator takes on the form of a bare propagator, but with the energy of its single pole shifted by the collective energies.

Finally, we write down the dressed photonic propagator in terms of the atomic

$$
\begin{aligned}
G^{(1)}_{\mathrm{ph}}(\boldsymbol{k}\nu,&\boldsymbol{k}'\nu',\omega) \\
&= (2\pi)^3\delta(\boldsymbol{k}-\boldsymbol{k}')\delta_{\nu\nu'}G^0_{\mathrm{ph}}(\boldsymbol{k}',\omega) \\
&\quad + (2\pi)^2\delta_{\mathrm{BZ}}(\boldsymbol{k}_\perp-\boldsymbol{k}'_\perp) \\
&\qquad \times G^0_{\mathrm{ph}}(\boldsymbol{k},\omega)g_{\boldsymbol{k}\nu\alpha}G^{(1)}_{\mathrm{at},\alpha\alpha'}(\boldsymbol{k}'_\perp,\omega)g^*_{\boldsymbol{k}'\nu'\alpha'}G^0_{\mathrm{ph}}(\boldsymbol{k}',\omega) \ ,
\end{aligned}
\tag{5.10}
$$

where the emission-absorption coupling strength in the even-odd basis is $g_{\boldsymbol{k}\nu\alpha} = i\sqrt{2\omega_{\boldsymbol{k}}\tilde{\Gamma}/\omega_a p(\boldsymbol{k}\nu)}\,\mathrm{trg}_\alpha(k_z L/2)$, with

$$
\mathrm{trg}_\alpha(x) = \frac{e^{-ix}+\alpha e^{ix}}{2} = \begin{cases} \cos(x), & \alpha = + \ , \\ -i\sin(x), & \alpha = - \ . \end{cases}
\tag{5.11}
$$

### 5.1.3 Two-photon propagator

Before writing down the two-photon propagator, we will consider the Hamiltonian in the even-odd basis (in particular the hardcore repulsion term), and the atomic $\mathcal{T}$-matrix that follows. A bit of algebra concerning the hardcore repulsion term leads to the Hamiltonian

$$
\begin{aligned}
H =& \sum_\nu \int \frac{\mathrm{d}^3k}{(2\pi)^3}\omega_{\boldsymbol{k}} b^\dagger_{\boldsymbol{k}\nu}b_{\boldsymbol{k}\nu} + \int_{\mathrm{BZ}} \frac{\mathrm{d}^2k_\perp}{(2\pi)^2}\omega_a a^\dagger_{\boldsymbol{k}_\perp,\alpha}a_{\boldsymbol{k}_\perp,\alpha} \\
&- \sum_\nu \int \frac{\mathrm{d}^3k}{(2\pi)^3}\left(g_{\boldsymbol{k}\nu\alpha}b^\dagger_{\boldsymbol{k}\nu}a_{\boldsymbol{k}_\perp,\alpha} + \mathrm{H.c.}\right) \\
&+ \frac{U^{\alpha\beta}_{\alpha'\beta'}a^2}{2}\int_{\mathrm{BZ}} \frac{\mathrm{d}^2k_\perp \mathrm{d}^2k'_\perp \mathrm{d}^2q_\perp}{(2\pi)^6} a^\dagger_{\boldsymbol{k}_\perp+\boldsymbol{q}_\perp,\alpha}a^\dagger_{\boldsymbol{k}'_\perp-\boldsymbol{q}_\perp,\beta}a_{\boldsymbol{k}_\perp,\alpha'}a_{\boldsymbol{k}'_\perp,\beta'} \ ,
\end{aligned}
\tag{5.12}
$$

where we have defined $a_{\boldsymbol{k}_\perp,\pm} = (a_{\boldsymbol{k}_\perp,0} \pm a_{\boldsymbol{k}_\perp,1})/\sqrt{2}$, and

$$
U^{\alpha\beta}_{\alpha'\beta'} = \begin{cases} \frac{U}{2}, & \text{for } \alpha=\beta, \alpha'=\beta' \text{ and } \alpha\neq\beta, \alpha'\neq\beta' \ , \\ 0, & \text{otherwise.} \end{cases}
\tag{5.13}
$$

In other words, the repulsion is still constant, but restricted to conserving the product of the parities of scattering excitations, i.e. two even or two odd excitations can only scatter to two even or two odd ones, while an odd and an even excitation will only scatter into an odd and an even.

As with the dressed propagators, we use the same diagrammatic equation as in the atomic mirror case, Eq. (4.29), to find the Bethe-Salpeter equation for the atomic $\mathcal{T}$-matrix, which here becomes

$$\mathcal{T}^{\alpha\beta}_{\alpha'\beta'}(Y) = U^{\alpha\beta}_{\alpha'\beta'} + iU^{\alpha\beta}_{\gamma\delta}a^2 \int \frac{\mathrm{d}^3 y}{(2\pi)^3} G^{(1)}_{\mathrm{at},\gamma\gamma'}(y) G^{(1)}_{\mathrm{at},\delta\delta'}(Y'-y)\mathcal{T}^{\gamma'\delta'}_{\alpha'\beta'}(Y) \quad , \tag{5.14}$$

where we have already written the $\mathcal{T}$-matrix as only a function of the total momentum-energy $Y$ (which is done with the same arguments as for the atomic mirror case), and we remind ourselves of the multi-index $y = (\boldsymbol{k}_\perp, \omega)$. Isolating the $\mathcal{T}$-matrix in this case, however, takes some matrix algebra, similar to what was necessary for the dressed atomic propagator. We will again skip the algebra, and simply quote the result in the limit of $U \to \infty$, which is

$$\begin{aligned}(\mathcal{T}^{\alpha\beta}_{\alpha'\beta'})^{-1}(Y) \\ = \begin{cases} -a^2 \int_{\mathrm{BZ}} \frac{\mathrm{d}^2 q_\perp}{(2\pi)^2} (\Pi_{++}(Y, \boldsymbol{q}_\perp) + \Pi_{--}(Y, \boldsymbol{q}_\perp)), & \text{for } \alpha = \beta, \alpha' = \beta' \quad , \\ -2a^2 \int_{\mathrm{BZ}} \frac{\mathrm{d}^2 q_\perp}{(2\pi)^2} \Pi_{+-}(Y, \boldsymbol{q}_\perp), & \text{for } \alpha \neq \beta, \alpha' \neq \beta' \quad , \\ 0, & \text{otherwise,} \end{cases}\end{aligned} \tag{5.15}$$

where the pair propagator is

$$\begin{aligned}\Pi_{\alpha\beta}(\boldsymbol{K}_\perp, \Omega, \boldsymbol{q}_\perp) &= i \int \frac{\mathrm{d}\omega}{2\pi} \frac{1}{\omega + i\eta - \omega_a - \Sigma_\alpha(\boldsymbol{q}_\perp)} \\ &\qquad \times \frac{1}{\Omega - \omega + i\eta - \omega_a - \Sigma_\beta(\boldsymbol{K}_\perp - \boldsymbol{q}_\perp)} \\ &= \frac{1}{\Omega + i\eta - 2\omega_a - \Sigma_\alpha(\boldsymbol{q}_\perp) - \Sigma_\beta(\boldsymbol{K}_\perp - \boldsymbol{q}_\perp)} \quad .\end{aligned} \tag{5.16}$$

Thus, the $\mathcal{T}$-matrix inherits the conservation of the product of scattering excitations' parities from the bare coupling, and its value depends on whether the pair of excitations have the same or opposite parities (and, of course, on the total momentum and energy). These two values we will refer to as $\mathcal{T}_{\mathrm{even}}$ and $\mathcal{T}_{\mathrm{odd}}$ (the first and second case of Eq. (5.15) respectively).

Fig. 5.1 shows an example of $\mathcal{T}_{\mathrm{even}}$ and $\mathcal{T}_{\mathrm{odd}}$. Equation (5.15) is evaluated numerically at the peak of a transmission resonance whose delay time, as calculated by Eq. (2.38), is equal to that of the "d" point of Fig. 3.1. In other words, we choose $\Delta$ and $L$ such that the delay time is the same as the "d" point we considered for the finite dual array. $\mathcal{T}_{\text{even}}/\tilde{\Gamma}$ and $\mathcal{T}_{\text{odd}}/\tilde{\Gamma}$ are plotted as a function of the total momentum. The dual array $\mathcal{T}$-matrix has a greater variation than that of the single array, Fig. 4.1, but it is still near flat. We will again be performing approximations based on the near-flatness of $\mathcal{T}_{\text{even}}$ and $\mathcal{T}_{\text{odd}}$ at small momenta.

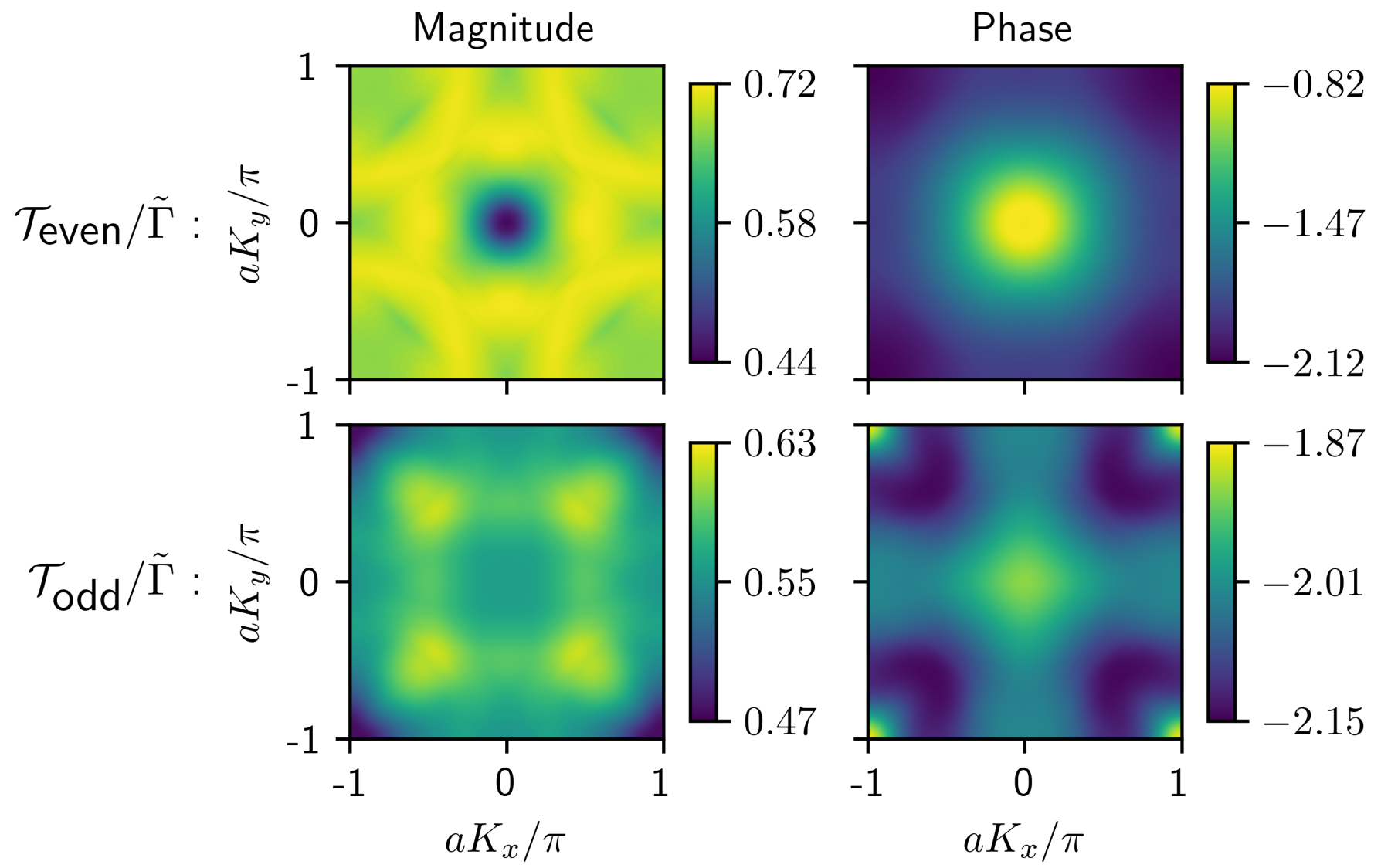

**Figure 5.1 | Example of atomic $\mathcal{T}$-matrix for the dual array.** We numerically evaluate Eq. (5.15) at a transmission resonance with a delay time corresponding to that of the "d" point of Fig. 3.1. $\mathcal{T}_{\text{even}}/\tilde{\Gamma}$ and $\mathcal{T}_{\text{odd}}/\tilde{\Gamma}$ are plotted as a function of $\boldsymbol{K}_{\perp}$.

We can now write down the two-photon propagator for the dual array,

using Eq. (4.26) again,

$$
\begin{aligned}
&G^{(2)}_{\text{ph}}(x_1, x_2; x'_1, x'_2) \\
&\quad = -G^{(1)}_{\text{ph}}(x_1, x'_1) G^{(1)}_{\text{ph}}(x_2, x'_2) - G^{(1)}_{\text{ph}}(x_1, x'_2) G^{(1)}_{\text{ph}}(x_2, x'_1) \\
&\qquad - 2ia^2 (2\pi)^3 \delta(\Omega - \Omega') \delta_{\text{BZ}}(\boldsymbol{K}_\perp - \boldsymbol{K}'_\perp) \\
&\qquad\quad \times G^0_{\text{ph}}(\boldsymbol{k}_1, \omega_1) g_{\boldsymbol{k}_1 \nu_1 \alpha} G^{(1)}_{\text{at},\alpha\gamma}(y_1) G^0_{\text{ph}}(\boldsymbol{k}_2, \omega_2) g_{\boldsymbol{k}_2 \nu_2 \beta} G^{(1)}_{\text{at},\beta\delta}(y_2) \\
&\qquad\quad \times \mathcal{T}^{\gamma\delta}_{\gamma'\delta'}(Y') \\
&\qquad\quad \times G^{(1)}_{\text{at},\gamma'\alpha'}(y'_1) g^*_{\boldsymbol{k}'_1 \nu'_1 \alpha'} G^0_{\text{ph}}(\boldsymbol{k}'_1, \omega'_1) G^{(1)}_{\text{at},\delta'\beta'}(y'_2) g^*_{\boldsymbol{k}'_2 \nu'_2 \beta'} G^0_{\text{ph}}(\boldsymbol{k}'_2, \omega'_2) \ .
\end{aligned}
\tag{5.17}
$$

with the multi-index $x = (\boldsymbol{k}, \nu, \omega)$. The propagator has exactly the same form as in the single array case, but now there are more terms due to the contractions of atomic excitation indices.

### 5.1.4 Two-photon steady state

We can now write down the two-photon steady state, and we will restrict ourselves to only write down the final expression for the steady state wave function after assuming the initial wave function continuous, specifically that it is proportional to a delta-function in the longitudinal momentum. We do so for each of the three partial wave functions, we considered previously, i.e. the wave function of a freely propagating photon, of a photon scattering off of the dual array, and of two photons scattering off of each other in the array. The expressions can be found either by performing the derivations performed in the previous chapter again, or simply by comparing the propagators, Eqs. (4.19) and (5.10), Eqs. (4.36) and (5.17), and then carrying the differences over to Eqs. (4.40), (4.46) and (4.63).

#### Free single-photon wave function

The free propagation is not affected by the atomic configuration and so it is identical to the single array case

$$
\Psi^{(1)}_0(\tilde{x}) = e^{ik_{z,\omega} z} \Psi^{(1)}_{\text{in}}(\boldsymbol{k}_\perp \nu) \ . \tag{5.18}
$$

with the multi-index $\tilde{x} = (\boldsymbol{k}_\perp, z, \nu)$.

### Scattered single-photon wave function

For the single photon scattering off of the dual array, we find

$$\begin{aligned}\Psi^{(1)}_{\text{sc}}(\tilde{x}) \\ = &-2i\frac{\tilde{\Gamma}\omega_a}{k_{z,\omega_a}}e^{ik_{z,\omega}|z|}\frac{\text{trg}_\alpha[\text{sgn}(z)k_{z,\omega_a}L/2]\,\text{trg}^*_\alpha[k_{z,\omega_a}L/2]}{\Delta-\tilde{\Delta}_{\boldsymbol{k}_\perp,\alpha}+i\tilde{\Gamma}_{\boldsymbol{k}_\perp,\alpha}} \\ &\times p_{\text{sgn}(z),\omega_a}(\boldsymbol{k}_\perp\nu)\sum_{\boldsymbol{q}_m,\nu'}p^*(\boldsymbol{k}'\nu')\Psi^{(1)}_{\text{in}}(\boldsymbol{k}'_\perp\nu')\ ,\end{aligned} \tag{5.19}$$

where we have performed the small detuning approximations. Carrying out the contractions[1], we would recognize the constellation of trigonometric functions in Eq. (2.13). Indeed, extracting the right-circular component of the above, as we did in the single array case, we would recognize the continuous Fourier transformed decay rate, Eqs. (2.9) and (2.10), as we did in Eq. (2.14) and the surrounding equations. Thus, we would arrive at a form of $\Psi^{(1)}_{\text{sc},+}$ similar to that of Eq. (4.48). In total, this expression would result in the same transmission and reflection coefficients that the previous analysis yielded (Section 2.3.2).

1: These now involve three indices, and as such do not follow the usual structure of contractions between tensors, but nonetheless the expression should simply be understood as having an implicit sum over $\alpha$.

**Interacting two-photon wave function**

Finally, for the photon-photon scattering part of the two-photon wave function we find

$$
\begin{aligned}
&\Psi^{(2)}_{\text{int}}(\tilde{x}_1, \tilde{x}_2) \\
&= -8\tilde{\Gamma}^2 \frac{\omega_a^2}{k_{1z,\omega_a} k_{2z,\omega_a}} p_{\text{sgn}(z_1),\omega_a}(k_{1\perp}\nu_1) p_{\text{sgn}(z_2),\omega_a}(k_{2\perp}\nu_2) \\
&\quad \times e^{i(k_{1z,\omega}|z_1| + k_{2z,\omega}|z_2|)} \\
&\quad \times \mathcal{T}^{\alpha\beta}_{\alpha'\beta'}(\boldsymbol{k}_{1\perp} + \boldsymbol{k}_{2\perp}, 2\omega) \\
&\quad \times \frac{\text{trg}_\alpha[\text{sgn}(z_1) k_{1z,\omega_a} L/2]\, \text{trg}_\beta[\text{sgn}(z_2) k_{2z,\omega_a} L/2]}{2\Delta - \tilde{\Delta}_{\boldsymbol{k}_{1\perp},\alpha} - \tilde{\Delta}_{\boldsymbol{k}_{2\perp},\beta} + i(\tilde{\Gamma}_{\boldsymbol{k}_{1\perp},\alpha} + \tilde{\Gamma}_{\boldsymbol{k}_{2\perp},\beta})} \\
&\quad \times \Bigg[\theta(\Delta z) e^{\left(i(\Delta - \tilde{\Delta}_{\boldsymbol{k}_{1\perp},\alpha}) - \tilde{\Gamma}_{\boldsymbol{k}_{1\perp},\alpha}\right)\Delta z} \\
&\qquad + \theta(-\Delta z) e^{\left(i(\Delta - \tilde{\Delta}_{\boldsymbol{k}_{2\perp},\beta}) - \tilde{\Gamma}_{\boldsymbol{k}_{2\perp},\beta}\right)(-\Delta z)}\Bigg] \\
&\quad \times a^2 \int \frac{\mathrm{d}^2 k'_{1\perp} \mathrm{d}^2 k'_{2\perp}}{(2\pi)^2} \delta_{\text{BZ}}(\boldsymbol{k}_{1\perp} + \boldsymbol{k}_{2\perp} - \boldsymbol{k}'_{1\perp} - \boldsymbol{k}'_{2\perp}) \\
&\qquad \times p(\boldsymbol{k}'_1, \nu'_1) p(\boldsymbol{k}'_2, \nu'_2) \Psi^{(2)}_{\text{in}}(\boldsymbol{k}'_{1\perp}\nu'_1, \boldsymbol{k}'_{2\perp}\nu'_2) \\
&\qquad \times \frac{\text{trg}^*_{\alpha'}[k_{1z,\omega_a} L/2]}{\Delta - \tilde{\Delta}_{\boldsymbol{k}'_{1\perp},\alpha'} + i\tilde{\Gamma}_{\boldsymbol{k}'_{1\perp},\alpha'}} \frac{\text{trg}^*_{\beta'}[k_{2z,\omega_a} L/2]}{\Delta - \tilde{\Delta}_{\boldsymbol{k}'_{2\perp},\beta'} + i\tilde{\Gamma}_{\boldsymbol{k}'_{2\perp},\beta'}} ,
\end{aligned}
\tag{5.20}
$$

where we again have contractions over multiply repeated indices. Up to the trigonometric factors pertaining to the propagation of light between the lattices and the contractions pertaining to multiple lattices, this is again identical to the single array case.

## 5.2 Correlations functions

The single- or two-photon wave functions for an incoming Gaussian beam, and those for detection in a Gaussian mode, are found completely analogously to what was done in Sections 4.6.1 and 4.6.3, and so we now skip forward to the final expressions for the correlators.

### 5.2.1 Momentum density

Having found the total two-photon wave function for the case of an incoming Gaussian beam, we proceed as in Section 4.6.2, and find the two-photon momentum density. This time we will write all combinations of transmitted and reflected photons in our expression for the density. To do so we introduce the variable $s_i$, which for $i = T, R$ takes on the values $s_T \equiv 1$ and $s_R \equiv 0$. We then write $\rho_{ij}$ for the density where the first (second) photon is transmitted or reflected according to $i = T, R$ ($j = T, R$). We furthermore have $\mathrm{sgn}(z_1) = (-1)^{s_i+1}$ and $\mathrm{sgn}(z_2) = (-1)^{s_j+1}$. With this we find

$$
\begin{aligned}
&\rho_{ij}(\boldsymbol{k}_{1\perp}, \boldsymbol{k}_{2\perp}, t)/(4\pi w_0^2)^2 \\
&= \Bigg| \Bigg( s_i e^{-k_{1\perp}^2 w_0^2/4} - 2\frac{i\Gamma^0_{\boldsymbol{k}_{1\perp}}}{a^2} \frac{\mathrm{trg}_\alpha[(-1)^{s_i+1} k_{1z,\omega_a} L/2]\, \mathrm{trg}^*_\alpha[k_{1z,\omega_a} L/2]}{\Delta - \tilde{\Delta}_\alpha + i\tilde{\Gamma}_\alpha} \\
&\qquad\qquad \times \sum_{\boldsymbol{q}_m} e^{-(\boldsymbol{k}_{1\perp}+\boldsymbol{q}_m)^2 w_0^2/4} \Bigg) \\
&\quad \times \Bigg( s_j e^{-k_{2\perp}^2 w_0^2/4} - 2\frac{i\Gamma^0_{\boldsymbol{k}_{2\perp}}}{a^2} \frac{\mathrm{trg}_\beta[(-1)^{s_j+1} k_{2z,\omega_a} L/2]\, \mathrm{trg}^*_\beta[k_{2z,\omega_a} L/2]}{\Delta - \tilde{\Delta}_\beta + i\tilde{\Gamma}_\beta} \\
&\qquad\qquad \times \sum_{\boldsymbol{q}'_m} e^{-(\boldsymbol{k}_{2\perp}+\boldsymbol{q}'_m)^2 w_0^2/4} \Bigg) \\
&\quad - \frac{2a^2}{\pi w_0^2} e^{\left(i(\Delta - \tilde{\Delta}_{\boldsymbol{k}_{2\perp},\beta}) - \tilde{\Gamma}_{\boldsymbol{k}_{2\perp},\beta}\right)t} \sum_{\boldsymbol{q}_m} e^{-(\boldsymbol{k}_{1\perp}+\boldsymbol{k}_{2\perp}+\boldsymbol{q}_m)^2 w_0^2/8} \\
&\qquad \times \frac{\Gamma^0_{\boldsymbol{k}_{1\perp}}\Gamma^0_{\boldsymbol{k}_{2\perp}}}{a^4} \frac{\mathrm{trg}^*_{\alpha'}[k_a L/2]\, \mathrm{trg}^*_{\beta'}[k_a L/2]}{(\Delta - \tilde{\Delta}_{\alpha'} + i\tilde{\Gamma}_{\alpha'})(\Delta - \tilde{\Delta}_{\beta'} + i\tilde{\Gamma}_{\beta'})} \\
&\qquad \times \mathcal{T}^{\alpha\beta}_{\alpha'\beta'}(\boldsymbol{k}_{1\perp} + \boldsymbol{k}_{2\perp}, 2\omega) \\
&\qquad \times \frac{\mathrm{trg}_\alpha[(-1)^{s_i+1} k_{1z,\omega_a} L/2]\, \mathrm{trg}_\beta[(-1)^{s_j+1} k_{2z,\omega_a} L/2]}{2\Delta - \tilde{\Delta}_{\boldsymbol{k}_{1\perp},\alpha} - \tilde{\Delta}_{\boldsymbol{k}_{2\perp},\beta} + i(\tilde{\Gamma}_{\boldsymbol{k}_{1\perp},\alpha} + \tilde{\Gamma}_{\boldsymbol{k}_{2\perp},\beta})} \Bigg|^2 ,
\end{aligned}
\tag{5.21}
$$

where we have again written $\tilde{\Gamma}_\pm = \tilde{\Gamma}_{0,\pm}$ and likewise for $\tilde{\Delta}_\pm$.

We can carry out the contractions in the above expression somewhat succinctly by introducing $C_{\boldsymbol{k}_\perp} = \cos(k_{z,\omega_a} L/2)$, $S_{\boldsymbol{k}_\perp} = \sin(k_{z,\omega_a} L/2)$, and $D_{\boldsymbol{k}_\perp,\pm} = \Delta - \tilde{\Delta}_{\boldsymbol{k}_\perp,\pm} + i\tilde{\Gamma}_{\boldsymbol{k}_\perp,\pm}$. Furthermore, for $\boldsymbol{k}_\perp = 0$, we suppress the momentum subscript such that $C = C_0$ and likewise for the others. For the sake of a simpler expression we will specialize to same-time correlations $t = 0$, and ignore Bragg scattering (keeping only $\boldsymbol{q}_0 = \boldsymbol{0}$ terms). With this we can find

$$
\begin{aligned}
&\rho_{ij}(\boldsymbol{k}_{1\perp}, \boldsymbol{k}_{2\perp}, t)/(4\pi w_0^2)^2 \\
&= \Bigg| S_i S_j e^{-k_{1\perp}^2 w_0^2/4} e^{-k_{2\perp}^2 w_0^2/4} \\
&\quad - \frac{2a^2}{\pi w_0^2} \frac{\Gamma^0_{\boldsymbol{k}_{1\perp}} \Gamma^0_{\boldsymbol{k}_{2\perp}}}{a^4} e^{-(\boldsymbol{k}_{1\perp}+\boldsymbol{k}_{2\perp})^2 w_0^2/8} \\
&\quad \times \Bigg[ \mathcal{T}_{\mathrm{even}}(0, 2\omega) \left( \frac{C^2}{D_+^2} - \frac{S^2}{D_-^2} \right) \\
&\qquad \times \left( \frac{C_{\boldsymbol{k}_{1\perp}} C_{\boldsymbol{k}_{2\perp}}}{D_{\boldsymbol{k}_{1\perp},+} + D_{\boldsymbol{k}_{2\perp},+}} - (-1)^{s_i+s_j} \frac{S_{\boldsymbol{k}_{1\perp}} S_{\boldsymbol{k}_{2\perp}}}{D_{\boldsymbol{k}_{1\perp},-} + D_{\boldsymbol{k}_{2\perp},-}} \right) \\
&\qquad - \mathcal{T}_{\mathrm{odd}}(0, 2\omega) \frac{4CS}{D_+ D_-} \\
&\qquad \times \left( (-1)^{s_j} \frac{C_{\boldsymbol{k}_{1\perp}} S_{\boldsymbol{k}_{2\perp}}}{D_{\boldsymbol{k}_{1\perp},+} + D_{\boldsymbol{k}_{2\perp},-}} + (-1)^{s_i} \frac{S_{\boldsymbol{k}_{1\perp}} C_{\boldsymbol{k}_{2\perp}}}{D_{\boldsymbol{k}_{1\perp},-} + D_{\boldsymbol{k}_{2\perp},+}} \right) \Bigg] \Bigg|^2 ,
\end{aligned}
\tag{5.22}
$$

where we have used the paraxial approximation in the linear terms, and introduced $S_i = s_i - 2i\tilde{\Gamma}\left(\frac{C^2}{D_+} - (-1)^{s_i}\frac{S^2}{D_-}\right)$, which for $i = T, R$ is equal to the normal-incidence ($\boldsymbol{k}_\perp = 0$) transmission or reflection amplitude respectively, given by Eq. (2.19). Furthermore, we have used the flatness of the $\mathcal{T}$-matrix at small momenta. We nicely see how the conservation of total momentum, up to the width of the driving beam, is imposed by the Gaussian factor. We can also see that any structure in the momentum correlations is given by the collective energies and the emission properties of the array. This, of course, follows from the assumption that the $\mathcal{T}$-matrix is flat for a small total momenta. To further understand this expression and the physics of it, one would need to study the $\mathcal{T}$-matrix and its dependence on the system parameters further. Likewise, one could study how the different momentum-dependent terms interfere to produce the variation along the antidiagonal of the momentum correlations we saw in the former analysis (see for example Fig. 3.2). This would give insight into the exchange of momentum between interacting photons in the system.

### 5.2.2 Temporal correlation

Having found the total two-photon wave function for the case of an incoming Gaussian beam and for detection in a matching Gaussian mode,

we proceed as in Section 4.6.4, and find the two-time correlation function, this time for all combinations of transmitted and reflected photons. We find

$$g_{ij}^{(2)}(t) = \Bigg| 1 + \frac{a^2}{2\pi w_0^2} e^{\left(i(\Delta-\tilde{\Delta}_\beta)-\tilde{\Gamma}_\beta\right)t} S_i^{-1} S_j^{-1} \times (2i\tilde{\Gamma})^2 \frac{\mathrm{trg}^*_{\alpha'}[k_a L/2]\,\mathrm{trg}^*_{\beta'}[k_a L/2]}{(\Delta-\tilde{\Delta}_{\alpha'}+i\tilde{\Gamma}_{\alpha'})(\Delta-\tilde{\Delta}_{\beta'}+i\tilde{\Gamma}_{\beta'})} \times \mathcal{T}_{\alpha'\beta'}^{\alpha\beta}(0,2\omega) \frac{\mathrm{trg}_\alpha[(-1)^{s_i+1}k_a L/2]\,\mathrm{trg}_\beta[(-1)^{s_j+1}k_a L/2]}{\Delta-(\tilde{\Delta}_\alpha+\tilde{\Delta}_\beta)/2+i(\tilde{\Gamma}_\alpha+\tilde{\Gamma}_\beta)/2} \Bigg|^2 , \tag{5.23}$$

using $s_i$ and $S_i$ introduced in the previous section. We can again carry out the contractions in the above expression using the notation of the previous section. With this we can find

$$g_{ij}^{(2)}(t) = \Bigg| 1 - \frac{2\tilde{\Gamma}^2 a^2}{\pi w_0^2} S_i^{-1} S_j^{-1} \times \Bigg[ \mathcal{T}_{\mathrm{even}} \left(\frac{C^2}{D_+^2} - \frac{S^2}{D_-^2}\right)\left(\frac{C^2}{D_+} e^{iD_+ t} - (-1)^{s_i+s_j}\frac{S^2}{D_-} e^{iD_- t}\right) - \mathcal{T}_{\mathrm{odd}} \frac{4C^2 S^2}{D_+ D_- (D_+ + D_-)} \left((-1)^{s_i} e^{iD_+ t} + (-1)^{s_j} e^{iD_- t}\right) \Bigg] \Bigg|^2 . \tag{5.24}$$

Let us now consider the regime of larger $L$, where the evanescent interaction between lattices can be neglected, such that the zero momentum collective energies are given by Eq. (2.23) without the evanescent term. Along the ridges of perfect transmission of normal-incidence light, defined by $\Delta - \tilde{\Delta} = -\tilde{\Gamma}\tan(k_a L)$ (Eq. (2.30)), we can find the following identities

$$S_T = -e^{-2ik_a L} , \tag{5.25a}$$

$$S_R = 0 , \tag{5.25b}$$

$$D_+ = 4i\tilde{\Gamma}\frac{C^2}{1+e^{-2ik_a L}} , \tag{5.25c}$$

$$D_- = 4i\tilde{\Gamma}\frac{S^2}{1+e^{-2ik_a L}} , \tag{5.25d}$$

where the first is identical to Eq. (2.31). If we then consider $g^{(2)}_{TT}$ along the transmission ridge (the normalized correlator is not defined for reflected light as the linearly reflected light vanishes), we can find with some algebra

$$
\begin{aligned}
g^{(2)}_{TT}(t) = \Bigg| 1 - i\frac{a^2}{2\pi w_0^2\tilde{\Gamma}}(1+e^{2ik_aL})\cos^2(k_aL) \\
\times \Bigg[ \mathcal{T}_{\text{even}}\frac{\cos(k_aL)}{\sin^2(k_aL)}\left(e^{iD_+t}-e^{iD_-t}\right) \\
+ \mathcal{T}_{\text{odd}}\left(e^{iD_+t}+e^{iD_-t}\right)\Bigg]\Bigg|^2 .
\end{aligned}
\tag{5.26}
$$

If we consider this near the diverging delay time, i.e. near $k_aL - n\pi \simeq 0$, specifically choosing an odd $n$, such that $\cos(k_aL) \simeq -1$ and $\Delta - \tilde{\Delta} \simeq \tilde{\Gamma}\sin(k_aL) \ll \tilde{\Gamma}$ , we can find

$$
\begin{aligned}
g^{(2)}_{TT}(t) = \Bigg| 1 - \frac{a^2}{2\pi w_0^2\tilde{\Gamma}^2}\left(2(\Delta-\tilde{\Delta}) + i\Gamma_{SR}\right) \\
\times \Bigg[ \left(\mathcal{T}_{\text{even}}\frac{\tilde{\Gamma}^2}{(\Delta-\tilde{\Delta})^2} + \mathcal{T}_{\text{odd}}\right)e^{-\Gamma_{sr}t} \\
+ \left(-\mathcal{T}_{\text{even}}\frac{\tilde{\Gamma}^2}{(\Delta-\tilde{\Delta})^2} + \mathcal{T}_{\text{odd}}\right)e^{\left(2i(\Delta-\tilde{\Delta})-\Gamma_{SR}\right)t}\Bigg]\Bigg|^2 ,
\end{aligned}
\tag{5.27}
$$

where the subradiant decay rate is $\Gamma_{sr} = 1/\tau$, with the atomic cavity delay time $\tau$ given by Eq. (2.38), and the superradiant decay rate is $\Gamma_{SR} = 2\tilde{\Gamma} - \Gamma_{sr} \simeq 2\tilde{\Gamma}$. The factor on $\mathcal{T}_{\text{even}}$ is proportional to $\tau$, but since we do not have an explicit expression for the $\mathcal{T}$-matrix factors, it is difficult to attempt an interpretation of the corresponding terms, as we do not know (without further, possibly numerical, analysis) how the $\mathcal{T}$-matrix behaves. However, we can see that this form of $g^{(2)}$ clearly matches the numerical results discussed in Section 3.4, in particular Fig. 3.7. Namely, there are two time scales that dictate the exponential time-dependence of the correlation function, one superradiant and one subradiant, the latter given by the delay time. We can also note that $g^{(2)}(0)$ is only affected by $\mathcal{T}_{\text{odd}}$

$$
g^{(2)}_{TT}(0) = \left| 1 - \frac{a^2}{\pi w_0^2\tilde{\Gamma}^2}\left(2(\Delta-\tilde{\Delta}) + i\Gamma_{SR}\right)\mathcal{T}_{\text{odd}}\right|^2 . \tag{5.28}
$$

This seems to imply that two photons can only be emitted simultaneously, if they inhabited collective states of opposite parity. Here, again, a further study of the $\mathcal{T}$-matrix would allow us to understand the initial value of the temporal correlation, which in the previous analysis was finite and implied bunching of the photons. Here, the parenthesis goes to zero for $\Delta - \tilde{\Delta} \to 0$, so it would be interesting to understand the $\mathcal{T}$-matrix' dependence on $\Delta - \tilde{\Delta}$.

# 6 Conclusion and Outlook

In this thesis, I have presented the work carried out during my Ph.D. studies concerning nonlinear optics in two-dimensional atomic arrays. The focus of my research has been the generation and analytical description of strong optical correlations, implying effective photon-photon interactions, choosing the platform of two-dimensional atomic arrays due to their strong and low-loss coupling to light at the linear level.

In the first part of the thesis, I presented the single- and two-excitation physics of the atomic cavity (the dual array) in itself and in comparison with the atomic mirror (the single array) and the single atom. In particular, at finite detuning of the incoming light and at finite inter-array distance, the atomic cavity shows repeated transmission resonances that become arbitrarily narrow as a consequence of hosting a perfectly subradiant state. This perfect subradiance is accompanied by a diverging delay time of the absorbed photons. While the individual atomic lattices are largely linear, exactly due to their collective behaviour which decreases the effect of the underlying saturable atoms, the confinement of photons to this weakly nonlinear environment results in a strong accumulated effect. This reveals the significance of the confinement time of photons with regard to generation of optical nonlinearities, and indicates an approach at the linear level to engineering nonlinearities.

I found that momentum correlations show an exchange of transverse momentum between pairs of photons, under conservation of total momentum, for both the single and the dual array. For the atomic mirror, the correlated output at a reflection resonance is completely dwarfed by the uncorrelated reflected light. Hence, while the transmitted photons have indeed interacted and are correlated, the signal of correlated light is comparatively very weak. For the atomic cavity, on the other hand, the momentum-correlated light dominates the output in both reflected and transmitted light, showing the strength of the effective photon-photon interaction of the system.

Finally, I looked at the temporal correlations, which told a similar story of the atomic cavity's ability to generate strongly correlated light. The shape of the temporal correlations were understood in terms of the post-detection single-excitation populations. A post-detection increase of population for the superradiant state results in bunching of photons, while the depletion of the subradiant state results in a following long period of anticorrelation, until the array is repopulated. The bunched light that is scattered into a different transverse mode than the driving Gaussian mode, together with the long period of anticorrelation indicates that the atomic cavity produces single photons in the Gaussian mode, and bunched pairs of photons in higher modes.

In the second part of the thesis, I re-derived many of the results from the first part, and expanded some, now using an analytical, Green's function-based approach, which allowed for the writing down of exact expressions for the two-photon steady state of propagating light. From these, correlation functions could be immediately written down. The resulting expressions tended to be large, but could nonetheless be understood at an intuitive level, if read factor by factor. This is a decided advantage over numerical approaches, where an intuitive understanding comes slowly.

I showed that the long confinement time of the atomic cavity not only explains the strength of the photon-photon scattering, but is also the origin of its momentum dependence. This follows from the approximately uniform exchange of momentum that takes place in the atomic mirror, corresponding to photons scattering off of each other at a single point, i.e. at a single atom. In other words, in the atomic mirror photons are emitted before having time to explore the array, and only interact if they immediately meet each other. The prolonged confinement in the atomic cavity, on the other hand, results in multiple scattering, yielding a momentum-dependent effective interaction.

The exact results came without the use of any additional approximations or restrictions compared to the numerical approach, but are in fact valid within the same regime, showing the power of the analytical approach. The only non-explicit components of my results are due to not having a closed analytical expression for the atomic self-energy. Thus, the self-energy, and also the atomic $\mathcal{T}$-matrix, were calculated numerically. However, due to the properties of these quantities and the considered choice of initial state, this was not a significant restriction to the analysis.

❉

From the photonic $\mathcal{T}$-matrix (essentially the third diagram in Eq. (4.26)) the effective interaction between photons could be extracted and an effective theory for the emerging quantum fluid of light could be written down and studied. In this way one could hope to study the behaviour of the photons in the presence of the atomic array, but without directly including the atoms, similar to how the photonic degrees of freedom are integrated out in the input-output formalism. Alternatively, one could work towards a description in terms of interacting polaritons, which have shown a rich variety of physics [29, 123, 129–132]. The momentum-dependent scattering, the overall strength of the correlations, and the long timescale associated with these, implies an environment where polaritons strongly interact in a spatially dependent manner, exchanging momentum, for a long period of time. Such a setup seems a strong candidate for the study of photonic many-body physics [4, 6]. This subject is notoriously difficult to tackle, but the present analytical description of the nonlinearity opens up the possibility for studying the many-body physics with existing theoretical tools.

A first step towards this, and an interesting exercise in itself, would be to derive an (approximate) closed expression for the atomic self-energy, and with it find an approximate closed expression for the atomic $\mathcal{T}$-matrix. This would eliminate the only part of the analytical approach which requires numerical calculation and, as such, is not completely explicit. With this, or otherwise numerically, it would also be insightful to study the value and behaviour of the atomic $\mathcal{T}$-matrix, to acquire greater understanding of how the nonlinearity of the atomic cavity depends on the system parameters.

The atomic cavity appears to produce both single photons and bunched pairs, depending on the transverse mode. Such production of highly nonclassical states of light ordered by mode motivates the exploration of application as a nonlinear quantum optical element to generate and process photonic quantum states [133–135], or to study the physics of propagating multiphoton quantum states through many of such nonlinear elements [35, 91, 136, 137]. It would be interesting to consider the decomposition of the described analytical wave functions in terms of Laguerre-Gaussian modes, or another basis, to see the distribution of single- and two-photon contributions among these, and whether some further physics could be derived. Higher Laguerre-Gaussian modes, for example, pertain to angular momentum, and it has already been shown that atomic arrays can generate light with orbital angular momentum at the linear level [30].

The presented Green's function-based approach could be applied to other atomic-optical systems, either simply different configurations of atoms in free space, atoms in waveguides (as done in Ref. [125]) or cavities, or other nonlinear systems used in quantum optics, such as layers of transition metal dichalcogenides [138]. Here, photons couple to excitons in atomically thin layers of semiconductor material. These have many similar properties to planar atomic arrays, like high single-photon reflectivity [27, 139], and have been shown to have nonlinear optical capabilities as well [93, 132, 140]. As such, the application of the analytical approach might prove fruitful.

Finally, an interesting extension of the atomic cavity would be to add topological effects by changing the level-structure of the atoms or the geometry of the lattices [141–145]. Topology is an evolving frontier of physics and offers many exciting and exotic phenomena. At the forefront are those that come about as a consequence turning on interactions among topological states [146]. A topological version of the atomic cavity should host topological states of the atomic excitations, coupled to photons that in turn interact through the saturability of the atoms, in total resulting in strongly interacting topological polaritons [147].

Here, at the end of all things, I will conclude by stating that this thesis wishes to contribute to the ongoing pursuit of understanding nonlinearities in quantum optics and how to generate them. The physics of the studied system gives a concrete example, while the presented analytical method is generally applicable in the field. The long photon confinement and the strong photon-photon interactions of the system lead to thoughts of many-body physics, while the analytical description opens up the possibility of an exact study. This paves the way for using the atomic cavity to study quantum fluids of light.

# List of Figures

# *Appendix*

# A

# Deriving the Feynman rules

We will perform the derivation that leads to the Feynman rules in a general setting, and then specialize to a system with two types of particles that have an emission-absorption type interaction, similar to our system (see Eq. (4.7)). The case that also includes the hard-core repulsion interaction (the last term of Eq. (4.7)) can be extrapolated from the case considered here. This derivation is included for completeness, as it is enlightening regarding the origin of Feynman diagrams, and it can be difficult to find it in a complete and unified version. The derivation, or parts of it, can be found in and is based on Refs. [116–118].

## A.1 Interaction picture time-evolution operator

Consider a Hamiltonian written as $H = H_0 + H_I$, where $H_0$ is some solvable Hamiltonian and $H_I$ is considered an interaction within this setting. We then work in the interaction picture, where the states and operators are given by

$$|\Psi(t)\rangle_I = e^{iH_0 t} |\Psi(t)\rangle_S \quad , \tag{A.1a}$$

$$\mathcal{O}_I(t) = e^{iH_0 t}\mathcal{O}_S(t)e^{-iH_0 t} \quad , \tag{A.1b}$$

where the subscript $S$ indicates the Schrödinger picture equivalent. Normally, Schrödinger picture operators will be time-independent, $\mathcal{O}_S(t) = \mathcal{O}_S$. In that case, we can find that these evolve according to

$$i\partial_t |\Psi(t)\rangle_I = H_I(t) |\Psi(t)\rangle_I \quad , \tag{A.2a}$$

$$i\partial_t \mathcal{O}_I(t) = [\mathcal{O}_I(t), H_0] \quad , \tag{A.2b}$$

such that the time-evolution of the operators is in principle known and the states only evolve according to the interaction. Notice that the interaction picture $H_I$ is generally time-dependent. If $H_0$ is of the diagonal form $\sum_i \epsilon_i a_i^\dagger a_i$, with $\epsilon_i$ an energy and $a_i$ an annihilation operator, then $H_I(t)$ is found by replacing $a_i$ in the Schrödinger picture interaction Hamiltonian with $a_i e^{-i\epsilon_i t}$. Note also, that for no interaction, i.e. $H_I = 0$, the interaction picture state is constant. The time-evolution operator of the states is $U(t, t_0)$, defined by

$$i\partial_t U(t,t_0) = H_I(t) U(t,t_0) \ , \tag{A.3}$$

which can be formally solved by

$$\begin{aligned} U(t,t_0) &= 1 - i\int_{t_0}^{t} \mathrm{d}t' H_I(t') U(t',t_0) \\ &= \sum_{n=0}^{\infty} (-i)^n \int_{t_0}^{t} \mathrm{d}t_1 \int_{t_0}^{t_1} \mathrm{d}t_2 \cdots \int_{t_0}^{t_{n-1}} \mathrm{d}t_n H_I(t_1)\cdots H_I(t_n) \\ &= \sum_{n=0}^{\infty} \frac{(-i)^n}{n!} \int_{t_0}^{t} \mathrm{d}t_1 \int_{t_0}^{t} \mathrm{d}t_2 \cdots \int_{t_0}^{t} \mathrm{d}t_n T[H_I(t_1)\cdots H_I(t_n)] \ , \end{aligned} \tag{A.4}$$

where the second line is the iteration of the first, and the third line uses a standard trick of extending the integration ranges which is compensated by the time-ordering operator $T[\cdot]$ and the combinatorial factor $1/n!$. This final expression is the Dyson series.

We note the following relations between the Schrödinger, Heisenberg, and interaction pictures

$$\mathcal{O}_H(t) = U(0,t)\mathcal{O}_I(t)U(t,0) \ , \tag{A.5a}$$
$$|\Psi\rangle_H = |\Psi(0)\rangle_S = |\Psi(0)\rangle_I \ , \tag{A.5b}$$
$$\mathcal{O}_S = \mathcal{O}_H(0) = \mathcal{O}_I(0) \ . \tag{A.5c}$$

## A.2 Gell-Mann and Low theorem

Let $|\Psi_0\rangle_H$ be an exact eigenstate of the full system in the Heisenberg picture. From the definition of the interaction picture, we have $|\Psi_0\rangle_H = |\Psi_0(0)\rangle_I = U(0,t)\,|\Psi_0(t)\rangle_I$.

Consider now the time-dependent Hamiltonian $H(t) = H_0 + e^{-\epsilon|t|} H_I$, where $\epsilon$ is a small positive number, and the corresponding time-evolution operator is $U_\epsilon(t, t_0)$. At times $t = \pm\infty$, the Hamiltonian is simply the non-interacting $H_0$, while at $t = 0$ it is the complete interacting Hamiltonian $H = H_0 + H_I$. Thus $|\Psi(t_0)\rangle_I$ for $t_0 \to \infty$ becomes time-independent (as the interaction is turned off in that limit).

If we say that the Schrödinger state of the system in this limit is an eigenstate, $|\Phi_0\rangle$, of $H_0$, the Heisenberg state $|\Psi\rangle_H$, which is always time-independent, must also be an eigenstate. Using the relation from above, $|\Psi_0\rangle_H = U(0,t)\,|\Psi_0(t)\rangle_I$, and the fact that the interaction state is time-independent in the limit $\lim_{t_0 \to -\infty} |\Psi(t_0)\rangle_I = |\Phi_0\rangle$, we can then write

$$|\Psi\rangle_H = \lim_{t_0 \to -\infty} U_\epsilon(0,t_0)\,|\Psi(t_0)\rangle_I = U_\epsilon(0,-\infty)\,|\Phi_0\rangle \quad , \tag{A.6}$$

which expresses a full eigenstate in terms of an eigenstate of $H_0$. The Gell-Mann and Low theorem [116] then states that if the following quantity exists to all orders in perturbation theory

$$\lim_{\epsilon \to 0} \frac{U_\epsilon(0,-\infty)\,|\Phi_0\rangle}{\langle\Phi_0|U_\epsilon(0,-\infty)|\Phi_0\rangle} \equiv \frac{|\Psi_0\rangle}{\langle\Phi_0|\Psi_0\rangle} \quad , \tag{A.7}$$

it is an eigenstate of the full Hamiltonian $H$. Note that neither the numerator nor denominator exist separately in the limit, even though the ratio does.

## A.3 Heisenberg matrix element in terms of interaction picture

Consider an operator in the Heisenberg picture $\mathcal{O}_H(t)$ and a matrix element of this operator

$$\frac{\langle\Psi_0|\mathcal{O}_H(t)|\Psi_0\rangle}{\langle\Psi_0|\Psi_0\rangle} \quad . \tag{A.8}$$

We wish to express this object in terms of the corresponding interaction picture operator and the interaction Hamiltonian. According to the Gell-Mann and Low theorem we can write

$$\frac{\langle\Psi_0|\Psi_0\rangle}{|\langle\Phi_0|\Psi_0\rangle|^2} = \frac{\langle\Phi_0|U_\epsilon^\dagger(0,\infty)U_\epsilon(0,-\infty)|\Phi_0\rangle}{|\langle\Phi_0|\Psi_0\rangle|^2} = \frac{\langle\Phi_0|U_\epsilon(\infty,-\infty)|\Phi_0\rangle}{|\langle\Phi_0|\Psi_0\rangle|^2} \quad , \tag{A.9}$$

where the limit $\epsilon \to 0$ is implicit. Likewise we can find

$$\begin{aligned}
\frac{\langle\Psi_0|\mathcal{O}_H(t)|\Psi_0\rangle}{|\langle\Phi_0|\Psi_0\rangle|^2} &= \frac{\langle\Phi_0|U_\epsilon^\dagger(0,\infty)\mathcal{O}_H(t)U_\epsilon(0,-\infty)|\Phi_0\rangle}{|\langle\Phi_0|\Psi_0\rangle|^2} \\
&= \frac{\langle\Phi_0|U_\epsilon(\infty,0)U_\epsilon(0,t)\mathcal{O}_I(t)U_\epsilon(t,0)U_\epsilon(0,-\infty)|\Phi_0\rangle}{|\langle\Phi_0|\Psi_0\rangle|^2} \\
&= \frac{\langle\Phi_0|U_\epsilon(\infty,t)\mathcal{O}_I(t)U_\epsilon(t,-\infty)|\Phi_0\rangle}{|\langle\Phi_0|\Psi_0\rangle|^2} \quad .
\end{aligned} \tag{A.10}$$

With this we write (exploiting the common denominator)

$$\frac{\langle\Psi_0|\mathcal{O}_H(t)|\Psi_0\rangle}{\langle\Psi_0|\Psi_0\rangle} = \frac{\langle\Phi_0|U_\epsilon(\infty,t)\mathcal{O}_I(t)U_\epsilon(t,-\infty)|\Phi_0\rangle}{\langle\Phi_0|U_\epsilon(\infty,-\infty)|\Phi_0\rangle} \quad . \tag{A.11}$$

Using the Dyson series expression for $U_\epsilon$, it is then a matter of combinatorics to show that

$$\begin{aligned}
&\frac{\langle\Psi_0|\mathcal{O}_H(t)|\Psi_0\rangle}{\langle\Psi_0|\Psi_0\rangle} \\
&\quad = \frac{1}{\langle\Phi_0|U(\infty,-\infty)|\Phi_0\rangle} \\
&\quad\quad \times \langle\Phi_0| \sum_{n=0}^{\infty} \frac{(-i)^n}{n!} \int_{-\infty}^{\infty} \mathrm{d}t_1 \cdots \int_{-\infty}^{\infty} \mathrm{d}t_n T[H_I(t_1)\cdots H_I(t_n)\mathcal{O}_I(t)]\, |\Phi_0\rangle \quad ,
\end{aligned} \tag{A.12}$$

where the limit $\epsilon \to 0$ has been taken. An equivalent expression can be found for time-ordered products of operators evaluated at different times, which we can use to calculate Green's functions.

## A.4 Wick's theorem

To evaluate the expectation value of the time-ordered product in Eq. (A.12), we use Wick's theorem. This states simply that a time-ordered product can be expressed in terms of normal-ordered contractions

$$\begin{aligned}
T[\mathcal{O}_1\cdots\mathcal{O}_n] &= \mathcal{N}[\mathcal{O}_1\cdots\mathcal{O}_n] + \sum_{\text{singles}} \mathcal{N}[\mathcal{O}_1^\bullet\mathcal{O}_2^\bullet\mathcal{O}_3\cdots\mathcal{O}_n] \\
&\quad + \sum_{\text{doubles}} \mathcal{N}[\mathcal{O}_1^\bullet\mathcal{O}_2^\bullet\mathcal{O}_3^{\bullet\bullet}\mathcal{O}_4^{\bullet\bullet}\mathcal{O}_5\cdots\mathcal{O}_n] + \ldots \\
&= \sum_{\text{all cont.}} \mathcal{N}[\mathcal{O}_1\cdots\mathcal{O}_n] \quad ,
\end{aligned} \tag{A.13}$$

where a contraction is defined as

$$\mathcal{O}^\bullet\mathcal{O}'^\bullet = T[\mathcal{O}\mathcal{O}'] - \mathcal{N}[\mathcal{O}\mathcal{O}'] \ , \tag{A.14}$$

and a sum over single, double, or all contractions indicates a sum over all possibilities for contracting a single pair or two pairs, or making all possible contractions, from singles to the fully contracted product. For bosonic or fermionic creation and annihilation operators $(a^\dagger, a)$ this becomes especially simple, as

$$a^\bullet(t)a^\bullet(t') = \begin{cases} a(t)a(t') - a(t)a(t') = 0, & t' \leq t, \\ \pm a(t')a(t) - a(t)a(t') = 0, & t < t', \end{cases} \tag{A.15a}$$

$$(a^\dagger(t))^\bullet(a^\dagger(t'))^\bullet = \begin{cases} a^\dagger(t)a^\dagger(t') - a^\dagger(t)a^\dagger(t') = 0, & t' \leq t, \\ \pm a^\dagger(t')a^\dagger(t) - a^\dagger(t)a^\dagger(t') = 0, & t < t', \end{cases} \tag{A.15b}$$

$$(a^\dagger(t))^\bullet a^\bullet(t') = \begin{cases} a^\dagger(t)a(t') - a^\dagger(t)a(t') = 0, & t' \leq t, \\ \pm a(t')a^\dagger(t) - a^\dagger(t)a(t') = \pm[a(t'), a^\dagger(t)]_{B,F}, & t < t', \end{cases} \tag{A.15c}$$

$$a(t)^\bullet(a^\dagger(t'))^\bullet = \begin{cases} a(t)a^\dagger(t') \mp a^\dagger(t')a(t) = [a(t), a^\dagger(t')]_{B,F}, & t' \leq t, \\ \pm a^\dagger(t')a(t) \mp a^\dagger(t')a(t) = 0, & t < t', \end{cases} \tag{A.15d}$$

where the $\pm$ and $B, F$ are each for bosons and fermions respectively. For $t \neq t'$ this can be summarized by

$$a^\bullet(t)a^\bullet(t') = (a^\dagger(t))^\bullet(a^\dagger(t'))^\bullet = 0 \ , \tag{A.16a}$$

$$a(t)^\bullet(a^\dagger(t'))^\bullet = \pm(a^\dagger(t'))^\bullet a^\bullet(t) = \theta(t - t')[a(t), a^\dagger(t')]_{B,F} \ . \tag{A.16b}$$

For $t = t'$, we have $a(t)^\bullet(a^\dagger(t))^\bullet = [a(t), a^\dagger(t)]_{B,F}$ and $(a^\dagger(t))^\bullet a^\bullet(t) = 0$. Furthermore, for creation and annihilation operators, the ground state expectation value of any normal-ordered product also vanishes. Thus, the expectation value of a time-ordered product of creation and annihilation operators is equal to the sum of fully contracted products, which will be a series of (anti-)commutators. This yields the Feynman diagrams and rules, as we shall see.

## A.5 Cancelling the disconnected contributions

With Wick's theorem we have

$$\frac{\langle\Psi_0|\mathcal{O}_H(t)|\Psi_0\rangle}{\langle\Psi_0|\Psi_0\rangle} = \frac{1}{\langle\Phi_0|U(\infty,-\infty)|\Phi_0\rangle}\sum_{n=0}^{\infty}\frac{(-i)^n}{n!}\int_{-\infty}^{\infty}\mathrm{d}t_1\cdots\int_{-\infty}^{\infty}\mathrm{d}t_n \times \langle\Phi_0|\sum_{\text{full cont.}}[H_I(t_1)\cdots H_I(t_n)\mathcal{O}_I(t)]|\Phi_0\rangle \quad , \tag{A.17}$$

where the sum is over all full contractions. The contractions of all the operators in the square bracket can be thought of in a graphical way. Specifically, we can define graphs, where the nodes are each of the time-coordinates, and these nodes are connected, when two operators evaluated at the corresponding times are contracted. We can then divide the contractions into whether they are connected to $\mathcal{O}_I(t)$ or not. If they are, we call the contractions *connected.* Any contraction may itself be a product of a smaller connected contraction and a number of smaller disconnected contractions. The terms of Eq. (A.17), which contain contractions not connected to $\mathcal{O}_I(t)$, factorize in such a way that the sum of all full contractions can be written as all full and connected contractions multiplied with all full and *disconnected* contractions. Performing the same steps as above, we can find that $\langle\Phi_0|U(\infty,-\infty)|\Phi_0\rangle$ can be written in the same way as $\langle\Psi_0|\mathcal{O}_H(t)|\Psi_0\rangle$, but without $\mathcal{O}_I(t)$. Hence the only contractions contributing to $\langle\Phi_0|U(\infty,-\infty)|\Phi_0\rangle$ will be identical to those disconnected from $\mathcal{O}_I(t)$ in $\langle\Psi_0|\mathcal{O}_H(t)|\Psi_0\rangle$. In total, the factor $1/\langle\Phi_0|U(\infty,-\infty)|\Phi_0\rangle$ serves to cancel the contribution from the disconnected contraction, and we can write

$$\frac{\langle\Psi_0|\mathcal{O}_H(t)|\Psi_0\rangle}{\langle\Psi_0|\Psi_0\rangle} = \sum_{n=0}^{\infty}\frac{(-i)^n}{n!}\int_{-\infty}^{\infty}\mathrm{d}t_1\cdots\int_{-\infty}^{\infty}\mathrm{d}t_n \times \langle\Phi_0|\sum_{\text{full conn. cont.}}[H_I(t_1)\cdots H_I(t_n)\mathcal{O}_I(t)]|\Phi_0\rangle \quad , \tag{A.18}$$

where the sum is over all full and connected contractions. The graphical picture we have alluded to is exactly that of Feynman diagrams. The cancelled disconnected contributions are those of diagrams consisting of a diagram connected to the incoming and outgoing particles plus any number of disconnected diagrams. As the behaviour of the incoming and outgoing particles should not be affected by vacuum fluctuations not connected to them, it makes good physical sense that the disconnected diagrams have been removed.

This result has been derived for an operator, $\mathcal{O}_H(t)$, depending on a single time, $t$, but as mentioned for time-ordered operators, $T[\mathcal{O}_1(t_1)\cdots\mathcal{O}_n(t_n)]$, an identical result can be derived, where we simply replace $\mathcal{O}_I(t)$ with $\mathcal{O}_{1I}(t_1)\cdots\mathcal{O}_{nI}(t_n)$, in the final expression.

## A.6 The time-ordered Green's function

As an example, we wish to calculate the time-ordered Green's function

$$G_{ij}(\boldsymbol{k}t,\boldsymbol{k}'t') = -i\frac{\langle\Psi_0|T[c^H_{i,\boldsymbol{k}}(t)c^{H\dagger}_{j,\boldsymbol{k}'}(t')]|\Psi_0\rangle}{\langle\Psi_0|\Psi_0\rangle} \tag{A.19}$$

of a Hamiltonian

$$H = \sum_{i=0,1}\int\frac{\mathrm{d}^3k}{(2\pi)^3}\epsilon_{i,\boldsymbol{k}}c^\dagger_{i,\boldsymbol{k}}c_{i,\boldsymbol{k}} + \int\frac{\mathrm{d}^3k}{(2\pi)^3}(g_{\boldsymbol{k}}c^\dagger_{0,\boldsymbol{k}}c_{1,\boldsymbol{k}} + \mathrm{H.c.}) \quad , \tag{A.20}$$

where the first term is considered to be $H_0$ and the second is $H_I$, and $c_{i,\boldsymbol{k}}$ is a bosonic annihilation operator of a particle with momentum $\boldsymbol{k}$ of the species $i$ (with two species $i = 0, 1$). We note that the time-dependence of the $c_{i,\boldsymbol{k}}$-operators in the definition of the Green's function, is the Heisenberg picture time-dependence as indicated here by the superscript, i.e. the operators evolve according to the full Hamiltonian $H$ (this will be the only place where it is the Heisenberg picture time-dependence). However, for interaction picture evolution, the time-dependence is simple, as it is only according to $H_0$, and we can write

$$c_{i,\boldsymbol{k}}(t) = c_{i,\boldsymbol{k}}e^{-i\epsilon_{i,\boldsymbol{k}}t} \quad , \tag{A.21a}$$

$$\begin{aligned}[c_{i,\boldsymbol{k}}(t), c^\dagger_{j,\boldsymbol{k}'}(t')] &= [c_{i,\boldsymbol{k}}, c^\dagger_{j,\boldsymbol{k}'}]e^{-i(\epsilon_{i,\boldsymbol{k}}t-\epsilon_{j,\boldsymbol{k}'}t')} \\ &= \delta_{ij}(2\pi)^3\delta(\boldsymbol{k}-\boldsymbol{k}')e^{-i\epsilon_{i,\boldsymbol{k}}(t-t')} \quad .\end{aligned} \tag{A.21b}$$

With this the *free* Green's function (i.e. time-evolution according to only $H_0$) is simply

$$\begin{aligned}G^0_{ij}(\boldsymbol{k}t,\boldsymbol{k}'t') &= -i\langle\Phi_0|T[c_{i,\boldsymbol{k}}(t)c^\dagger_{j,\boldsymbol{k}'}(t')]|\Phi_0\rangle \\ &= -i\theta(t-t')\langle\Phi_0|c_{i,\boldsymbol{k}}(t)c^\dagger_{j,\boldsymbol{k}'}(t')|\Phi_0\rangle \\ &= -i\theta(t-t')\langle\Phi_0|c_{i,\boldsymbol{k}}c^\dagger_{j,\boldsymbol{k}'}|\Phi_0\rangle e^{-i\epsilon_{i,\boldsymbol{k}}(t-t')} \\ &= -i\theta(t-t')\langle i,\boldsymbol{k}|j,\boldsymbol{k}'\rangle e^{-i\epsilon_{i,\boldsymbol{k}}(t-t')} \\ &= -i(2\pi)^3\theta(t-t')\delta_{ij}\delta(\boldsymbol{k}-\boldsymbol{k}')e^{-i\epsilon_{i,\boldsymbol{k}}(t-t')} \quad ,\end{aligned} \tag{A.22}$$

where we have used the normalization of the states $|i, \boldsymbol{k}\rangle = c^\dagger_{i,\boldsymbol{k}} |\Phi_0\rangle$. Note that, up to a suppressed identity operator,

$$\begin{aligned} G^0_{ij}(\boldsymbol{k}t, \boldsymbol{k}'t') &= -i\theta(t-t')[c_{i,\boldsymbol{k}}(t), c^\dagger_{j,\boldsymbol{k}'}(t')] \\ &= -ic^\bullet_{i,\boldsymbol{k}}(t)c^{\dagger\bullet}_{j,\boldsymbol{k}'}(t') \\ &= -ic^{\dagger\bullet}_{j,\boldsymbol{k}'}(t')c^\bullet_{i,\boldsymbol{k}}(t) \end{aligned} \tag{A.23}$$

for $t \neq t'$. Fourier transforming this with respect to time, yields the fully Fourier transformed free Green's function

$$\begin{aligned} G^0_{ij}(\boldsymbol{k}\omega, \boldsymbol{k}'\omega') &= \int_{-\infty}^{\infty} \mathrm{d}t \int_{-\infty}^{\infty} \mathrm{d}t' e^{i(\omega t - \omega' t')} G^0_{ij}(\boldsymbol{k}t, \boldsymbol{k}'t') \\ &= -i(2\pi)^3 \delta_{ij}\delta(\boldsymbol{k}-\boldsymbol{k}') \\ &\qquad \times \int_{-\infty}^{\infty} \mathrm{d}t \int_{-\infty}^{\infty} \mathrm{d}t' e^{i(\omega t - \omega' t')}\theta(t-t')e^{-i\epsilon_{i,\boldsymbol{k}}(t-t')} \\ &= -i(2\pi)^3 \delta_{ij}\delta(\boldsymbol{k}-\boldsymbol{k}') \\ &\qquad \int_{-\infty}^{\infty} \mathrm{d}t \int_{-\infty}^{\infty} \mathrm{d}t' e^{i\omega(t-t')} e^{i(\omega-\omega')t'}\theta(t-t')e^{-i\epsilon_{i,\boldsymbol{k}}(t-t')} \\ &= -i(2\pi)^3 \delta_{ij}\delta(\boldsymbol{k}-\boldsymbol{k}') \int_{-\infty}^{\infty} \mathrm{d}t e^{i\omega t}\theta(t)e^{-i\epsilon_{i,\boldsymbol{k}}t} \int_{-\infty}^{\infty} \mathrm{d}t' e^{i(\omega-\omega')t'} \\ &= (2\pi)^4 \delta_{ij}\delta(\boldsymbol{k}-\boldsymbol{k}')\delta(\omega-\omega')\frac{1}{\omega + i\eta - \epsilon_{i,\boldsymbol{k}}} \quad . \end{aligned} \tag{A.24}$$

We have chosen a sign convention for the Fourier transform of the two time-arguments of the Green's function, which is consistent with the Hermitian conjugate of the second operator in the definition of it. We have used an identity from complex contour integration for the Fourier transform of $\theta(t)$ ($\eta = 0^+$ is a positive infinitesimal).

To calculate the full propagator, we proceed by using the result of the previous sections

$$\begin{aligned} G_{ij}(\boldsymbol{k}t, \boldsymbol{k}'t') = &-i\sum_{n=0}^{\infty} \frac{(-i)^n}{n!} \int \mathrm{d}t_1 \cdots \int \mathrm{d}t_n \int \frac{\mathrm{d}^3k_1}{(2\pi)^3} \cdots \int \frac{\mathrm{d}^3k_n}{(2\pi)^3} \\ &\times \langle\Phi_0| \sum_{\text{full conn. cont.}} [(g_{\boldsymbol{k}_1} c^\dagger_{0,\boldsymbol{k}_1}(t_1)c_{1,\boldsymbol{k}_1}(t_1) + \text{H.c.}) \cdots \\ &\times (g_{\boldsymbol{k}_n} c^\dagger_{0,\boldsymbol{k}_n}(t_n)c_{1,\boldsymbol{k}_n}(t_n) + \text{H.c.})c_{i,\boldsymbol{k}}(t)c^\dagger_{j,\boldsymbol{k}'}(t')] |\Phi_0\rangle \quad . \end{aligned} \tag{A.25}$$

To get an intuition for this rather large expression, we explicitly write the first two terms of the above. We have

$$\begin{aligned} &G_{ij}(\boldsymbol{k}t, \boldsymbol{k}'t') \\ &\quad = -i\Bigg[ \langle\Phi_0|\, c^{\bullet}_{i,\boldsymbol{k}}(t) c^{\dagger\bullet}_{j,\boldsymbol{k}'}(t')\, |\Phi_0\rangle \\ &\qquad\quad - i \int \mathrm{d}t_1 \int \frac{\mathrm{d}^3 k_1}{(2\pi)^3} \langle\Phi_0| \sum_{\text{full conn. cont.}} \\ &\qquad\qquad \times \left[(g_{\boldsymbol{k}_1} c^{\dagger}_{0,\boldsymbol{k}_1}(t_1) c_{1,\boldsymbol{k}_1}(t_1) + \text{H.c.}) c_{i,\boldsymbol{k}}(t) c^{\dagger}_{j,\boldsymbol{k}'}(t')\right] |\Phi_0\rangle + \dots \Bigg] \; . \end{aligned} \tag{A.26}$$

The first term is simply $G^0_{ij}(\boldsymbol{k}t, \boldsymbol{k}'t')$. The contractions in the second term are

$$\begin{aligned} &\langle\Phi_0| \sum_{\text{full conn. cont.}} [(g_{\boldsymbol{k}_1} c^{\dagger}_{0,\boldsymbol{k}_1}(t_1) c_{1,\boldsymbol{k}_1}(t_1) \\ &\qquad\qquad + g^*_{\boldsymbol{k}_1} c^{\dagger}_{1,\boldsymbol{k}_1}(t_1) c_{0,\boldsymbol{k}_1}(t_1)) c_{i,\boldsymbol{k}}(t) c^{\dagger}_{j,\boldsymbol{k}'}(t')] |\Phi_0\rangle \\ &\quad = \langle\Phi_0| \Big[ g_{\boldsymbol{k}_1} (c^{\dagger}_{0,\boldsymbol{k}_1}(t_1))^{\bullet} c^{\bullet\bullet}_{1,\boldsymbol{k}_1}(t_1) c^{\bullet}_{i,\boldsymbol{k}}(t) (c^{\dagger}_{j,\boldsymbol{k}'}(t'))^{\bullet\bullet} \\ &\qquad\qquad + g^*_{\boldsymbol{k}_1} (c^{\dagger}_{1,\boldsymbol{k}_1}(t_1))^{\bullet} c^{\bullet\bullet}_{0,\boldsymbol{k}_1}(t_1) c^{\bullet}_{i,\boldsymbol{k}}(t) (c^{\dagger}_{j,\boldsymbol{k}'}(t'))^{\bullet\bullet} \Big] |\Phi_0\rangle \\ &\quad = -g_{\boldsymbol{k}_1} G^0_{i0}(\boldsymbol{k}t, \boldsymbol{k}_1 t_1) G^0_{1j}(\boldsymbol{k}_1 t_1, \boldsymbol{k}'t') - g^*_{\boldsymbol{k}_1} G^0_{i1}(\boldsymbol{k}t, \boldsymbol{k}_1 t_1) G^0_{0j}(\boldsymbol{k}_1 t_1, \boldsymbol{k}'t') \; , \end{aligned} \tag{A.27}$$

such that the full second term of Eq. (A.26) is

$$\begin{aligned} \int \mathrm{d}t_1 \int \frac{\mathrm{d}^3 k_1}{(2\pi)^3} \Big[ & g_{\boldsymbol{k}_1} G^0_{i0}(\boldsymbol{k}t, \boldsymbol{k}_1 t_1) G^0_{1j}(\boldsymbol{k}_1 t_1, \boldsymbol{k}'t') \\ & + g^*_{\boldsymbol{k}_1} G^0_{i1}(\boldsymbol{k}t, \boldsymbol{k}_1 t_1) G^0_{0j}(\boldsymbol{k}_1 t_1, \boldsymbol{k}'t') \Big] \; . \end{aligned} \tag{A.28}$$

Note that this is only non-zero if $i \neq j$, with each term corresponding to one of two options for this. A clearer picture can be found Fourier transforming these terms. The first is again simply $G^0_{ij}(\boldsymbol{k}\omega, \boldsymbol{k}'\omega')$. The first part of the second term without the coupling strength, and the

$\boldsymbol{k}_1$-integral is

$$\begin{aligned}
&\int \mathrm{d}t \int \mathrm{d}t' e^{i(\omega t-\omega' t')} \int \mathrm{d}t_1 \int \frac{\mathrm{d}\omega_0}{2\pi} \int \frac{\mathrm{d}\omega_1}{2\pi} \int \frac{\mathrm{d}\omega_1'}{2\pi} \int \frac{\mathrm{d}\omega_0'}{2\pi} \\
&\qquad \times e^{-i(\omega_0 t-\omega_1 t_1+\omega_1' t_1-\omega_0' t')} G^0_{i0}(\boldsymbol{k}\omega_0, \boldsymbol{k}_1\omega_1) G^0_{1j}(\boldsymbol{k}_1\omega_1', \boldsymbol{k}'\omega_0') \\
&= \int \frac{\mathrm{d}\omega_0}{2\pi} \int \frac{\mathrm{d}\omega_1}{2\pi} \int \frac{\mathrm{d}\omega_1'}{2\pi} \int \frac{\mathrm{d}\omega_0'}{2\pi} (2\pi)^3 \delta(\omega_1-\omega_1') \\
&\qquad \times \delta(\omega-\omega_0)\delta(\omega'-\omega_0') G^0_{i0}(\boldsymbol{k}\omega_0, \boldsymbol{k}_1\omega_1) G^0_{1j}(\boldsymbol{k}_1\omega_1', \boldsymbol{k}'\omega_0') \\
&= \int \frac{\mathrm{d}\omega_1}{2\pi} G^0_{i0}(\boldsymbol{k}\omega, \boldsymbol{k}_1\omega_1) G^0_{1j}(\boldsymbol{k}_1\omega_1, \boldsymbol{k}'\omega') \ .
\end{aligned} \tag{A.29}$$

With this we can write

$$\begin{aligned}
G_{ij}(\boldsymbol{k}\omega, \boldsymbol{k}'\omega') &= G^0_{ij}(\boldsymbol{k}\omega, \boldsymbol{k}'\omega') \\
&\quad + \int \frac{\mathrm{d}\omega_1}{2\pi} \int \frac{\mathrm{d}^3 k_1}{(2\pi)^3} \Big[ g_{\boldsymbol{k}_1} G^0_{i0}(\boldsymbol{k}\omega, \boldsymbol{k}_1\omega_1) G^0_{1j}(\boldsymbol{k}_1\omega_1, \boldsymbol{k}'\omega') \\
&\qquad + g^*_{\boldsymbol{k}_1} G^0_{i1}(\boldsymbol{k}\omega, \boldsymbol{k}_1\omega_1) G^0_{0j}(\boldsymbol{k}_1\omega_1, \boldsymbol{k}'\omega') \Big] \\
&\quad + \ldots
\end{aligned} \tag{A.30}$$

In this particular example both the full momentum and energy are conserved throughout as shown by the delta-functions in the bare propagators, Eq. (A.24). We therefore define $G_{ij}(\boldsymbol{k}\omega, \boldsymbol{k}'\omega') = (2\pi)^4\delta(\omega-\omega')\delta(\boldsymbol{k}-\boldsymbol{k}')G_{ij}(\boldsymbol{k}\omega)$ and likewise for the bare propagator. This results in the above reducing to

$$\begin{aligned}
G_{ij}(\boldsymbol{k}\omega) &= G^0_{ij}(\boldsymbol{k}\omega) \\
&\quad + g_{\boldsymbol{k}} G^0_{i0}(\boldsymbol{k}\omega) G^0_{1j}(\boldsymbol{k}\omega) + g^*_{\boldsymbol{k}} G^0_{i1}(\boldsymbol{k}\omega) G^0_{0j}(\boldsymbol{k}\omega) \\
&\quad + \ldots
\end{aligned} \tag{A.31}$$

where we have removed an overall factor of $(2\pi)^4\delta(\omega-\omega')\delta(\boldsymbol{k}-\boldsymbol{k}')$ representing energy-momentum conservation. From this we can glean the structure of Feynman diagrams and the rules that define them. The full propagator is sum of terms, with the first being easily interpreted as simple bare propagation, the second as propagation interrupted by a single interaction, and so on. This line of thinking leads us to the Feynman rules.

## A.7 General Feynman rules

Due to the conservation of all momenta and energy and the one-to-one interaction in our example, the Feynman diagrams and rules also reduce quite a bit. Nonetheless, we can glean off the general rules for calculating the full propagator representing a certain particle coming in and a particle (not necessarily of the same type) coming out, and they are as follows. First, we define the components of the diagrams

1. Assign to each type of particle, represented by a momentum-space annihilation operator $a_{i,\boldsymbol{k}}$ and a diagonal energy term $\epsilon_{i,\boldsymbol{k}} a^\dagger_{i,\boldsymbol{k}} a_{i,\boldsymbol{k}}$ in $H_0$, an oriented line, representing the propagation of this particle. The bare propagator of each type of particle is $G^0_i(\boldsymbol{k}\omega) = [\omega + i\eta - \epsilon_{i,\boldsymbol{k}}]^{-1}$ with $\eta = 0^+$.

2. Assign to each type of interaction, represented by an interaction term $g a^\dagger_{i,\boldsymbol{k}} a^\dagger_{j,\boldsymbol{q}} \cdots a_{i',\boldsymbol{k}'} a_{j',\boldsymbol{q}'}$ in $H_0$ (where the coupling strength may depend on the indices or momenta), a vertex, representing a single interaction of this type. Each vertex must in the end have an incoming line for each annihilation operator in the interaction term, and an outgoing line for each creation operator.

Then we define the approach to draw the $n$'th order (in the coupling strengths) contribution to the full propagator

1. Draw all topologically distinct, fully connected diagrams consisting of $n$ vertices of the types defined by the interactions in the system, connected with lines according to which types of particles are involved in the interaction, with appropriate loose-ended lines representing the incoming and outgoing particles defined by the quantity being calculated.

2. Assign to each line a momentum and energy (and other variables if needed, e.g. polarization), $\boldsymbol{k}\omega$.

3. Reduce these momenta and energies by conserving whatever is conserved at each vertex.

4. Write for each line in the resulting diagram the appropriate propagator, evaluated at the appropriate momenta and energy.

5. Write for each vertex the appropriate coupling strength, evaluated at the appropriate momentum and energy.

6. Multiply with the number of permutations of the incoming and outgoing lines which leave the topology of the diagram unchanged.

7. Integrate and sum over all remaining free variables (i.e. variables other than those pertaining to the incoming and outgoing particles).

8. Multiply with $-i$ for each interaction vertex, and $i$ for each propagator. Furthermore, multiply with constants from the quantity you are calculating.

The factor $1/n!$ in the expression of the full propagator cancels with the number of permutations of labels in each diagram that leave the contribution of that diagram invariant. The $i$'s generally cancel, except for those stemming from the quantity being calculated (either constants from it or contractions of its operators). These are the rules we have used in Section 4.2.